\documentclass[pra,floatfix,a4paper,superscriptaddress]{revtex4}
\usepackage{bm,color,graphicx,amsmath,txfonts}
\usepackage{here}
\usepackage{array}
\usepackage{graphicx}
\usepackage[colorlinks=true,
citecolor=blue,
linkcolor=blue,
urlcolor=blue]{hyperref}
\usepackage{braket}
\usepackage{dsfont}
\usepackage[left=	2cm,top=2cm,right=1.2cm,bottom=2cm]{geometry}
\usepackage{tikz}
\usetikzlibrary{decorations.pathmorphing, positioning}

\everymath{\displaystyle}

\makeatletter
\renewcommand{\fnum@figure}{\textbf{Fig.~\thefigure:}}
\makeatother

\makeatletter
\renewcommand{\@makecaption}[2]{%
	\textbf{#1} #2\par
}
\makeatother
\makeatletter
\renewcommand{\fnum@figure}{\normalsize\textbf{Fig.~\thefigure:}}
\makeatother
\makeatletter
\renewcommand{\@makecaption}[2]{%
	\begin{flushleft}
		\textbf{#1} #2
	\end{flushleft}
}
\makeatother

\begin{document}
\makeatletter
\renewcommand{\@biblabel}[1]{%
	\makebox[2.1em][l]{\fontsize{10}{13}\selectfont[#1]}}
\makeatother

\title{Environmental Memory Effects and Quantum Resource Hierarchies in Polarized Hyperon--Antihyperon Systems
}

\author{Omar Bachain}
\address{LPHE-Modeling and Simulation, Faculty of Sciences, Mohammed V University in Rabat, Rabat, Morocco}

\author{Mohamed \surname{Amazioug} }
\email{m.amazioug@uiz.ac.ma}
\address{LPTHE-Department of Physics, Faculty of Sciences, Ibnou Zohr University, Agadir 80000, Morocco}

\author{Rachid Ahl Laamara}
\address{LPHE-Modeling and Simulation, Faculty of Sciences, Mohammed V University in Rabat, Rabat, Morocco}
\address{Centre of Physics and Mathematics, CPM, Faculty of Sciences, Mohammed V University in Rabat, Rabat, Morocco}
\date{\today}
\begin{abstract}
Hyperon--antihyperon pairs produced in $e^{+}e^{-}\rightarrow J/\psi\rightarrow Y\bar{Y}$ ($Y=\Lambda,\Sigma^{+},\Xi^{-},\Xi^{0}$) constitute a unique high-energy platform for probing quantum correlations through experimentally accessible spin observables. We investigate the impact of correlated dephasing environments on the
stationary and dynamical properties of logarithmic negativity, geometric quantum discord, and $l_{1}$-norm quantum coherence under both longitudinal and transverse beam polarizations. Our analysis reveals that environmental memory plays a crucial role in preserving quantum resources. In the non-Markovian regime, information
backflow generates recurrent revivals of quantum correlations and significantly delays decoherence, whereas Markovian evolution drives the system toward asymptotic stationary states through an irreversible loss of quantum information. The influence of beam polarization is found to be strongly channel dependent and can substantially enhance the amount of accessible quantum correlations. A comparative investigation of different quantifiers uncovers a clear hierarchy of quantum resources. Quantum coherence remains robust over the widest parameter range, geometric quantum discord survives even in regions where entanglement is strongly reduced, while logarithmic
negativity is the most sensitive to environmental degradation. This hierarchy persists for all considered hyperon channels and under both polarization configurations. The dependence of quantum resources on the production angle, azimuthal angle, polarization degree, and memory parameter is examined using
experimental inputs from BESIII. The predicted effects are found to be
compatible with the precision expected at BESIII and future high-luminosity facilities such as STCF and CEPC. Our results establish polarized hyperon--antihyperon systems as a realistic high-energy laboratory for investigating the interplay between quantum correlations, open-system dynamics, and environmental memory effects.
\end{abstract}

\maketitle
\section{Introduction}

Quantum correlations constitute one of the most remarkable features of quantum mechanics and play a central role in modern quantum information science. Entanglement, quantum steering, Bell nonlocality, quantum discord, and quantum coherence represent distinct manifestations of nonclassical correlations and provide essential resources for quantum communication, quantum computation, quantum metrology, and quantum cryptography \cite{Einstein1935,Schrodinger1935,Bell1964,Horodecki2009,Nielsen2010}. Over the last decades, considerable efforts have been devoted to understanding the generation, quantification, and manipulation of these quantum resources in a wide variety of physical systems, including atomic ensembles, photonic platforms, trapped ions, solid-state devices, and superconducting circuits \cite{Baumgratz2014,Streltsov2017}. Beyond their fundamental significance, quantum correlations constitute indispensable ingredients for the development of emerging quantum technologies.

In realistic situations, however, no quantum system can be completely isolated from its surrounding environment. The unavoidable interaction between a quantum system and external degrees of freedom leads to decoherence, dissipation, and information leakage, which progressively suppress quantum features and induce the quantum-to-classical transition \cite{Zurek2003,Breuer2016}. Consequently, the theory of open quantum systems has become an indispensable framework for investigating the behavior of quantum resources under realistic conditions. While Markovian environments are characterized by a continuous loss of information from the system to the environment, non-Markovian dynamics may induce information backflow, correlation revivals, and enhanced robustness of quantum properties \cite{Breuer2009,Rivas2014,deVega2017}. Understanding the interplay between environmental effects and quantum correlations therefore remains one of the central challenges in contemporary quantum physics \cite{Bachain2026,Bachain2026NPB}.

Recently, the investigation of quantum phenomena has expanded beyond
traditional low-energy platforms toward the domain of high-energy
particle physics. Elementary particles produced in scattering and decay
processes provide natural multipartite quantum systems whose spin,
flavor, and momentum degrees of freedom can exhibit genuine quantum
correlations \cite{Fano1983}. In particular, quantum entanglement has
been explored in neutral mesons, neutrino oscillations, positronium
systems, top-quark production, and hyperon--antihyperon pairs, opening
a promising bridge between quantum information science and particle
physics \cite{Lisi2000,Mavromatos2010,Afik2022,Fetscher2023,Tandean2024}.
Recent studies have further shown that beam polarization can be used to
manipulate Bell nonlocality and entanglement in electron--positron
annihilation processes \cite{Zhang2026,Jaloum2026,Fabbrichesi2021,Fabbrichesi2022}.

Among the various high-energy systems currently under investigation, hyperon--antihyperon pairs produced in electron--positron annihilation are especially attractive. Owing to angular-momentum conservation and the spin structure of the production mechanism, the hyperon pair emerges in an entangled spin state that can be fully characterized through its spin-density matrix \cite{Perotti2019}. Moreover, the weak and self-analyzing decays of hyperons provide direct access to their spin degrees of freedom, making it possible to reconstruct polarization observables and spin-correlation parameters experimentally. In recent years, the BESIII Collaboration has achieved remarkable precision in measurements of hyperon polarization and spin correlations in reactions such as
$
e^{+}e^{-}\rightarrow J/\psi\rightarrow \Lambda\bar{\Lambda},
$
thereby establishing hyperon systems as realistic laboratories for investigating entanglement, Bell nonlocality, and other manifestations of quantum correlations in high-energy experiments \cite{Ablikim2019NatPhys,BESIII2019,BESIII2022,BESIII2024}. Future facilities such as the Super Tau-Charm Facility (STCF) and the Circular Electron--Positron Collider (CEPC) are expected to provide significantly larger data samples and improved detector capabilities, opening unprecedented opportunities for precision studies of quantum resources in particle collisions \cite{STCF2024,CEPC2023}.

Despite these important advances, most existing studies have focused primarily on idealized isolated systems and on the characterization of static quantum properties. In realistic experimental conditions, however, the produced hyperon--antihyperon pair cannot be regarded as perfectly isolated. Interactions with surrounding degrees of freedom, finite detector resolution, production uncertainties, finite experimental resolutions, and effective noise mechanisms may alter the quantum state and affect the observable correlations. Consequently, assessing the robustness of quantum resources against environmental disturbances is essential for establishing reliable quantum-information protocols in high-energy experiments and for evaluating the feasibility of observing genuine quantum effects in realistic collider environments.

Furthermore, several important questions remain largely unexplored.
First, while entanglement and Bell nonlocality have been extensively
investigated in hyperon systems, much less attention has been devoted
to the behavior of the broader hierarchy of quantum resources under
open-system dynamics. Second, the combined influence of beam
polarization and environmental decoherence on the generation and
preservation of quantum correlations remains poorly understood.
Since beam polarization constitutes one of the most powerful
experimental tools for controlling spin dynamics in
electron--positron collisions, understanding its interplay with
environmental effects is of both fundamental and practical
importance. Third, the role of memory effects in protecting quantum
resources against decoherence has not yet been systematically
investigated in experimentally relevant hyperon channels. Finally,
a comprehensive comparison among Bell nonlocality, entanglement,
logarithmic negativity, quantum steering, geometric quantum discord,
and quantum coherence is still lacking for polarized
hyperon--antihyperon systems. Moreover, the combined impact of
environmental memory effects, beam polarization, and the hierarchy
of quantum resources has not yet been investigated within a unified
framework for experimentally relevant hyperon channels.

\begin{figure}[H]
	\centering
	\includegraphics[
	width=0.64\textwidth,
	trim=0 40 0 40,
	clip
	]{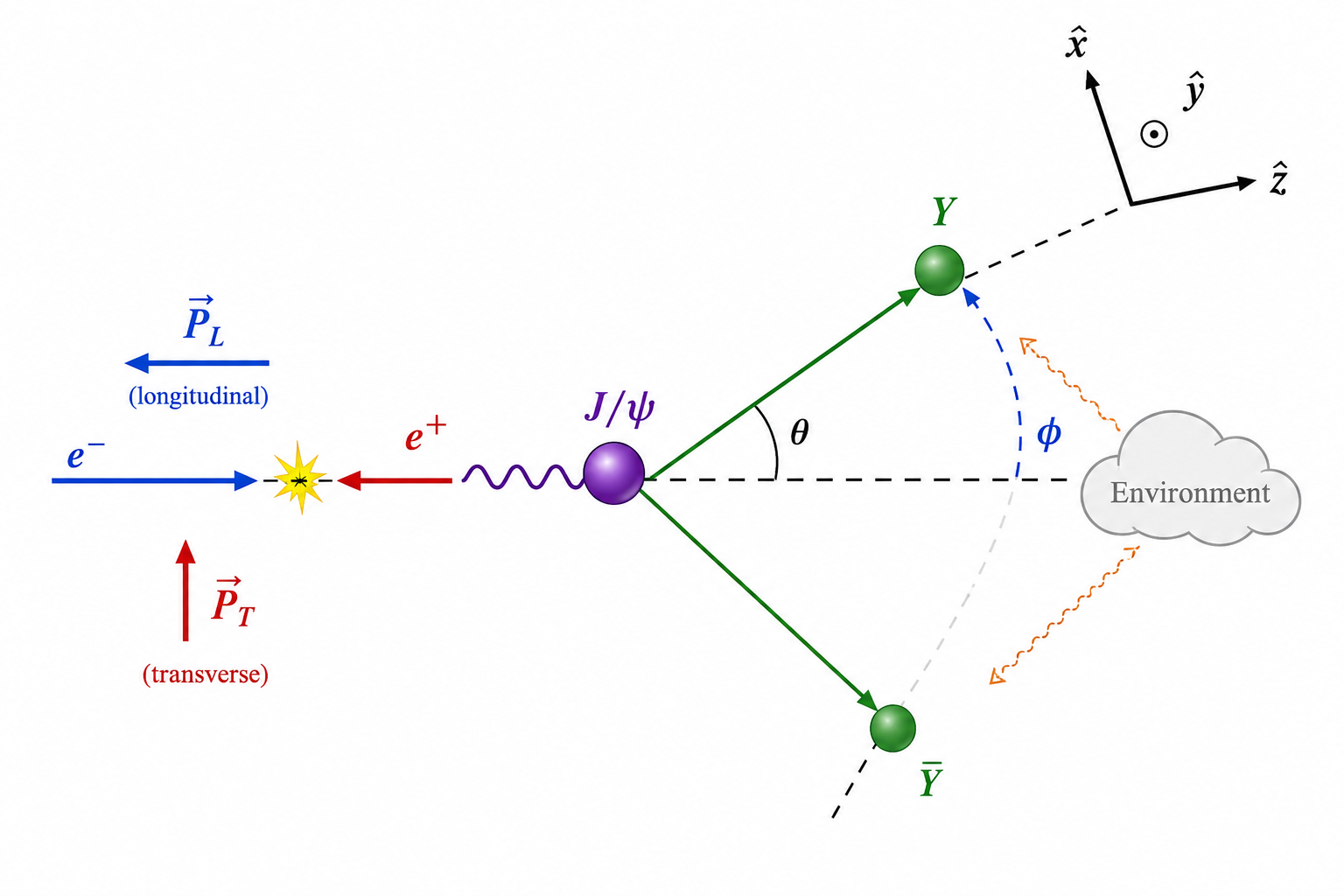}
\caption{
	Schematic illustration of the physical framework considered in this
	work. Polarized electron--positron beams produce the process
	$e^{+}e^{-}\rightarrow J/\psi\rightarrow Y\bar{Y}$, where
	$Y=\Lambda,\Sigma^{+},\Xi^{-},\Xi^{0}$.
	The longitudinal ($P_L$) and transverse ($P_T$) beam polarizations,
	together with the production angles $(\theta,\phi)$, determine the
	spin state of the hyperon pair. Environmental interactions are modeled
	through correlated dephasing channels with memory effects, allowing the
	investigation of the robustness of quantum correlations and coherence
	under both Markovian and non-Markovian dynamics.
}
	\label{fig1}
\end{figure}

Motivated by these considerations, in the present work we investigate the static and dynamical behavior of quantum resources in polarized hyperon--antihyperon systems produced through the reaction
$
e^{+}e^{-}\rightarrow J/\psi\rightarrow Y\bar{Y},
$
where $(Y=\Lambda,\Sigma^{+},\Xi^{-},\Xi^{0})$. The physical scenario considered throughout this work is schematically illustrated in Fig.~\ref{fig1}. The environmental influence is modeled through correlated dephasing channels incorporating memory effects, which allows us to investigate both Markovian and non-Markovian dynamics within a unified framework. Using experimentally measured production parameters reported by the BESIII Collaboration, we systematically analyze Bell nonlocality, logarithmic negativity, quantum steering, geometric quantum discord, and the $(l_{1})$-norm quantum coherence under both longitudinal and transverse beam polarizations. Particular attention is devoted to the role of polarization and environmental memory in protecting quantum resources against decoherence, as well as to the hierarchy among different forms of quantum correlations. In addition, we discuss the experimental feasibility of observing the predicted effects in current and future high-luminosity electron--positron facilities.

The remainder of this paper is organized as follows. In Sec.~\ref{sec2}, we introduce the polarized hyperon--antihyperon system produced in the process $(e^{+}e^{-}\rightarrow J/\psi\rightarrow Y\bar{Y})$ and present the corresponding spin-density-matrix formalism. Section~\ref{sec3} describes the correlated dephasing model and the dynamical framework employed to investigate environmental effects under both Markovian and non-Markovian conditions. In Sec.~\ref{sec4}, we review the quantum resources considered throughout this work, including Bell nonlocality, logarithmic negativity, quantum steering, geometric quantum discord, and the $(l_{1})$-norm quantum coherence. The static and dynamical properties of these resources under longitudinal and transverse beam polarizations are analyzed in Sec.~\ref{sec5}. The influence of environmental memory effects and the hierarchy among different quantum resources are also discussed in this section. In Sec.~\ref{sec6}, we address the experimental feasibility of observing the predicted phenomena in current and future high-luminosity $(e^{+}e^{-})$ facilities. Finally, Sec.~\ref{sec7} summarizes the main results and presents concluding remarks.

\section{Hyperon--antihyperon density matrix}\label{sec2}

\label{sec:theory}

The production of a spin-$1/2$ hyperon--antihyperon pair in the reaction
$
	e^{+}e^{-}\rightarrow \gamma^{*}/\psi \rightarrow Y\bar{Y},
$
provides a natural realization of a bipartite quantum system composed of two massive qubits. Since both the hyperon and antihyperon carry spin-$1/2$, the complete spin information of the pair is encoded in a $4\times4$ density operator acting on the Hilbert space
\begin{equation}
	\mathcal{H}=\mathcal{H}_{Y}\otimes\mathcal{H}_{\bar Y}.
\end{equation}
The spin density matrix contains all measurable information concerning polarization effects, spin correlations, quantum entanglement and Bell nonlocality.
A convenient representation of an arbitrary two-qubit state is provided by the Bloch--Fano decomposition \cite{Fano1983},
\begin{align}
	\rho_{Y\bar Y}
	=
	\frac14
	\Big(
	I^{(4)}
	+
	\vec B^{+}\cdot\boldsymbol{\sigma}\otimes I^{(2)}
	+
	I^{(2)}\otimes
	\vec B^{-}\cdot\boldsymbol{\sigma}
	+
	\sum_{i,j=1}^{3}
	C_{ij}
	\sigma_i\otimes\sigma_j
	\Big),
	\label{eq:Fano_rep}
\end{align}
where $I^{(2)}$ and $I^{(4)}$ denote the identity operators in the single- and two-qubit Hilbert spaces, respectively. The vectors
\begin{equation}
	\vec B^{+}=
	(B_x^{+},B_y^{+},B_z^{+}),
	\qquad
	\vec B^{-}=
	(B_x^{-},B_y^{-},B_z^{-}),
\end{equation}
represent the polarization of the hyperon and antihyperon, while
$
	C=
	\left(C_{ij}\right)
$
is the spin-correlation tensor describing the mutual quantum correlations between both subsystems.
Introducing the extended Pauli basis
$
	\sigma_0=I^{(2)},
$
the density matrix can alternatively be written in the compact tensorial form
\begin{equation}
	\rho_{Y\bar Y}
	=
	\frac14
	\Theta_{\mu\nu}
	\sigma_\mu\otimes\sigma_\nu ,
	\qquad
	\mu,\nu=0,1,2,3.
	\label{eq:Theta_def}
\end{equation}
The coefficient matrix $\Theta_{\mu\nu}$ is defined through
\begin{equation}
	\Theta_{\mu\nu}
	=
	\mathrm{Tr}
	\left[
	\rho_{Y\bar Y}
	(\sigma_\mu\otimes\sigma_\nu)
	\right],
\end{equation}
with
\begin{equation}
	\Theta_{00}=1,
	\qquad
	\Theta_{i0}=B_i^{+},
	\qquad
	\Theta_{0j}=B_j^{-},
	\qquad
	\Theta_{ij}=C_{ij}.
\end{equation}
Therefore, the matrix $\Theta_{\mu\nu}$ provides a complete description of the spin structure of the produced baryon pair.
The kinematics is formulated in the center-of-mass frame of the reaction.
The hyperon and antihyperon are emitted back-to-back as a consequence of momentum conservation.
The scattering angle $\theta$ is defined through
\begin{equation}
	\cos\theta
	=
	\hat{\mathbf p}_{e}
	\cdot
	\hat{\mathbf p}_{Y},
\end{equation}
where $\hat{\mathbf p}_{e}$ and $\hat{\mathbf p}_{Y}$ denote the unit vectors along the electron and hyperon momenta, respectively.
For polarized lepton beams, the spin-density matrix depends on the production dynamics through the decay parameter $\alpha_\psi$ and the relative phase $\Delta\Phi$ between the electric and magnetic form factors.
It is convenient to define
\begin{equation}
	\beta_\psi
	=
	\sqrt{1-\alpha_\psi^{2}}
	\sin\Delta\Phi ,
	\label{eq:beta}
\end{equation}
and
\begin{equation}
	\gamma_\psi
	=
	\sqrt{1-\alpha_\psi^{2}}
	\cos\Delta\Phi .
	\label{eq:gamma}
\end{equation}
The quantity $\alpha_\psi$ characterizes the angular distribution of the produced hyperons, whereas $\Delta\Phi$ controls the interference between helicity amplitudes and is responsible for the emergence of transverse polarization effects.
The complete coefficient matrix describing the hyperon pair can then be written as
\begin{equation}
	\Theta_{\mu\nu}
	=
	\Theta_{\mu\nu}^{(0)}
	+
	P_T^2\,
	\Theta_{\mu\nu}^{(T)}
	+
	P_L\,
	\Theta_{\mu\nu}^{(L)},
	\label{eq:Theta_general}
\end{equation}
where $P_L$ and $P_T$ denote the longitudinal and transverse beam polarizations.
The explicit expressions of the three matrices are those reported in Eq.~(4) of the original formulation and contain the full dynamical information of the reaction.
A remarkable feature of hyperon production is that the generated baryons may acquire transverse polarization even in the absence of polarized beams. Such a phenomenon originates from the non-vanishing relative phase $\Delta\Phi$ and therefore reflects the complex structure of the production amplitudes \cite{Perotti2019}. This behavior has no analogue in elementary lepton production and represents one of the most distinctive characteristics of hyperon physics.
For a longitudinally polarized beam, the normalized polarization vectors are obtained from Eq.~(\ref{eq:Theta_general}) as
\begin{equation}
	\vec B_L^{+}
	=
	\frac{1}{\chi_L}
	\begin{pmatrix}
		P_L\gamma_\psi\sin\theta\\
		-\beta_\psi\sin\theta\cos\theta\\
		-P_L(1+\alpha_\psi)\cos\theta
	\end{pmatrix},\qquad
	\vec B_L^{-}
	=
	\frac{1}{\chi_L}
	\begin{pmatrix}
		P_L\gamma_\psi\sin\theta\\
		\beta_\psi\sin\theta\cos\theta\\
		P_L(1+\alpha_\psi)\cos\theta
	\end{pmatrix},
\end{equation}
with
\begin{equation}
	\chi_L
	=
	1+\alpha_\psi\cos^2\theta.
\end{equation}

The associated correlation matrix is

\begin{equation}
	C_L
	=
	\frac1{\chi_L}
	\begin{pmatrix}
		\sin^2\theta &
		0 &
		\gamma_\psi\sin\theta\cos\theta
		\\
		0 &
		\alpha_\psi\sin^2\theta &
		-P_L\beta_\psi\sin\theta
		\\
		-\gamma_\psi\sin\theta\cos\theta &
		-P_L\beta_\psi\sin\theta &
		-\alpha_\psi-\cos^2\theta
	\end{pmatrix}.
	\label{eq:CL_new}
\end{equation}

Equation (\ref{eq:CL_new}) shows that the spin correlations are highly anisotropic.
The off-diagonal terms proportional to $\gamma_\psi$ and $\beta_\psi$ encode interference effects between different helicity amplitudes, while the diagonal elements determine the strength of spin-spin correlations along the principal axes.

For transversely polarized beams one obtains

\begin{equation}
	\vec B_T^{+}
	=
	\frac1{\chi_T}
	\begin{pmatrix}
		-P_T^2\beta_\psi\sin\theta\sin2\phi
		\\
		\beta_\psi\sin\theta\cos\theta
		(P_T^2\cos2\phi-1)
		\\
		0
	\end{pmatrix},\qquad
	\vec B_T^{-}
	=
	\frac1{\chi_T}
	\begin{pmatrix}
		-P_T^2\beta_\psi\sin\theta\sin2\phi
		\\
		\beta_\psi\sin\theta\cos\theta
		(1-P_T^2\cos2\phi)
		\\
		0
	\end{pmatrix},
\end{equation}

where

\begin{equation}
	\chi_T
	=
	1+\alpha_\psi\cos^2\theta
	+
	P_T^2\alpha_\psi
	\sin^2\theta\cos2\phi .
\end{equation}

The corresponding correlation tensor is

\begin{equation}
	C_T=
	\frac1{\chi_T}
	\begin{pmatrix}
		c_{11}&c_{12}&c_{13}
		\\
		c_{21}&c_{22}&c_{23}
		\\
		c_{31}&c_{32}&c_{33}
	\end{pmatrix},
\end{equation}

with

\begin{align}
	c_{11}
	&=
	\sin^2\theta
	+
	P_T^2
	(\alpha_\psi+\cos^2\theta)
	\cos2\phi,
\qquad\qquad\,
	c_{12}
	=
	-P_T^2(1+\alpha_\psi)
	\cos\theta
	\sin2\phi,
	\\
	c_{13}
	&=
	\gamma_\psi
	\sin\theta\cos\theta
	(1-P_T^2\cos2\phi),
\qquad\qquad\quad\,\,
	c_{21}
	=
	P_T^2(1+\alpha_\psi)
	\cos\theta
	\sin2\phi,
	\\
	c_{22}
	&=
	\alpha_\psi\sin^2\theta
	+
	P_T^2
	(1+\alpha_\psi\cos^2\theta)
	\cos2\phi,
\qquad\
	c_{23}
	=
	-P_T^2\gamma_\psi
	\sin\theta\sin2\phi,
	\\
	c_{31}
	&=
	-\gamma_\psi
	\sin\theta\cos\theta
	(1-P_T^2\cos2\phi),
\qquad \qquad\,\,\,\,
	c_{32}
	=
	-P_T^2\gamma_\psi
	\sin\theta\sin2\phi,
	\\
	c_{33}
	&=
	-\alpha_\psi
	-\cos^2\theta
	-P_T^2\sin^2\theta\cos2\phi .
\end{align}

The physical interpretation of these expressions is particularly transparent. The beam polarization modifies both the local polarization vectors and the correlation tensor, thereby affecting simultaneously the local and nonlocal quantum properties of the produced state. Consequently, entanglement measures and Bell parameters become controllable through external polarization settings.

To investigate these quantum correlations analytically, it is advantageous to perform local unitary transformations that diagonalize the correlation tensor. Since local unitary operations preserve both entanglement and Bell nonlocality \cite{Dur2000}, they do not alter the physical quantum resources contained in the state.

After suitable rotations of the local reference frames, the coefficient matrix can be reduced to the canonical form

\begin{equation}
	\Theta'
	=
	\begin{pmatrix}
		1&0&0&B_z
		\\
		0&\lambda_1&0&0
		\\
		0&0&\lambda_2&0
		\\
		B_z&0&0&\lambda_3
	\end{pmatrix},
	\label{eq:Theta_canonical}
\end{equation}

where $\lambda_1$, $\lambda_2$, and $\lambda_3$ are the eigenvalues of the correlation tensor and $B_z$ denotes the remaining non-vanishing polarization component.
Substituting Eq.~(\ref{eq:Theta_canonical}) into the Bloch-Fano expansion yields the corresponding $X$ state

\begin{equation}
	\rho^X
	=
	\frac14
	\begin{pmatrix}
		1+2B_z+\lambda_3
		&
		0
		&
		0
		&
		\lambda_1-\lambda_2
		\\
		0
		&
		1-\lambda_3
		&
		\lambda_1+\lambda_2
		&
		0
		\\
		0
		&
		\lambda_1+\lambda_2
		&
		1-\lambda_3
		&
		0
		\\
		\lambda_1-\lambda_2
		&
		0
		&
		0
		&
		1-2B_z+\lambda_3
	\end{pmatrix}.\label{rhoX}
\end{equation}

The importance of the $X$ representation lies in the fact that most entanglement measures, Bell inequalities and quantum Fisher information quantities can be derived analytically.
At the special scattering angles
$
	\theta=0,
	\theta=\frac{\pi}{2},
	\theta=\pi,
$
the hyperon-antihyperon state naturally reduces to an $X$ state. For longitudinal polarization one obtains

\begin{equation}
	\theta=\frac{\pi}{2}:
	\quad
	B_z=P_L\gamma_\psi,\,\,\,
	\lambda_{1,2}
	=
	\pm
	\sqrt{\alpha_\psi^2
		+
		P_L^2\beta_\psi^2},\,\,\,
	\lambda_3=1,
\end{equation}

and

\begin{align}
	\theta=0,\pi:
	\quad
	B_z=-P_L,
	\qquad
	\lambda_{1,2}=0,
	\qquad
	\lambda_3=1.
\end{align}

Similarly, for transverse polarization

\begin{equation}
	\theta=\frac{\pi}{2}:
	\quad
	B_z=
	-\frac{P_T^2\beta_\psi\sin2\phi}
	{1+P_T^2\alpha_\psi\cos2\phi},\,\,\,
	\lambda_{1,2}
	=
	\pm
	\frac{
		\sqrt{
			\alpha_\psi^2
			+
			2P_T^2\alpha_\psi\cos2\phi
			+
			P_T^4\eta
		}
	}
	{
		1+P_T^2\alpha_\psi\cos2\phi
	},\,\,\,
	\lambda_3=1,
\end{equation}

where

\begin{equation}
	\eta=
	\gamma_\psi^2\sin^22\phi
	+
	\cos^22\phi .
\end{equation}

In the limiting case $B_z=0$, the state becomes Bell diagonal,

\begin{equation}
	\rho_{\rm BDS}
	=
	\frac14
	\left(
	I^{(4)}
	+
	\sum_{i=1}^{3}
	\lambda_i
	\sigma_i\otimes\sigma_i
	\right).
\end{equation}

Bell-diagonal states occupy a central role in quantum information theory because they can be expressed as convex combinations of the four maximally entangled Bell states \cite{Bennett1996}. Consequently, the hyperon-antihyperon system produced in $e^+e^-$ annihilation provides a unique laboratory in which entanglement, Bell nonlocality and quantum metrology can be investigated within a relativistic particle-physics environment.
\section{Correlated dephasing dynamics}\label{sec3}
\label{sec:channel}

In realistic quantum systems, interactions with the surrounding environment are unavoidable and generally lead to decoherence, resulting in the degradation of quantum superposition and quantum correlations. Understanding the influence of environmental noise on quantum resources is therefore essential for both quantum information science and high-energy quantum systems. In the present work, we investigate the dynamical evolution of the hyperon--antihyperon spin state generated in the process
$
e^{+}e^{-}\rightarrow J/\psi \rightarrow Y\bar Y ,
$
when the system is subjected to correlated dephasing noise. Particular attention is devoted to the behavior of quantum entanglement, geometric quantum discord, and quantum coherence in both transient and asymptotic regimes.
The dynamics of an open quantum system can be described within the operator-sum representation introduced by Kraus \cite{Kraus1983,Nielsen2010}. Let $\rho_{Y\bar Y}(0)$ denote the initial density operator of the hyperon--antihyperon pair. The evolved state at time $t$ is given by

\begin{equation}
	\rho_{Y\bar Y}(t)
	=
	\mathcal{E}
	\left[
	\rho_{Y\bar Y}(0)
	\right]
	=
	\sum_{i,j}
	L_{ij}
	\rho_{Y\bar Y}(0)
	L_{ij}^{\dagger},
	\label{krausmap}
\end{equation}

where $L_{ij}$ are Kraus operators satisfying the completeness relation
$
\sum_{i,j}
L_{ij}^{\dagger}L_{ij}
=
I_4,
$

which guarantees that the quantum map $\mathcal{E}$ is completely positive and trace preserving.

To incorporate memory effects between successive uses of the channel, we adopt the correlated quantum-channel model introduced by Macchiavello and Palma \cite{Macchiavello2002}. In this framework, the Kraus operators take the form

\begin{equation}
	L_{ij}
	=
	\sqrt{p_{ij}}
	\,\sigma_i\otimes\sigma_j,
	\label{kraus}
\end{equation}

where
$
\sigma_0=I,
\sigma_1=\sigma_x,
\sigma_2=\sigma_y,
\sigma_3=\sigma_z
$
are the Pauli matrices.
The joint probabilities are defined by
\begin{equation}
	p_{ij}
	=
	(1-\mu)p_i p_j
	+
	\mu p_i \delta_{ij},
	\label{joint}
\end{equation}
where $\delta_{ij}$ is the Kronecker delta and
$
0\leq \mu \leq 1
$
is the memory parameter characterizing the strength of classical correlations between consecutive applications of the channel.
The limiting cases possess a clear physical interpretation:
\begin{itemize}
	\item $\mu=0$: completely uncorrelated channel (memoryless or Markovian limit),
	\item $\mu=1$: fully correlated channel,
	\item $0<\mu<1$: partially correlated channel.
\end{itemize}

Following Refs.~\cite{Daffer2004,Mazzola2010,Nielsen2010}, we consider a pure dephasing process for which

\begin{equation}
	p_0=1-p,
	\qquad
	p_x=p_y=0,
	\qquad
	p_z=p.
	\label{dephasing}
\end{equation}

The environmental fluctuations are modeled through a random telegraph noise process governed by the stochastic Hamiltonian

\begin{equation}
	H(t)
	=
	\hbar \Gamma(t)\sigma_z ,
	\label{Hamiltonian}
\end{equation}

where
$
\Gamma(t)=a\,n(t).
$
Here, $a$ is a random variable taking the values $\pm1$ with equal probability and $n(t)$ follows a Poisson process characterized by the average number of jumps
$
\langle n(t)\rangle
=
\frac{t}{2\tau}.
$
The dephasing probability can be expressed as

\begin{equation}
	p
	=
	\frac{1-K(t)}{2},
	\label{prob}
\end{equation}

where $K(t)$ is the decoherence function.
For the non-Markovian regime ($4\tau>1$), one obtains \cite{Daffer2004}

\begin{equation}
	K(t)
	=
	e^{-ut}
	\left[
	\cos(vt)
	+
	\frac{u}{v}\sin(vt)
	\right],
	\label{KNM}
\end{equation}

where
$
u=\frac{1}{2\tau},
v=\sqrt{1-u^2}.
$
The oscillatory behavior of Eq.~(\ref{KNM}) reflects the presence of memory effects and information backflow from the environment to the quantum system.
In contrast, for the Markovian regime ($4\tau<1$),

\begin{equation}
	K(t)
	=
	e^{-ut}
	\left[
	\cosh(vt)
	+
	\frac{u}{v}\sinh(vt)
	\right],
	\label{KM}
\end{equation}
with
$
v=\sqrt{u^2-1}.
$

In this case, the information continuously leaks into the environment, resulting in a monotonic decay of quantum resources.
Applying the correlated dephasing channel to the initial $X$ state given in Eq.~(\ref{rhoX}) yields

\begin{equation}
	\rho_{Y\bar Y}(t)
	=
	\begin{pmatrix}
		\rho_{11} & 0 & 0 & \eta\rho_{14}\\
		0 & \rho_{22} & \eta\rho_{23} & 0\\
		0 & \eta\rho_{32} & \rho_{33} & 0\\
		\eta\rho_{41} & 0 & 0 & \rho_{44}
	\end{pmatrix},
	\label{rhot}
\end{equation}

where the decoherence factor is

\begin{equation}
	\eta(t)
	=
	K^2(t)
	+
	\mu\left[1-K^2(t)\right].
	\label{eta}
\end{equation}

Equation~(\ref{eta}) explicitly shows the competition between decoherence and memory effects. While the function $K(t)$ tends to suppress the off-diagonal elements of the density matrix, the memory parameter $\mu$ counteracts this degradation and contributes to preserving quantum coherence and correlations.
The long-time behavior of the system is characterized by the stationary state

\begin{equation}
	\rho_{Y\bar Y}^{(\infty)}
	=
	\lim_{t\rightarrow\infty}
	\rho_{Y\bar Y}(t).
	\label{steady}
\end{equation}

For Markovian environments, the system approaches the stationary state monotonically. In contrast, non-Markovian dynamics may exhibit revivals and oscillatory behavior before eventually converging to the asymptotic regime. The stationary density matrix provides valuable information regarding the residual quantum resources surviving environmental decoherence.

In the following sections, logarithmic negativity, geometric quantum discord, and quantum coherence will be evaluated for both the dynamical state $\rho_{Y\bar Y}(t)$ and the stationary state $\rho_{Y\bar Y}^{(\infty)}$. This analysis enables a direct comparison between transient quantum effects and the long-time robustness of quantum resources in hyperon--antihyperon systems.
\subsection{Experimental input parameters}

The quantum-correlation properties of the hyperon--antihyperon system are fully determined by the spin-correlation matrix introduced in the previous subsection. In the present work, the latter is constructed using the experimentally measured weak-decay parameters $\alpha_{\psi}$ and $\Delta\Phi$ obtained by the BESIII Collaboration from the analysis of the reactions
$J/\psi\rightarrow Y\bar Y$.
These quantities characterize the polarization and spin-correlation structure of the produced hyperon pairs and therefore play a central role in determining the associated quantum resources.

The values of $\alpha_{\psi}$ and $\Delta\Phi$ employed throughout this work are summarized in Table~\ref{tab:BESIII}. They correspond to the channels
$J/\psi\rightarrow\Lambda\bar{\Lambda}$,
$J/\psi\rightarrow\Sigma^{+}\bar{\Sigma}^{-}$,
$J/\psi\rightarrow\Sigma^{0}\bar{\Sigma}^{0}$,
$J/\psi\rightarrow\Xi^{-}\bar{\Xi}^{+}$,
and
$J/\psi\rightarrow\Xi^{0}\bar{\Xi}^{0}$.
The quoted uncertainties are taken directly from the BESIII measurements.
\begin{table}[H]
	\caption{\label{tab:BESIII}
		Experimental values of the weak-decay parameters $\alpha_{\psi}$ and
		$\Delta\Phi$ for the processes
		$J/\psi\rightarrow Y\bar Y$
		measured by the BESIII Collaboration.}
	\centering
	\resizebox{\textwidth}{!}{
	\begin{tabular}{lccc}
		\hline\hline
		Decay channel &
		$\alpha_{\psi}$ &
		$\Delta\Phi$ (rad) &
		Refs.
		\\
		\hline
		$J/\psi\rightarrow\Lambda\bar{\Lambda}$
		&
		$0.4748\pm0.0022\pm0.0031$
		&
		$0.748\pm0.006\pm0.004$
		&
		\cite{Ablikim2025CP,Ablikim2022PRL129,Ablikim2019NatPhys,Ablikim2017PRD95}
		\\
		
		$J/\psi\rightarrow\Sigma^{+}\bar{\Sigma}^{-}$
		&
		$-0.5047\pm0.0018\pm0.0010$
		&
		$-0.2744\pm0.0033\pm0.0010$
		&
		\cite{Ablikim2019NatPhys,Ablikim2008PRD78,Ablikim2020PRL125}
		\\
		
		$J/\psi\rightarrow\Sigma^{0}\bar{\Sigma}^{0}$
		&
		$-0.4133\pm0.0035\pm0.0077$
		&
		$-0.0828\pm0.0068\pm0.0033$
		&
		\cite{Ablikim2024PRL133}
		\\
		
		$J/\psi\rightarrow\Xi^{-}\bar{\Xi}^{+}$
		&
		$0.5851\pm0.0044\pm0.0034$
		&
		$1.2205\pm0.0159\pm0.0056$
		&
		\cite{Ablikim2026PRL136,Ablikim2022Nature606,PDG2022}
		\\
		
		$J/\psi\rightarrow\Xi^{0}\bar{\Xi}^{0}$
		&
		$0.514\pm0.006\pm0.015$
		&
		$1.168\pm0.019\pm0.018$
		&
		\cite{Ablikim2017PLB770,Ablikim2023PRD108}
		\\
		\hline\hline
	\end{tabular}}
\end{table}
As shown in Table~\ref{tab:BESIII}, the decay channels exhibit markedly different values of the asymmetry parameter $\alpha_{\psi}$ and the relative phase $\Delta\Phi$. Such variations lead to distinct eigenvalue spectra of the spin-correlation matrix and consequently generate different behaviors of entanglement, quantum discord, and quantum coherence. In particular, the $\Xi^{-}\bar{\Xi}^{+}$ and $\Xi^{0}\bar{\Xi}^{0}$ channels possess comparatively large phase parameters, suggesting stronger spin correlations than those observed in the $\Lambda\bar{\Lambda}$ and $\Sigma^{+}\bar{\Sigma}^{-}$ modes. In the following sections, these experimentally determined parameters will be used to investigate and compare the stationary and dynamical quantum resources associated with each hyperon pair.
\section{Quantum resource quantifiers}\label{sec4}
\subsection{Logarithmic Negativity}

To quantify the entanglement shared by the hyperon--antihyperon pair, we employ the logarithmic negativity, which is a computable entanglement monotone based on the positive partial transpose (PPT) criterion \cite{Peres1996,Horodecki1996,Vidal2002,Plenio2005}. This measure is defined from the negative eigenvalues of the partially transposed density matrix and provides a reliable characterization of bipartite quantum entanglement.

For a bipartite mixed state $\rho^X$, the logarithmic negativity is defined as

\begin{equation}
	E_N(\rho^X)
	=
	\frac{
		\left\|
		(\rho^X)^{T_B}
		\right\|_1
		-1
	}{2}
	=
	-\sum_i \mu_i,
	\label{LN1}
\end{equation}

where $\mu_i$ denote the negative eigenvalues of the partially transposed density matrix $(\rho^X)^{T_B}$ and $T_B$ denotes partial transposition with respect to subsystem $B$. The trace norm is given by

\begin{equation}
	\left\|
	(\rho^X)^{T_B}
	\right\|_1
	=
	{\rm Tr}
	\sqrt{
		\left[
		(\rho^X)^{T_B}
		\right]^\dagger
		(\rho^X)^{T_B}
	}.
	\label{LN2}
\end{equation}

For two-qubit systems, Eq.~(\ref{LN1}) can be rewritten as

\begin{equation}
	E_N(\rho^X)
	=
	\max
	\left\{
	0,-2\mu_{\min}
	\right\},
	\label{LN3}
\end{equation}

with

\begin{equation}
	\mu_{\min}
	=
	\min
	\{e_1,e_2,e_3,e_4\}.
\end{equation}

For the state (\ref{rhoX}), the partial transpose reads

\begin{equation}
	(\rho^X)^{T_B}
	=
	\frac14
	\begin{pmatrix}
		1+2B_z+\lambda_3
		&
		0
		&
		0
		&
		\lambda_1+\lambda_2
		\\
		0
		&
		1-\lambda_3
		&
		\lambda_1-\lambda_2
		&
		0
		\\
		0
		&
		\lambda_1-\lambda_2
		&
		1-\lambda_3
		&
		0
		\\
		\lambda_1+\lambda_2
		&
		0
		&
		0
		&
		1-2B_z+\lambda_3
	\end{pmatrix}.
	\label{PT}
\end{equation}

The eigenvalues of Eq.~(\ref{PT}) are obtained as

\begin{align}
	e_{1,2}
	&=
	\frac14
	\left[
	1+\lambda_3
	\pm
	\sqrt{
		4B_z^2
		+
		(\lambda_1+\lambda_2)^2
	}
	\right],
\\
	e_{3,4}
	&=
	\frac14
	\left[
	1-\lambda_3
	\pm
	(\lambda_1-\lambda_2)
	\right].
	\label{eig34}
\end{align}

Substituting Eq.~(\ref{eig34})  into Eq.~(\ref{LN3}) yields

\begin{equation}
	E_N(\rho^X)
	=
	\max
	\Big[
	0,
	-\min
	\{
	e_1,e_2,e_3,e_4
	\}
	\Big].
	\label{LNfinal}
\end{equation}

A nonzero value of $E_N$ signals the presence of quantum entanglement between the hyperon and antihyperon spins, while $E_N=0$ corresponds to a separable state. Under correlated dephasing dynamics, the quantities $\lambda_1$ and $\lambda_2$ become time dependent through the decoherence factor $\eta(t)$, allowing the investigation of both dynamical and steady-state entanglement properties.

\subsection{Geometric Quantum Discord}
\label{sec:GQD}

Quantum discord constitutes a broader measure of quantum correlations than entanglement, as it can remain nonzero even for separable mixed states \cite{Ollivier2001,Henderson2001,Luo2010}. Although the entropic quantum discord provides a rigorous characterization of nonclassical correlations, its evaluation generally requires a challenging optimization procedure. To overcome this difficulty, Dakić \textit{et al.} introduced the geometric formulation of quantum discord \cite{Modi2012,Dakic2010,Ali2010,Hu2018}, while Paula \textit{et al.} later proposed the Schatten one-norm geometric quantum discord, which satisfies the fundamental requirements of a bona fide correlation measure \cite{Paula2013,Nakano2013}.

The one-norm geometric quantum discord is defined as the minimum trace distance between the quantum state and the set of classical-quantum states,

\begin{equation}
	Q_G(\rho_{AB})
	=
	\min_{\chi\in\Omega}
	\left\|
	\rho_{AB}-\chi
	\right\|_1,
	\label{GQD1}
\end{equation}

where $\Omega$ denotes the set of classical states and

\begin{equation}
	\|X\|_1
	=
	{\rm Tr}
	\sqrt{X^\dagger X}
\end{equation}

is the Schatten one-norm.

To evaluate Eq.~(\ref{GQD1}), it is convenient to express the density matrix in the Bloch--Fano representation \cite{Ciccarello2014,Fano1957},

\begin{equation}
	\rho_{AB}
	=
	\frac14
	\sum_{\mu,\nu=0}^{3}
	R_{\mu\nu}
	\sigma_\mu
	\otimes
	\sigma_\nu,
	\label{Fano}
\end{equation}

where

\begin{equation}
	R_{\mu\nu}
	=
	{\rm Tr}
	\Big[
	\rho_{AB}
	(\sigma_\mu\otimes\sigma_\nu)
	\Big].
\end{equation}

For the hyperon--antihyperon X state (\ref{rhoX}), the nonvanishing
Bloch--Fano coefficients are

\begin{equation}
	R_{11}=\lambda_1,
	\quad
	R_{22}=\lambda_2,
	\quad
	R_{33}=\lambda_3,
		\quad
	R_{30}=R_{03}=B_z.\label{GQDgeneral}
\end{equation}

Substituting these coefficients into Eq.~(\ref{GQDgeneral}),
the geometric quantum discord becomes

\begin{equation}
	Q_G(\rho^X)
	=
	\frac12
	\sqrt{
		\frac{
			\lambda_1^{2}
			\max
			\left(
			\lambda_2^{2}+B_z^{2},
			\lambda_3^{2}
			\right)
			-
			\lambda_2^{2}
			\min
			\left(
			\lambda_1^{2},
			\lambda_3^{2}
			\right)
		}{
			\max
			\left(
			\lambda_2^{2}+B_z^{2},
			\lambda_3^{2}
			\right)
			-
			\min
			\left(
			\lambda_1^{2},
			\lambda_3^{2}
			\right)
			+
			\lambda_1^{2}
			-
			\lambda_2^{2}
		}
	}.
	\label{GQDHyperon}
\end{equation}
The geometric quantum discord of the dynamical state is obtained by
substituting the time-dependent Bloch--Fano coefficients into
Eq.~(\ref{GQDHyperon}). Since the correlated dephasing channel affects
only the transverse correlations, one has

\begin{equation}
	R_{11}(t)=\eta(t)\lambda_1,
	\quad
	R_{22}(t)=\eta(t)\lambda_2,
\quad
	R_{33}(t)=\lambda_3,
	\quad
	R_{30}(t)=R_{03}(t)=B_z.
\end{equation}

Consequently, the geometric quantum discord becomes a function of the
decoherence factor $\eta(t)$, allowing us to investigate the temporal
evolution of quantum correlations as well as their asymptotic stationary
behavior.
\subsection{Quantum Coherence}
\label{sec:coherence}

Quantum coherence originates from the superposition principle and constitutes a fundamental resource in quantum information processing \cite{Baumgratz2014,Streltsov2017,Aberg2006,Winter2016}. Following the resource-theoretic framework introduced in Ref.~\cite{Baumgratz2014}, the coherence of a quantum state is quantified by the $l_1$-norm, defined as

\begin{equation}
	C_{l_1}(\rho)
	=
	\sum_{i\neq j}
	|\rho_{ij}|.
	\label{coh1}
\end{equation}

For the hyperon--antihyperon $X$ state given in Eq.~(\ref{rhoX}), only the off-diagonal elements $\rho_{14}$ and $\rho_{23}$ contribute to the coherence. Therefore,

\begin{equation}
	C_{l_1}(\rho^X)
	=
	2
	\left(
	|\rho_{14}|+|\rho_{23}|
	\right).
	\label{coh2}
\end{equation}

Using Eq.~(\ref{rhoX}), one has

\begin{equation}
	\rho_{14}
	=
	\frac{\lambda_1-\lambda_2}{4},
	\qquad
	\rho_{23}
	=
	\frac{\lambda_1+\lambda_2}{4},
\end{equation}

which yields

\begin{equation}
	C_{l_1}(\rho^X)
	=
	\frac12
	\left(
	|\lambda_1-\lambda_2|
	+
	|\lambda_1+\lambda_2|
	\right).
	\label{coh3}
\end{equation}

Equation~(\ref{coh3}) shows that the quantum coherence is entirely determined by the transverse spin-correlation coefficients $\lambda_1$ and $\lambda_2$, whereas the local polarization $B_z$ and the longitudinal correlation coefficient $\lambda_3$ do not contribute to the coherence measure.

When the system is subjected to the correlated dephasing channel introduced in Sec.~III, the off-diagonal elements evolve according to

\begin{equation}
	\rho_{14}(t)
	=
	\eta(t)\rho_{14},
	\qquad
	\rho_{23}(t)
	=
	\eta(t)\rho_{23},
\end{equation}

where $\eta(t)$ denotes the decoherence factor. Consequently, the time-dependent coherence becomes

\begin{equation}
	C_{l_1}(t)
	=
	2
	\left(
	|\rho_{14}(t)|
	+
	|\rho_{23}(t)|
	\right),
\end{equation}

or equivalently,

\begin{equation}
	C_{l_1}(t)
	=
	\frac{\eta(t)}{2}
	\left(
	|\lambda_1-\lambda_2|
	+
	|\lambda_1+\lambda_2|
	\right).
	\label{coht}
\end{equation}

The stationary coherence is obtained from the asymptotic limit

\begin{equation}
	C_{l_1}^{(\infty)}
	=
	\lim_{t\rightarrow\infty}
	C_{l_1}(t)
	=
	\frac{\eta_\infty}{2}
	\left(
	|\lambda_1-\lambda_2|
	+
	|\lambda_1+\lambda_2|
	\right),
	\label{cohsteady}
\end{equation}

with
$
\eta_\infty
=
\lim_{t\rightarrow\infty}\eta(t).
$
The quantities $C_{l_1}(t)$ and $C_{l_1}^{(\infty)}$ will be analyzed in the following sections to characterize the robustness of coherence in the dynamical and steady-state regimes.

\section{Results and Discussion}\label{sec5}
In this section, we present a detailed analysis of the quantum resources associated with hyperon--antihyperon pairs produced in the decay processes $J/\psi\rightarrow Y\bar{Y}$, where $Y=\Lambda$, $\Sigma^{+}$, $\Xi^{-}$, and $\Xi^{0}$. Using the experimentally measured decay parameters summarized in Table~\ref{tab:BESIII}, we investigate the behavior of logarithmic negativity, geometric quantum discord, and $l_1$-norm quantum coherence under both longitudinal and transverse beam polarizations. We first examine the stationary properties of these quantum resources and their dependence on the kinematic variables and polarization parameters. We then explore their dynamical evolution in correlated dephasing environments, emphasizing the influence of memory effects and the distinction between Markovian and non-Markovian regimes. Finally, we analyze the hierarchy among the considered quantum resources and discuss their relative robustness against decoherence.

\subsection{Dynamical quantum resources}
\begin{figure}[H]
	\centering
	\includegraphics[width=0.24\linewidth]{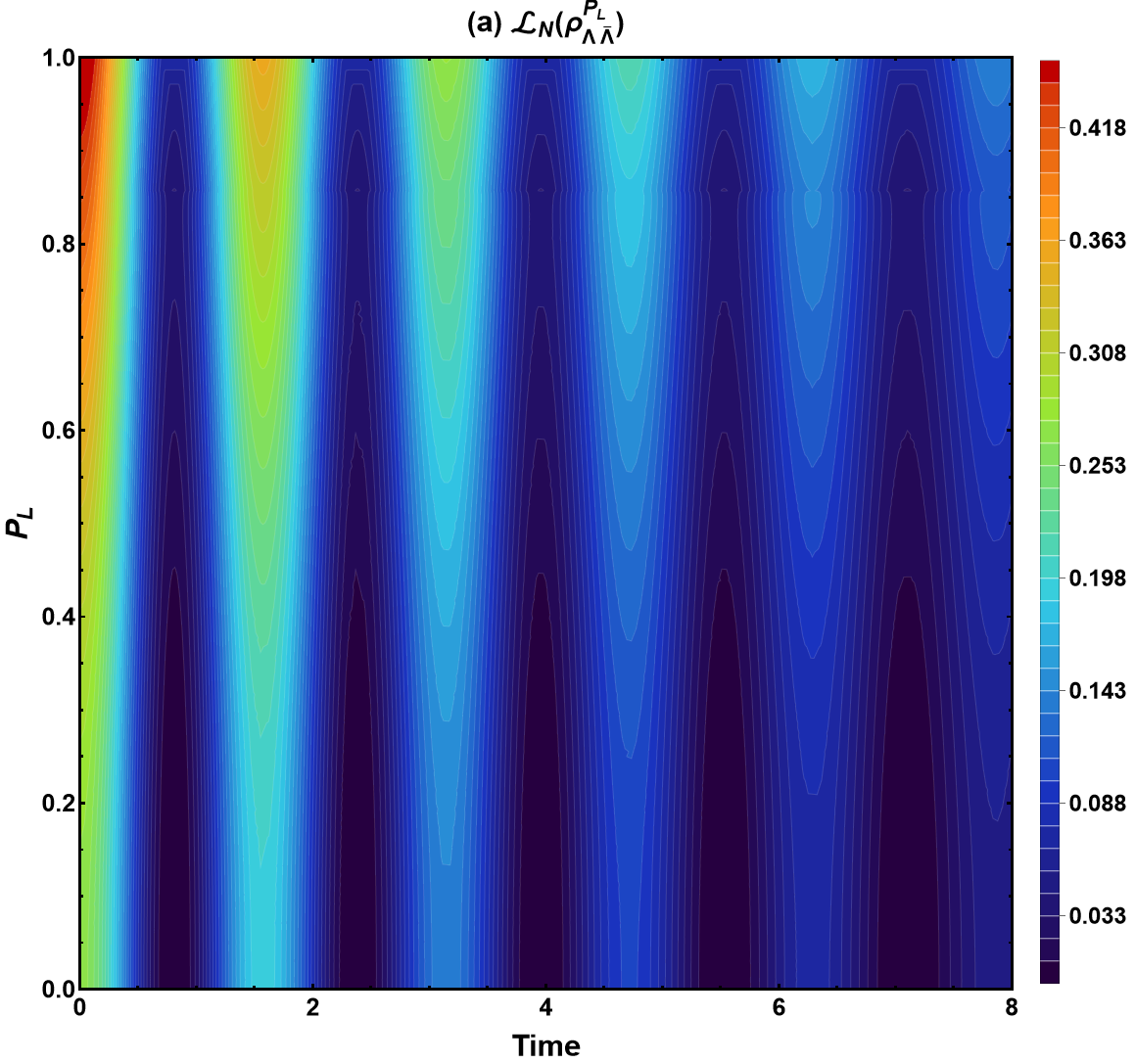}
	\includegraphics[width=0.24\linewidth]{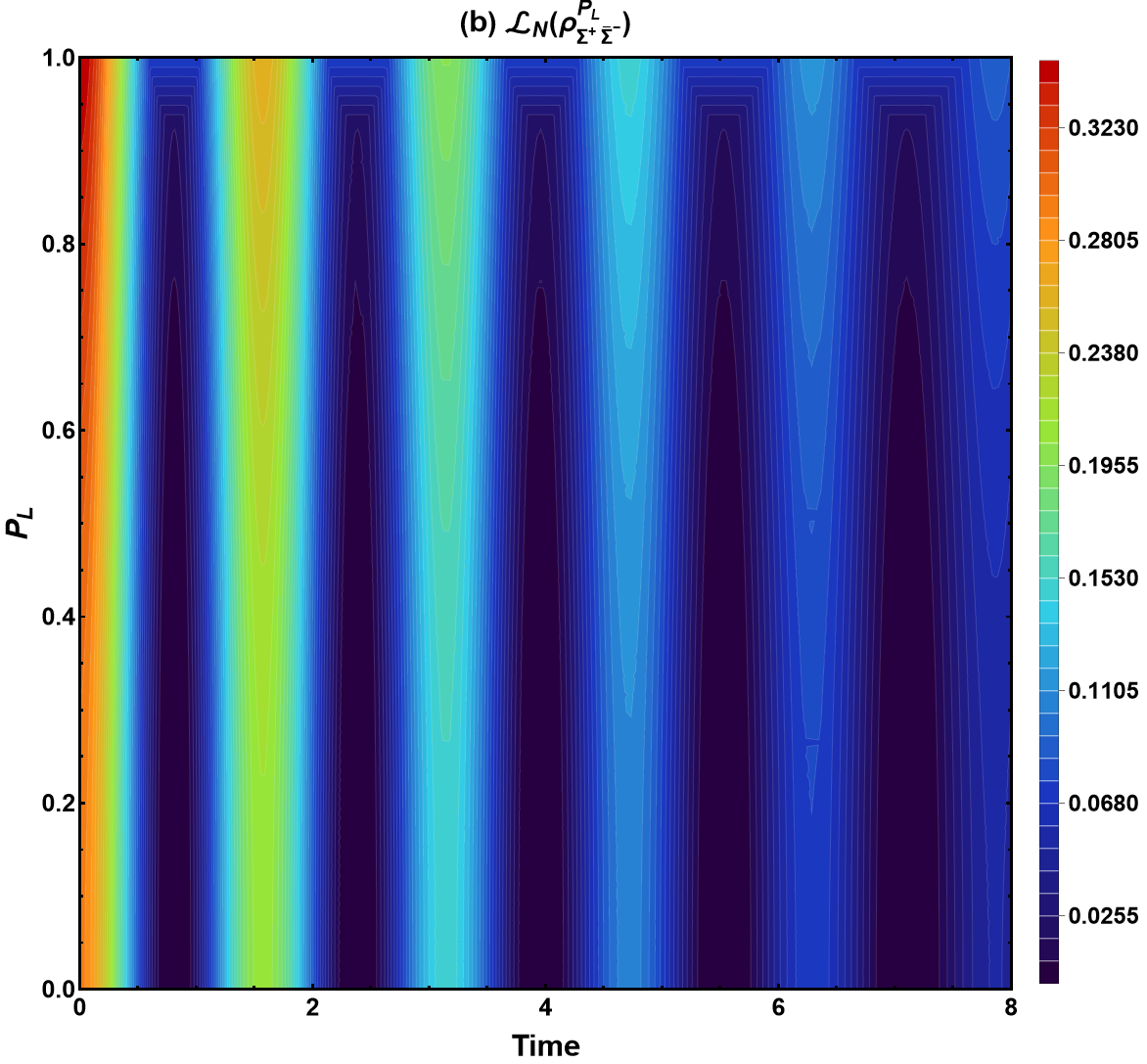}
	\includegraphics[width=0.24\linewidth]{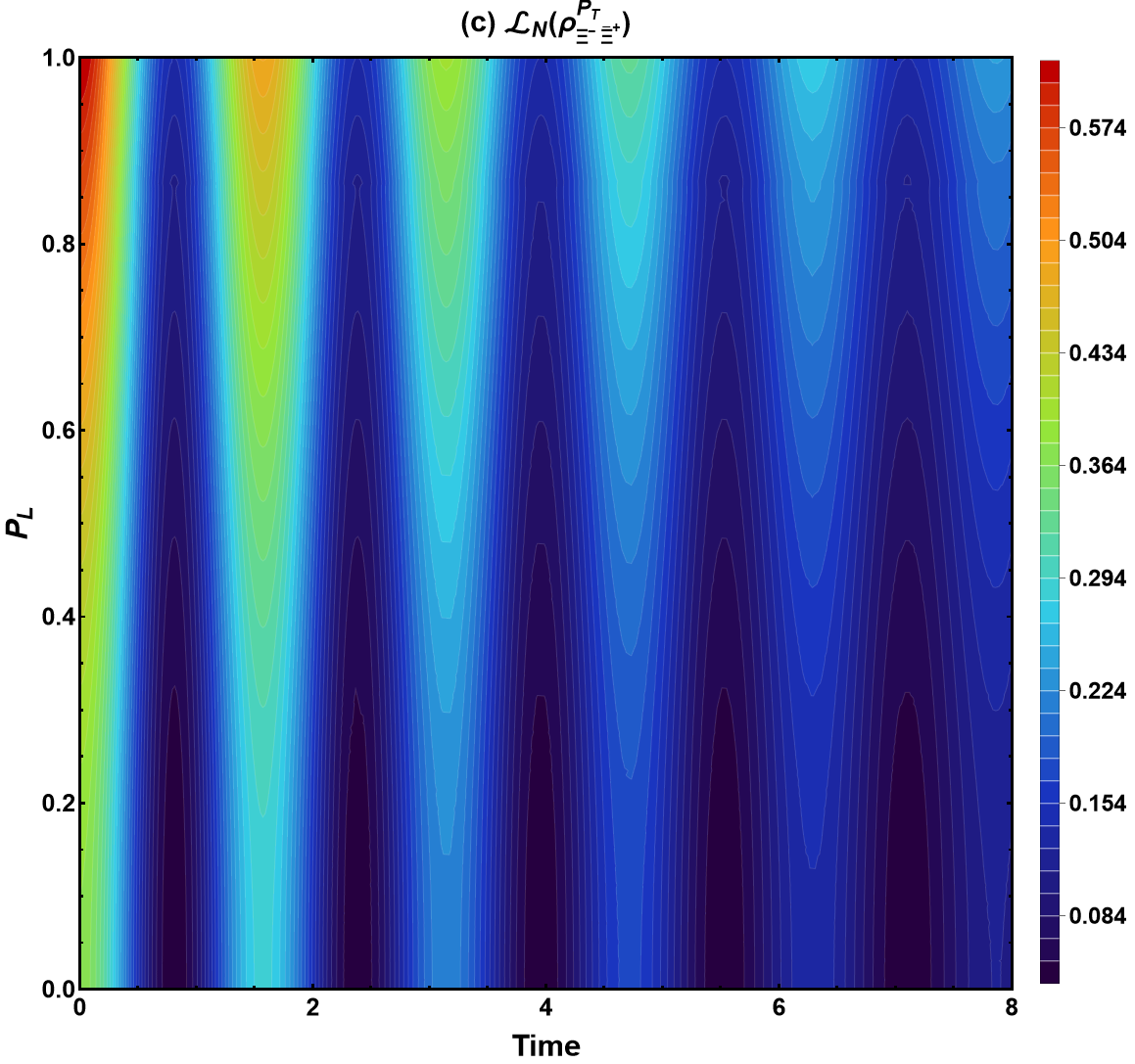}
	\includegraphics[width=0.24\linewidth]{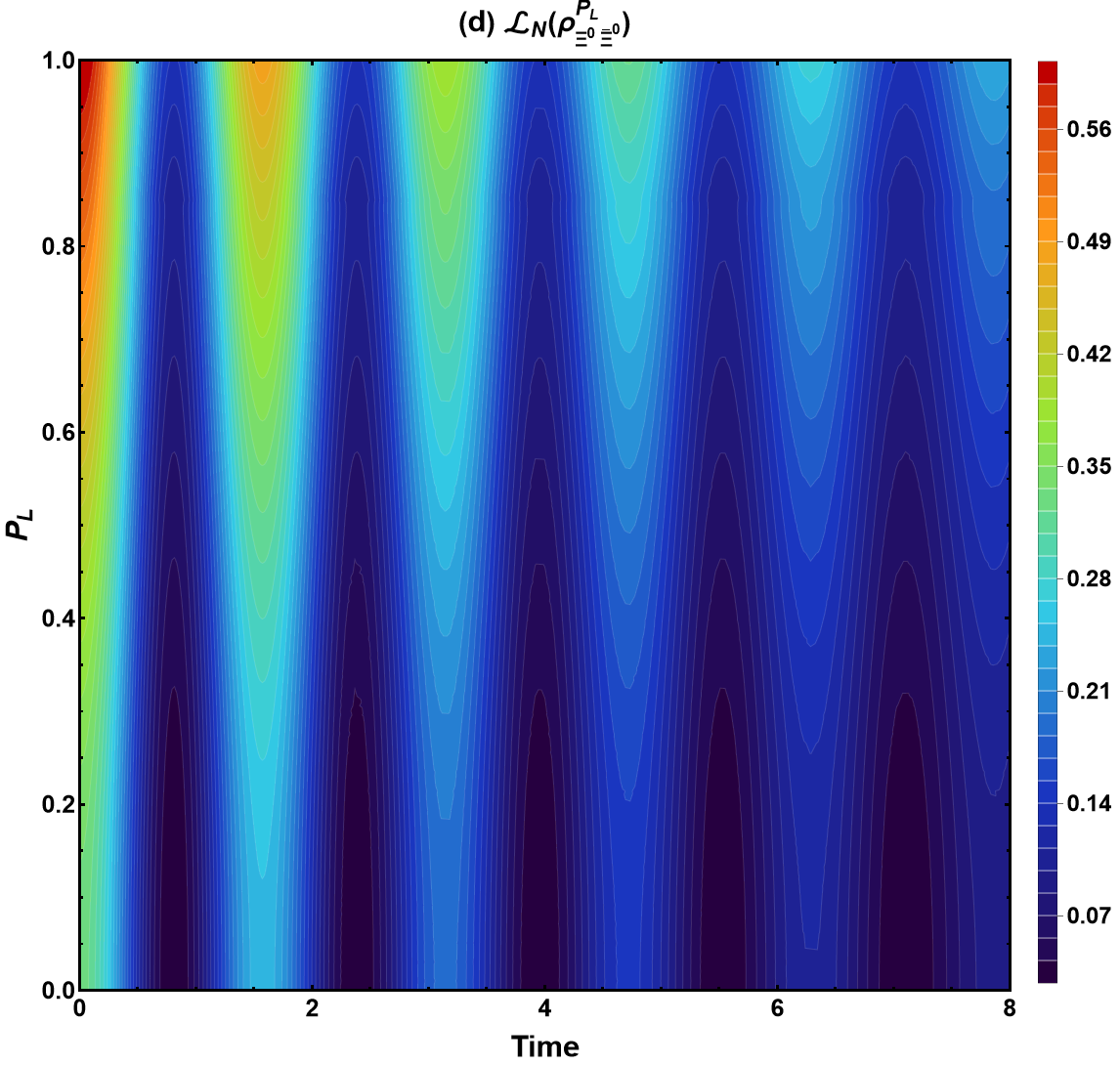}
		\includegraphics[width=0.24\linewidth]{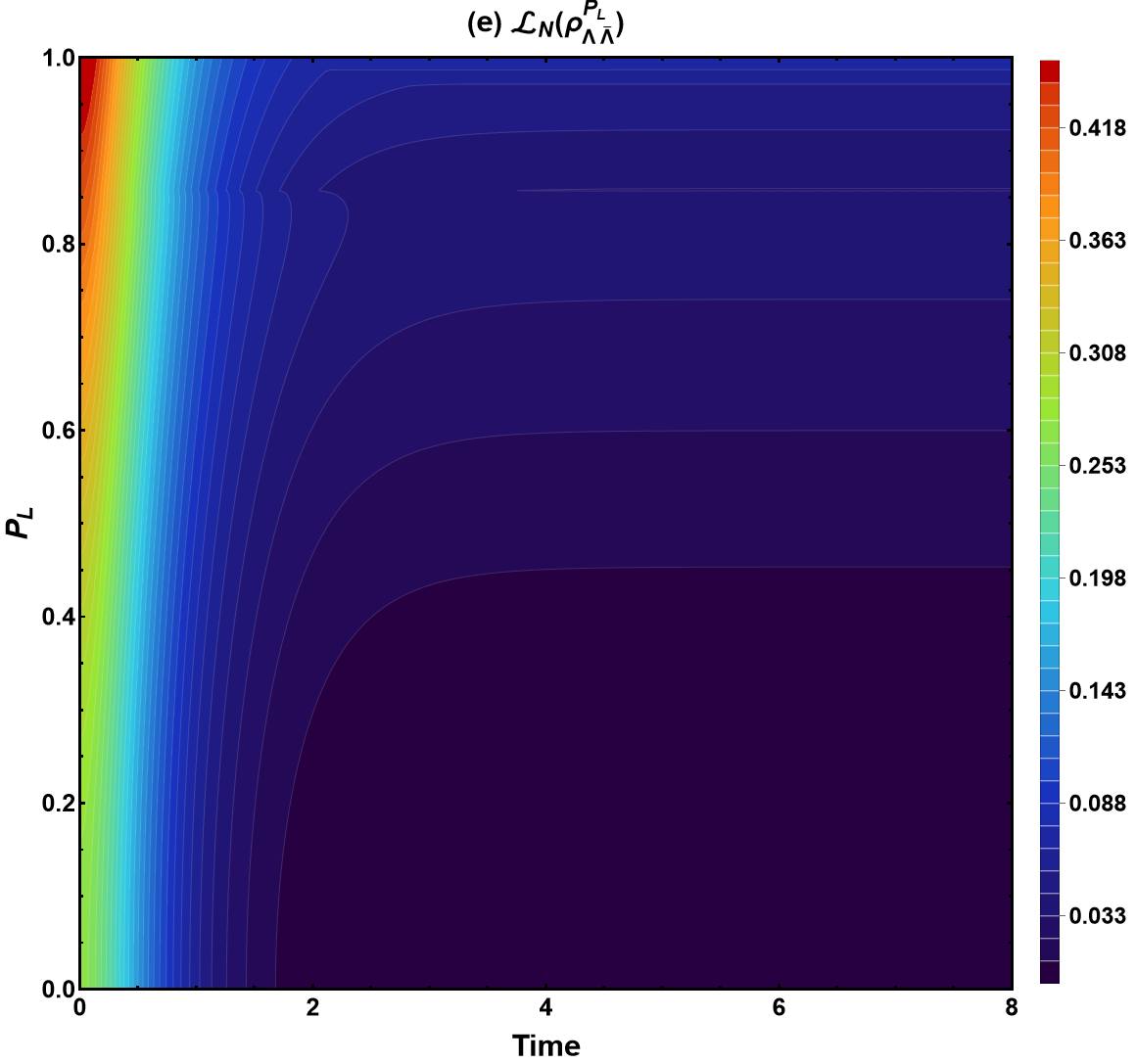}
	\includegraphics[width=0.24\linewidth]{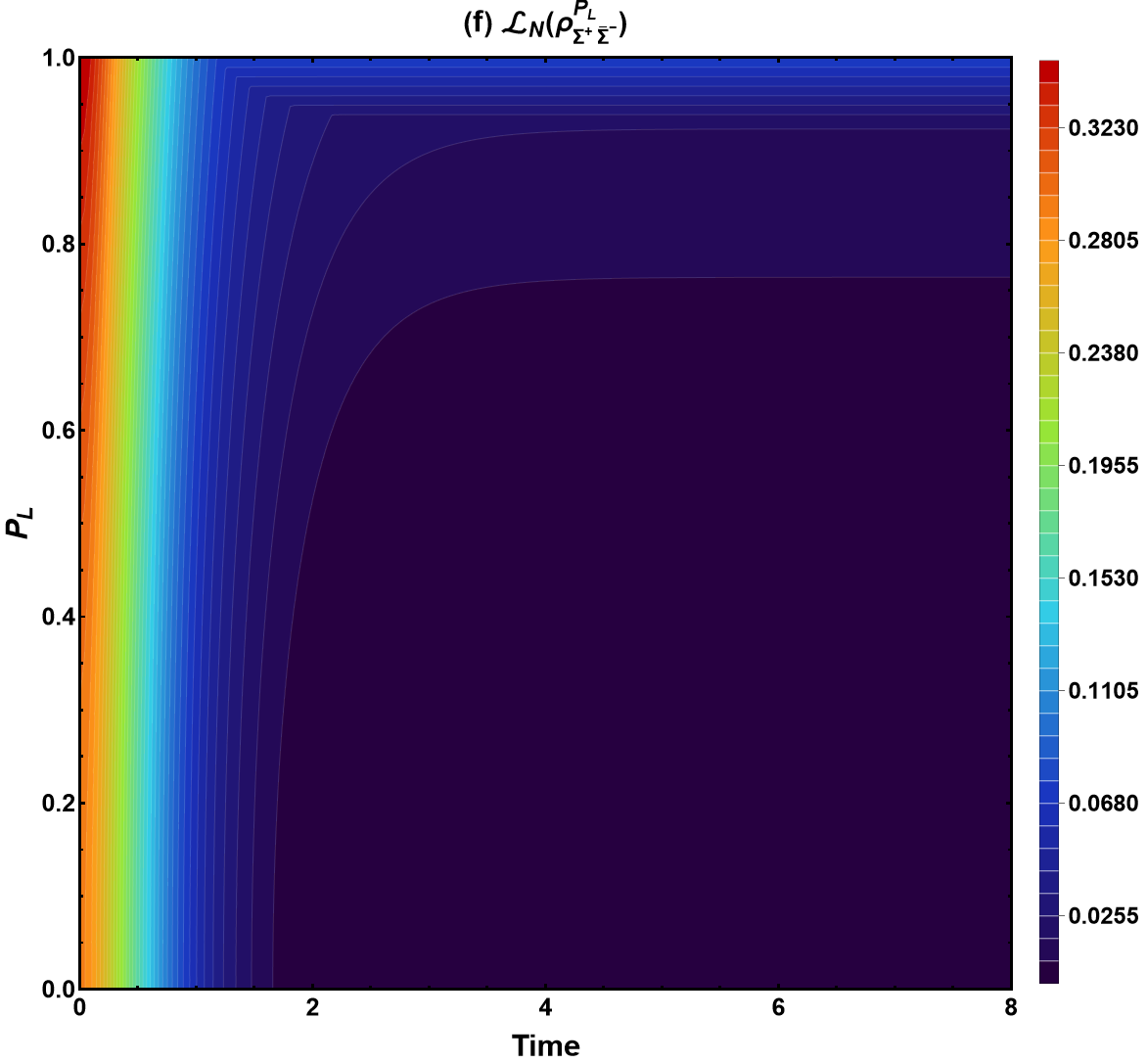}
	\includegraphics[width=0.24\linewidth]{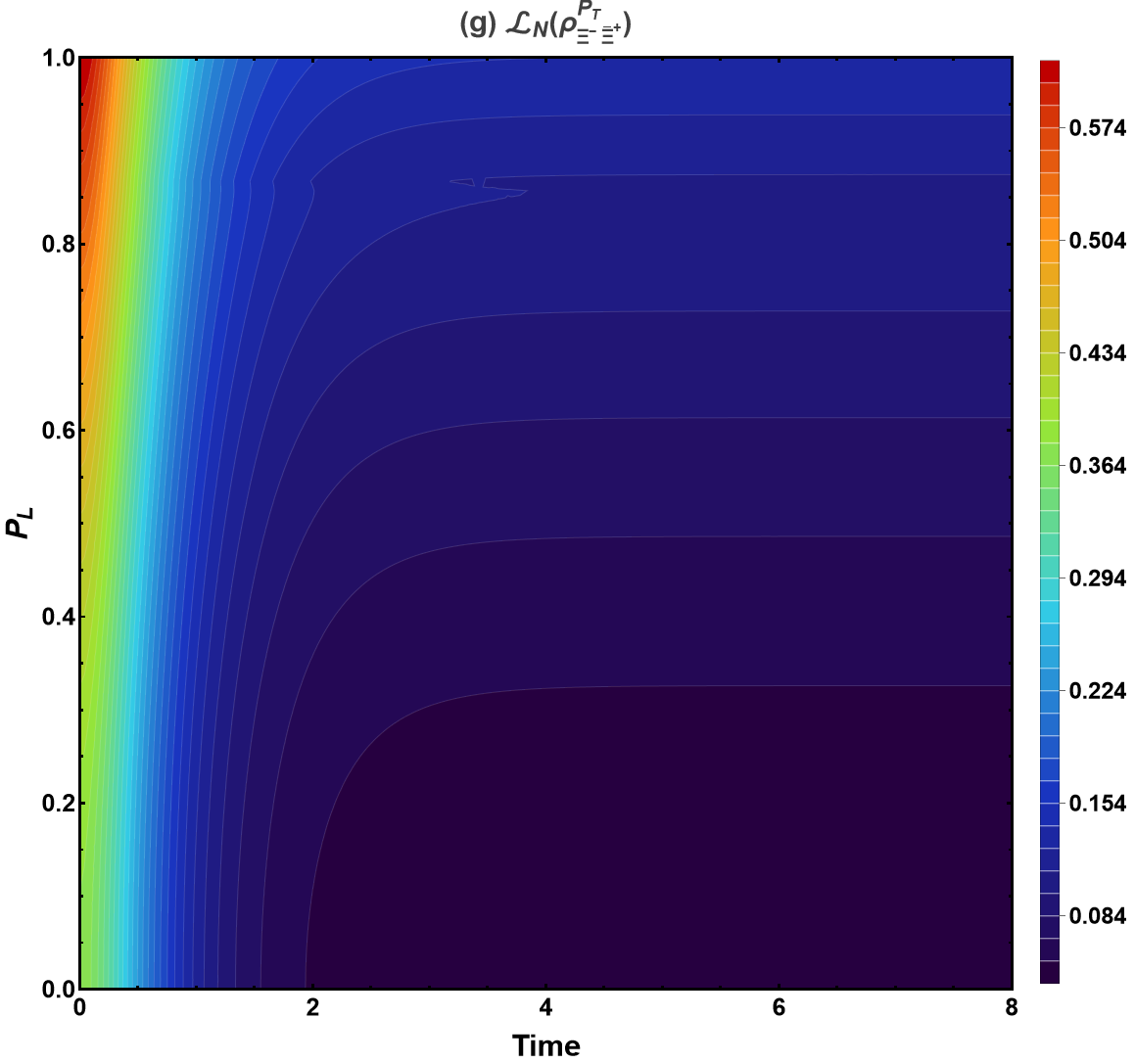}
	\includegraphics[width=0.24\linewidth]{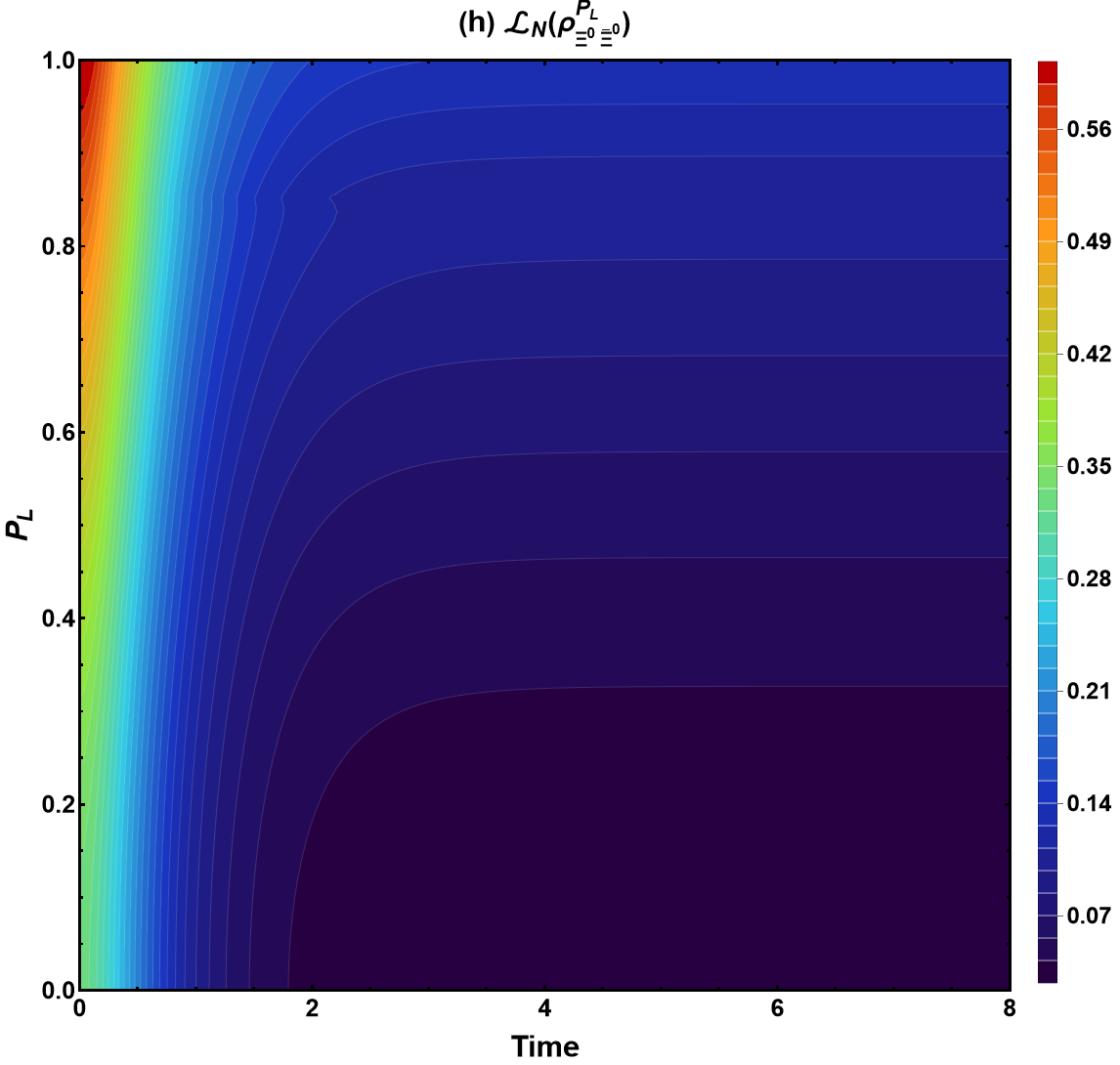}
\caption{
	Dynamical evolution of the logarithmic negativity
	$\mathcal{L}_N(\rho^{P_L}_{Y\bar{Y}})$ as a function of time and the
	longitudinal polarization degree $P_L$ for
	$J/\psi\rightarrow Y\bar{Y}$ with
	$Y=\Lambda$, $\Sigma^{+}$, $\Xi^{-}$, and $\Xi^{0}$ at
	$\cos\theta=0.5$. Panels (a)--(d) [(e)--(h)] correspond to the
	non-Markovian (Markovian) regime with $\tau=5$ ($\tau=0.2$) and
	$\mu=0.4$. The experimental parameters are taken from
	Table~\ref{tab:BESIII}.
}
	\label{fig11}
\end{figure}

A prominent feature emerging from Fig.~\ref{fig11} is the strong
dependence of the logarithmic negativity
$\mathcal{L}_{N}(\rho^{P_L}_{Y\bar{Y}})$ on both the evolution time and
the longitudinal polarization degree $P_L$.
The non-Markovian dynamics displayed in
Figs.~\ref{fig11}(a)--(d) are characterized by a sequence of damped
entanglement oscillations. The periodic alternation between high- and
low-entanglement regions indicates a continuous exchange of quantum
information between the hyperon pair and its environment. Although the
overall amplitude decreases with time, the recurrent maxima demonstrate
that a significant fraction of the initial quantum correlations remains
accessible over extended evolution times.
The influence of the longitudinal polarization is clearly visible in all
channels. Larger values of $P_L$ strengthen the entanglement peaks and
extend the temporal intervals over which nonvanishing quantum
correlations persist. This behavior suggests that spin alignment along
the beam direction enhances the coherence of the production process and
therefore improves the stability of the entangled hyperon state.
Quantitatively, the strongest entanglement is observed for the
$\Xi^{-}\bar{\Xi}^{+}$ and $\Xi^{0}\bar{\Xi}^{0}$ channels, whose peak
values exceed those of the $\Lambda\bar{\Lambda}$ and
$\Sigma^{+}\bar{\Sigma}^{-}$ systems throughout the evolution. This
difference reflects the channel dependence of the spin-correlation
structure encoded in the production amplitudes.
A markedly different picture emerges in the Markovian regime
[Figs.~\ref{fig11}(e)--(h)].
\begin{figure}[H]
	\centering
	\includegraphics[width=0.24\linewidth]{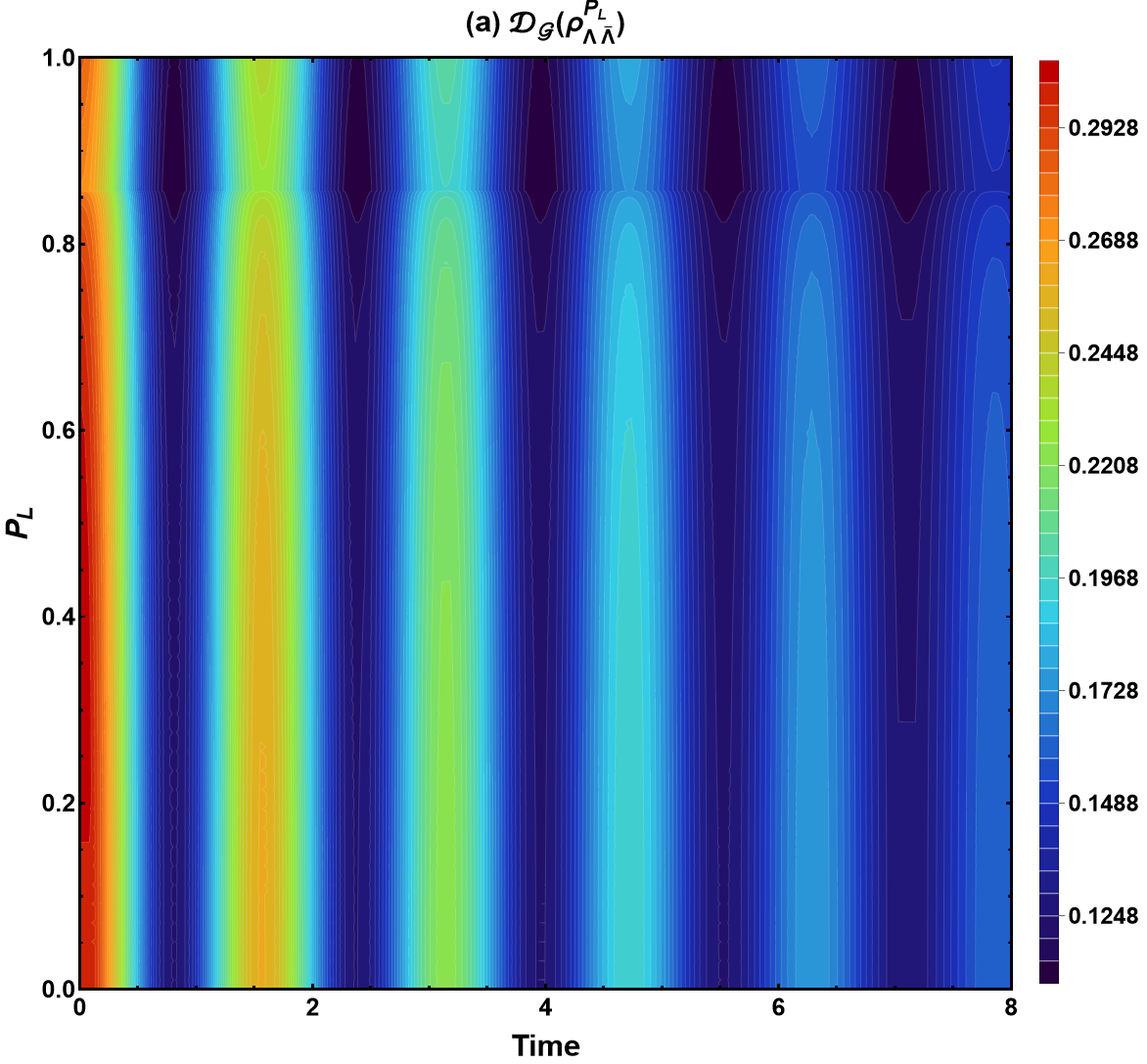}
	\includegraphics[width=0.24\linewidth]{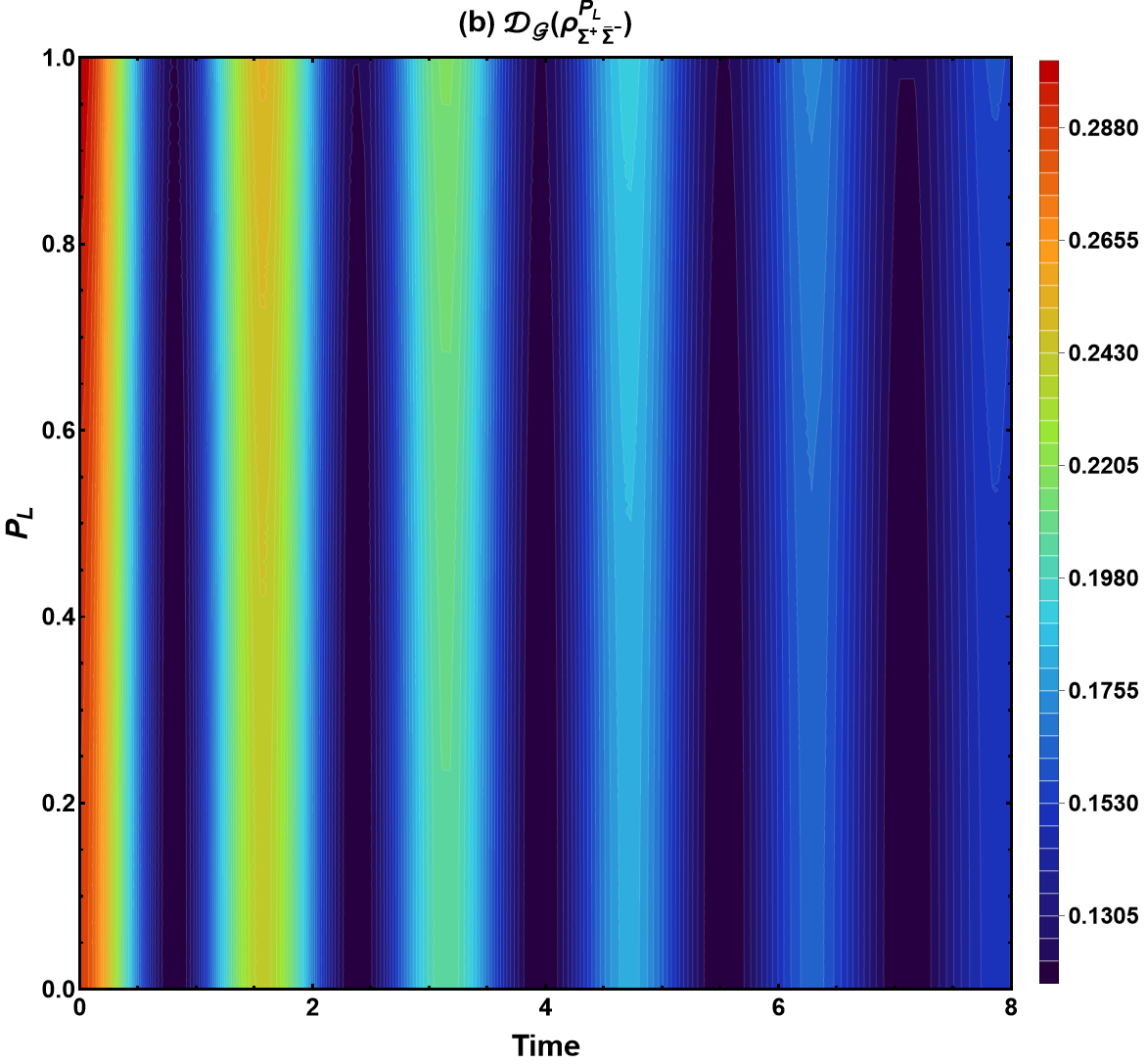}
	\includegraphics[width=0.24\linewidth]{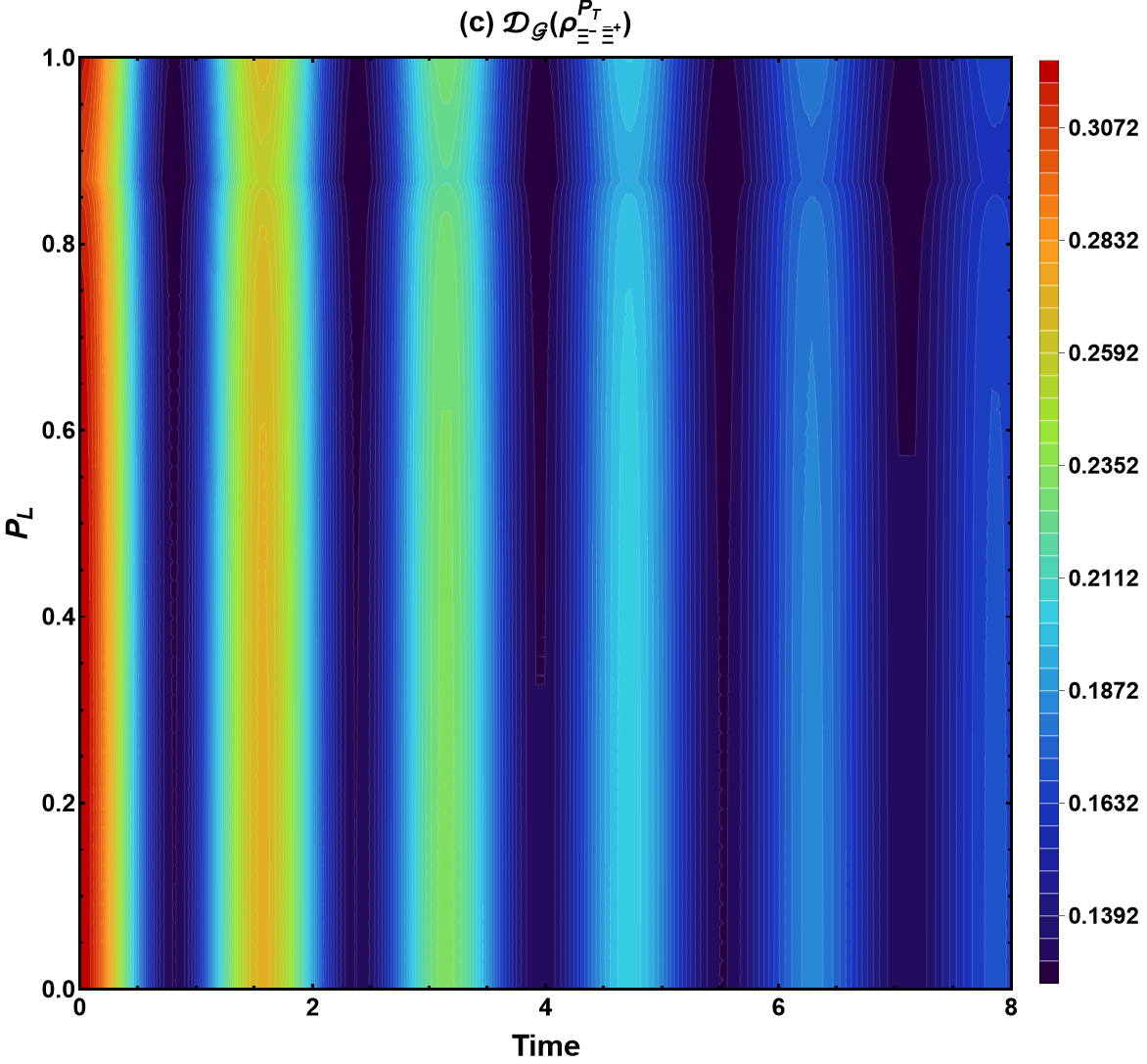}
	\includegraphics[width=0.24\linewidth]{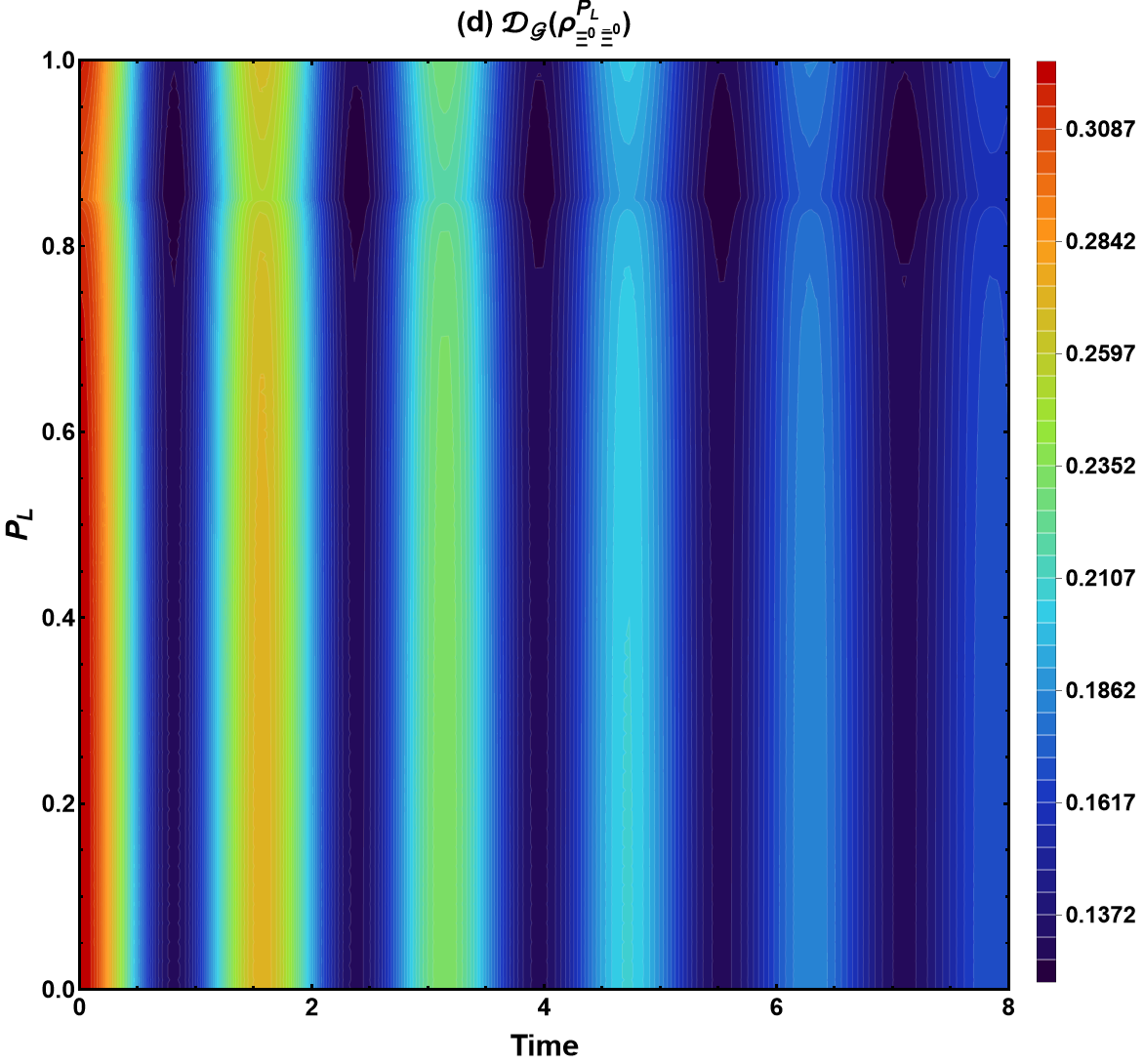}
	\includegraphics[width=0.24\linewidth]{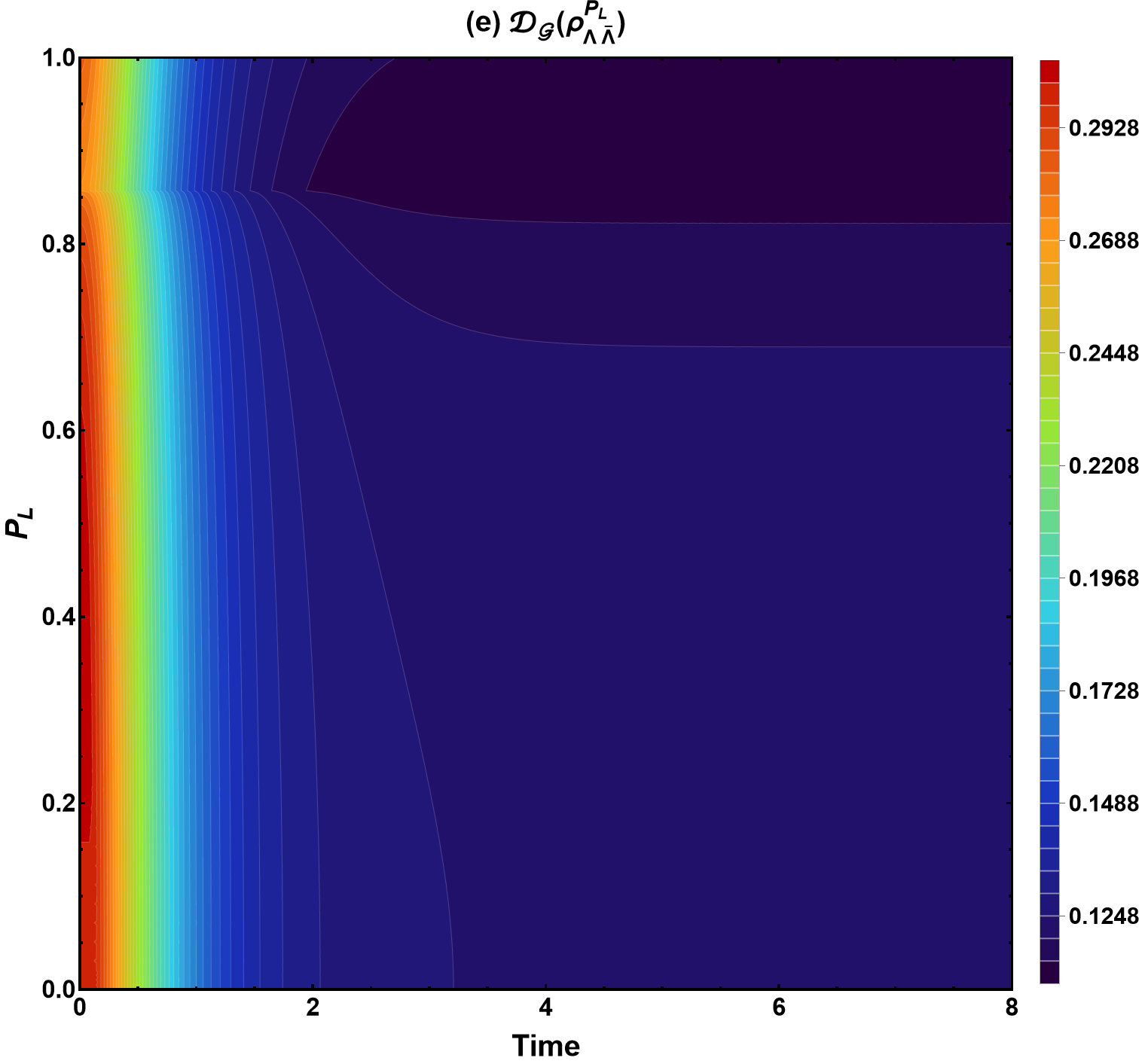}
	\includegraphics[width=0.24\linewidth]{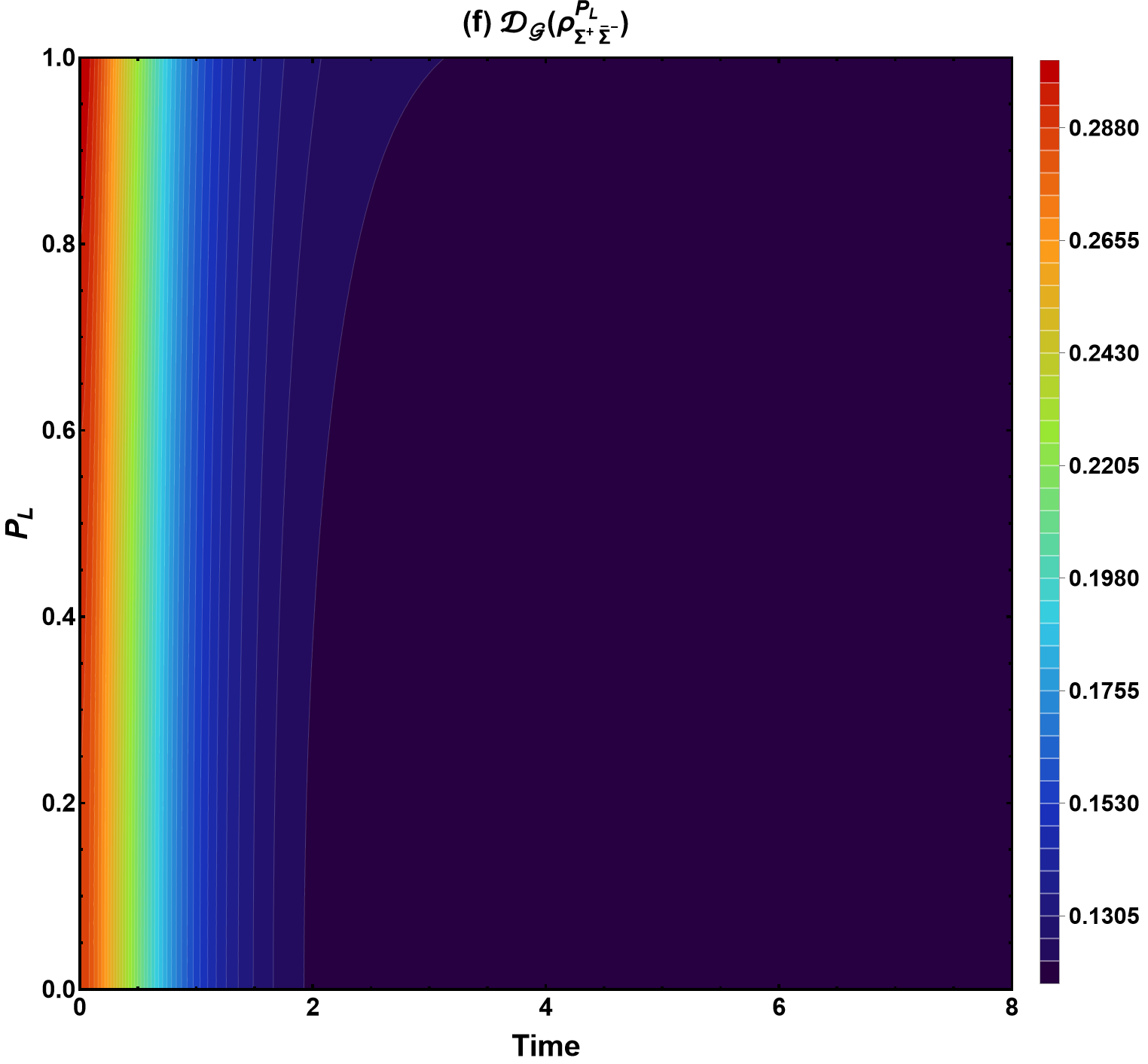}
	\includegraphics[width=0.24\linewidth]{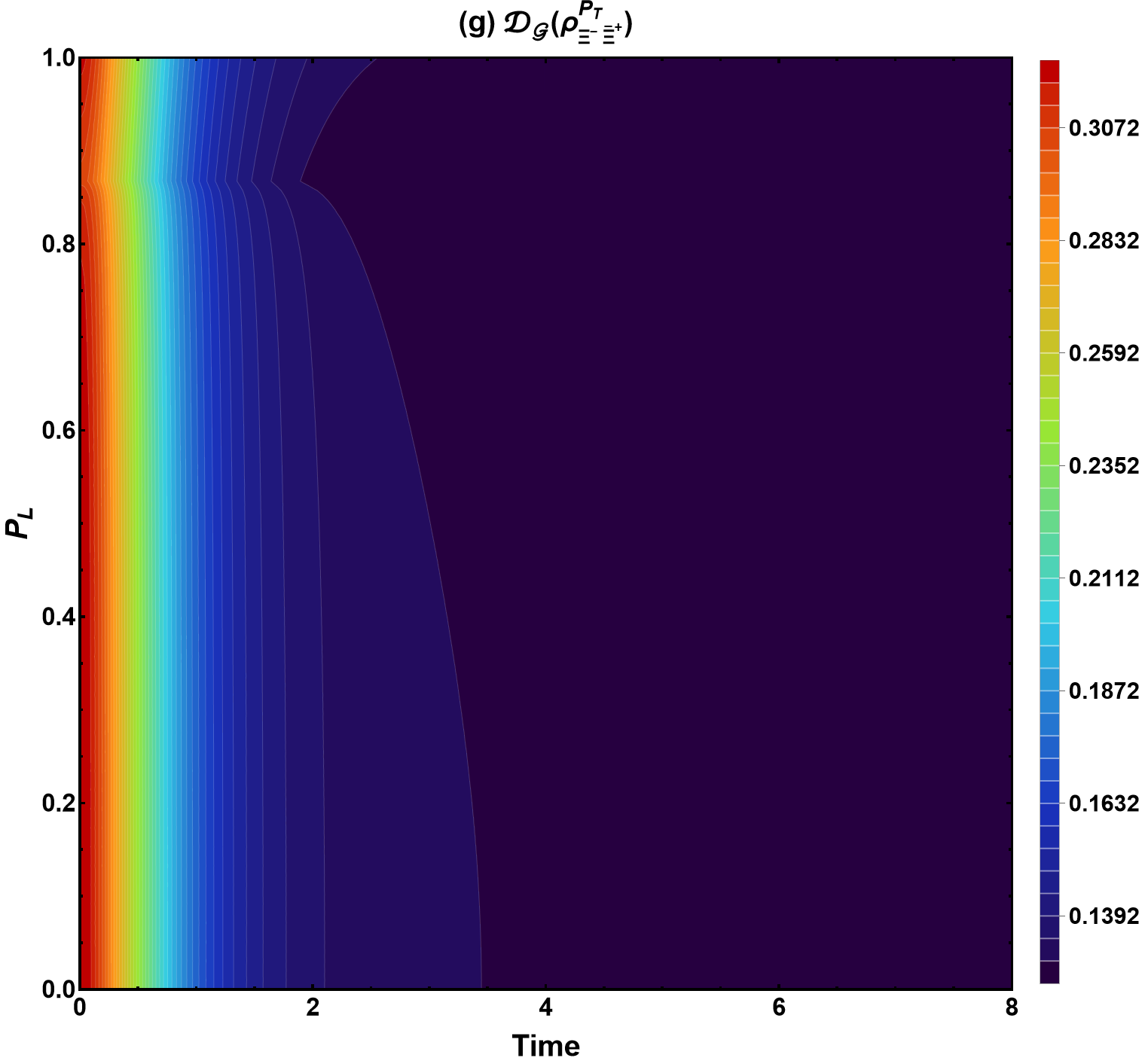}
	\includegraphics[width=0.24\linewidth]{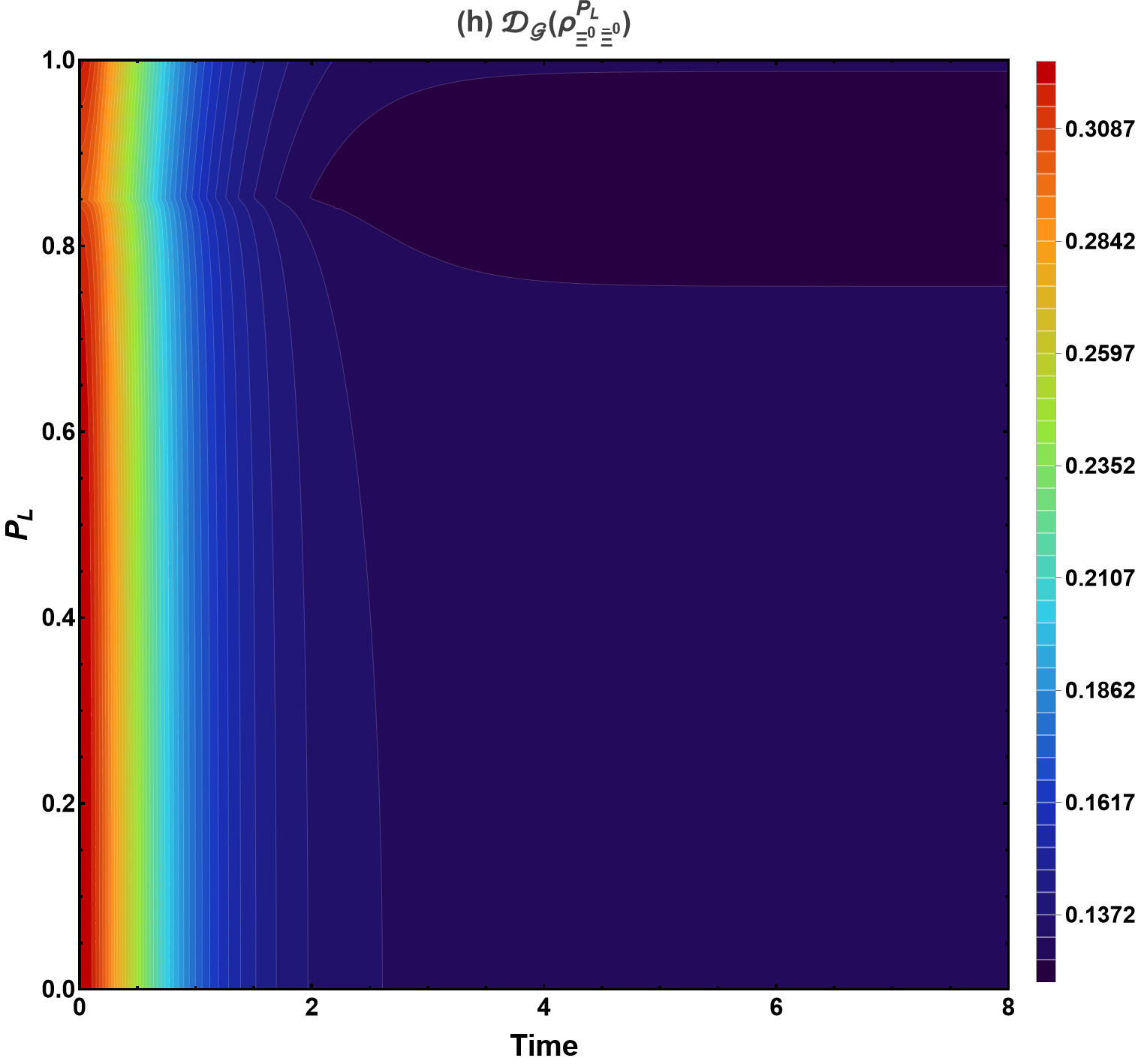}
\caption{
	Dynamical evolution of the geometric quantum discord
	$\mathcal{D}_G(\rho^{P_L}_{Y\bar{Y}})$ as a function of time and the
	longitudinal polarization degree $P_L$ for
	$J/\psi\rightarrow Y\bar{Y}$ with
	$Y=\Lambda$, $\Sigma^{+}$, $\Xi^{-}$, and $\Xi^{0}$ at
	$\cos\theta=0.5$. Panels (a)--(d) [(e)--(h)] correspond to the
	non-Markovian (Markovian) regime with $\tau=5$ ($\tau=0.2$) and
	$\mu=0.4$. The experimental parameters are taken from
	Table~\ref{tab:BESIII}.
}
	\label{fig12}
\end{figure}
 The oscillatory pattern disappears and the
entanglement undergoes a rapid monotonic decay toward a weak stationary
regime. The absence of revivals indicates that the lost quantum
correlations are no longer recovered during the evolution. In this
situation, the effect of $P_L$ is mainly restricted to short times,
where it slightly delays the decay rate, while the long-time dynamics
remain governed by irreversible decoherence.
These results demonstrate that the preservation of entanglement is
substantially enhanced when environmental memory and longitudinal
polarization act simultaneously. The most favorable conditions are
obtained for highly polarized beams in the non-Markovian regime, whereas
Markovian evolution leads to a much faster degradation of quantum
correlations.

Figure~\ref{fig12} displays the dynamical behavior of the geometric
quantum discord $\mathcal{D}_{G}(\rho^{P_L}_{Y\bar{Y}})$ as a function
of the evolution time and the longitudinal beam polarization $P_L$ for
the four hyperon-antihyperon channels.
In the non-Markovian regime [Figs.~\ref{fig12}(a)--(d)], the discord
exhibits an oscillatory dynamics similar to that observed for the
logarithmic negativity. However, the minima remain significantly above
zero during the entire evolution, indicating that a substantial portion
of the nonclassical correlations survives even when entanglement is
strongly suppressed. The oscillations gradually decrease in amplitude,
yet the correlation revivals remain visible over the whole time
interval, revealing the persistent action of environmental memory.
The longitudinal polarization enhances the discord for all channels.
Larger values of $P_L$ lead to broader high-correlation regions and
increase the magnitude of the revival peaks. This effect is most
pronounced close to full polarization, where the discord remains
appreciable even at relatively long times. Such behavior indicates that
spin alignment along the beam direction favors the preservation of
quantum correlations generated during the production process.
Compared with the entanglement dynamics, the channel dependence is
considerably weaker. The $\Xi^{-}\bar{\Xi}^{+}$ and
$\Xi^{0}\bar{\Xi}^{0}$ systems attain the largest values, approaching
$0.31$, while the $\Lambda\bar{\Lambda}$ and
$\Sigma^{+}\bar{\Sigma}^{-}$ channels remain only slightly lower.
This small quantitative variation suggests that geometric discord is
less sensitive to the details of the hyperon spin structure and
provides a more universal characterization of quantum correlations.
A different evolution is observed in the Markovian regime
[Figs.~\ref{fig12}(e)--(h)], where the oscillatory pattern disappears
and the discord decreases smoothly with time. After a rapid initial
reduction, the system approaches a stationary regime characterized by a
finite residual discord. The persistence of this asymptotic value
contrasts with the stronger degradation observed for logarithmic
negativity and demonstrates that nonclassical correlations can survive
even after a significant loss of entanglement.
These results confirm the remarkable robustness of geometric quantum
discord against dephasing effects. While non-Markovian memory effects
promote recurrent correlation recoveries, even purely Markovian
evolution preserves a finite amount of quantum correlations at long
times. Consequently, geometric discord constitutes a particularly
reliable resource for probing quantum features in polarized
hyperon-antihyperon systems.
\begin{figure}[H]
	\centering
	\includegraphics[width=0.24\linewidth]{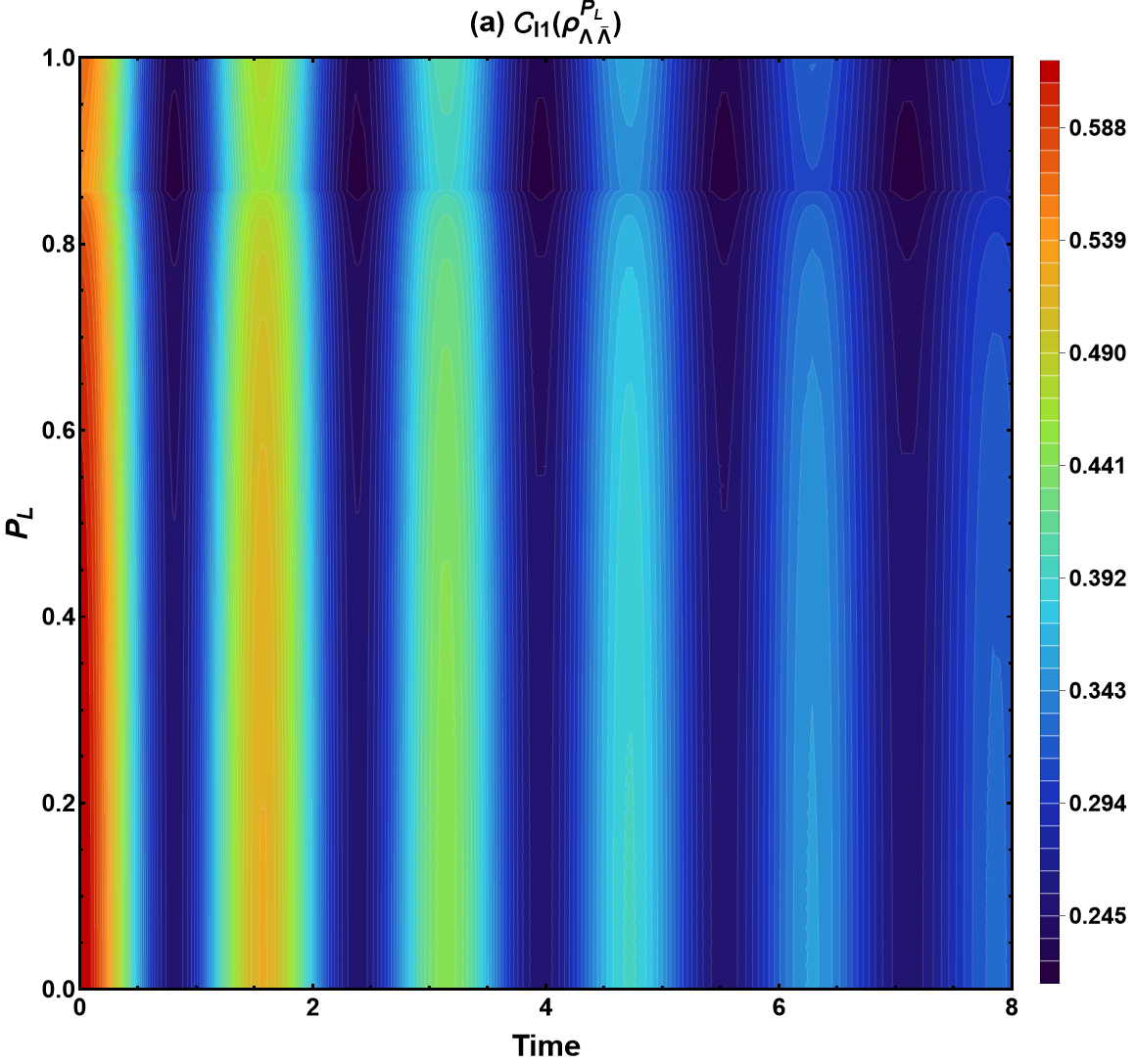}
	\includegraphics[width=0.24\linewidth]{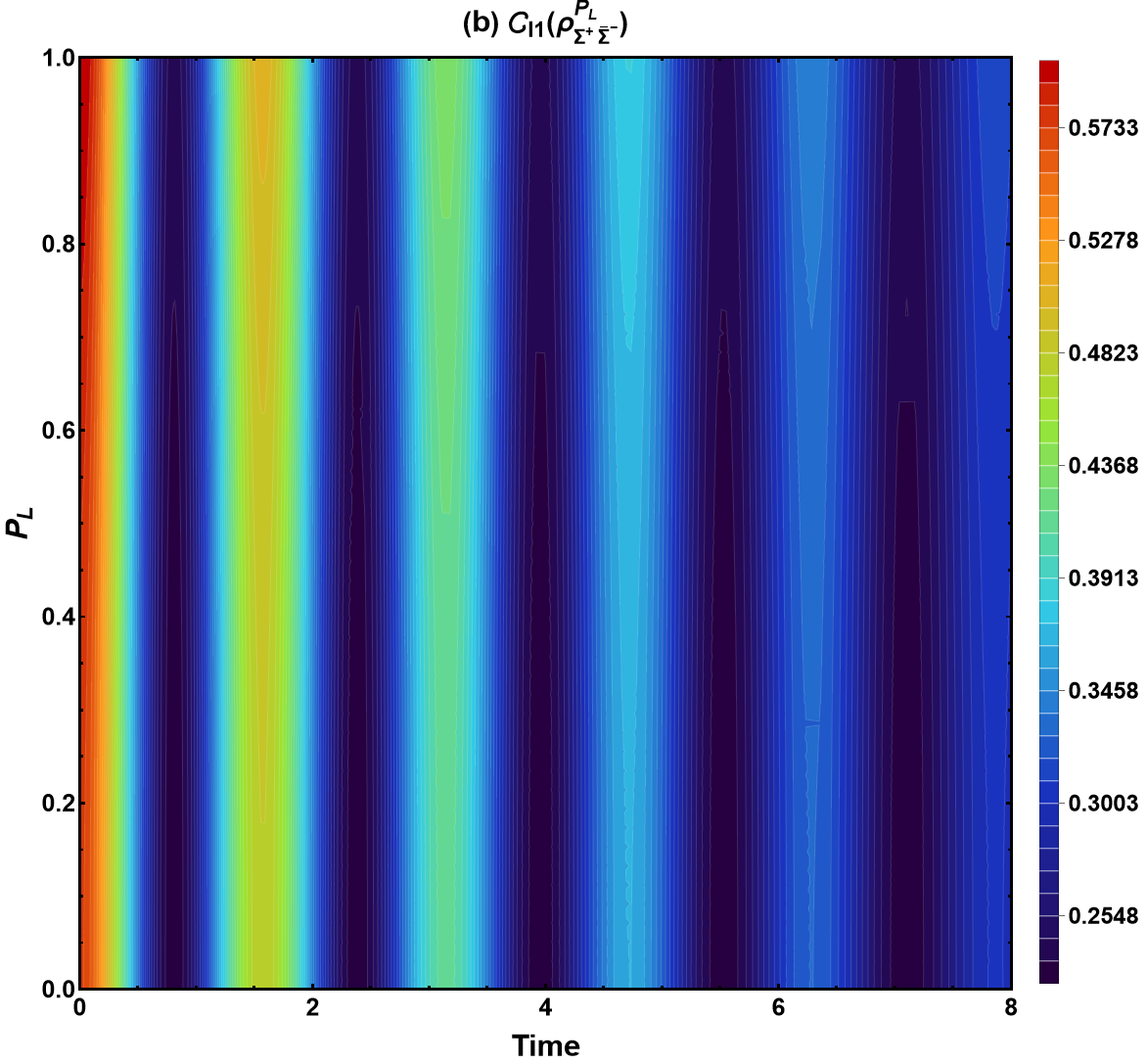}
	\includegraphics[width=0.24\linewidth]{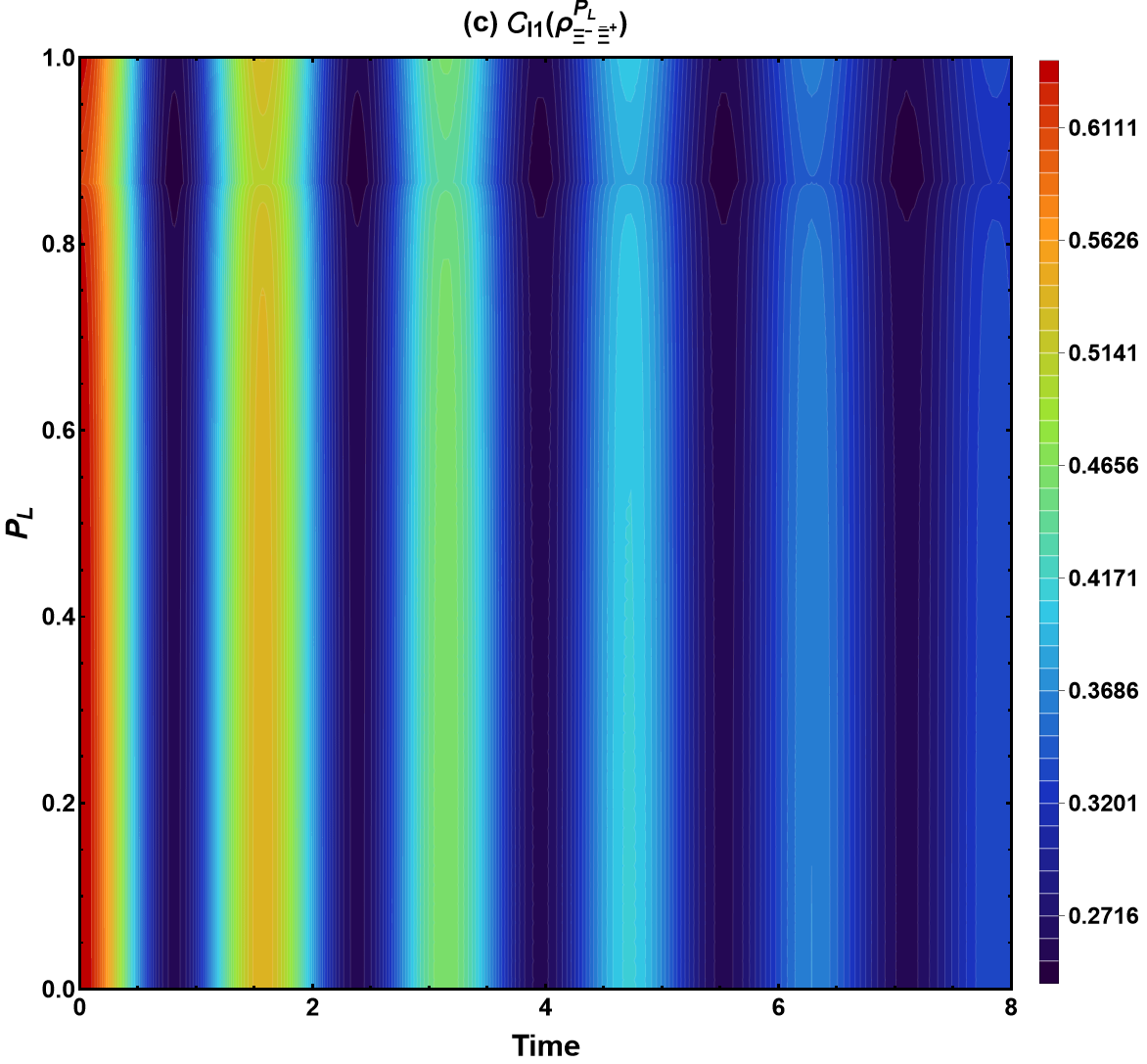}
	\includegraphics[width=0.24\linewidth]{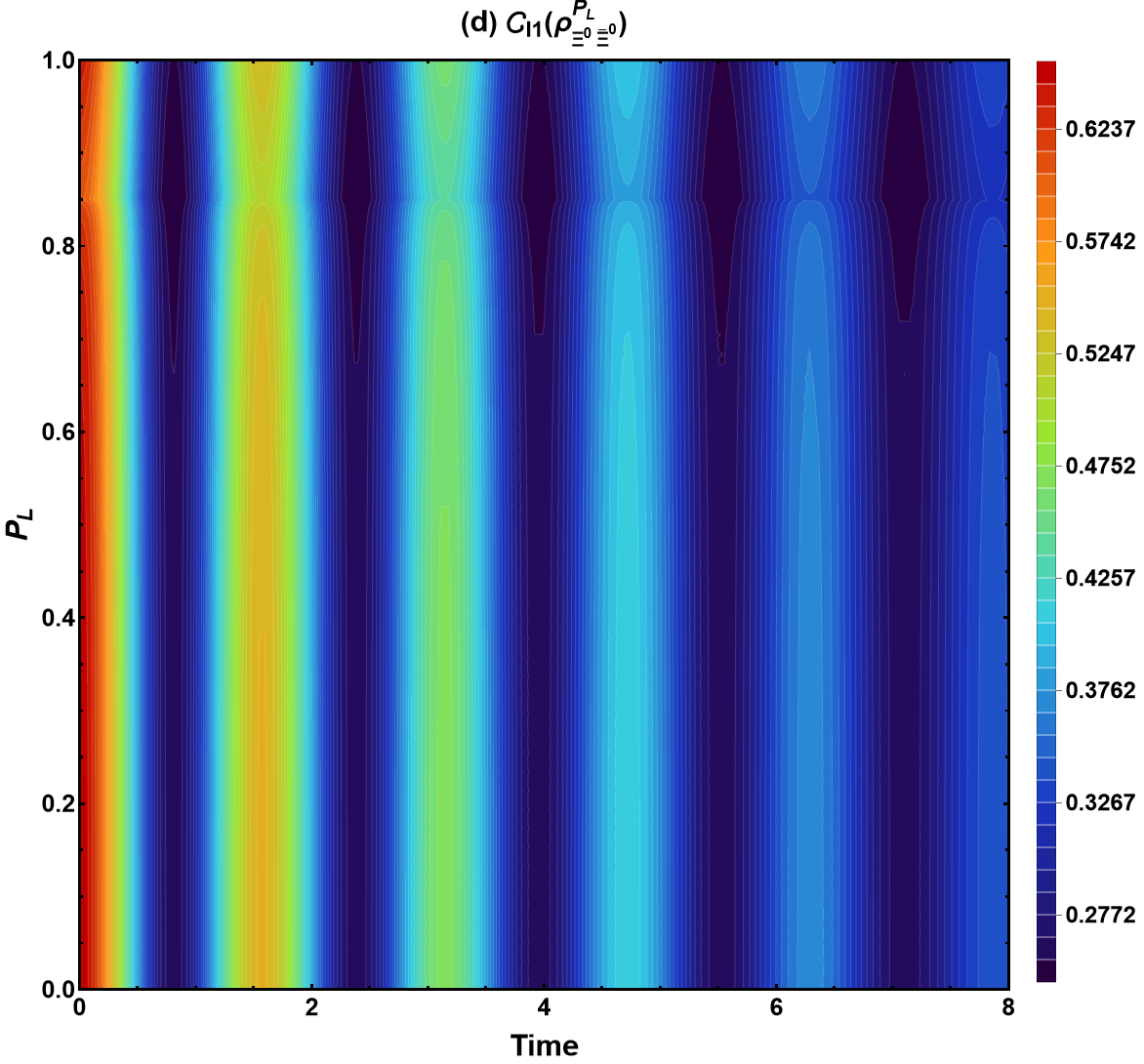}
	\includegraphics[width=0.24\linewidth]{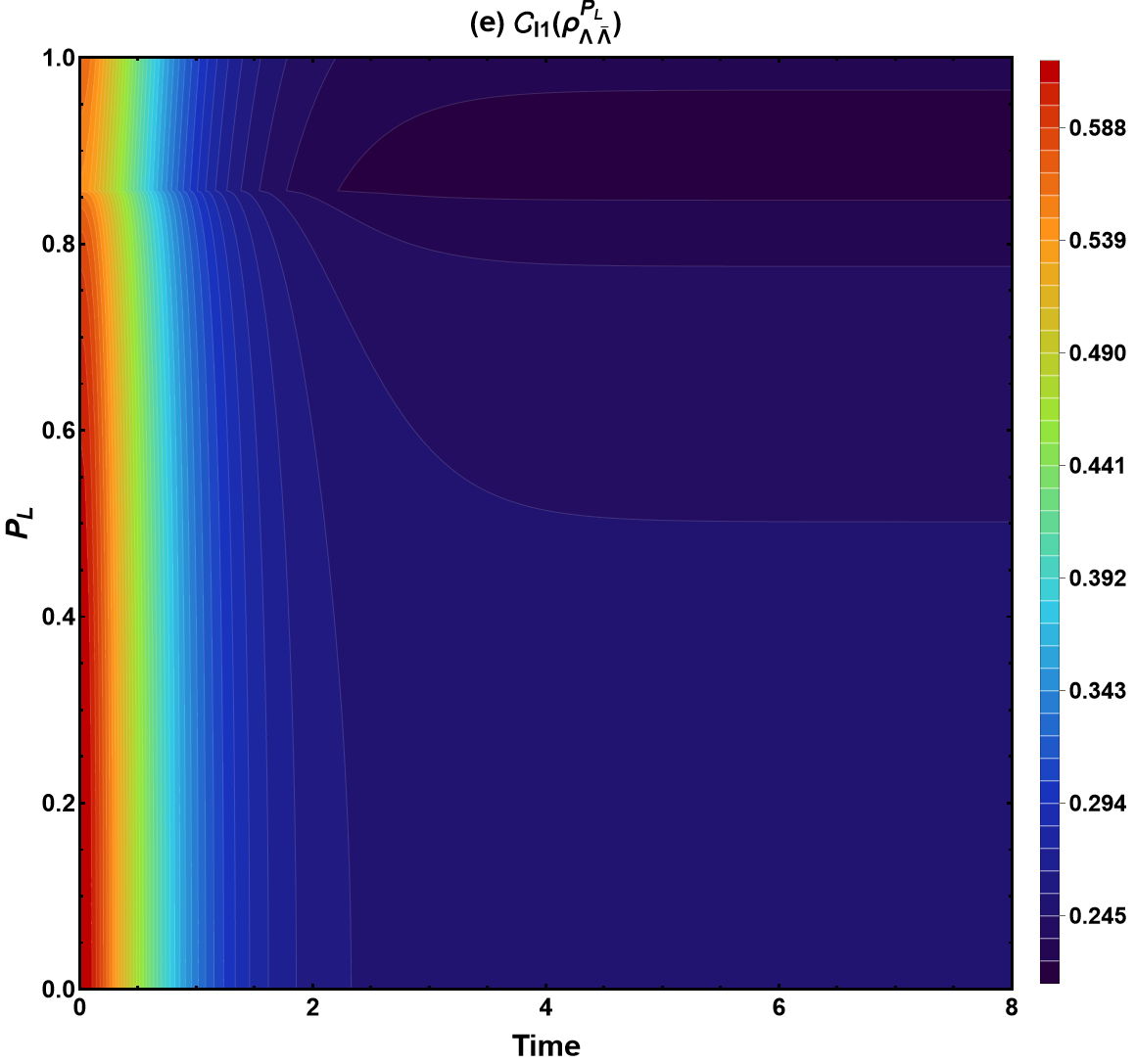}
	\includegraphics[width=0.24\linewidth]{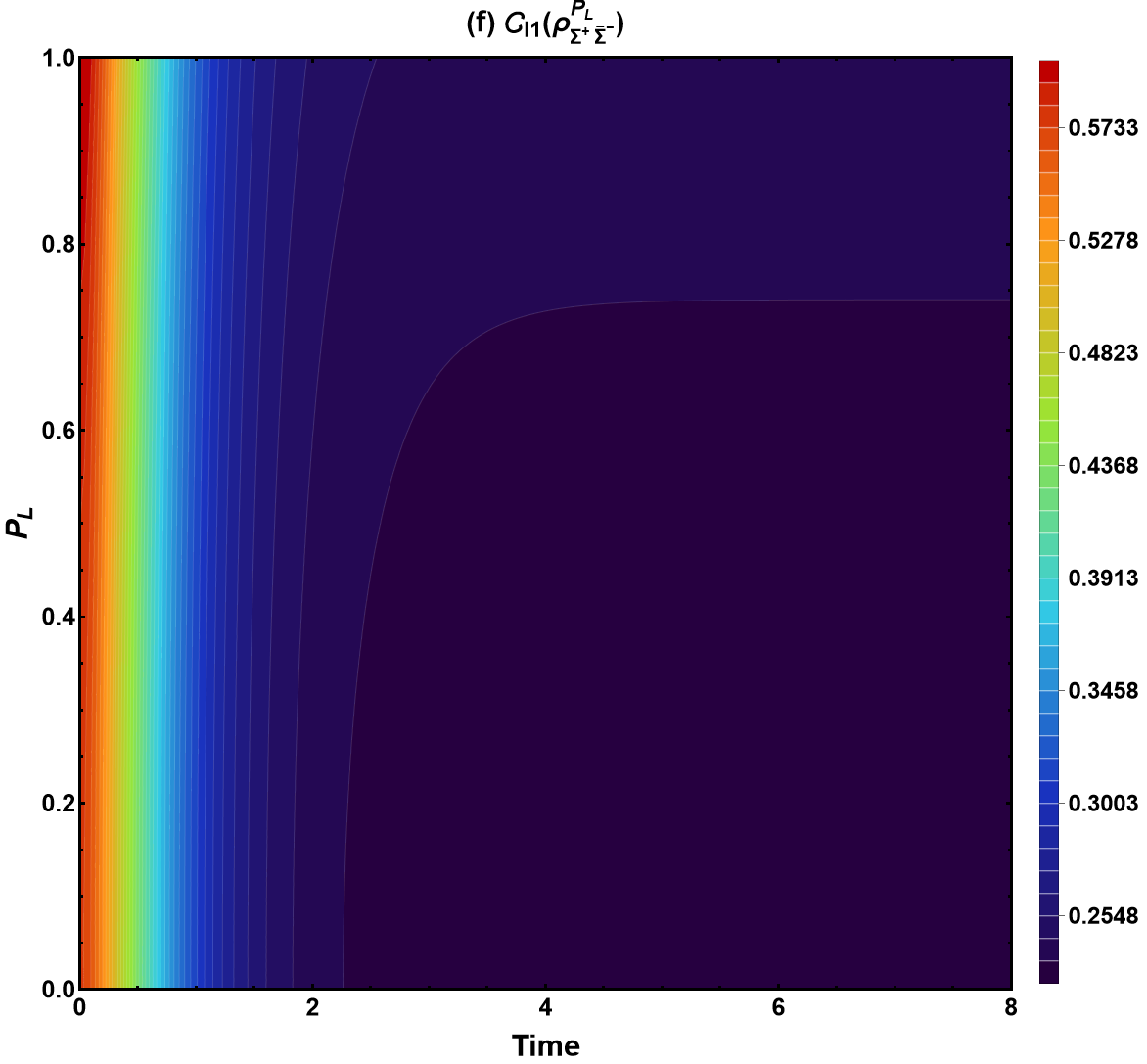}
	\includegraphics[width=0.24\linewidth]{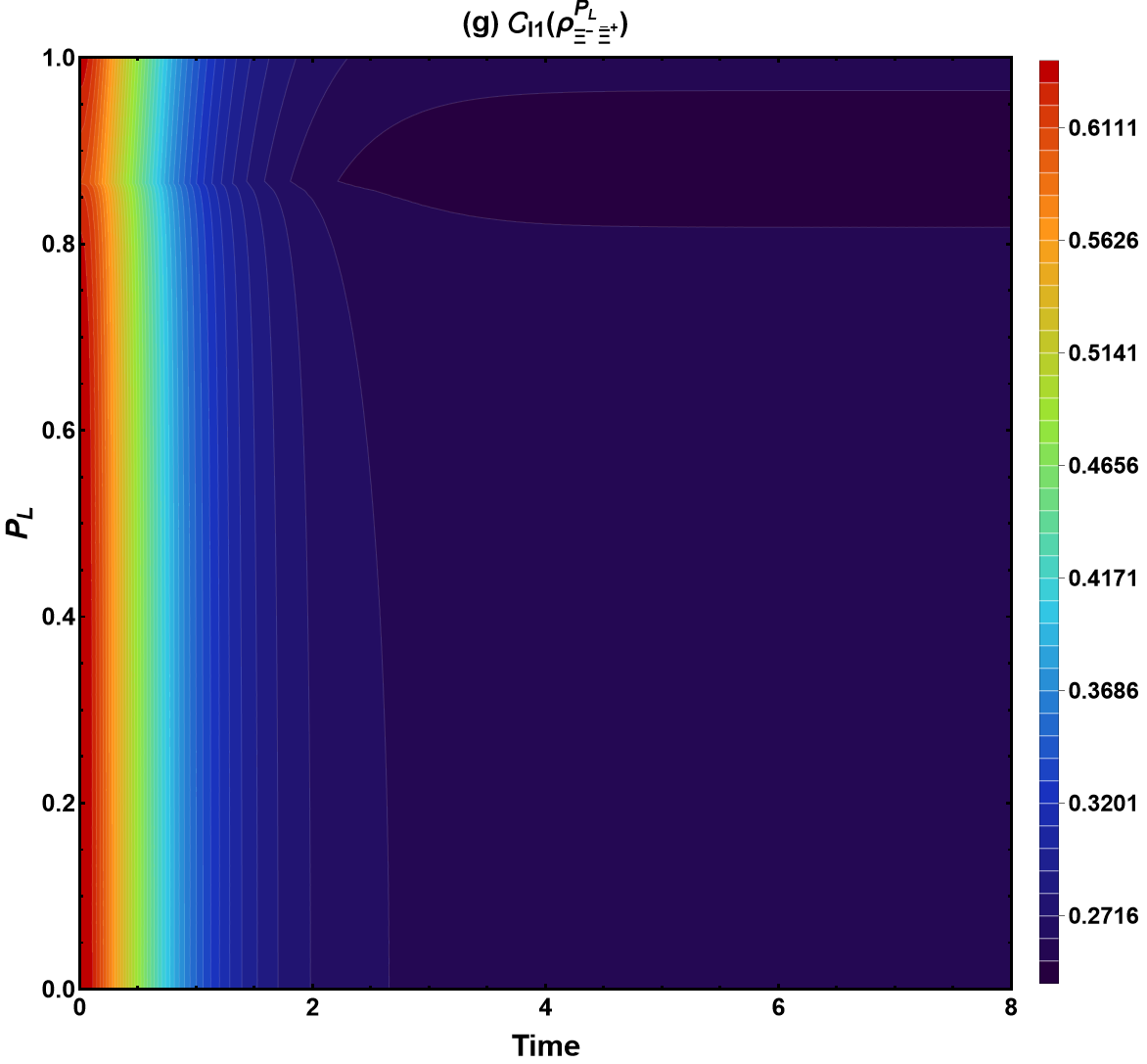}
	\includegraphics[width=0.24\linewidth]{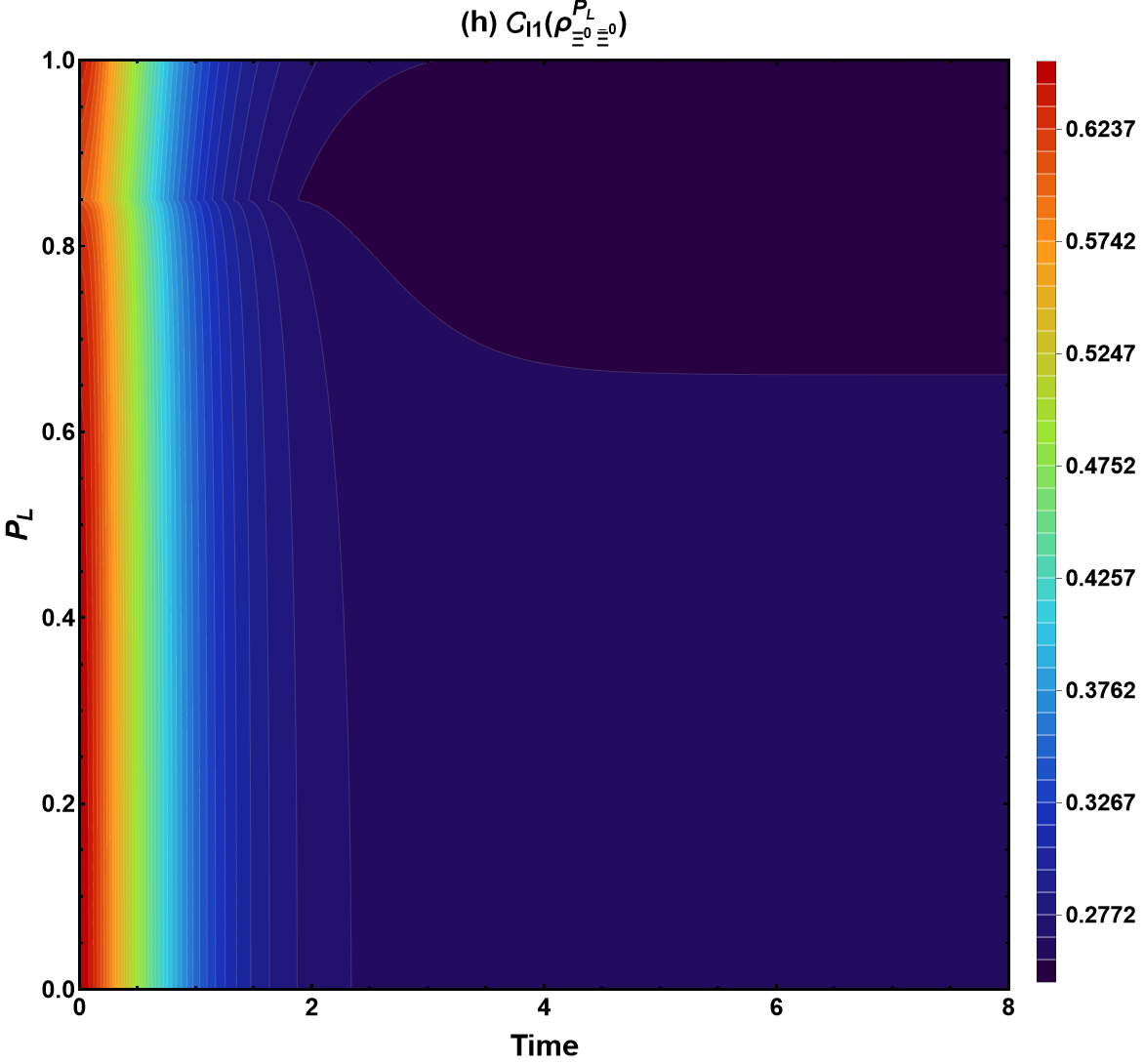}
\caption{
	Dynamical evolution of the $l_1$-norm quantum coherence
	$C_{l_1}(\rho^{P_L}_{Y\bar{Y}})$ as a function of time and the
	longitudinal polarization degree $P_L$ for
	$J/\psi\rightarrow Y\bar{Y}$ with
	$Y=\Lambda$, $\Sigma^{+}$, $\Xi^{-}$, and $\Xi^{0}$ at
	$\cos\theta=0.5$. Panels (a)--(d) [(e)--(h)] correspond to the
	non-Markovian (Markovian) regime  with $\tau=5$ ($\tau=0.2$) and
	$\mu=0.4$. The experimental parameters are taken from
	Table~\ref{tab:BESIII}.
}
	\label{fig13}
\end{figure}
The dynamics of the $l_{1}$-norm quantum coherence
$C_{l_{1}}(\rho^{P_L}_{Y\bar{Y}})$, illustrated in
Fig.~\ref{fig13}, is investigated as a function of the evolution time
$t$ and the longitudinal beam polarization degree $P_L$ for the four
hyperon--antihyperon channels. In the non-Markovian regime
[Figs.~\ref{fig13}(a)--(d)], the coherence exhibits pronounced damped
oscillations accompanied by recurrent revival phenomena. These revivals
originate from the backflow of information from the environment to the
system and constitute a clear signature of memory effects. Although the
oscillation amplitude gradually decreases with time, the coherence
remains finite throughout the evolution. Increasing the longitudinal
polarization enhances the coherence and enlarges the regions where
strong quantum superpositions survive. Among the considered channels,
the $\Xi^{0}\bar{\Xi}^{0}$ and $\Xi^{-}\bar{\Xi}^{+}$ systems display
the largest coherence values, reaching approximately $0.62$ and $0.61$,
respectively, while slightly smaller maxima are observed for the
$\Lambda\bar{\Lambda}$ and $\Sigma^{+}\bar{\Sigma}^{-}$ channels.
In contrast, the Markovian dynamics shown in
Figs.~\ref{fig13}(e)--(h) are characterized by a monotonic decay of the
coherence without any revival structures. The absence of environmental
memory leads to an irreversible loss of quantum information, causing the
coherence to rapidly decrease before approaching a stationary asymptotic
value. Nevertheless, a finite amount of coherence survives even at long
times, demonstrating its remarkable robustness against dephasing noise.
Compared with logarithmic negativity and geometric quantum discord,
$C_{l_{1}}$ remains nonzero over a much broader parameter region and
maintains significantly larger values during the evolution. These
results indicate that quantum coherence constitutes the most resilient
quantum resource in hyperon-antihyperon systems, while the combined
action of strong longitudinal polarization and non-Markovian memory
provides the most favorable conditions for its preservation.
\begin{figure}[H]
	\centering
	\includegraphics[width=0.24\linewidth]{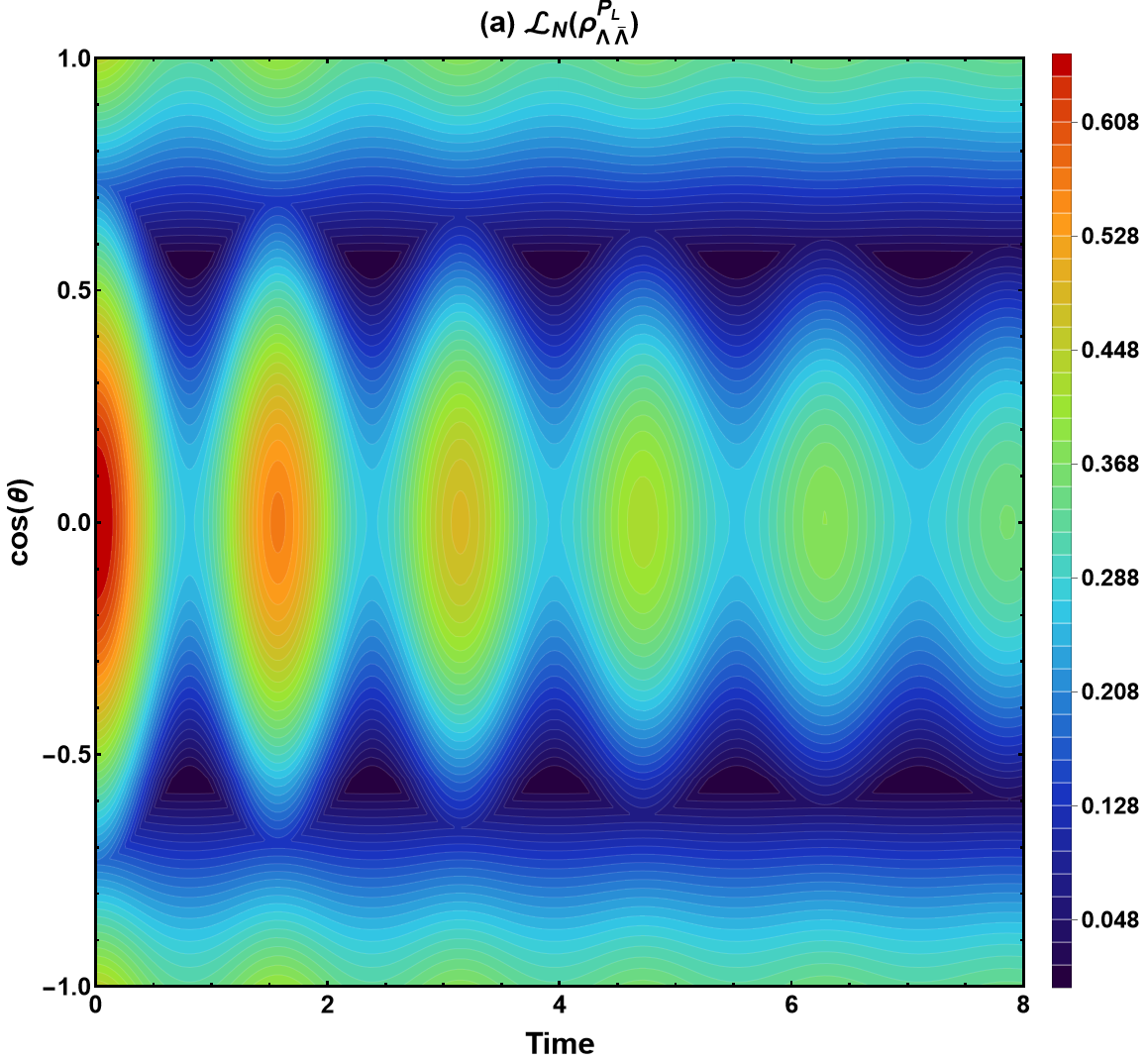}
	\includegraphics[width=0.24\linewidth]{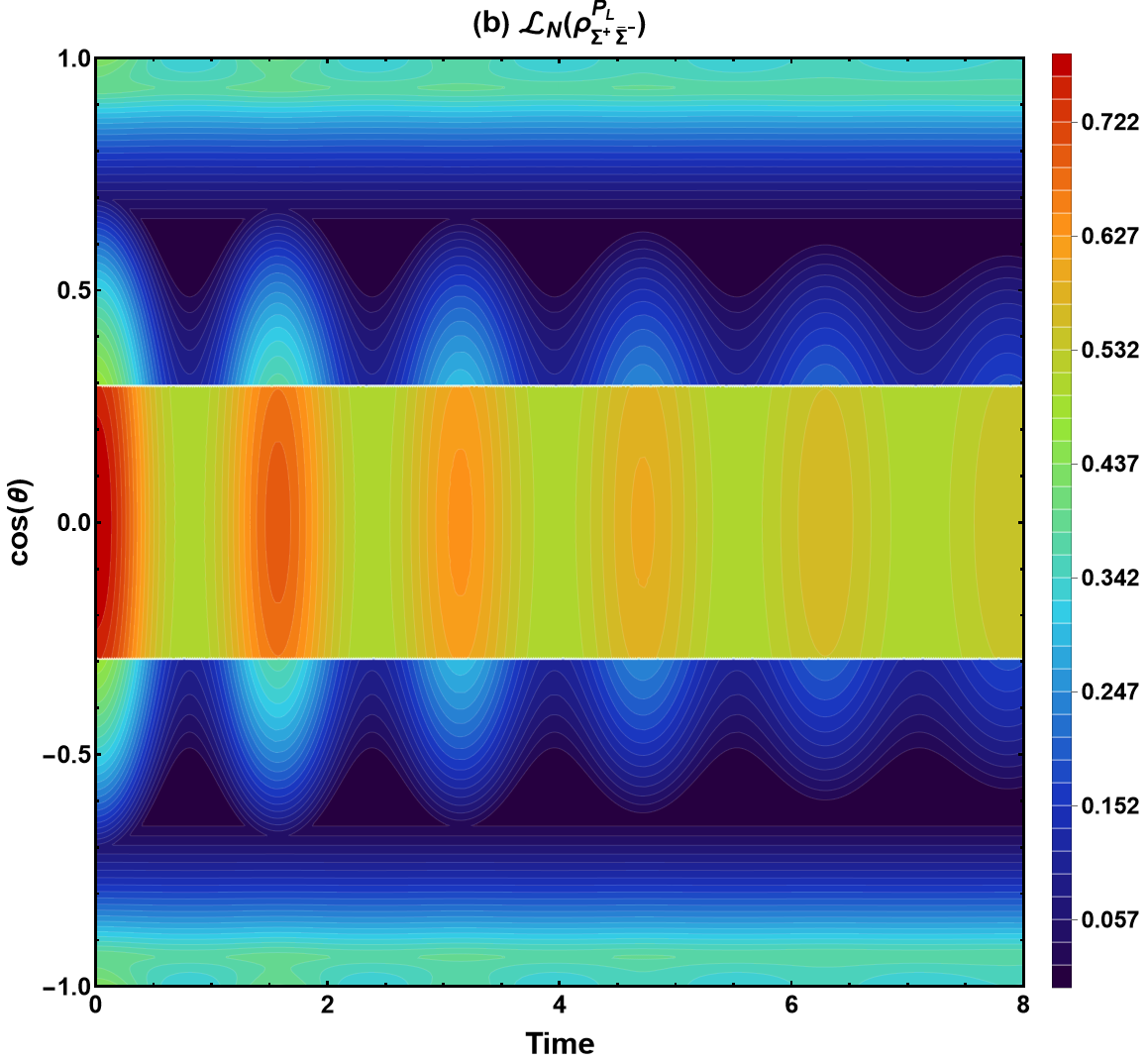}
	\includegraphics[width=0.24\linewidth]{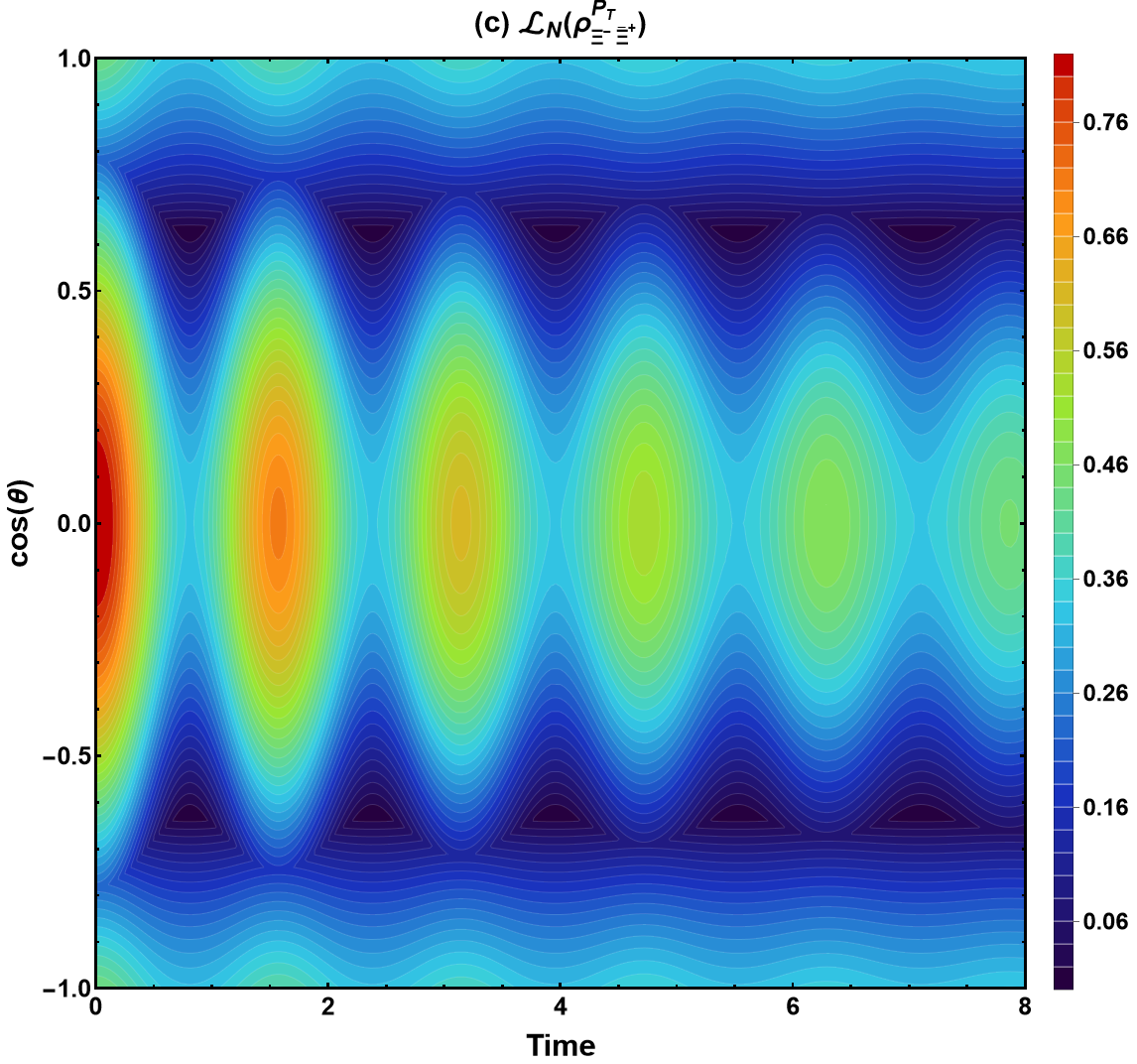}
	\includegraphics[width=0.24\linewidth]{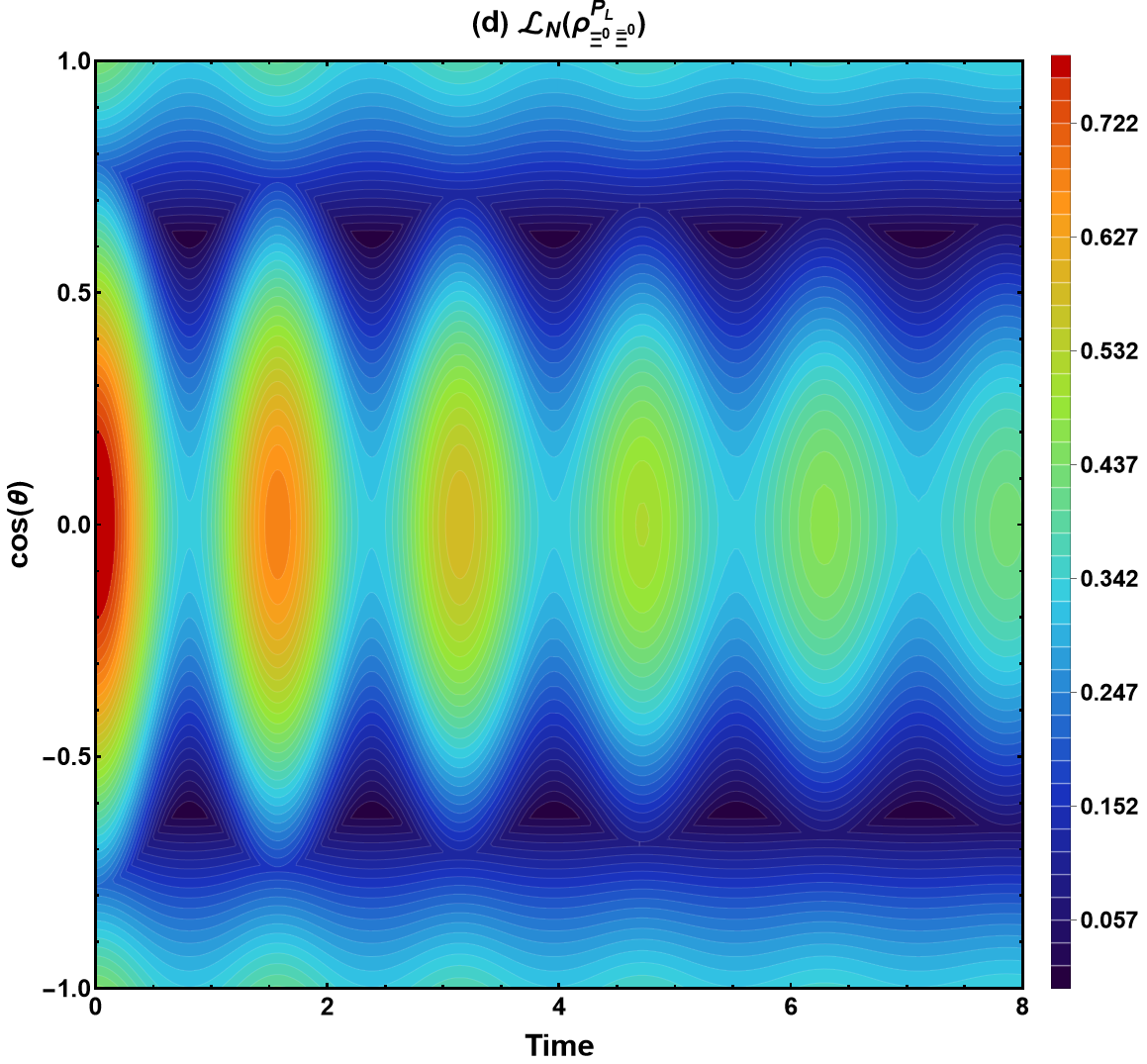}
	\includegraphics[width=0.24\linewidth]{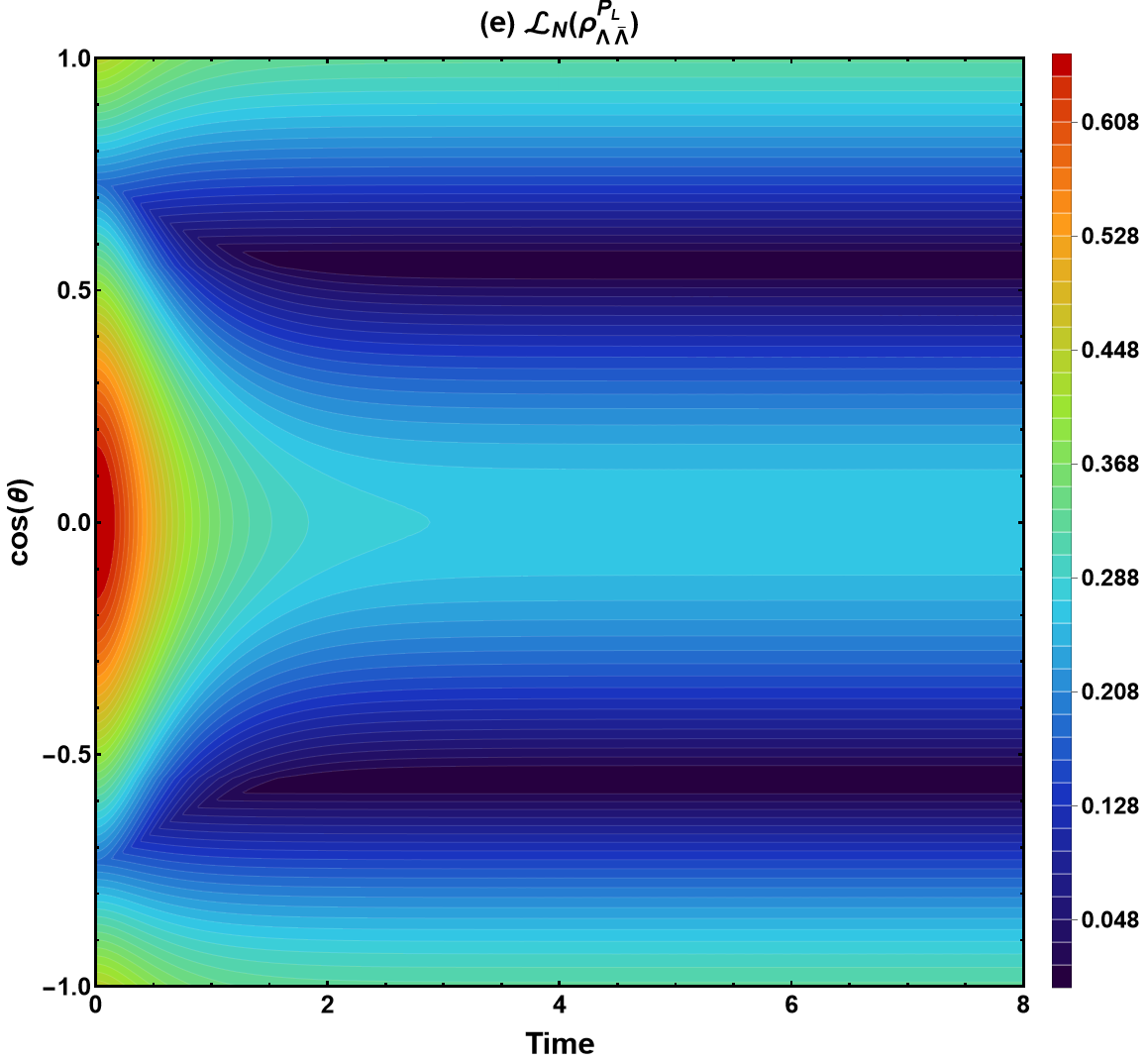}
	\includegraphics[width=0.24\linewidth]{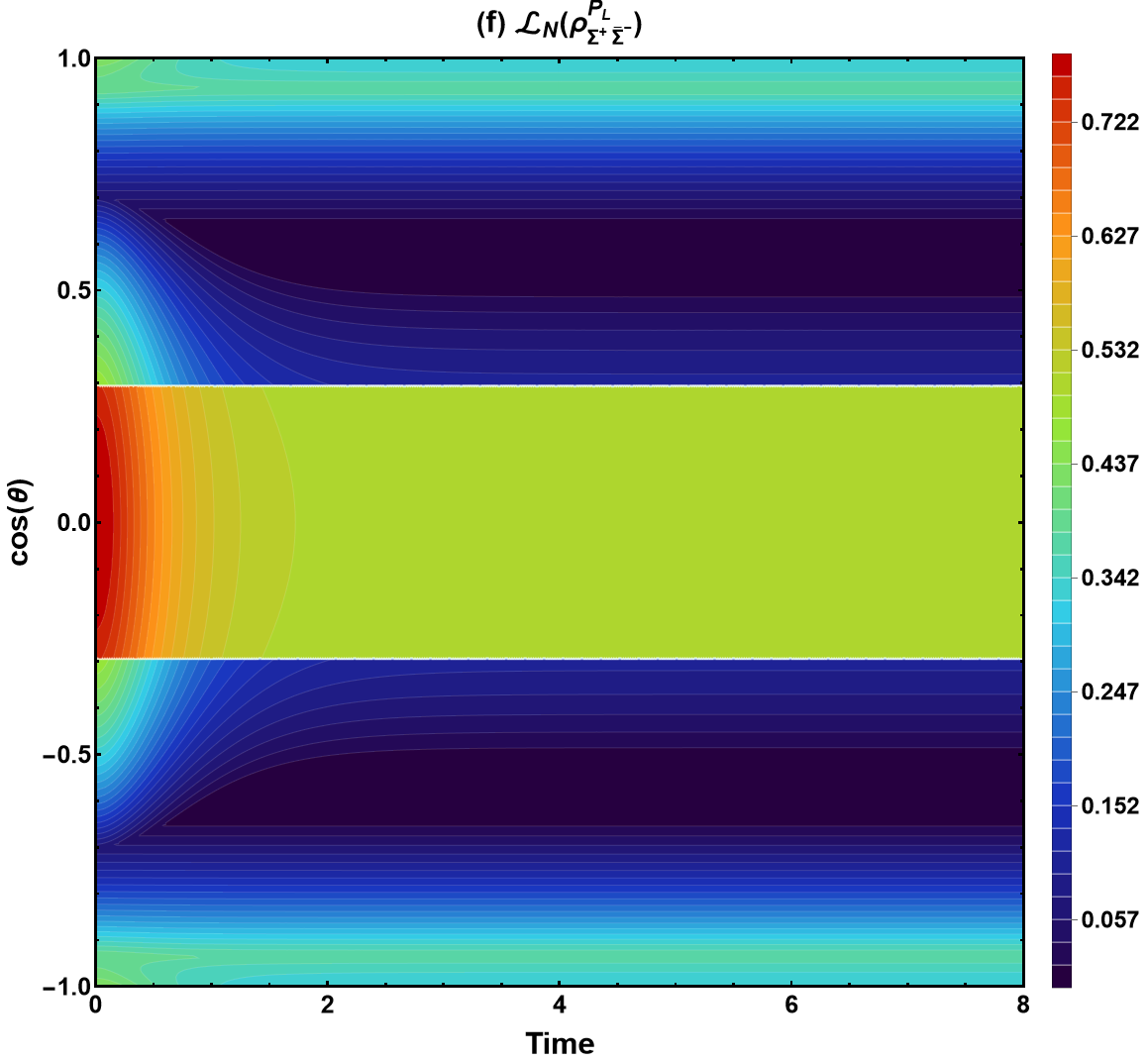}
	\includegraphics[width=0.24\linewidth]{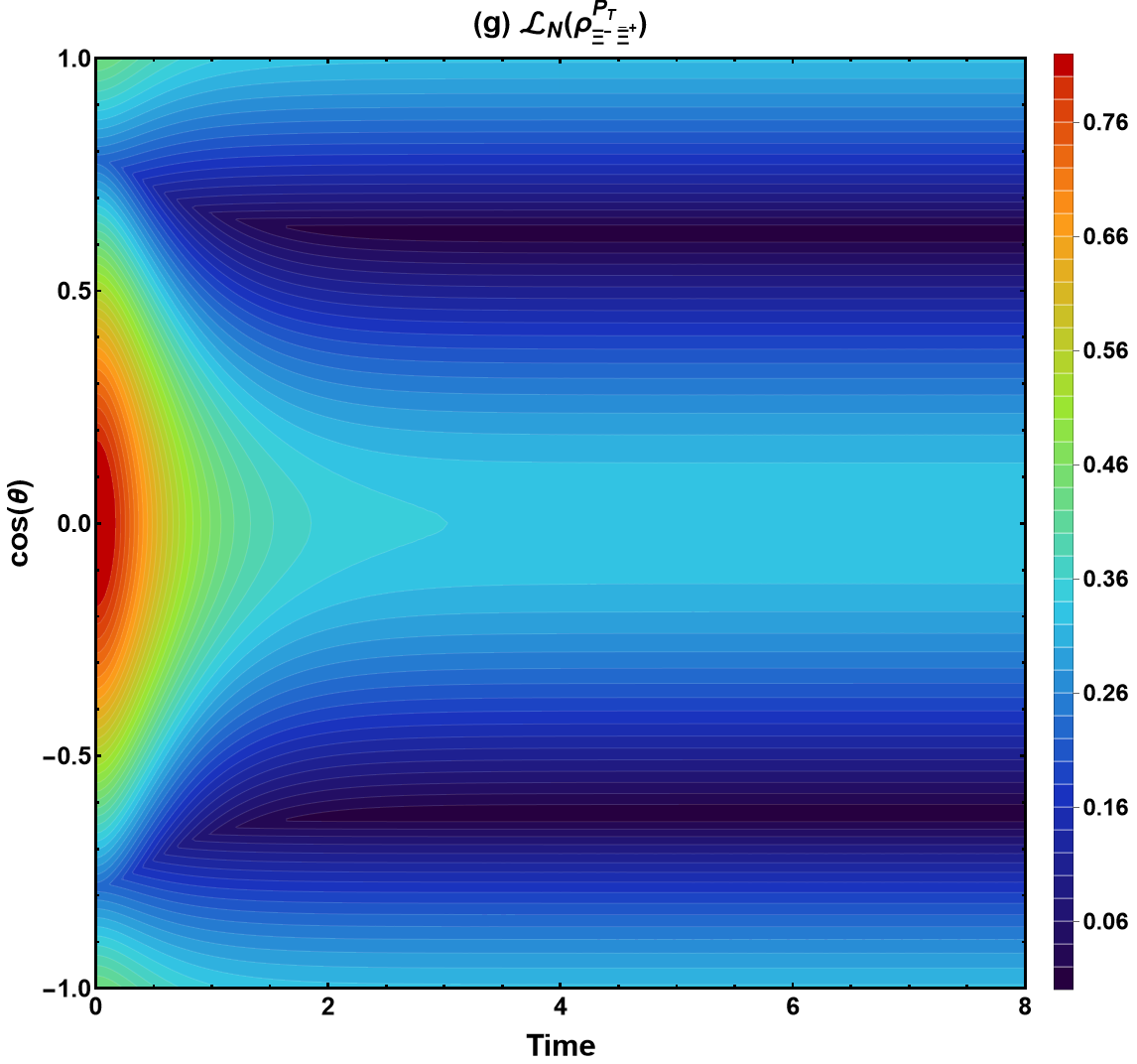}
	\includegraphics[width=0.24\linewidth]{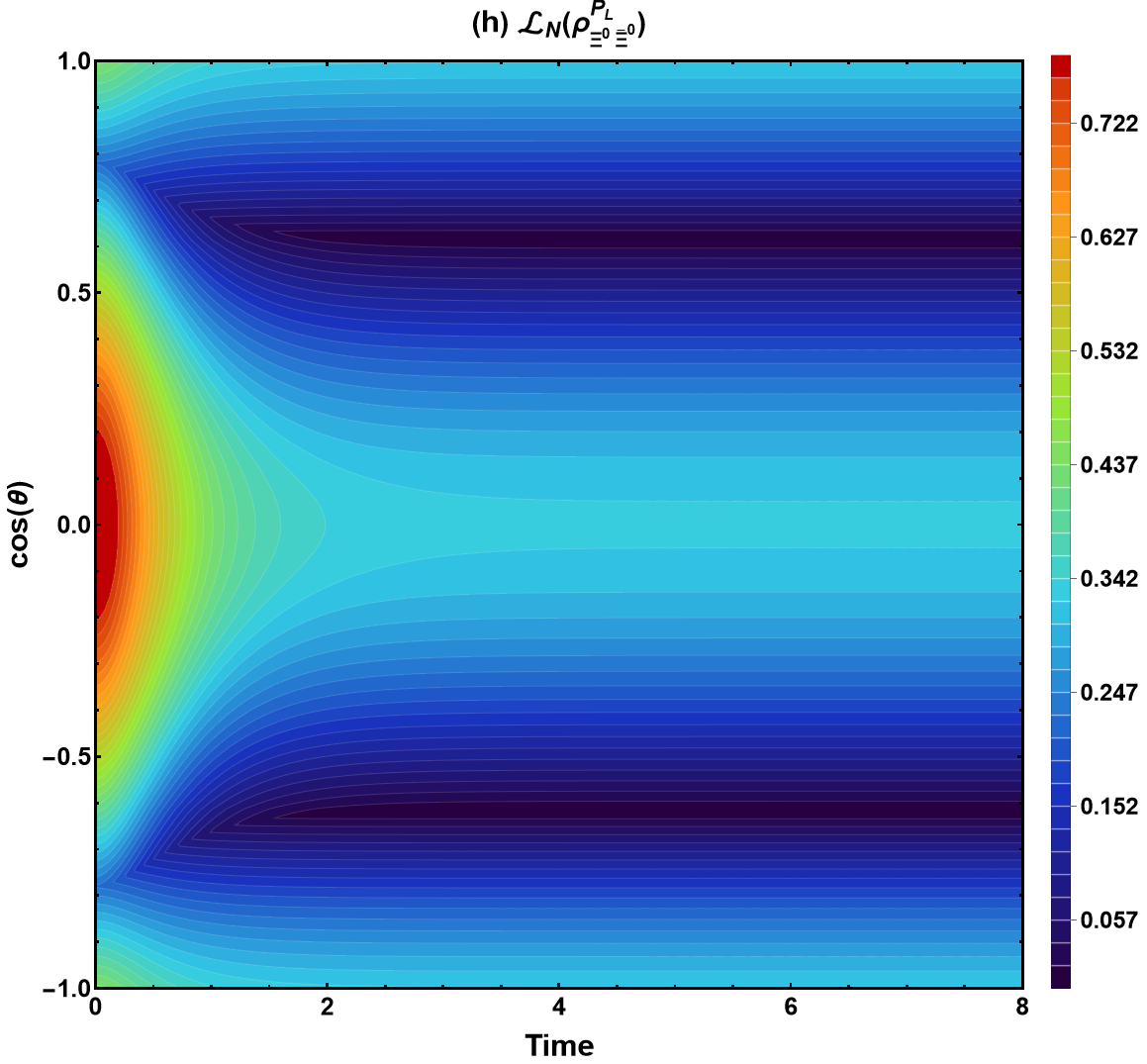}
\caption{
	Dynamical evolution of the logarithmic negativity
	$\mathcal{L}_N(\rho^{P_L}_{Y\bar{Y}})$ as a function of time and the
	production angle $\cos\theta$ for
	$J/\psi\rightarrow Y\bar{Y}$ with
	$Y=\Lambda$, $\Sigma^{+}$, $\Xi^{-}$, and $\Xi^{0}$ at
	$P_L=0.8$. Panels (a)--(d) [(e)--(h)] correspond to the
	non-Markovian (Markovian) regime  with $\tau=5$ ($\tau=0.2$) and
	$\mu=0.4$. The experimental parameters are taken from
	Table~\ref{tab:BESIII}.
}
	\label{fig14}
\end{figure}
The dynamical behavior of the logarithmic negativity
$\mathcal{L}_{N}(\rho^{P_T}_{Y\bar{Y}})$, shown in
Fig.~\ref{fig14}, is examined as a function of the production angle
$\cos\theta$ and the evolution time $t$ for the four
hyperon--antihyperon channels. In the
non-Markovian regime [Figs.~\ref{fig14}(a)--(d)], the entanglement
displays pronounced oscillatory revivals along the time axis, which are
a direct consequence of environmental memory effects. The largest values
are localized around the central angular region
$\cos\theta \simeq 0$, while two pronounced minima appear near
$\cos\theta \approx \pm 0.6$, where the logarithmic negativity becomes
strongly suppressed. Although the revival amplitudes decrease
progressively with time, the oscillatory structure remains visible over
the entire evolution range, indicating a persistent backflow of quantum
information from the environment to the hyperon system.
The four channels exhibit the same angular structure but differ in the
amount of generated entanglement. The $\Xi^{-}\bar{\Xi}^{+}$ channel
shows the strongest quantum correlations, reaching values close to
$\mathcal{L}_{N}\simeq0.76$, followed by the
$\Sigma^{0}\bar{\Sigma}^{0}$ and $\Xi^{0}\bar{\Xi}^{0}$ channels,
whereas the $\Lambda\bar{\Lambda}$ channel presents comparatively lower
values. These differences reflect the distinct spin-correlation
coefficients characterizing each production process.
A markedly different behavior emerges in the Markovian regime
[Figs.~\ref{fig14}(e)--(h)]. The oscillatory revivals disappear and the
entanglement rapidly approaches a stationary angular distribution. The
suppression regions around $\cos\theta\approx\pm0.6$ become time
independent, while the central region around $\cos\theta=0$ retains the
largest amount of entanglement. The absence of temporal revivals
indicates an irreversible loss of quantum information into the
environment. Comparing both dynamical regimes clearly demonstrates that
non-Markovian memory effects substantially enhance the preservation of
entanglement, whereas Markovian evolution drives the system toward a
stable asymptotic state with reduced quantum correlations.

\begin{figure}[H]
	\centering
	\includegraphics[width=0.24\linewidth]{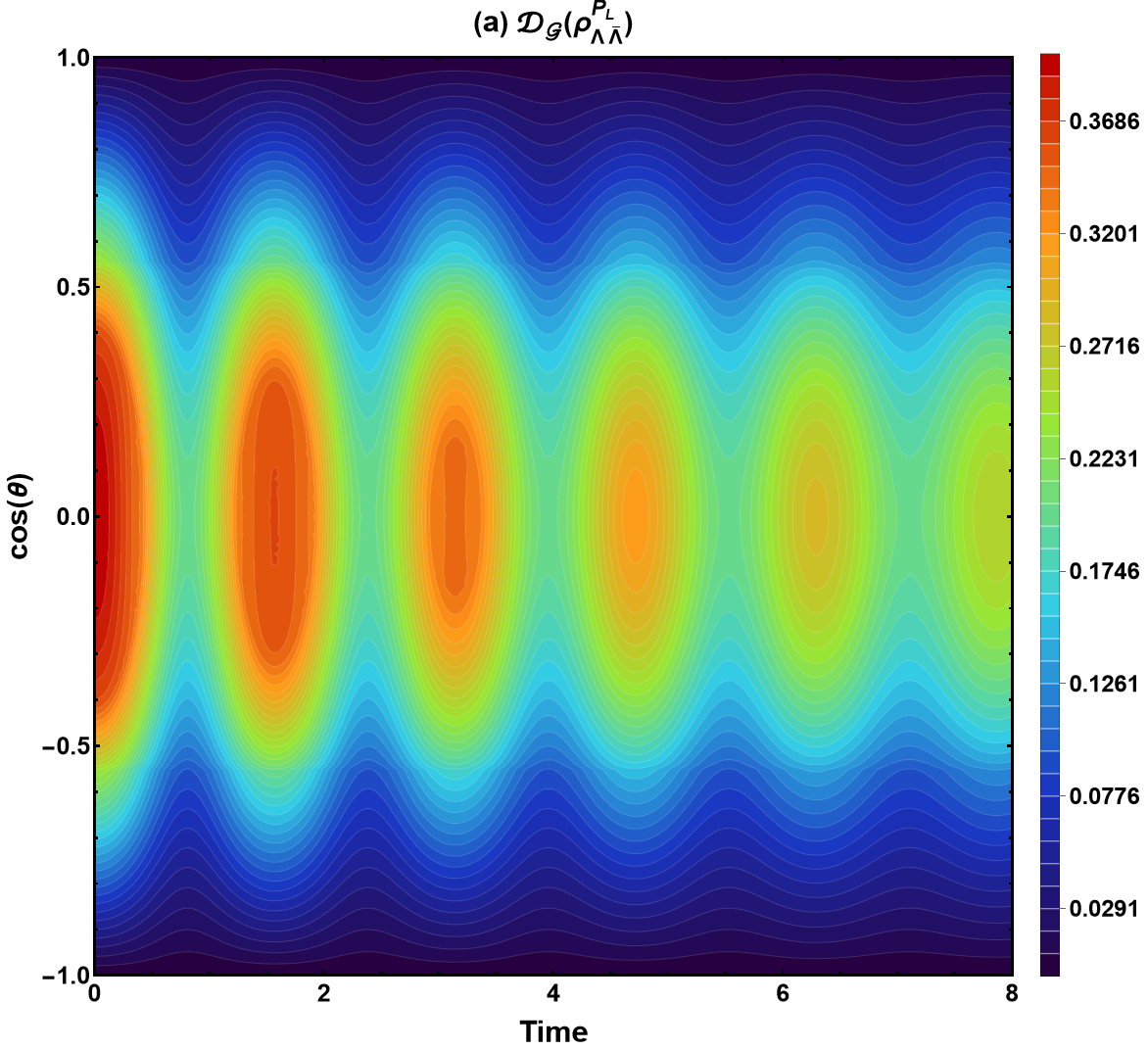}
	\includegraphics[width=0.24\linewidth]{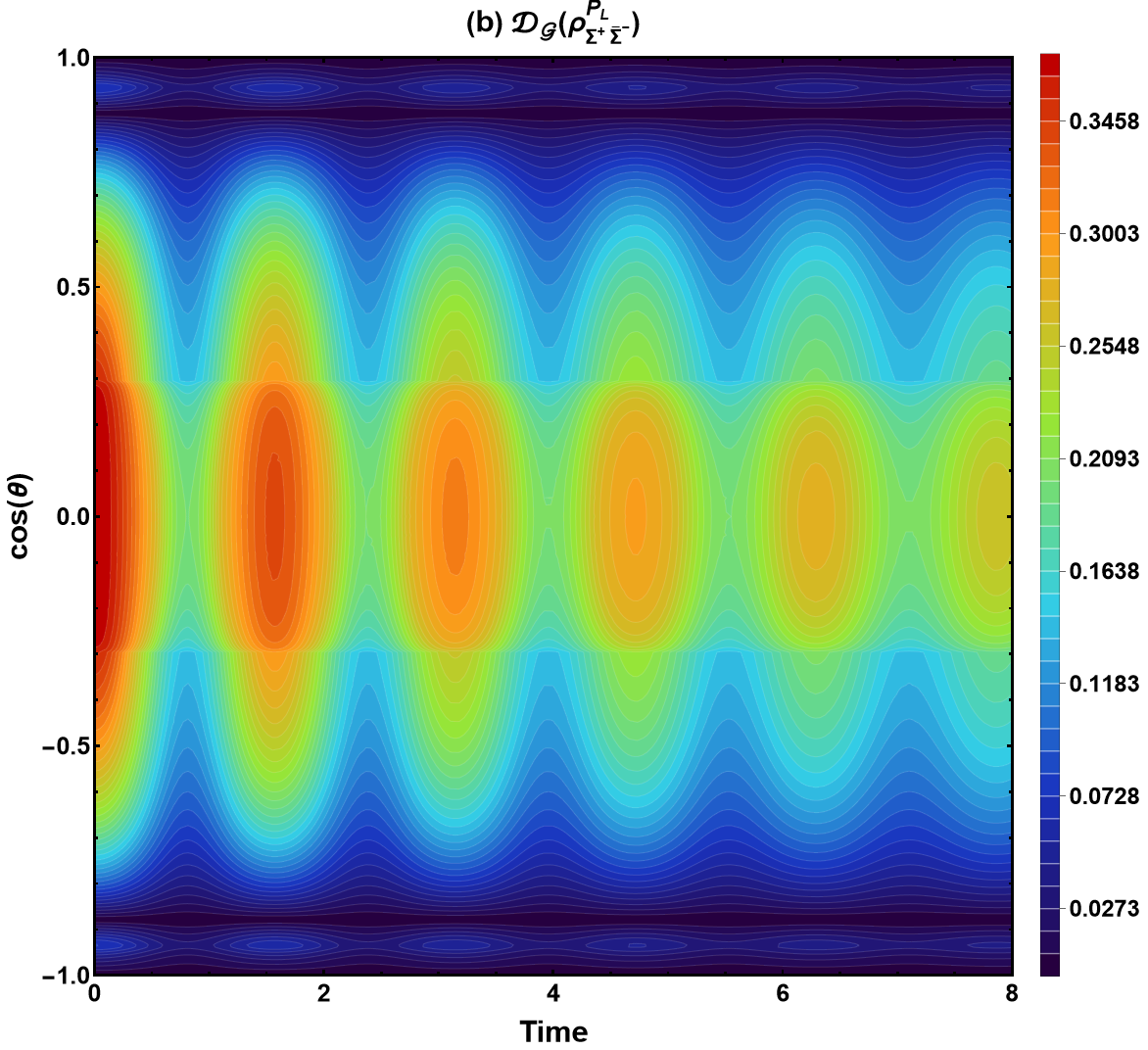}
	\includegraphics[width=0.24\linewidth]{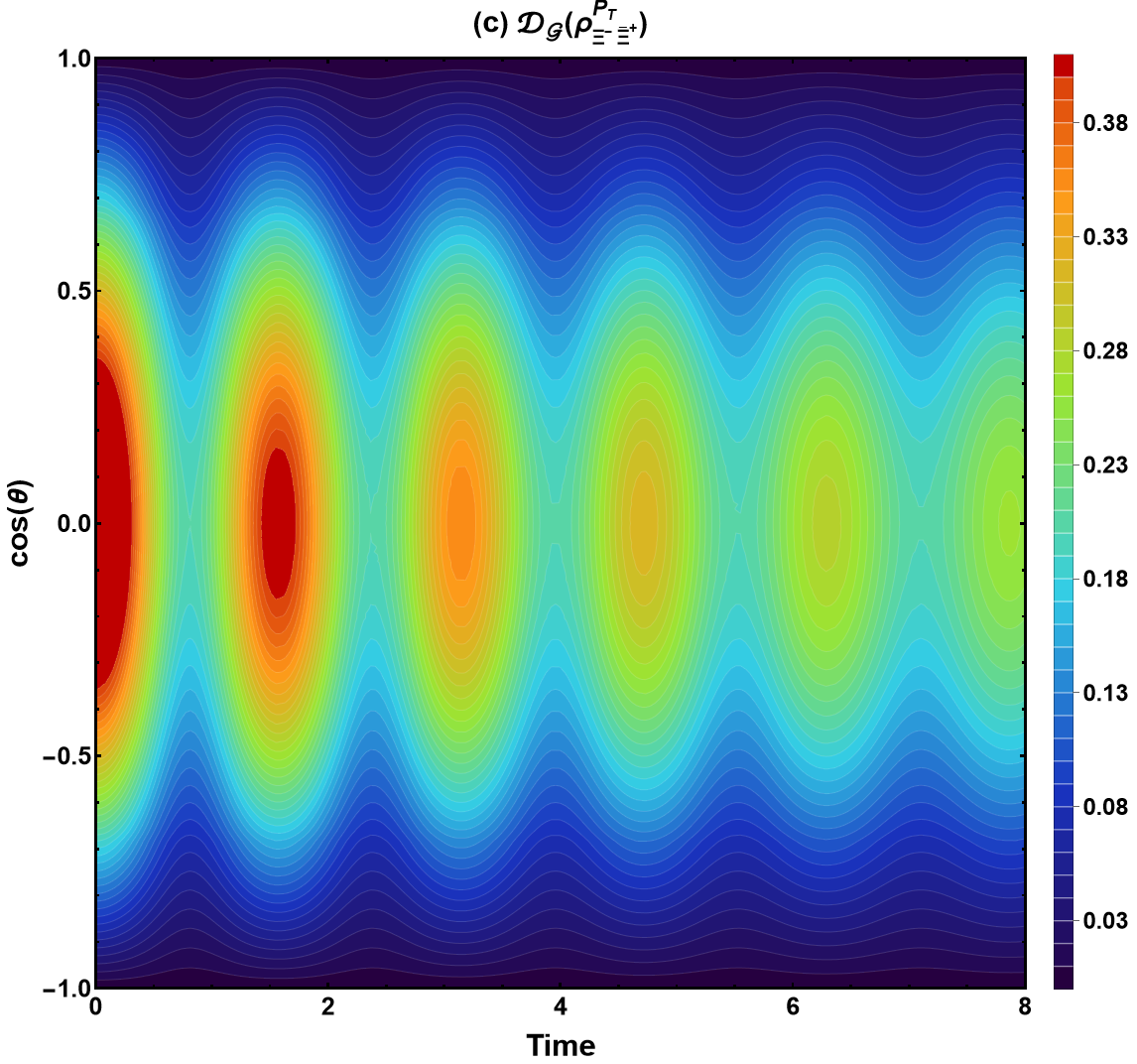}
	\includegraphics[width=0.24\linewidth]{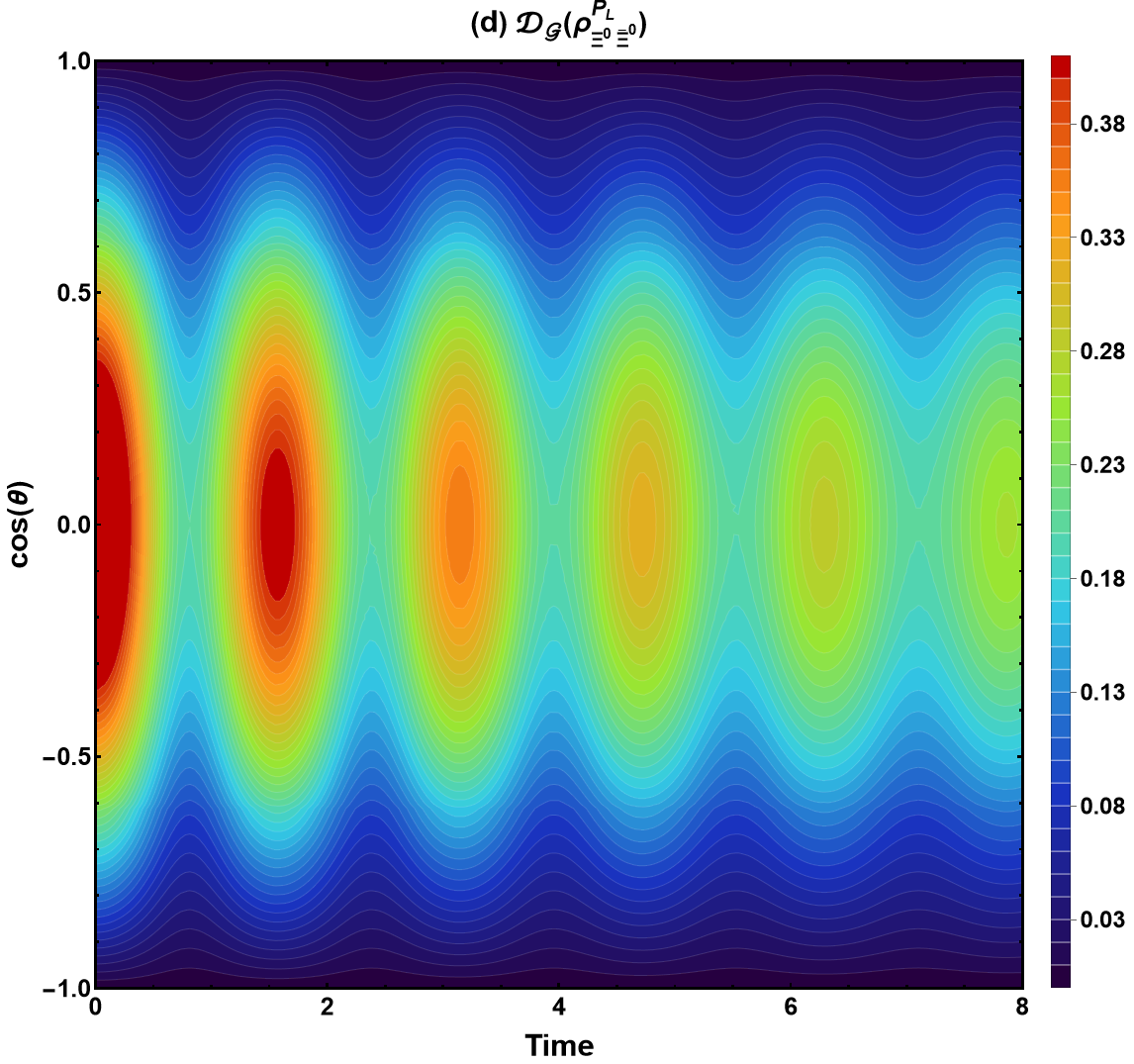}
	\includegraphics[width=0.24\linewidth]{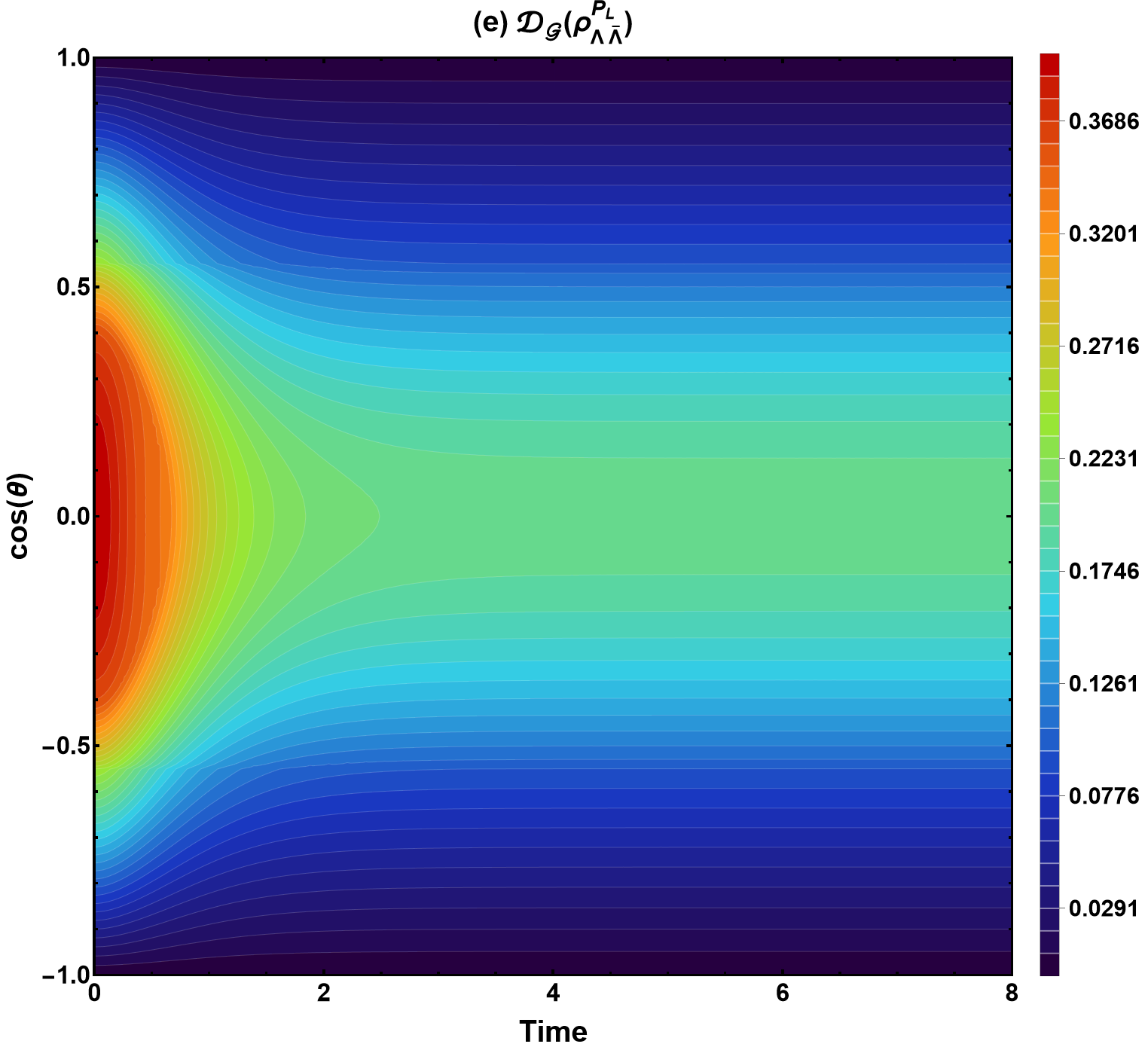}
	\includegraphics[width=0.24\linewidth]{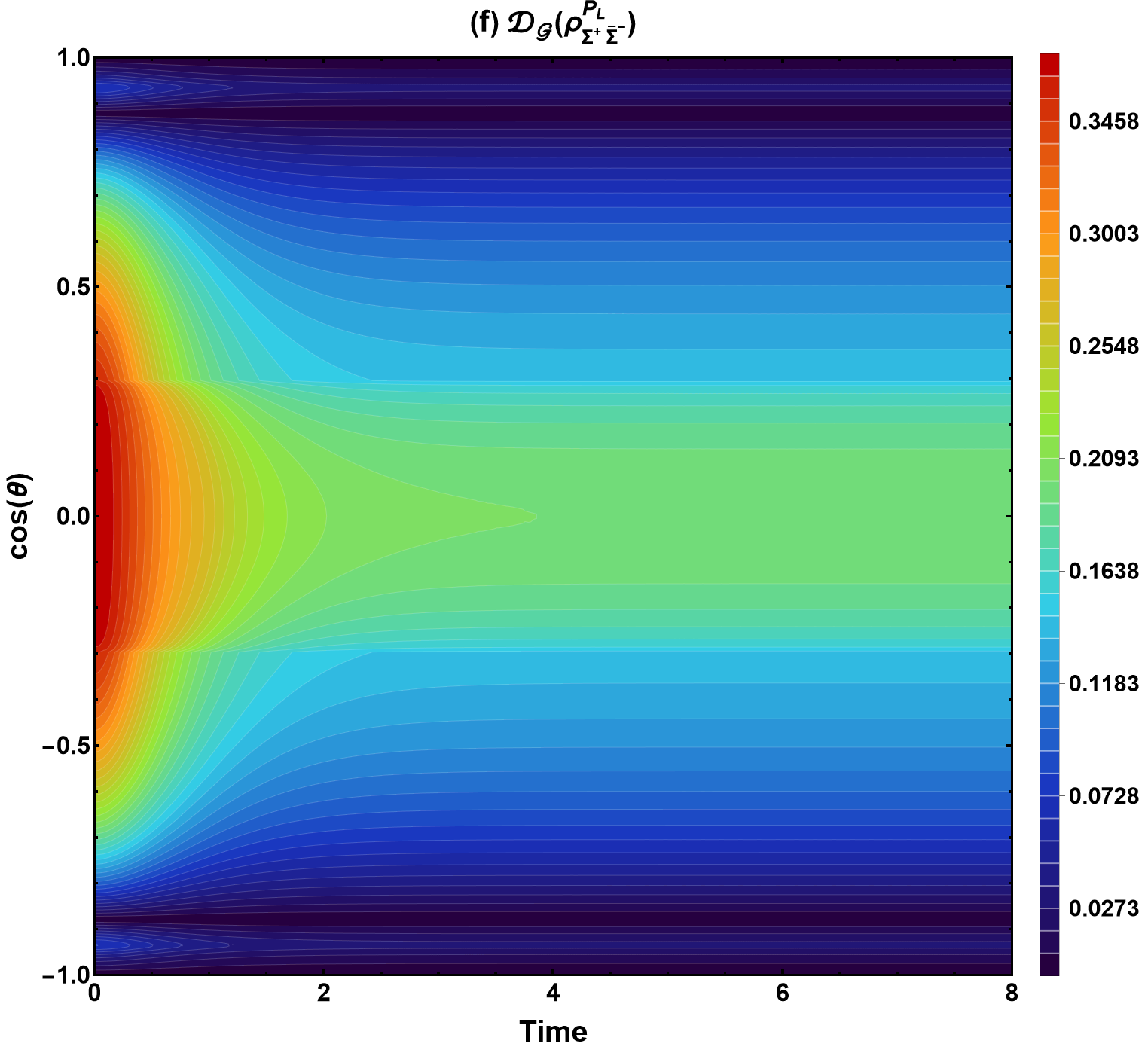}
	\includegraphics[width=0.24\linewidth]{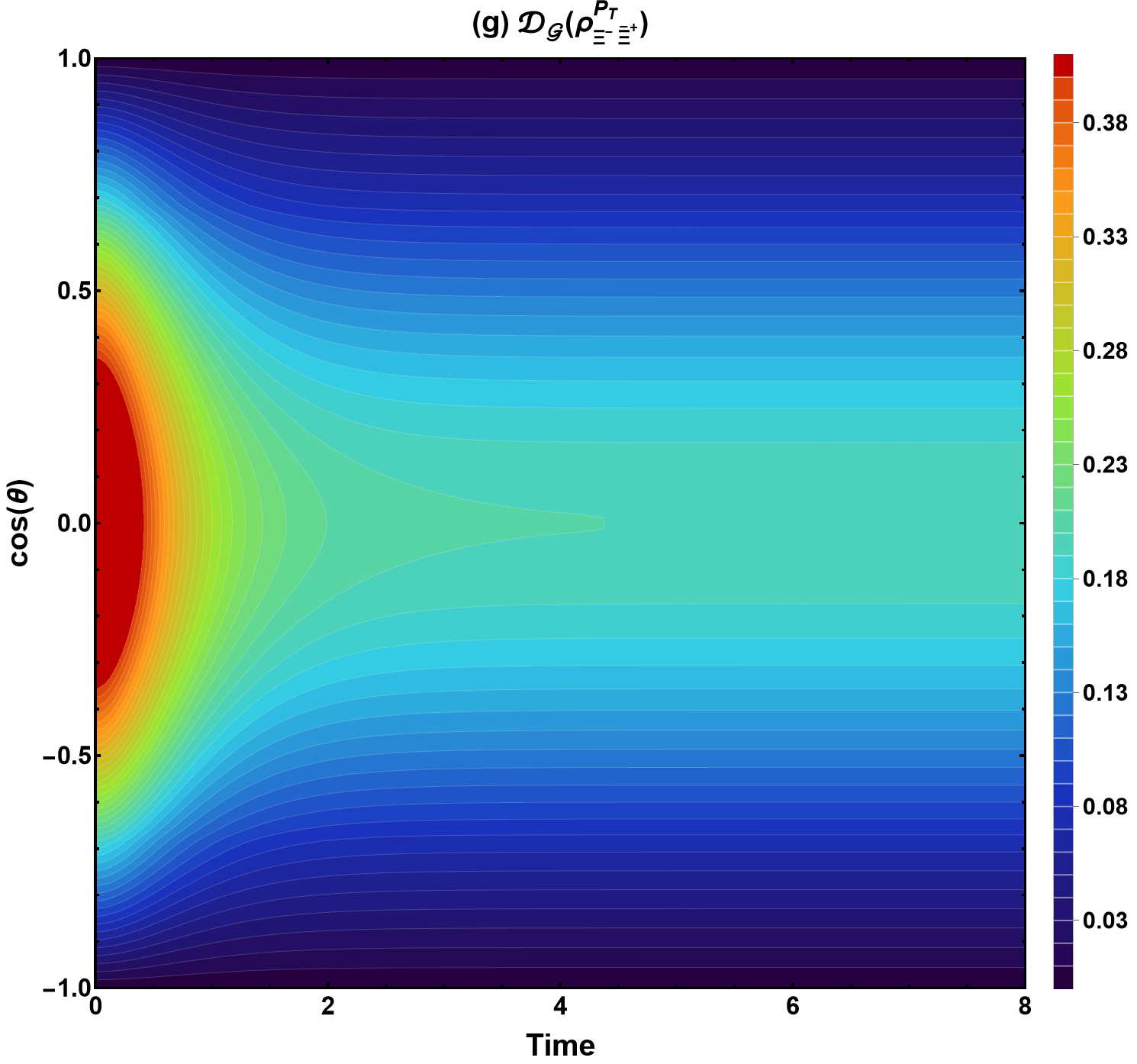}
	\includegraphics[width=0.24\linewidth]{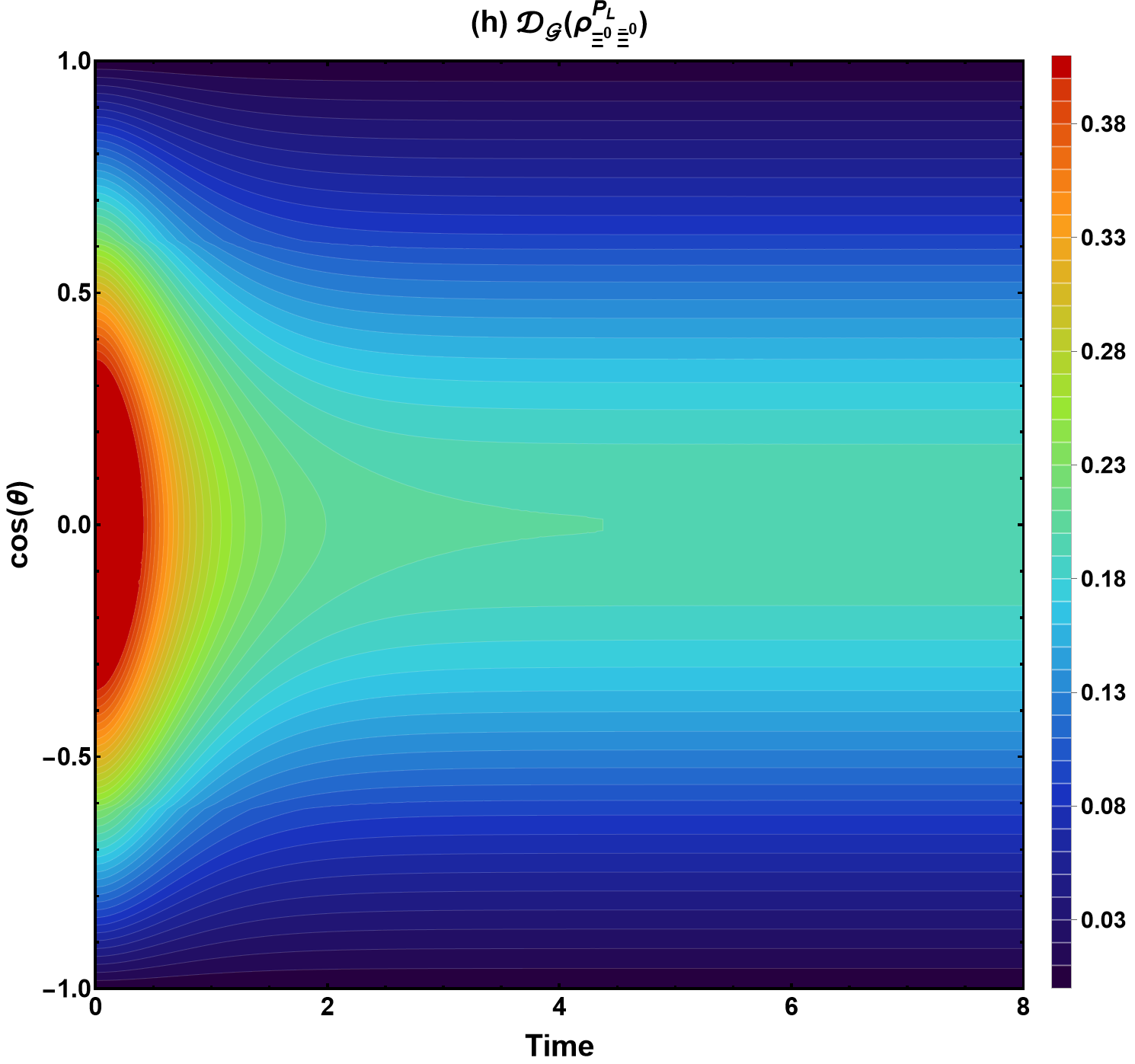}
\caption{
	Dynamical evolution of the geometric quantum discord
	$\mathcal{D}_G(\rho^{P_L}_{Y\bar{Y}})$ as a function of time and the
	production angle $\cos\theta$ for
	$J/\psi\rightarrow Y\bar{Y}$ with
	$Y=\Lambda$, $\Sigma^{+}$, $\Xi^{-}$, and $\Xi^{0}$ at
	$P_L=0.8$. Panels (a)--(d) [(e)--(h)] correspond to the
	non-Markovian (Markovian) regime  with $\tau=5$ ($\tau=0.2$) and
	$\mu=0.4$. The experimental parameters are taken from
	Table~\ref{tab:BESIII}.
}
	\label{fig15}
\end{figure}

Figure~\ref{fig15} displays the evolution of the geometric quantum
discord $\mathcal{D}_G(\rho^{P}_{Y\bar{Y}})$ as a function of the production angle
$\cos\theta$ and time for the four hyperon-antihyperon channels. In the
non-Markovian regime [Figs.~\ref{fig15}(a)--(d)], the discord exhibits
pronounced oscillatory patterns together with successive revival
structures, revealing the recurrent exchange of information between the
hyperon system and its environment. The maxima are systematically
localized around the central angular region $\cos\theta\simeq0$, where
the quantum correlations are strongest. As the evolution proceeds, the
oscillation amplitudes decrease gradually due to dissipative effects,
yet the revivals remain clearly visible throughout the considered time
interval.
The four channels possess similar dynamical characteristics but differ
quantitatively in the magnitude of the generated discord. The
$\Xi^{-}\bar{\Xi}^{+}$ and $\Xi^{0}\bar{\Xi}^{0}$ channels attain the
largest values, with $\mathcal{D}_G$ approaching $0.39$, while the
$\Lambda\bar{\Lambda}$ and $\Sigma^{+}\bar{\Sigma}^{-}$ channels exhibit
slightly smaller correlations. For the
$\Sigma^{+}\bar{\Sigma}^{-}$ channel, additional suppression regions are
observed near $|\cos\theta|\simeq0.9$, leading to a more structured
angular dependence than in the other channels.

A markedly different behavior appears in the Markovian regime
[Figs.~\ref{fig15}(e)--(h)]. The revival patterns disappear completely
and the discord evolves monotonically toward a stationary distribution.
After a short transient period, the contours become nearly time
independent, indicating the absence of information backflow and the
irreversible leakage of quantum coherence into the environment. Despite
this damping, finite values of geometric discord survive over the entire
angular domain. The largest correlations remain concentrated around
$\cos\theta=0$, whereas the discord decreases progressively toward the
forward and backward scattering regions $|\cos\theta|\rightarrow1$.
Comparing Figs.~\ref{fig15}(a)--(d) with
Figs.~\ref{fig15}(e)--(h) demonstrates that non-Markovian memory effects
significantly enhance the persistence of quantum correlations by
generating recurrent revivals, while Markovian dynamics drive the system
toward a stable asymptotic configuration. Moreover, unlike the
logarithmic negativity discussed previously, the geometric quantum
discord remains finite throughout the evolution in both dynamical
regimes, confirming its greater robustness against environmental
decoherence and highlighting the presence of nonclassical correlations
even in parameter regions where entanglement becomes strongly
suppressed.
\begin{figure}[H]
	\centering
	\includegraphics[width=0.24\linewidth]{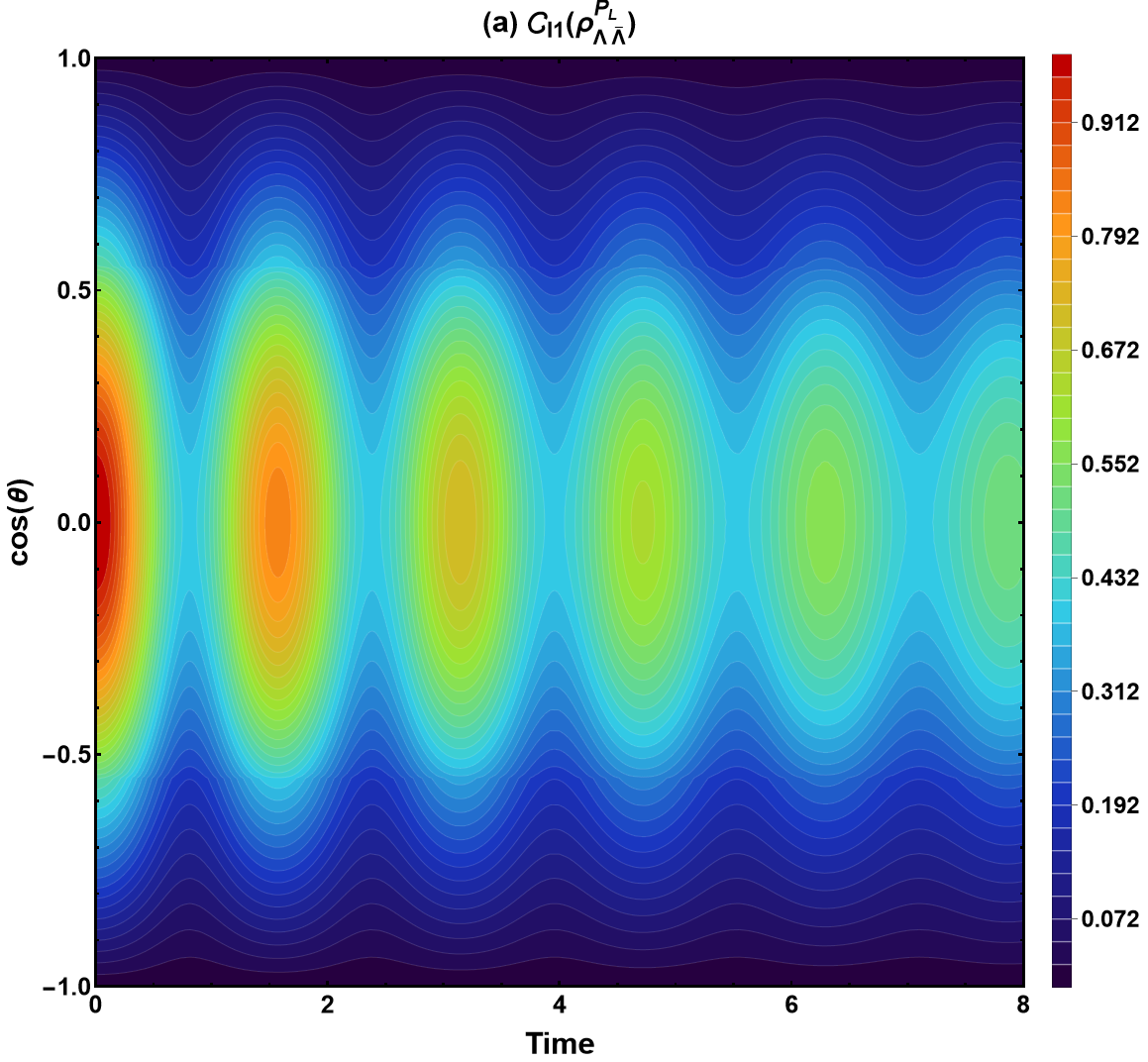}
	\includegraphics[width=0.24\linewidth]{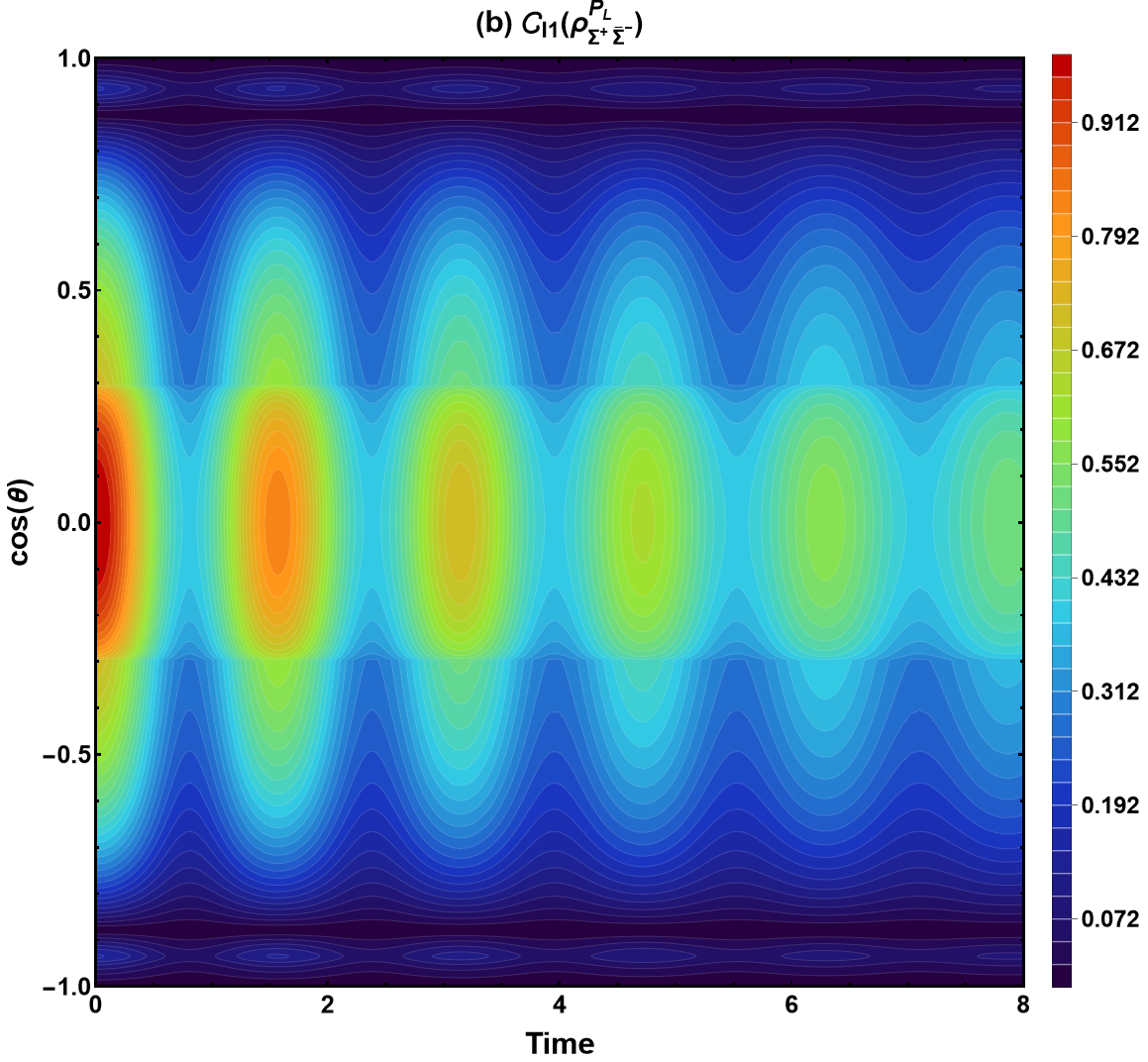}
	\includegraphics[width=0.24\linewidth]{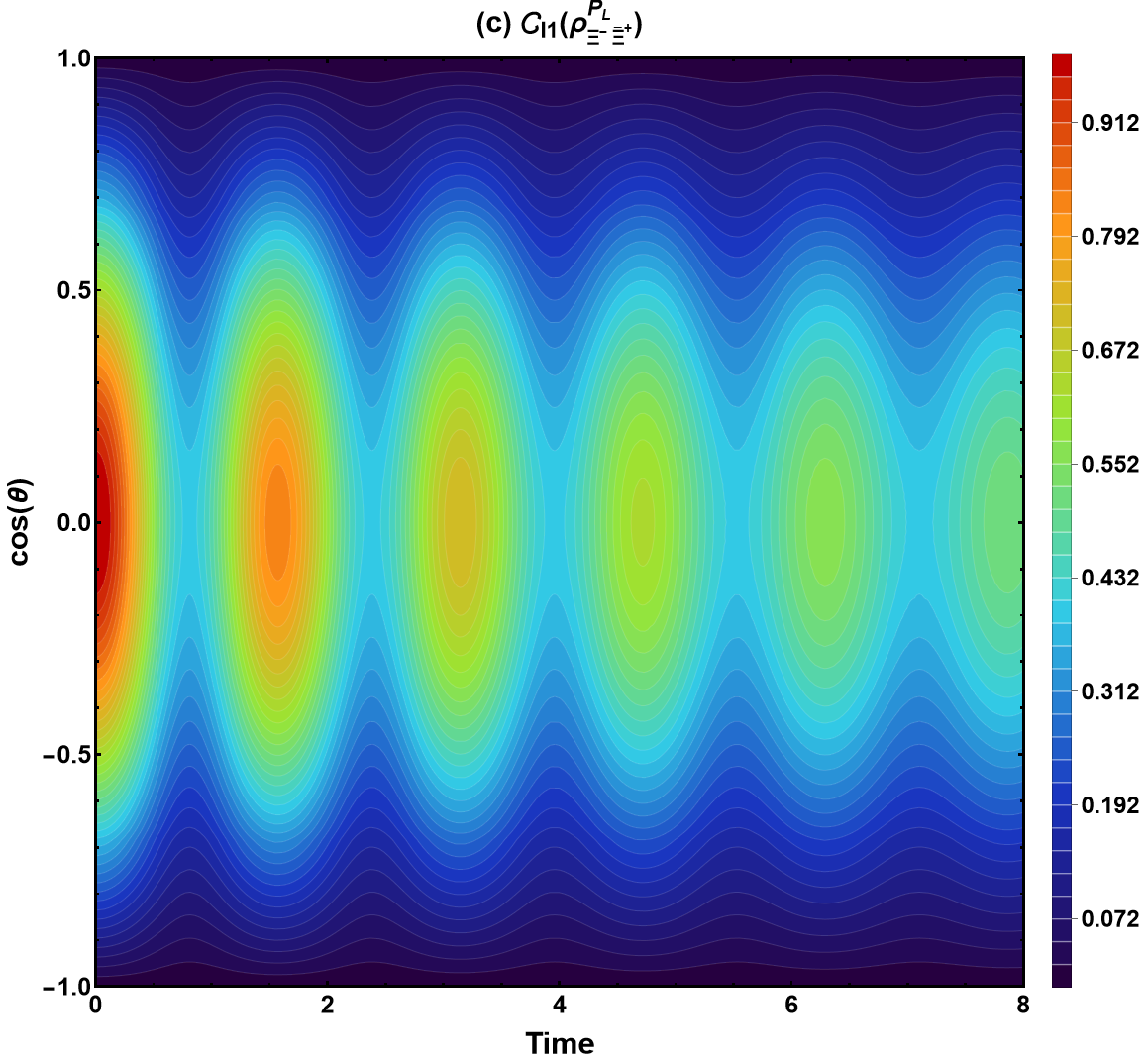}
	\includegraphics[width=0.24\linewidth]{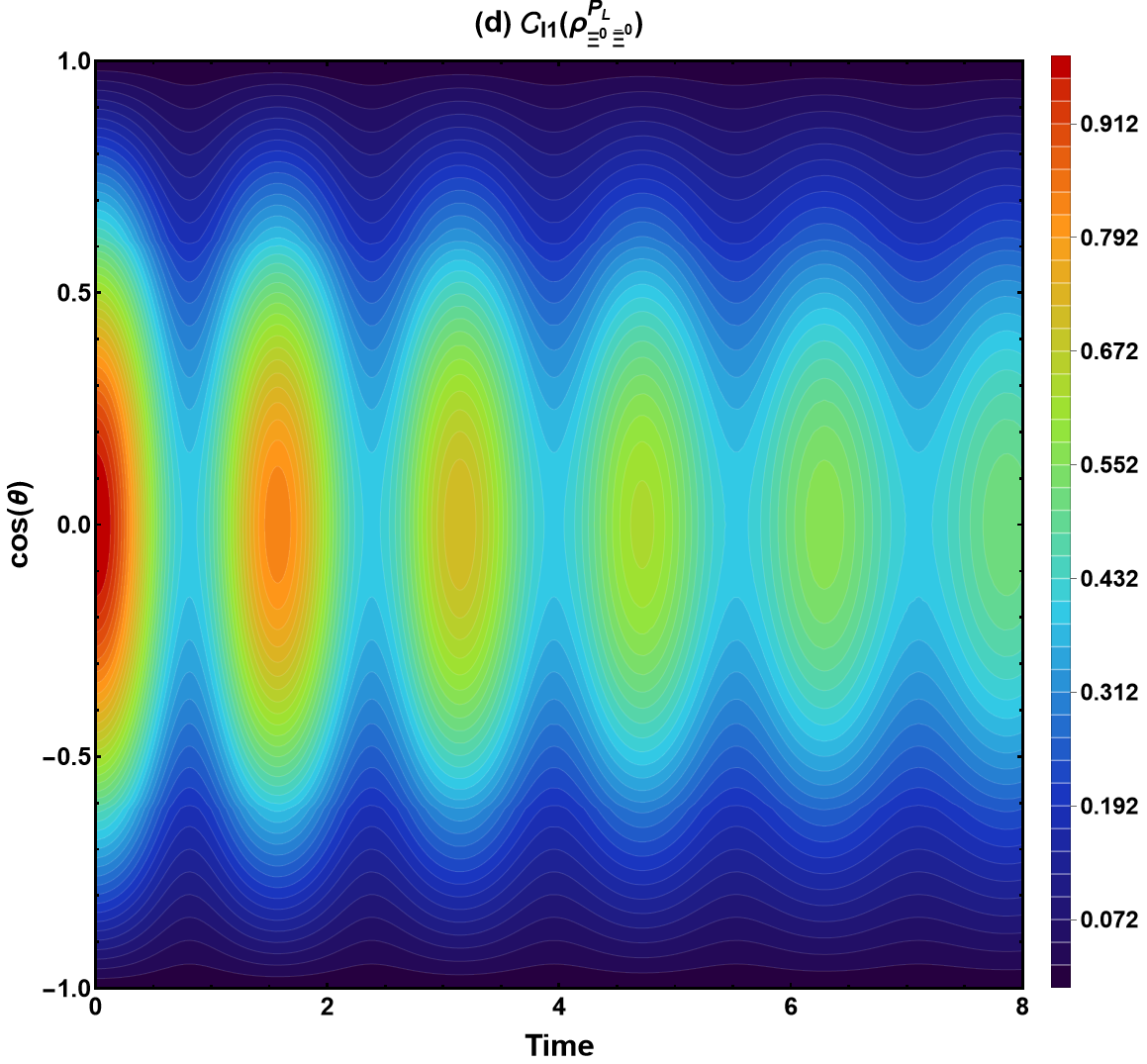}
	\includegraphics[width=0.24\linewidth]{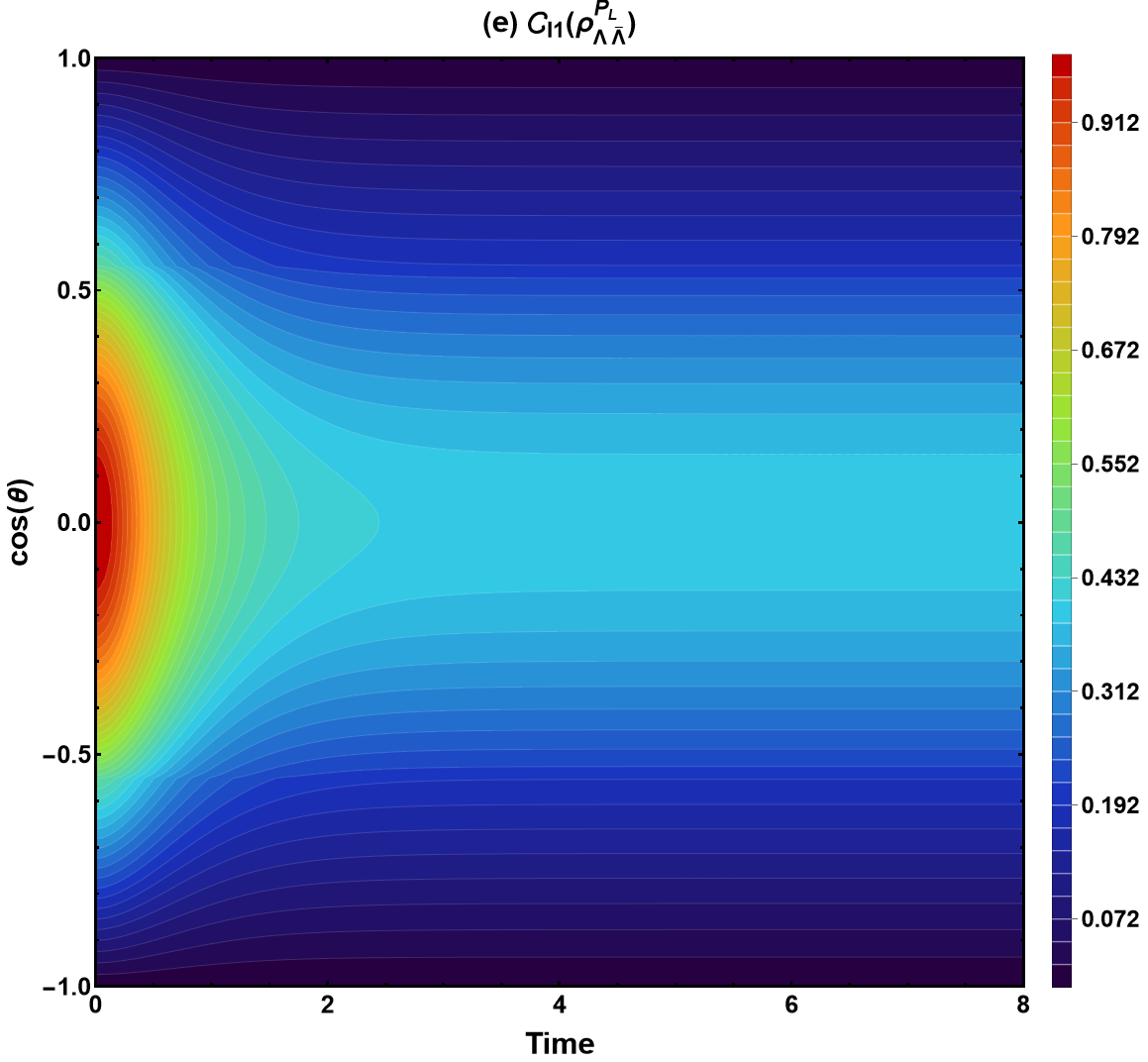}
	\includegraphics[width=0.24\linewidth]{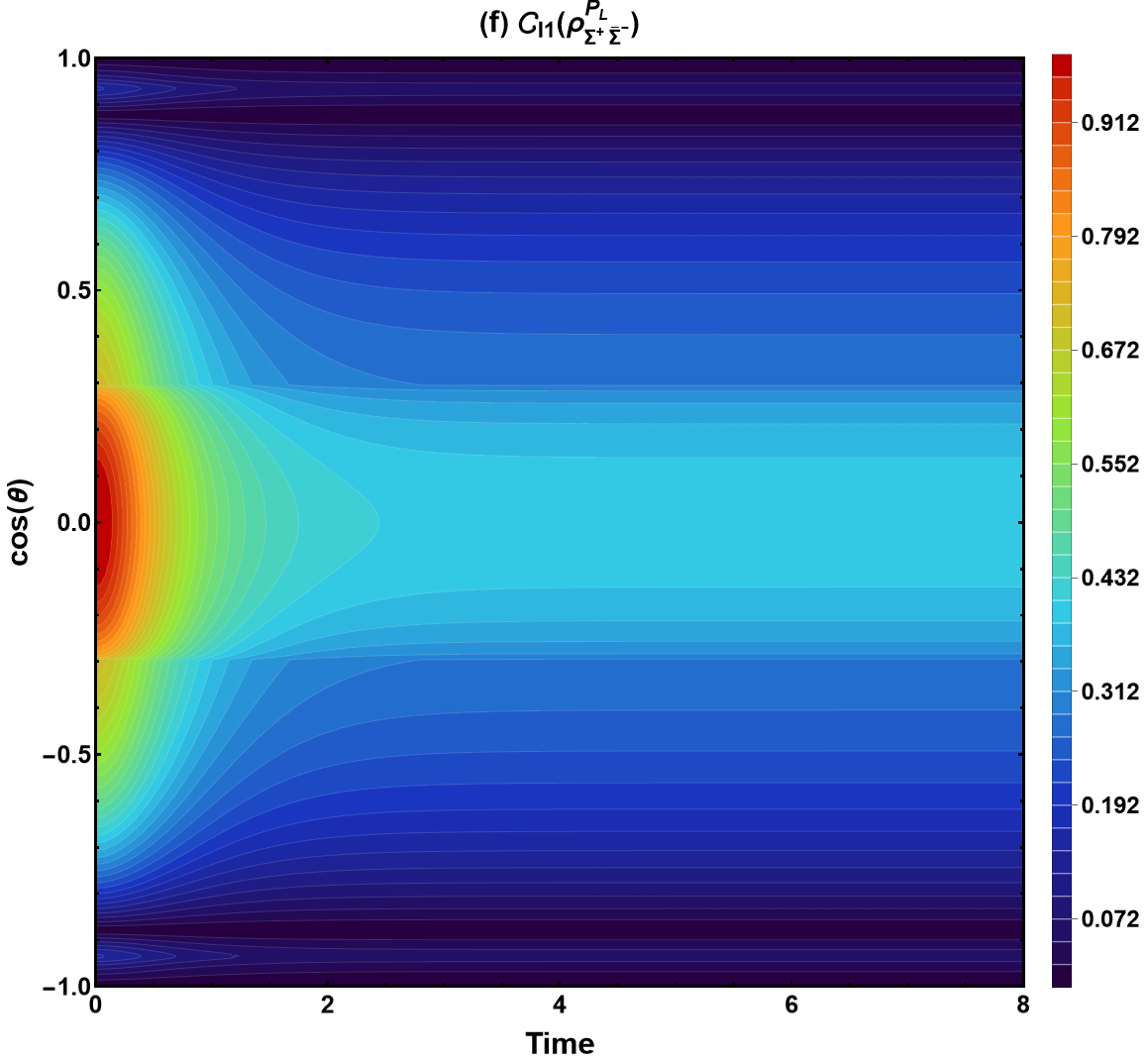}
	\includegraphics[width=0.24\linewidth]{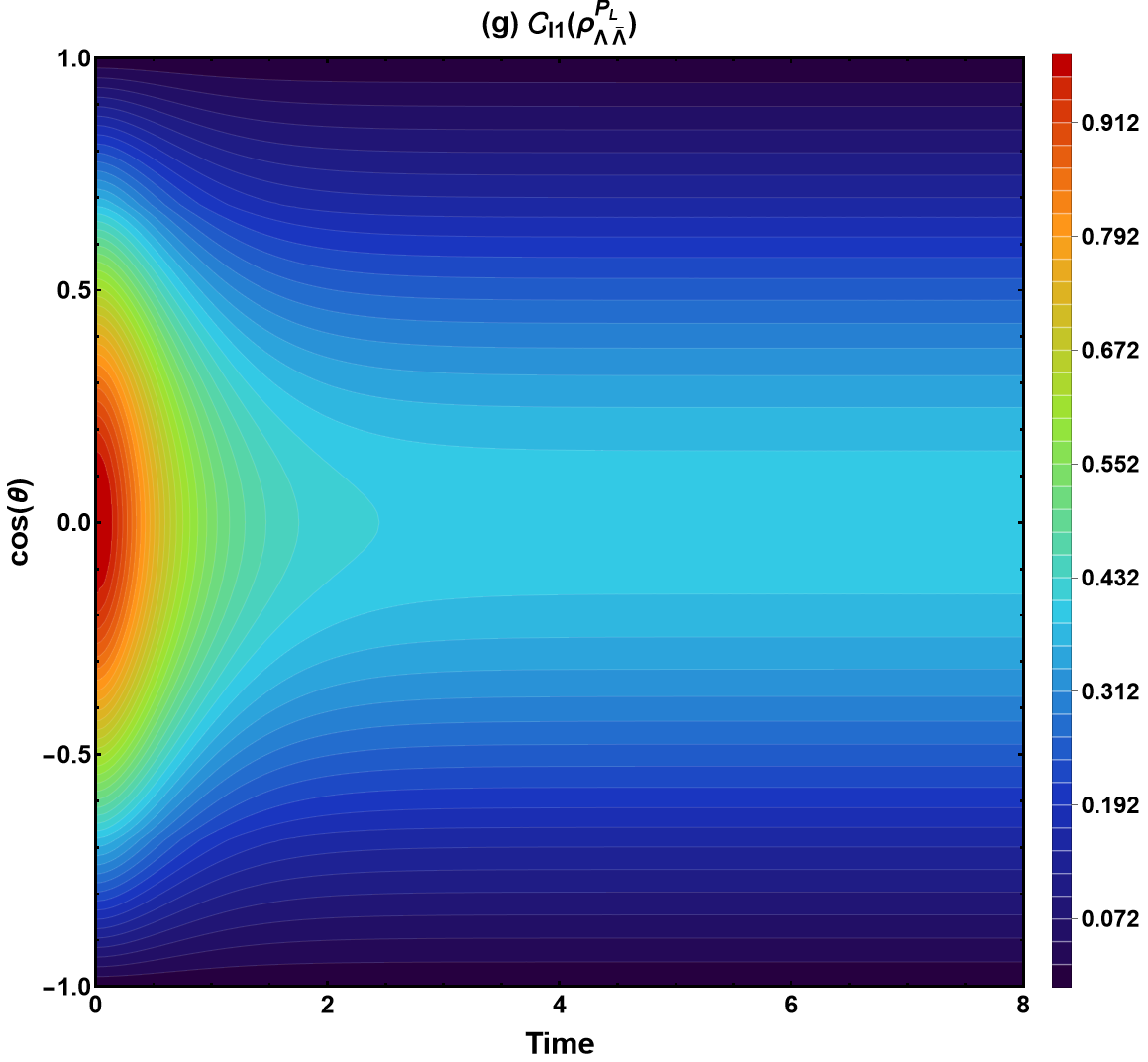}
	\includegraphics[width=0.24\linewidth]{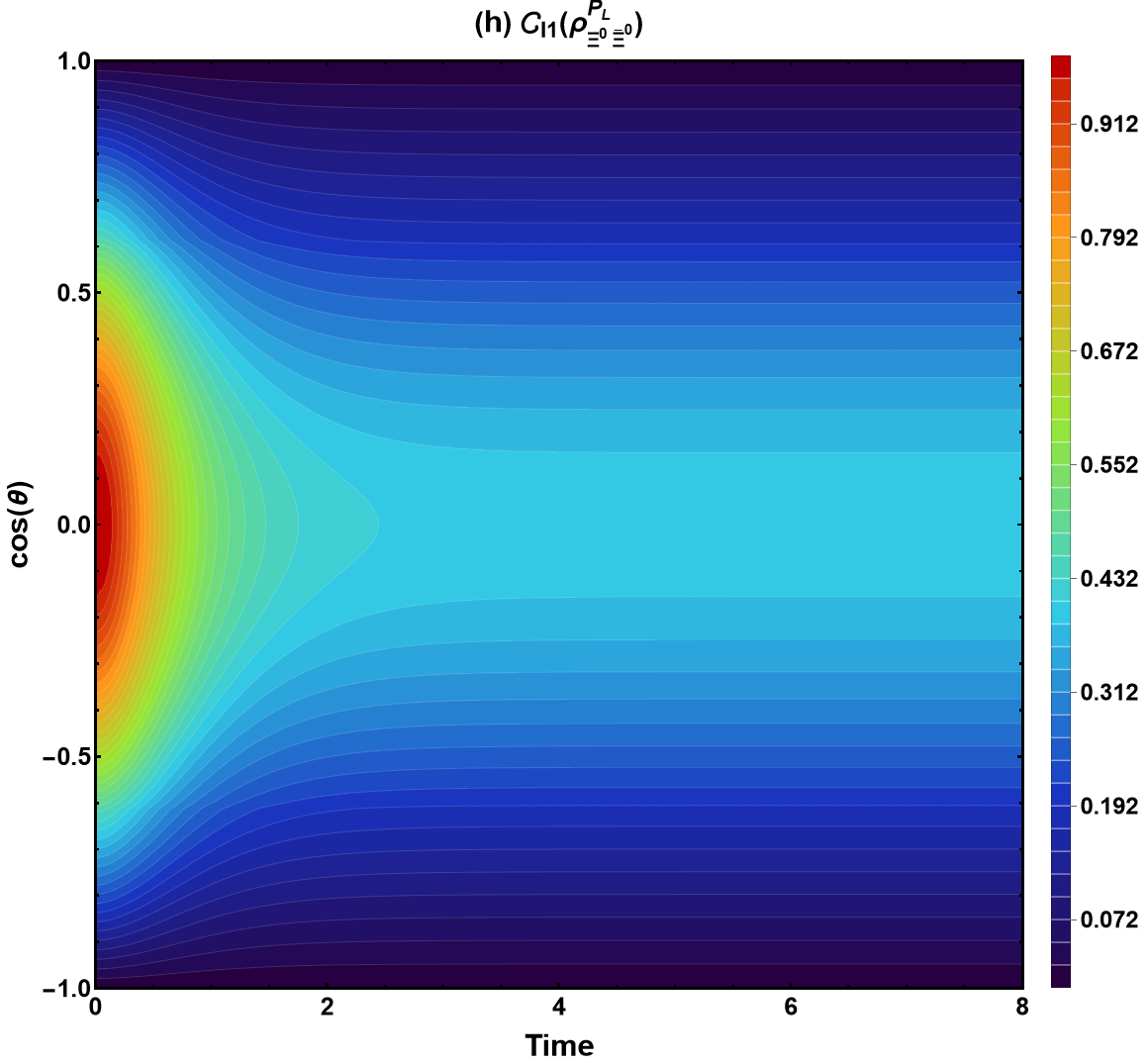}
\caption{
	Dynamical evolution of the $l_1$-norm quantum coherence
	$C_{l_1}(\rho^{P_L}_{Y\bar{Y}})$ as a function of time and the
	production angle $\cos\theta$ for
	$J/\psi\rightarrow Y\bar{Y}$ with
	$Y=\Lambda$, $\Sigma^{+}$, $\Xi^{-}$, and $\Xi^{0}$ at
	$P_L=0.8$. Panels (a)--(d) [(e)--(h)] correspond to the
	non-Markovian (Markovian) regime  with $\tau=5$ ($\tau=0.2$) and
	$\mu=0.4$. The experimental parameters are taken from
	Table~\ref{tab:BESIII}.
}
	\label{fig16}
\end{figure}
As illustrated in Fig.~\ref{fig16}, the $l_{1}$-norm quantum coherence $C_{l_1}$ emerges as the most resilient quantum resource among the various correlation measures examined. The coherence exhibits a
strong dependence on both the production angle $\cos\theta$ and the
environmental dynamics, with its largest values consistently localized
around the central scattering region $\cos\theta \simeq 0$.
In the non-Markovian regime
[Figs.~\ref{fig16}(a)--(d)], pronounced oscillations and recurrent
revivals dominate the dynamics. These features originate from the
backflow of information from the environment to the hyperon-antihyperon
system, allowing part of the lost coherence to be temporarily restored.
The revival amplitudes gradually decrease with time, reflecting the
competition between memory-induced recovery processes and irreversible
decoherence. Although all four channels display a similar revival
structure, the $\Xi^{-}\bar{\Xi}^{+}$ and $\Xi^{0}\bar{\Xi}^{0}$
channels maintain slightly higher coherence values over the entire
evolution, indicating a greater resistance to environmental
disturbances.
A qualitatively different behavior emerges in the Markovian regime
[Figs.~\ref{fig16}(e)--(h)]. In the absence of memory effects, the
oscillatory revivals disappear and the coherence undergoes a smooth
relaxation toward a stationary configuration. After a short transient
period, the contour profiles become nearly time independent,
demonstrating the irreversible loss of quantum information into the
environment. Nevertheless, a substantial amount of coherence survives at
long times, particularly in the vicinity of $\cos\theta =0$, where the
largest asymptotic values are observed.
An important feature of Fig.~\ref{fig16} is that the angular dependence
of coherence remains remarkably stable throughout the evolution. In both
dynamical regimes, the coherence decreases monotonically as
$|\cos\theta|$ approaches unity, indicating that quantum superposition
effects are strongest in the central production region and become
progressively suppressed toward the forward and backward scattering
directions. The $\Sigma^{+}\bar{\Sigma}^{-}$ channel exhibits a slightly
more structured angular profile than the other channels, with additional
suppression regions appearing near $|\cos\theta|\simeq0.9$.
The comparison between the non-Markovian and Markovian dynamics clearly
demonstrates the beneficial role of environmental memory in preserving
quantum coherence. More importantly, unlike entanglement and quantum
steering, which may undergo substantial degradation under decoherence,
the $l_{1}$-norm coherence remains finite throughout the evolution and
retains significant residual values even in the asymptotic limit. This
behavior confirms that coherence constitutes the most persistent quantum
resource in hyperon-antihyperon systems and provides a robust signature
of quantumness even in parameter regions where stronger forms of quantum
correlations are strongly suppressed.
\begin{figure}[H]
	\centering
	\includegraphics[width=0.24\linewidth]{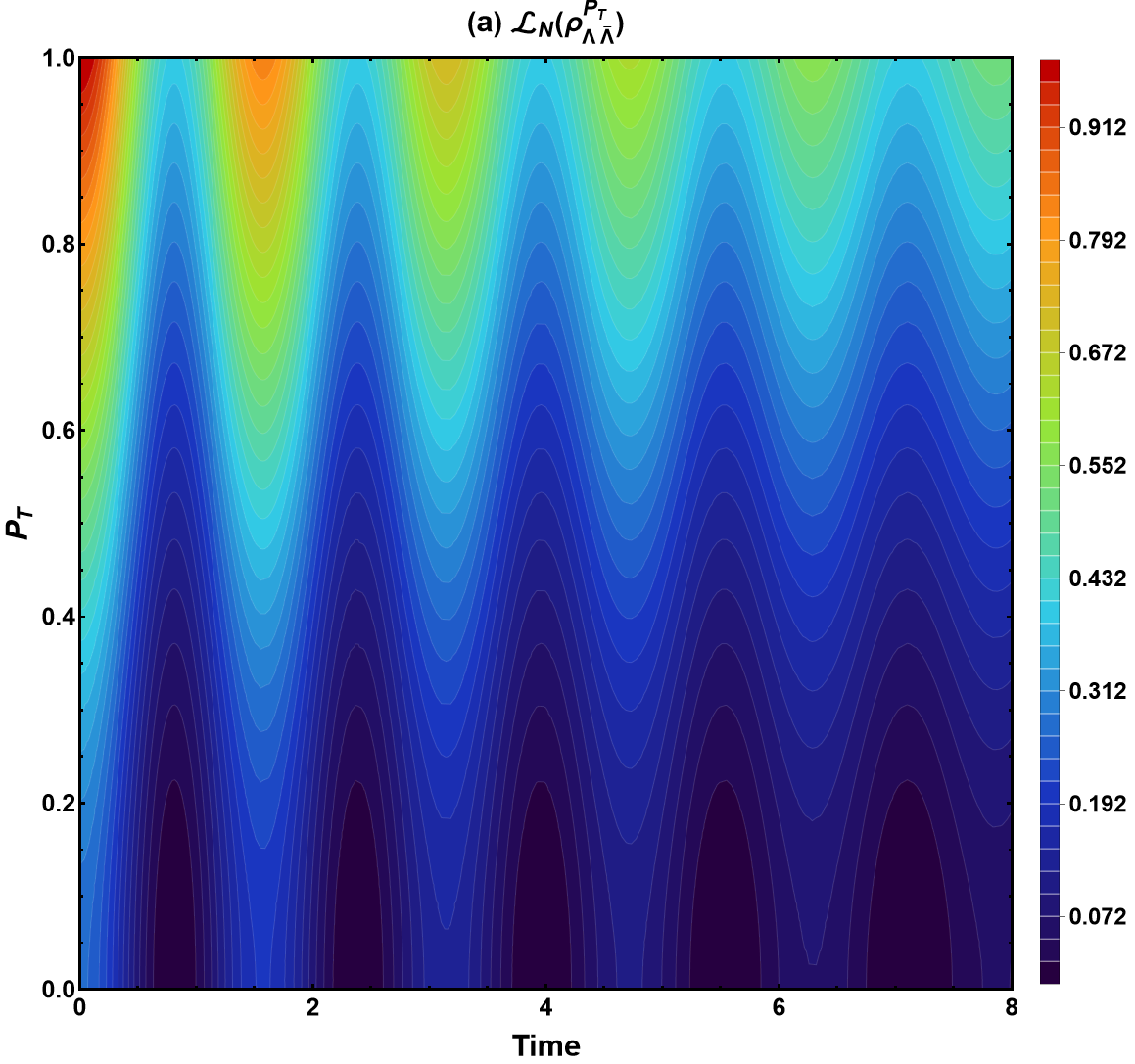}
	\includegraphics[width=0.24\linewidth]{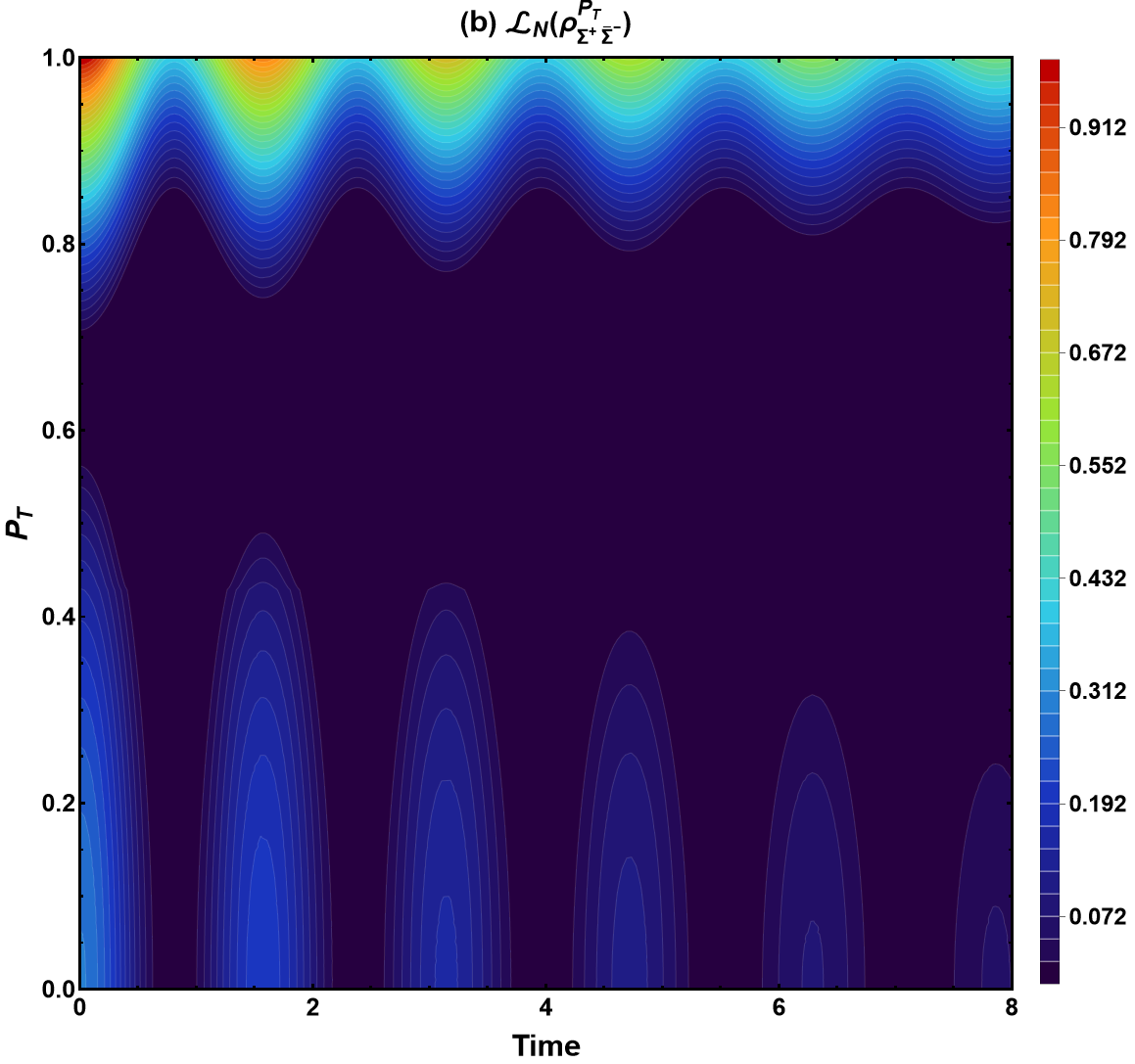}
	\includegraphics[width=0.24\linewidth]{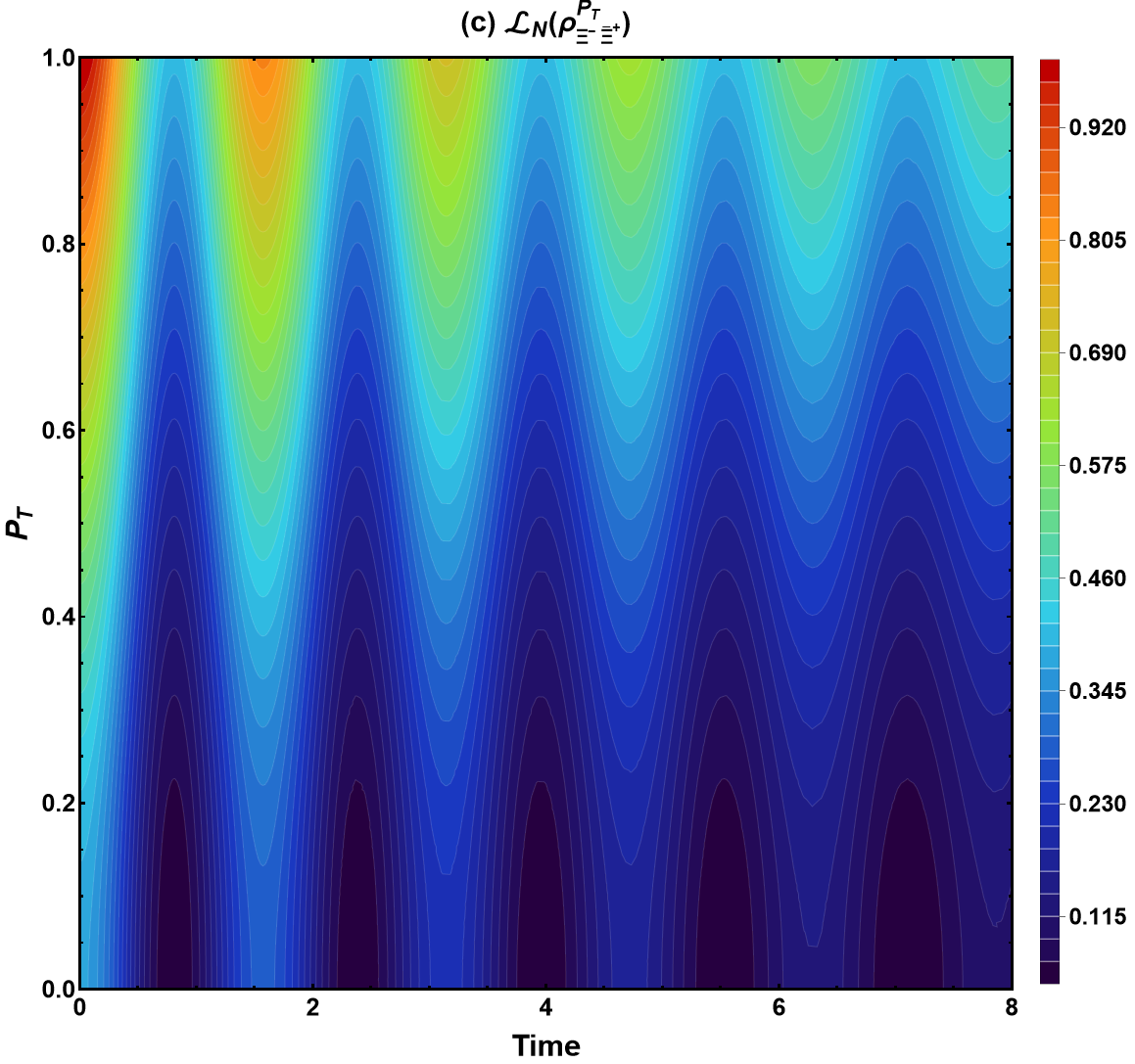}
	\includegraphics[width=0.24\linewidth]{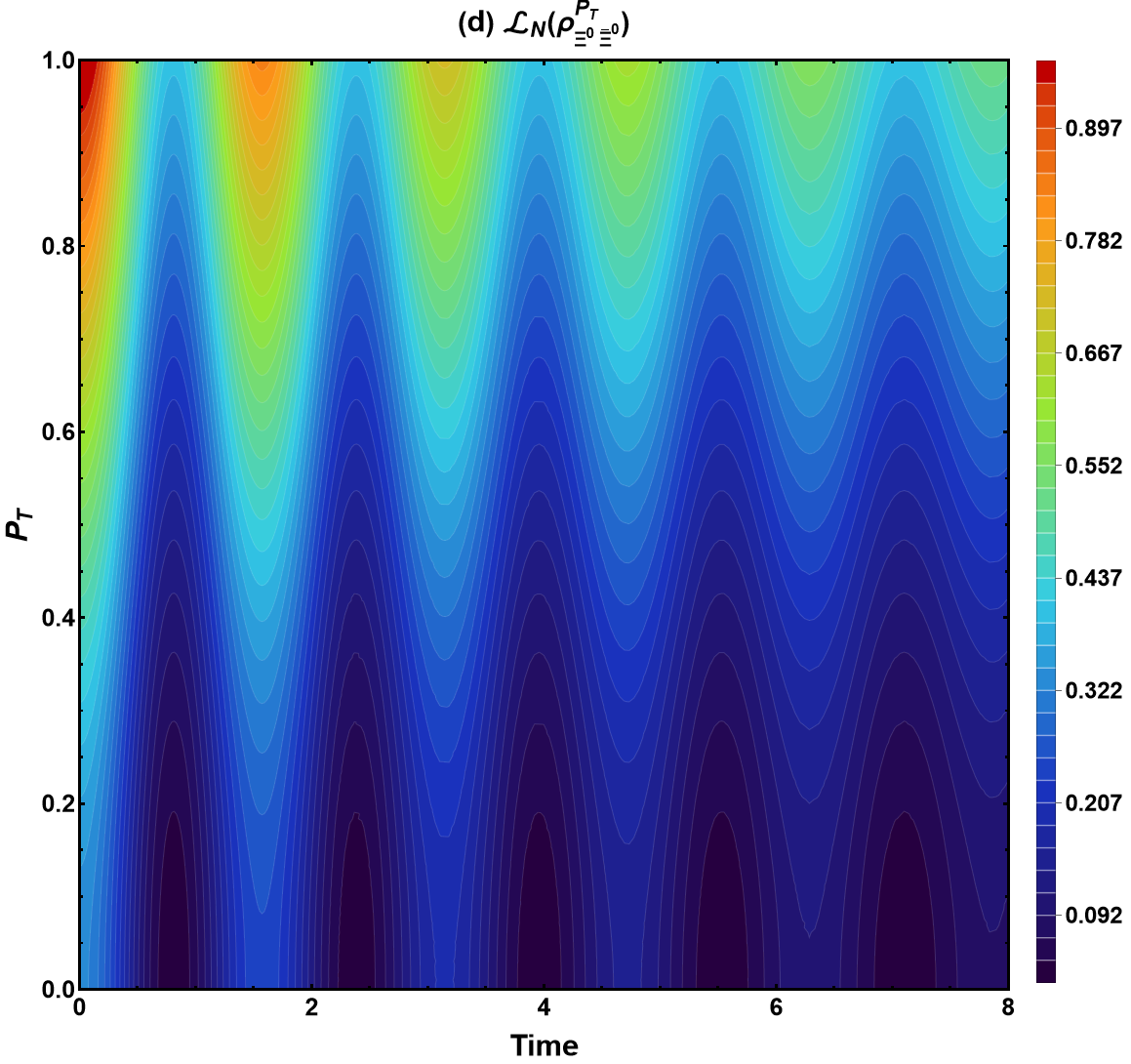}
	\includegraphics[width=0.24\linewidth]{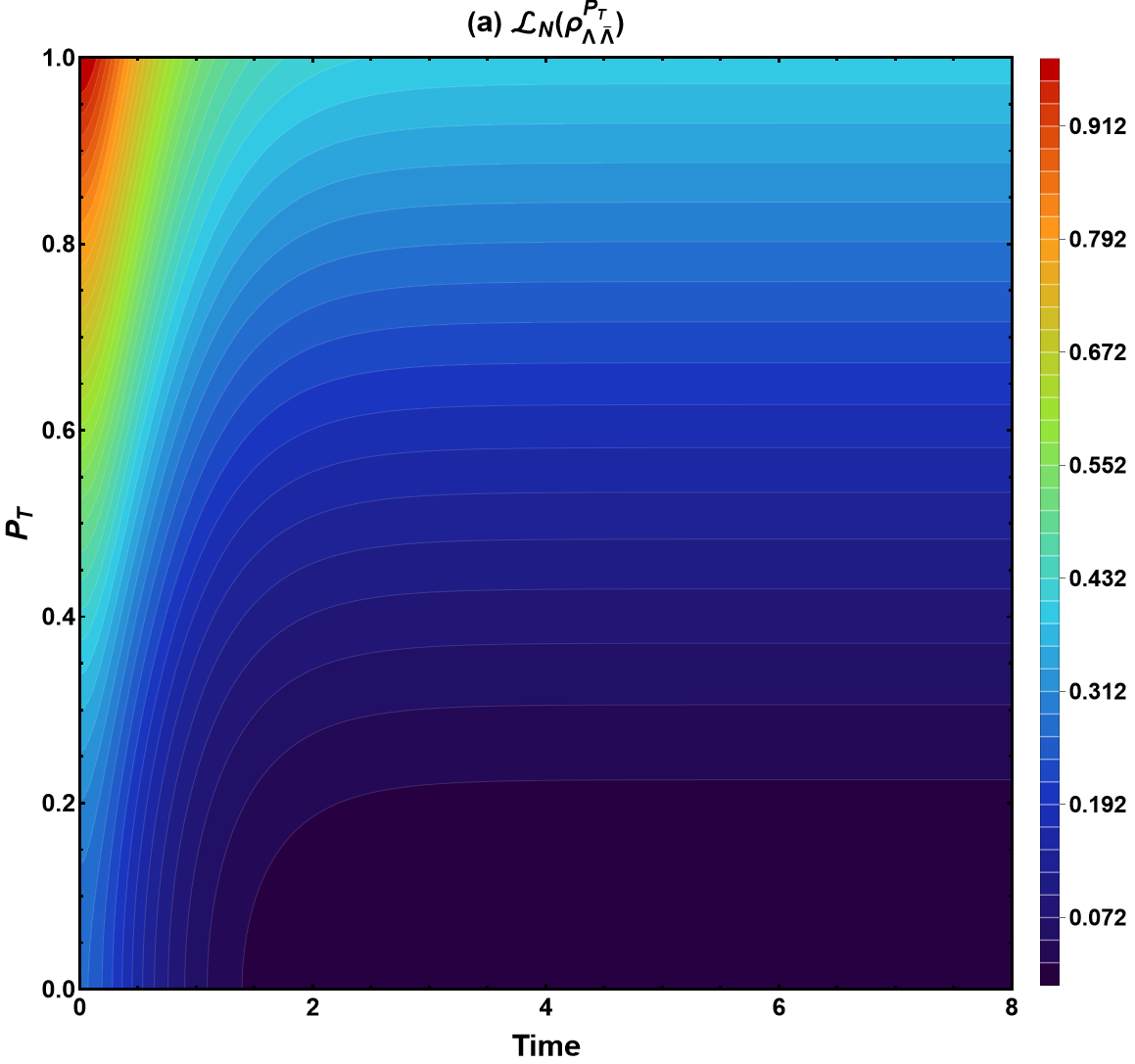}
	\includegraphics[width=0.24\linewidth]{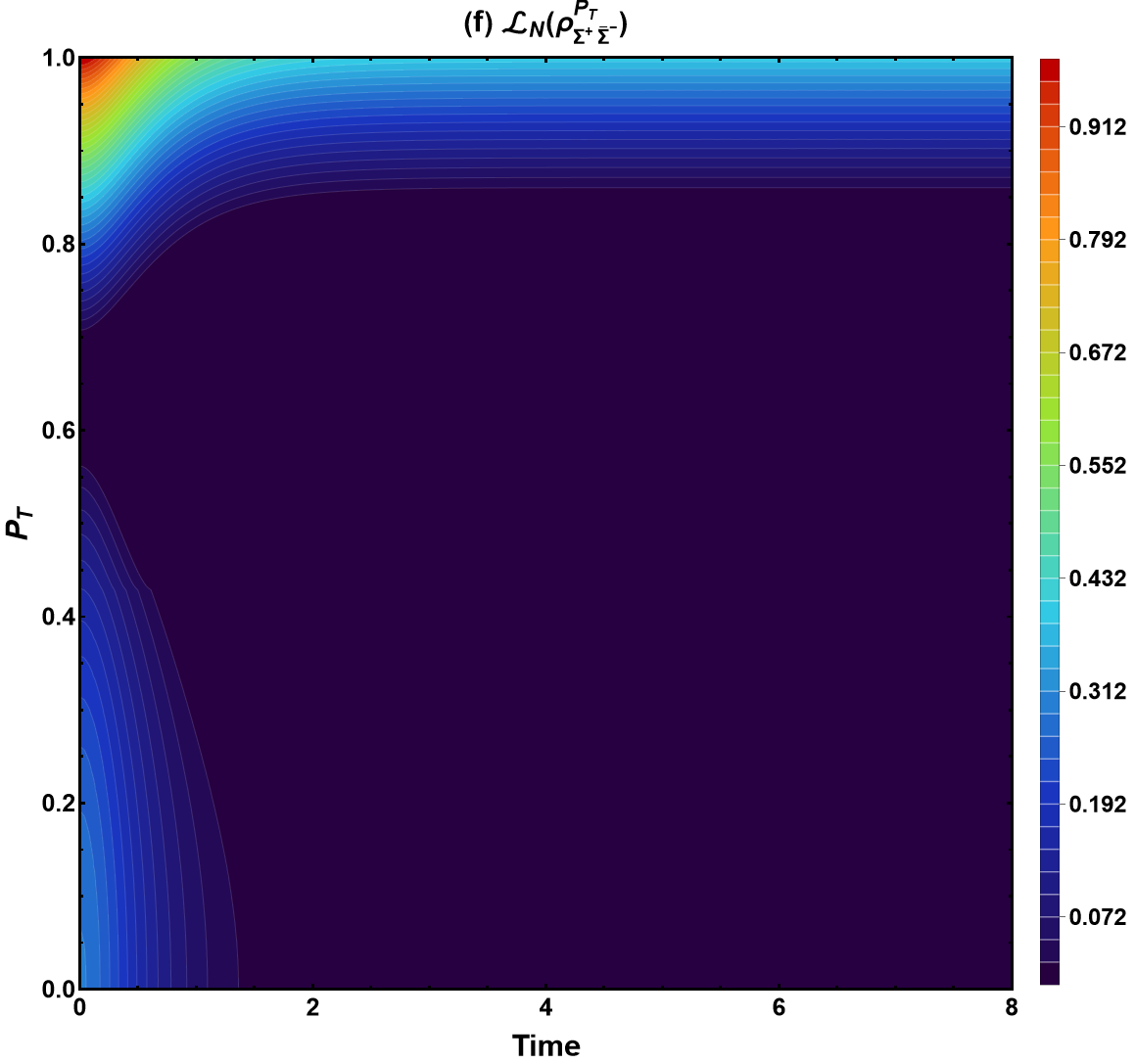}
	\includegraphics[width=0.24\linewidth]{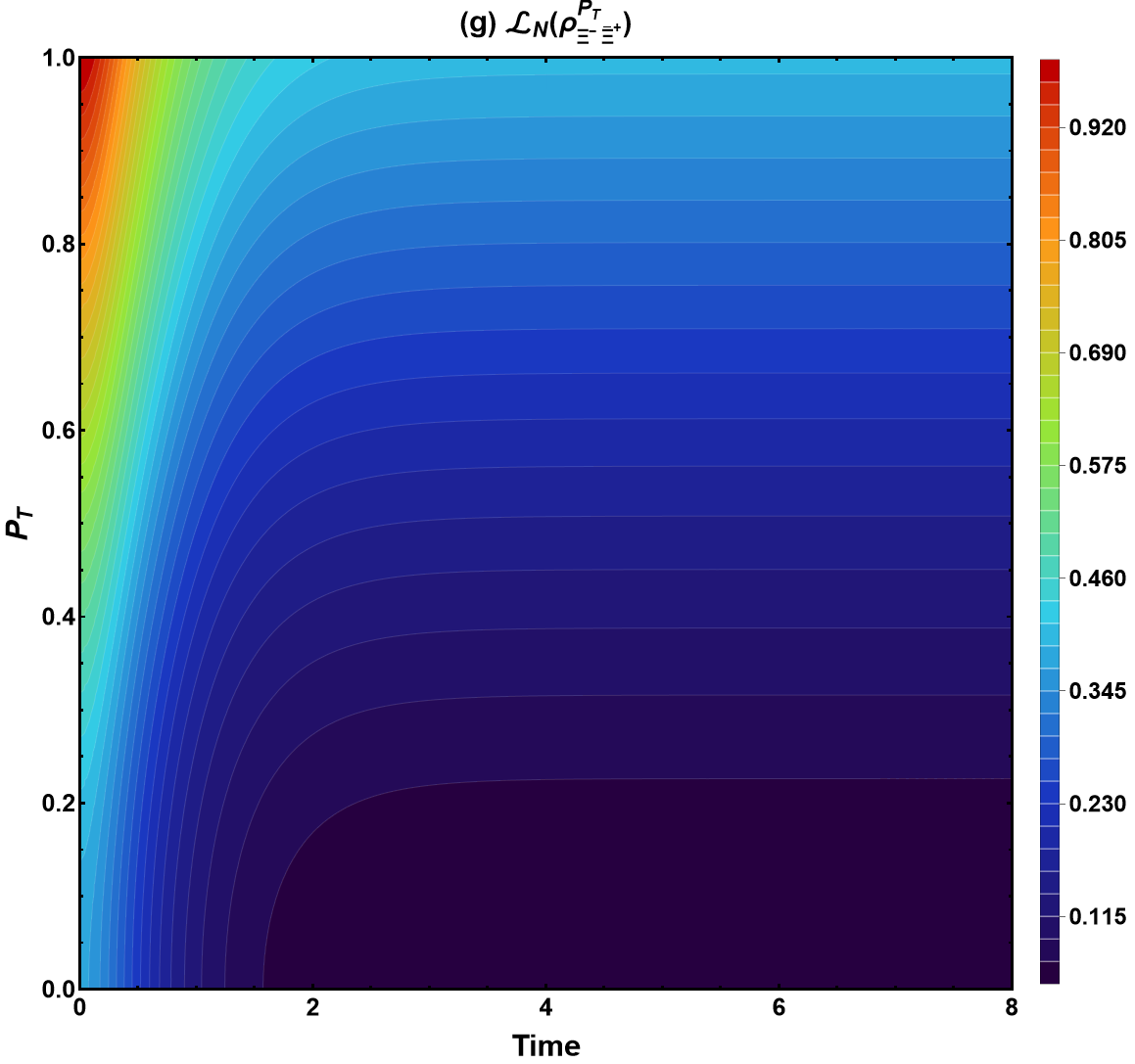}
	\includegraphics[width=0.24\linewidth]{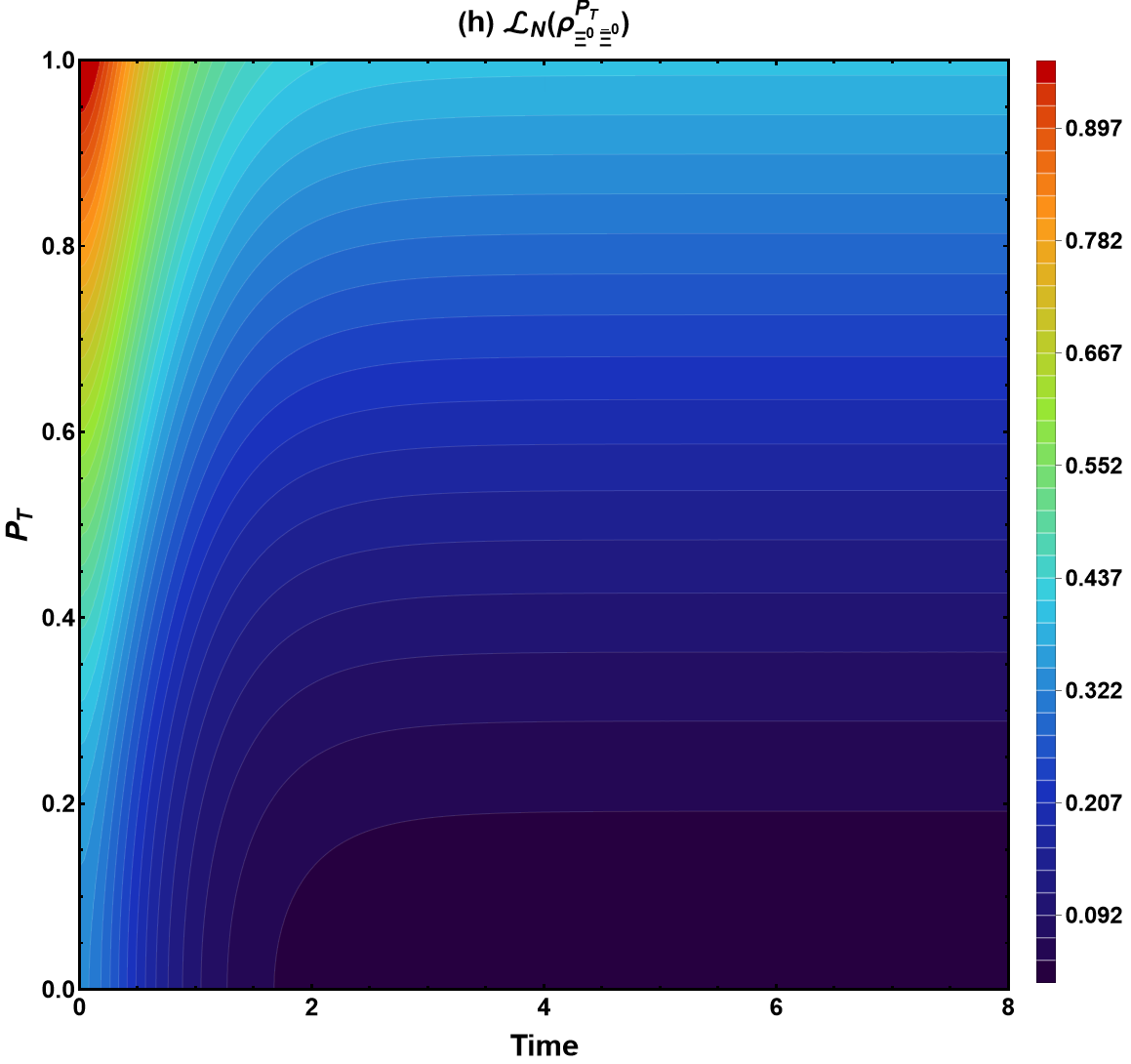}
\caption{
	Dynamical evolution of the logarithmic negativity
	$\mathcal{L}_N(\rho^{P_T}_{Y\bar{Y}})$ as a function of time and the
	transverse polarization degree $P_T$ for
	$J/\psi\rightarrow Y\bar{Y}$ with
	$Y=\Lambda$, $\Sigma^{+}$, $\Xi^{-}$, and $\Xi^{0}$ at
	$\cos\theta=0.5$ and $\phi=0$. Panels (a)--(d) [(e)--(h)] correspond
	to the non-Markovian (Markovian) regime  with $\tau=5$ ($\tau=0.2$) and
	$\mu=0.4$. The experimental parameters are taken from
	Table~\ref{tab:BESIII}.
}
	\label{fig17}
\end{figure}
The evolution of the logarithmic negativity $\mathcal{L}_{N}$ with time and transverse polarization degree $P_T$ is presented in Fig.~\ref{fig17} for the four hyperon--antihyperon channels. The results reveal a strong interplay between polarization,
environmental memory effects, and entanglement preservation.
In the non-Markovian regime
[Figs.~\ref{fig17}(a)--(d)], the entanglement exhibits pronounced
oscillatory dynamics accompanied by successive revivals. These
oscillations are a direct manifestation of information backflow from the
environment to the hyperon-antihyperon system, allowing part of the lost
quantum correlations to be temporarily restored. The revival amplitudes
gradually decrease with time, indicating that decoherence remains active
despite the presence of memory effects. For the
$\Lambda\bar{\Lambda}$, $\Xi^{-}\bar{\Xi}^{+}$, and
$\Xi^{0}\bar{\Xi}^{0}$ channels, the entanglement increases
monotonically with the transverse polarization degree and reaches its
maximum values near $P_T=1$. This behavior demonstrates that transverse
beam polarization acts as an efficient resource for enhancing quantum
correlations in hyperon production processes.
A qualitatively different structure emerges for the
$\Sigma^{+}\bar{\Sigma}^{-}$ channel [Fig.~\ref{fig17}(b)]. In this
case, entanglement survives only within a narrow region of large
polarization degrees, approximately $P_T\gtrsim0.8$, whereas a broad
intermediate domain remains almost completely disentangled throughout
the evolution. Although weak revivals are still visible at small
polarizations, the overall magnitude of $\mathcal{L}_{N}$ is strongly
suppressed compared with the other channels. This behavior indicates
that the $\Sigma^{+}\bar{\Sigma}^{-}$ state is considerably more
sensitive to environmental decoherence and requires a high degree of
transverse polarization to sustain nonclassical correlations.
The Markovian dynamics
[Figs.~\ref{fig17}(e)--(h)] exhibits a markedly different behavior. The
oscillatory revivals disappear completely and the entanglement decays
smoothly toward a stationary configuration. The absence of memory
effects prevents any recovery of quantum correlations once they are
transferred to the environment. As a result, the temporal evolution is
governed by a monotonic relaxation process characterized by a rapid
initial decay followed by saturation at long times.
Despite the stronger decohering action of the Markovian environment,
substantial entanglement remains preserved at large polarization
degrees. In particular, the $\Lambda\bar{\Lambda}$,
$\Xi^{-}\bar{\Xi}^{+}$, and $\Xi^{0}\bar{\Xi}^{0}$ channels retain
significant residual logarithmic negativity for $P_T\approx1$, whereas
the $\Sigma^{+}\bar{\Sigma}^{-}$ channel again displays a restricted
entangled region confined to the vicinity of maximal polarization.
Therefore, the transverse beam polarization plays a dual role: it not
only enhances the initial amount of entanglement but also increases its
robustness against environmental noise.
Overall, Fig.~\ref{fig17} demonstrates that environmental memory
primarily affects the temporal persistence of entanglement through
revivals, while the transverse polarization degree controls its overall
magnitude. The coexistence of these two mechanisms leads to the most
favorable conditions for entanglement preservation in the highly
polarized non-Markovian regime, where the logarithmic negativity
maintains large values over extended time intervals.
\begin{figure}[H]
	\centering
	\includegraphics[width=0.24\linewidth]{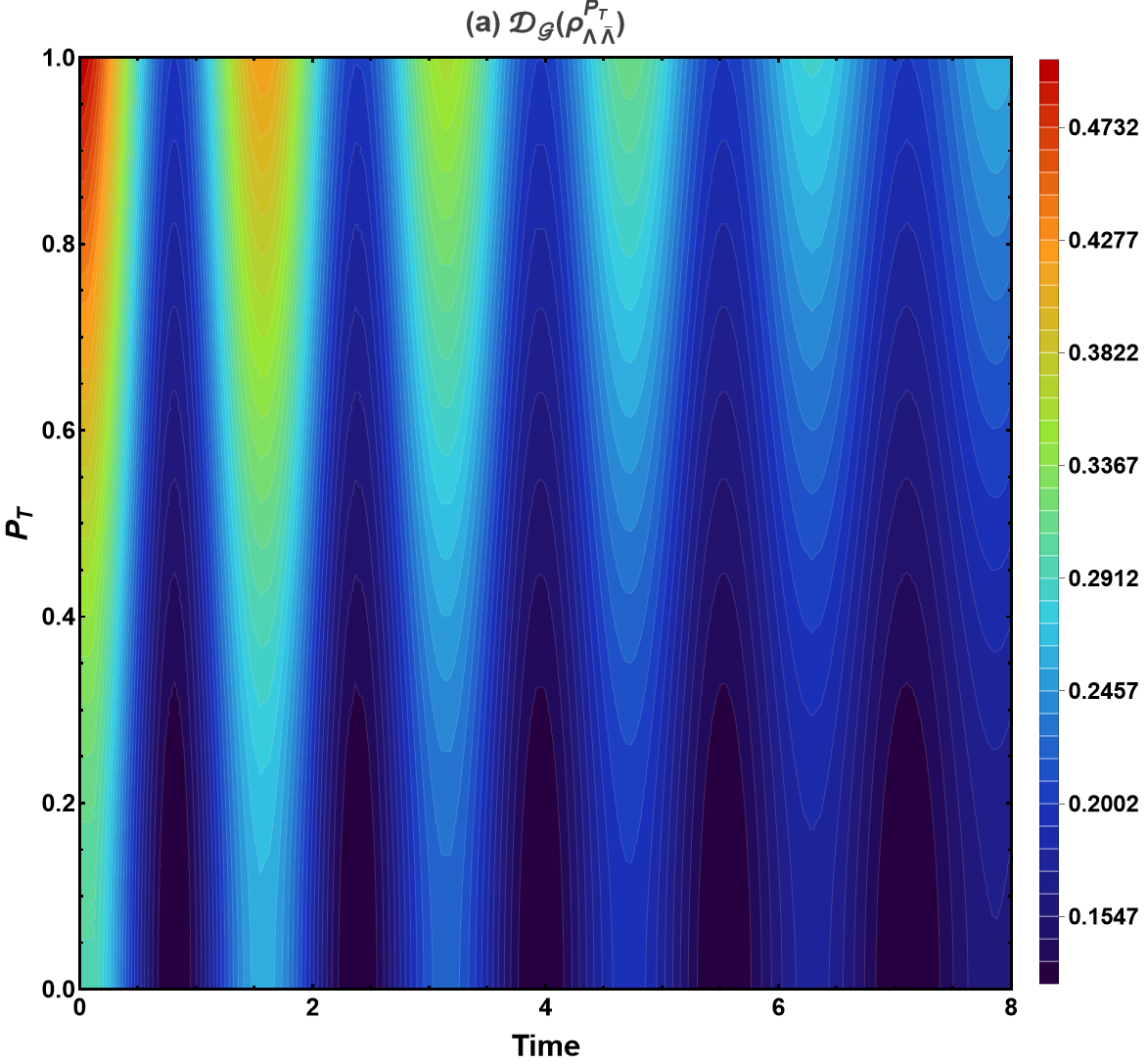}
	\includegraphics[width=0.24\linewidth]{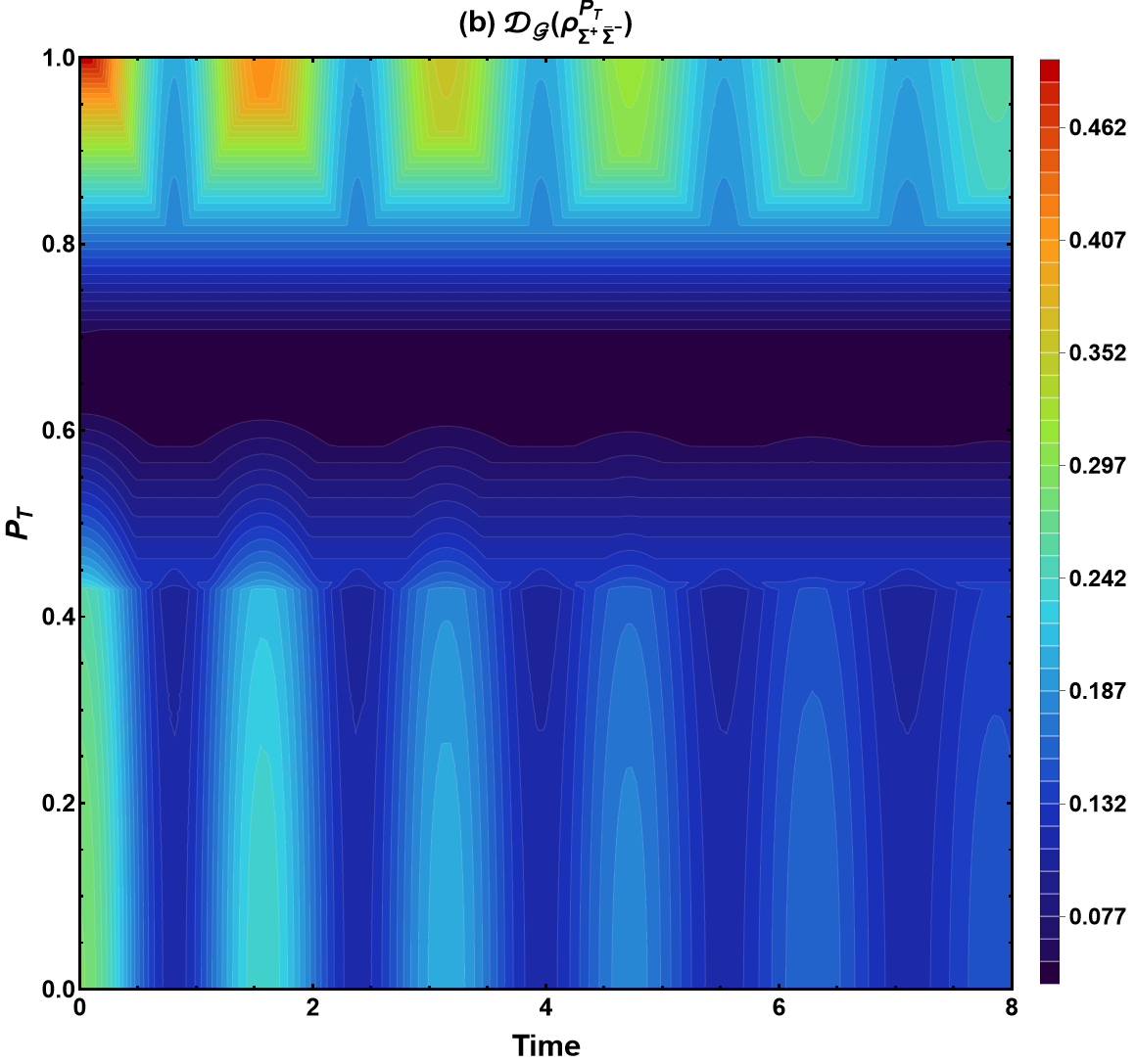}
	\includegraphics[width=0.24\linewidth]{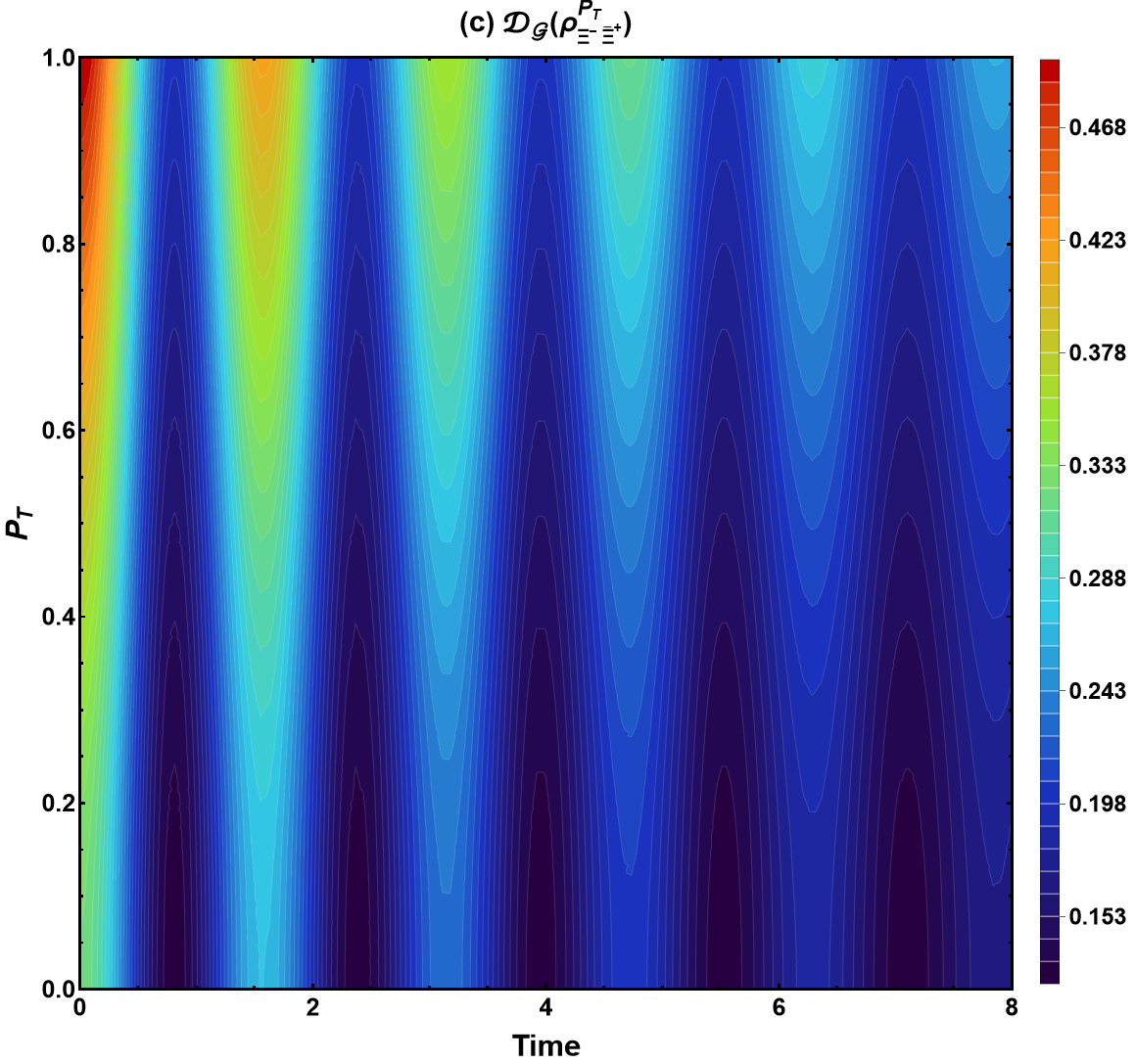}
	\includegraphics[width=0.24\linewidth]{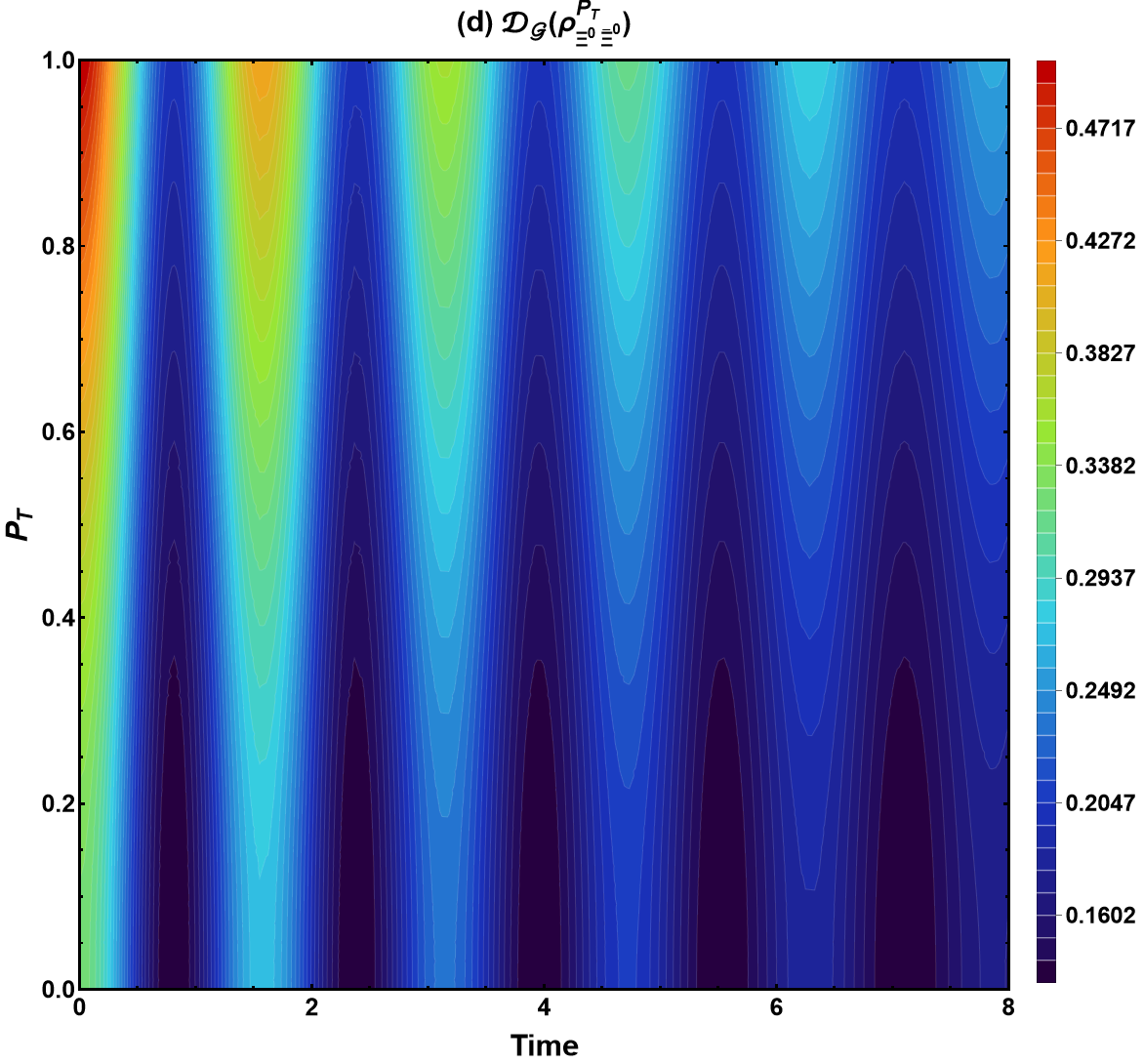}
	\includegraphics[width=0.24\linewidth]{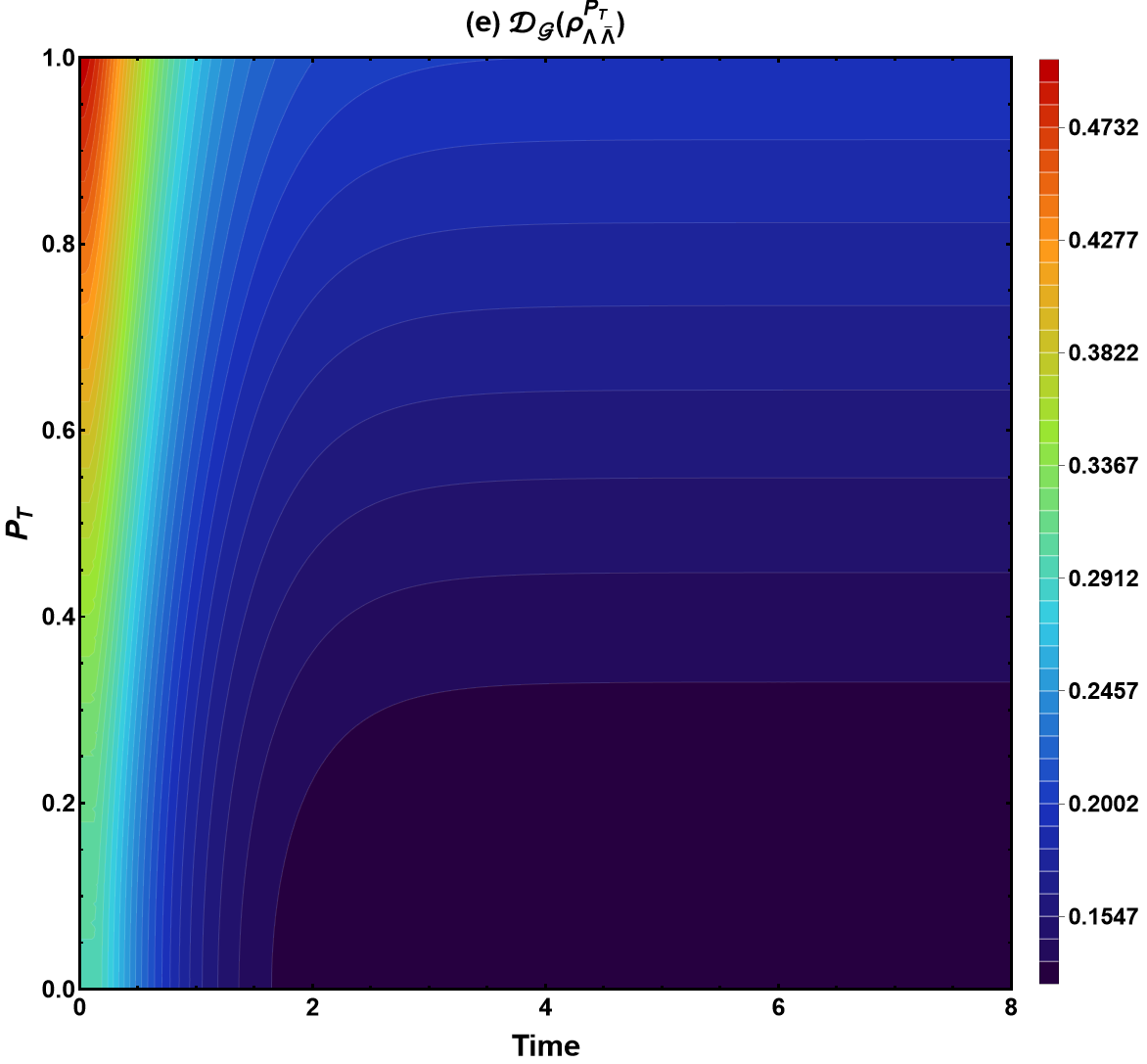}
	\includegraphics[width=0.24\linewidth]{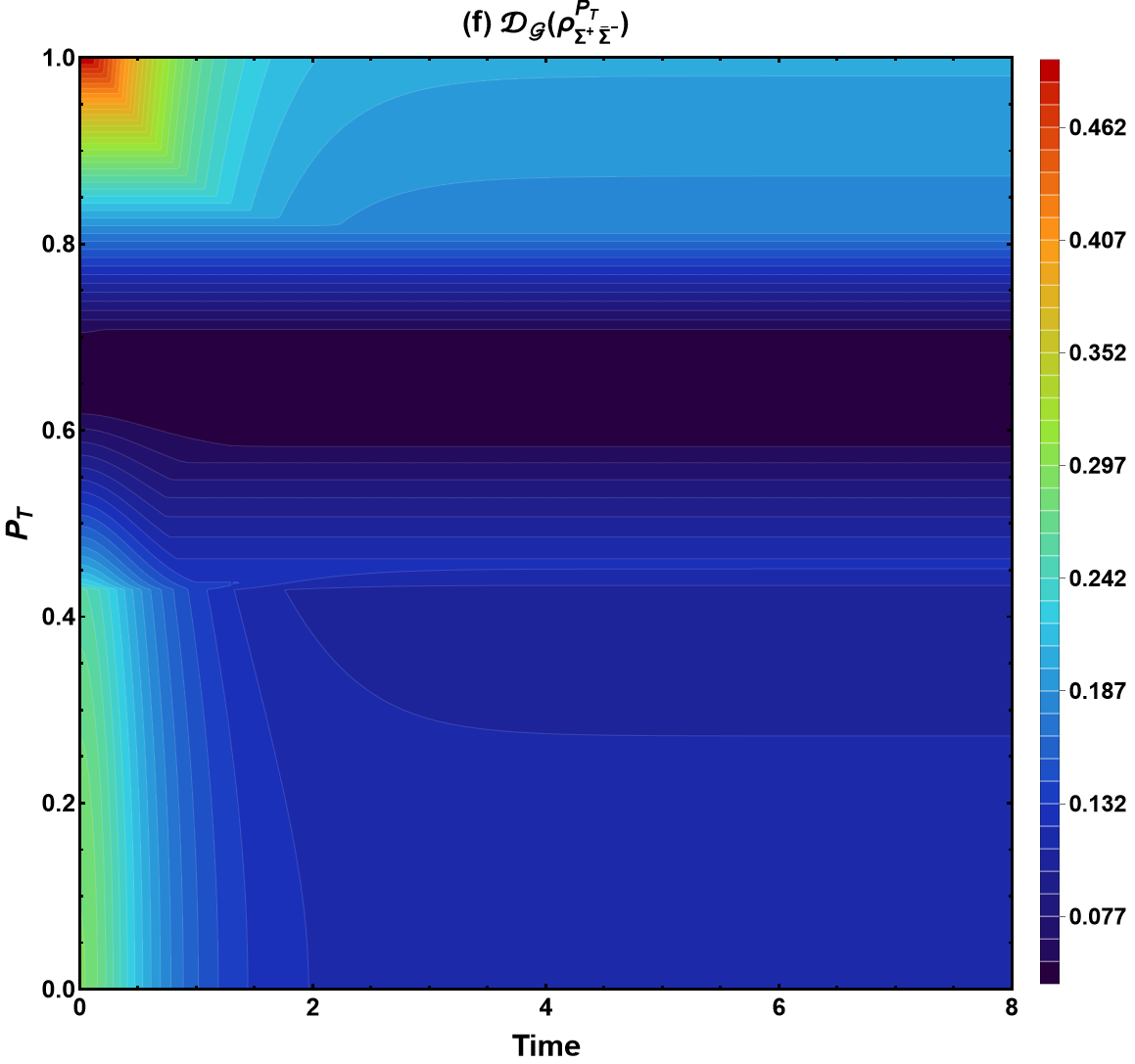}
	\includegraphics[width=0.24\linewidth]{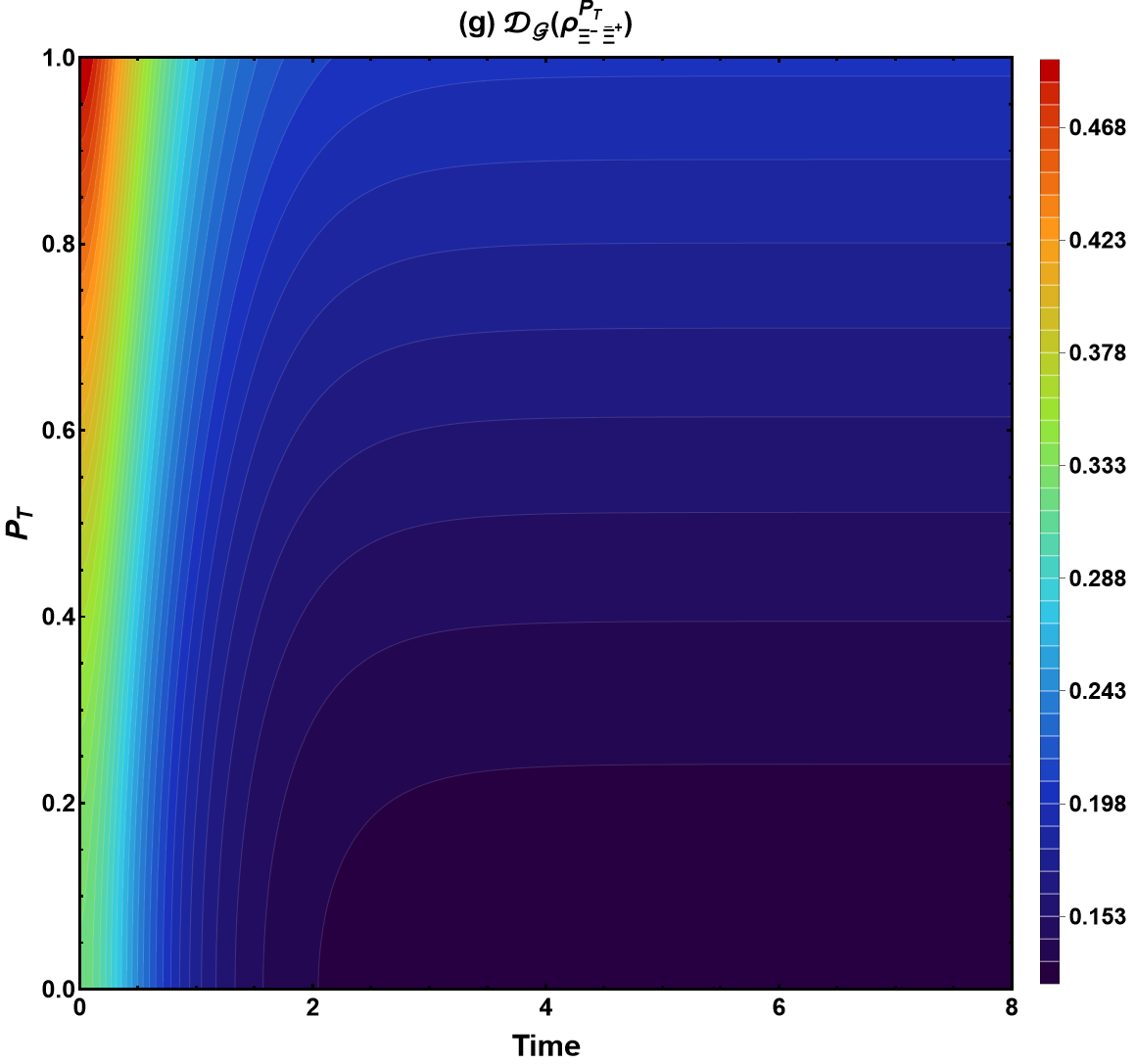}
	\includegraphics[width=0.24\linewidth]{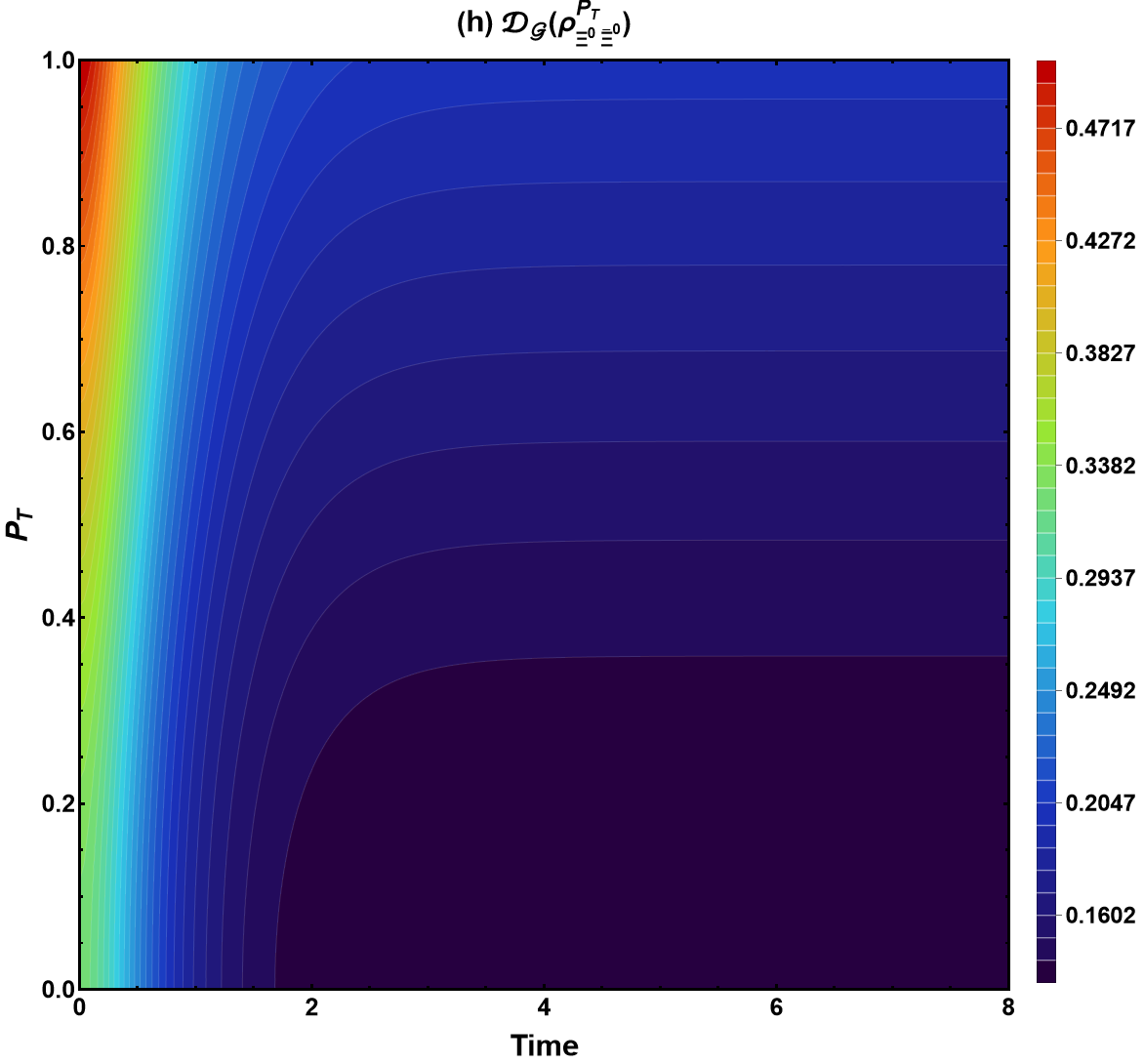}
\caption{
	Dynamical evolution of the geometric quantum discord
	$\mathcal{D}_G(\rho^{P_T}_{Y\bar{Y}})$ as a function of time and the
	transverse polarization degree $P_T$ for
	$J/\psi\rightarrow Y\bar{Y}$ with
	$Y=\Lambda$, $\Sigma^{+}$, $\Xi^{-}$, and $\Xi^{0}$ at
	$\cos\theta=0.5$ and $\phi=0$. Panels (a)--(d) [(e)--(h)] correspond
	to the non-Markovian (Markovian) regime  with $\tau=5$ ($\tau=0.2$) and
	$\mu=0.4$. The experimental parameters are taken from
	Table~\ref{tab:BESIII}.
}
	\label{fig20}
\end{figure}
The behavior of the geometric quantum discord $\mathcal{D}_{G}(\rho^{P_T}_{Y\bar{Y}})$ as a function of time and the transverse polarization degree $P_T$ for the four hyperon channels is presented in Fig.~\ref{fig20}. In the
non-Markovian regime [panels (a)--(d)], the discord exhibits damped
oscillations associated with the temporary backflow of information from
the environment to the hyperon subsystem. The oscillation amplitude
gradually decreases with time, indicating the progressive suppression of
memory effects. In contrast, the Markovian regime [panels (e)--(h)]
shows a monotonic relaxation toward stationary values, reflecting the
irreversible loss of coherence in the absence of environmental memory.
For all channels, $\mathcal{D}_{G}$ increases with the transverse
polarization degree, demonstrating that transverse polarization
enhances the nonclassical correlations generated during the production
process. A comparison among the different channels reveals that
$\Lambda\bar{\Lambda}$, $\Xi^{-}\bar{\Xi}^{+}$, and
$\Xi^{0}\bar{\Xi}^{0}$ display very similar behaviors, whereas the
$\Sigma^{+}\bar{\Sigma}^{-}$ channel exhibits a broader low-discord
region, indicating a stronger sensitivity to decoherence. Most
importantly, the discord remains finite over the entire parameter space,
even in regions where entanglement is strongly suppressed, confirming
the higher robustness of quantum correlations beyond entanglement under
correlated dephasing dynamics.

\begin{figure}[H]
	\centering
	\includegraphics[width=0.24\linewidth]{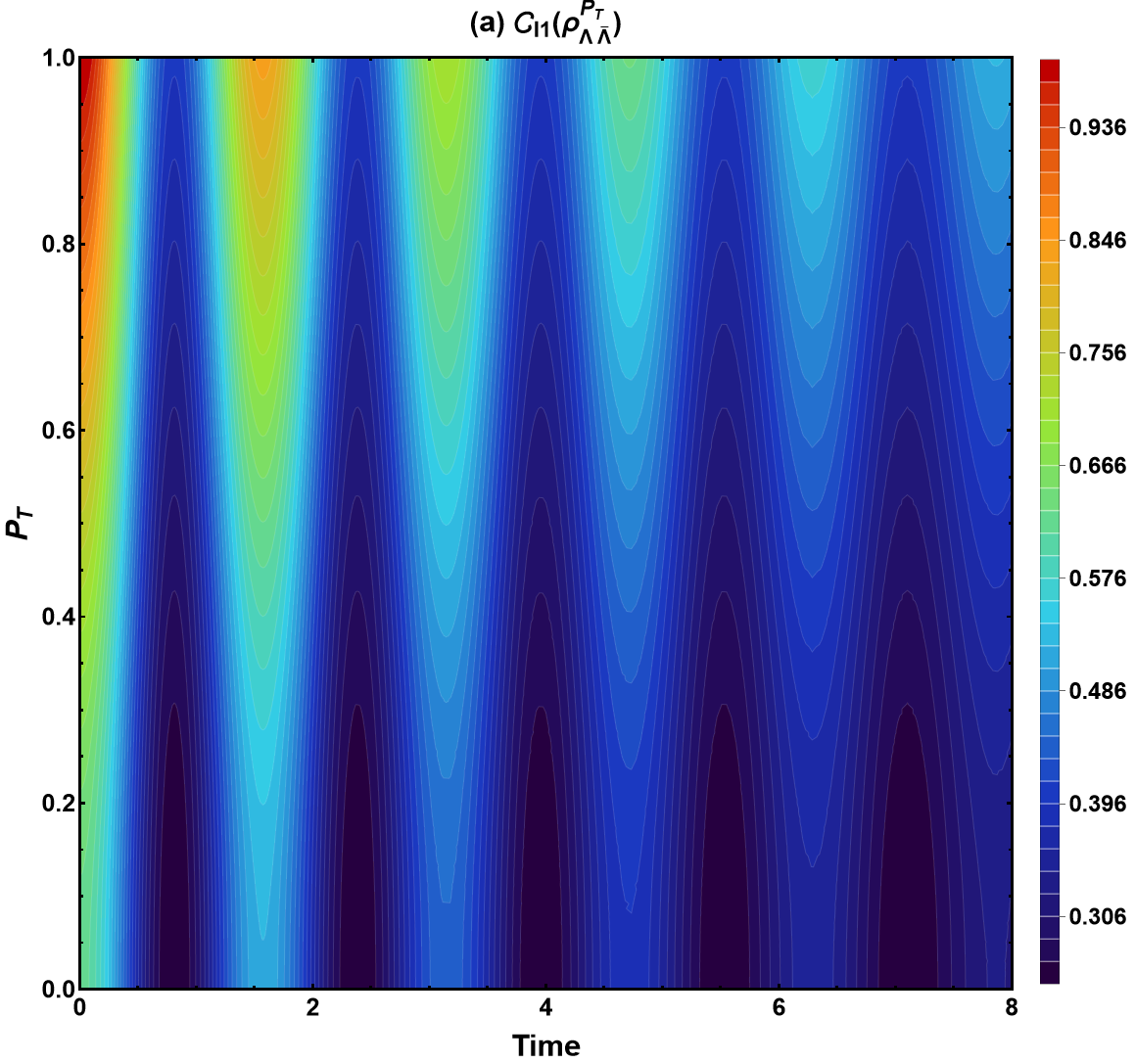}
	\includegraphics[width=0.24\linewidth]{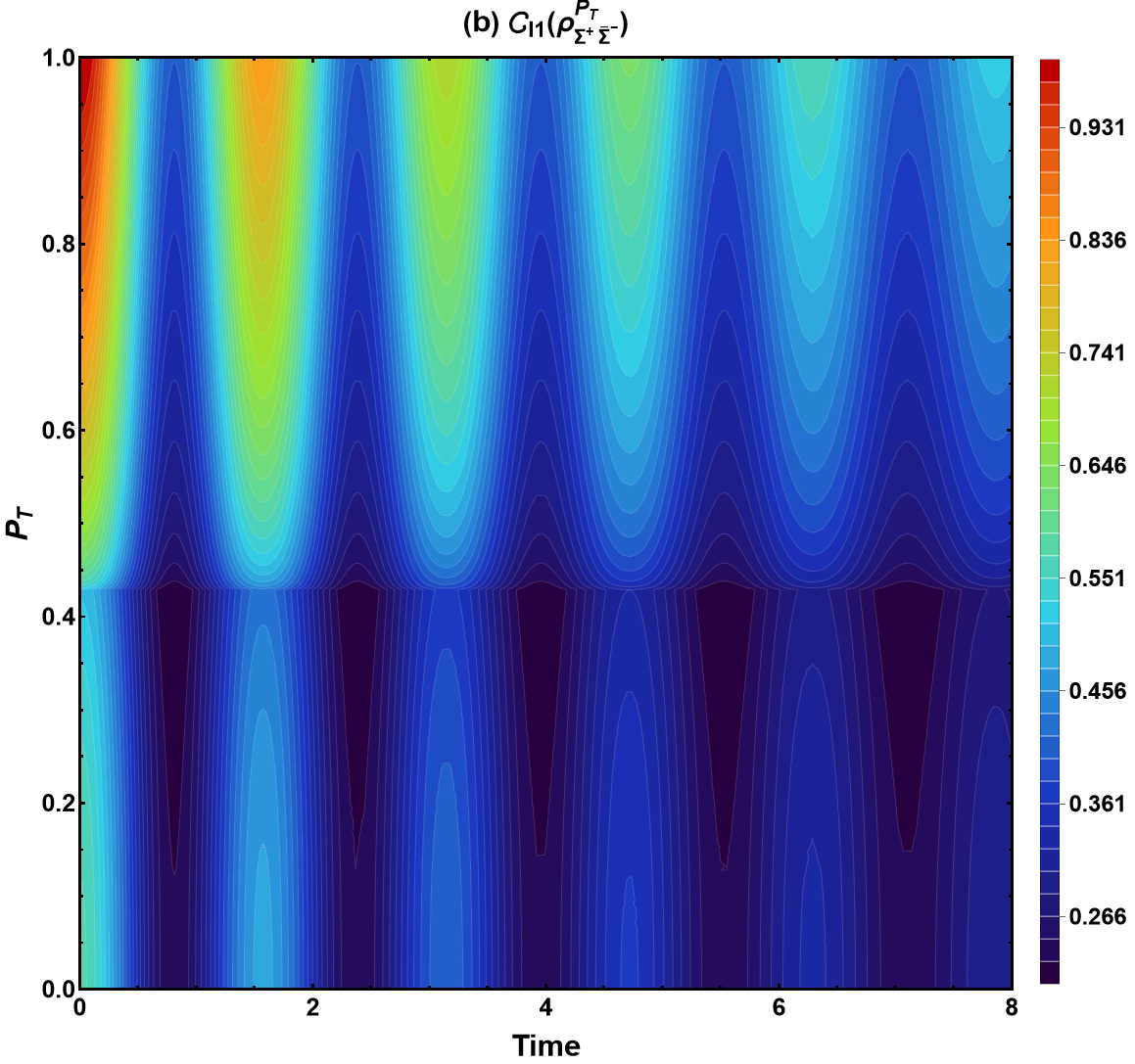}
	\includegraphics[width=0.24\linewidth]{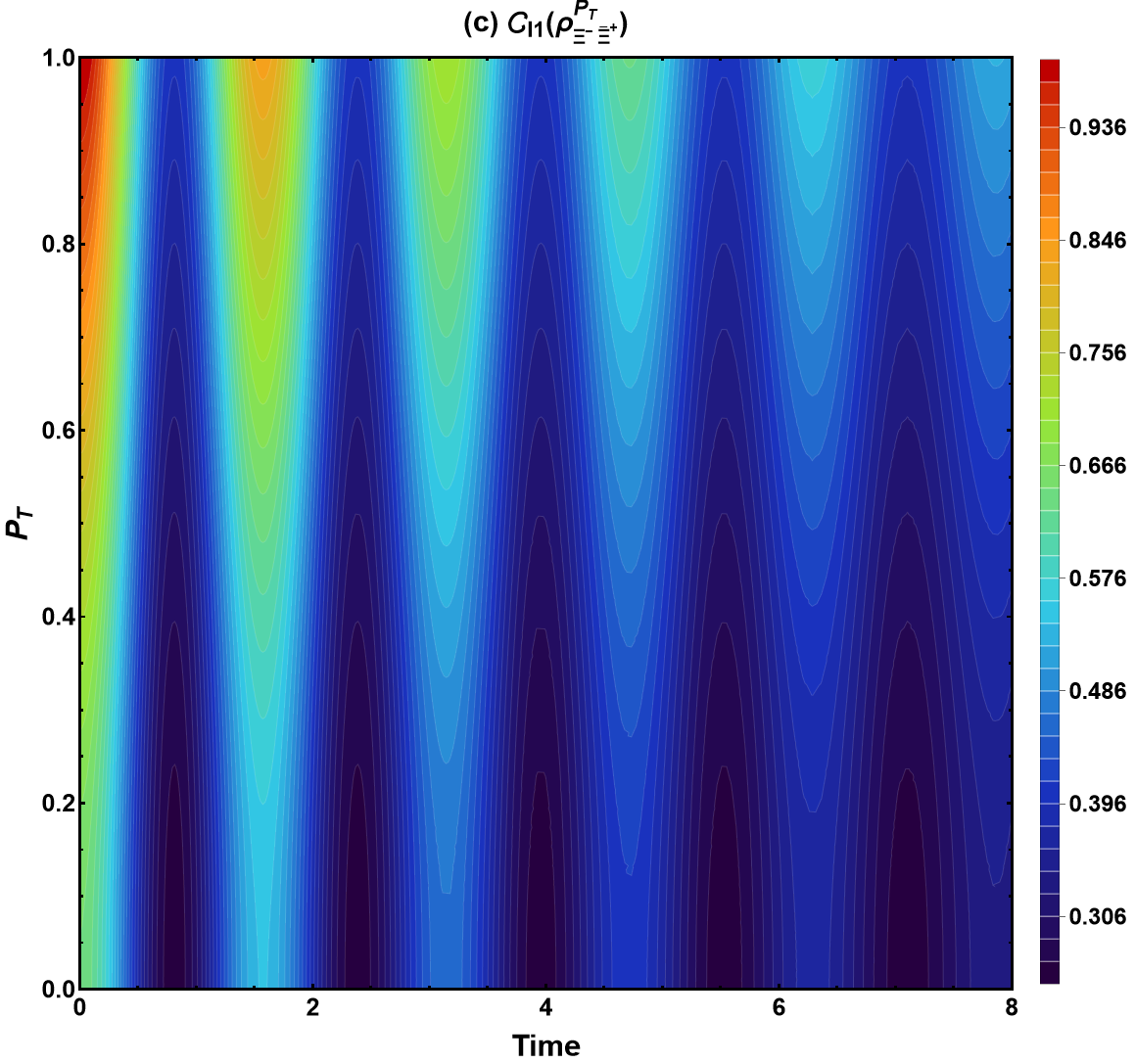}
	\includegraphics[width=0.24\linewidth]{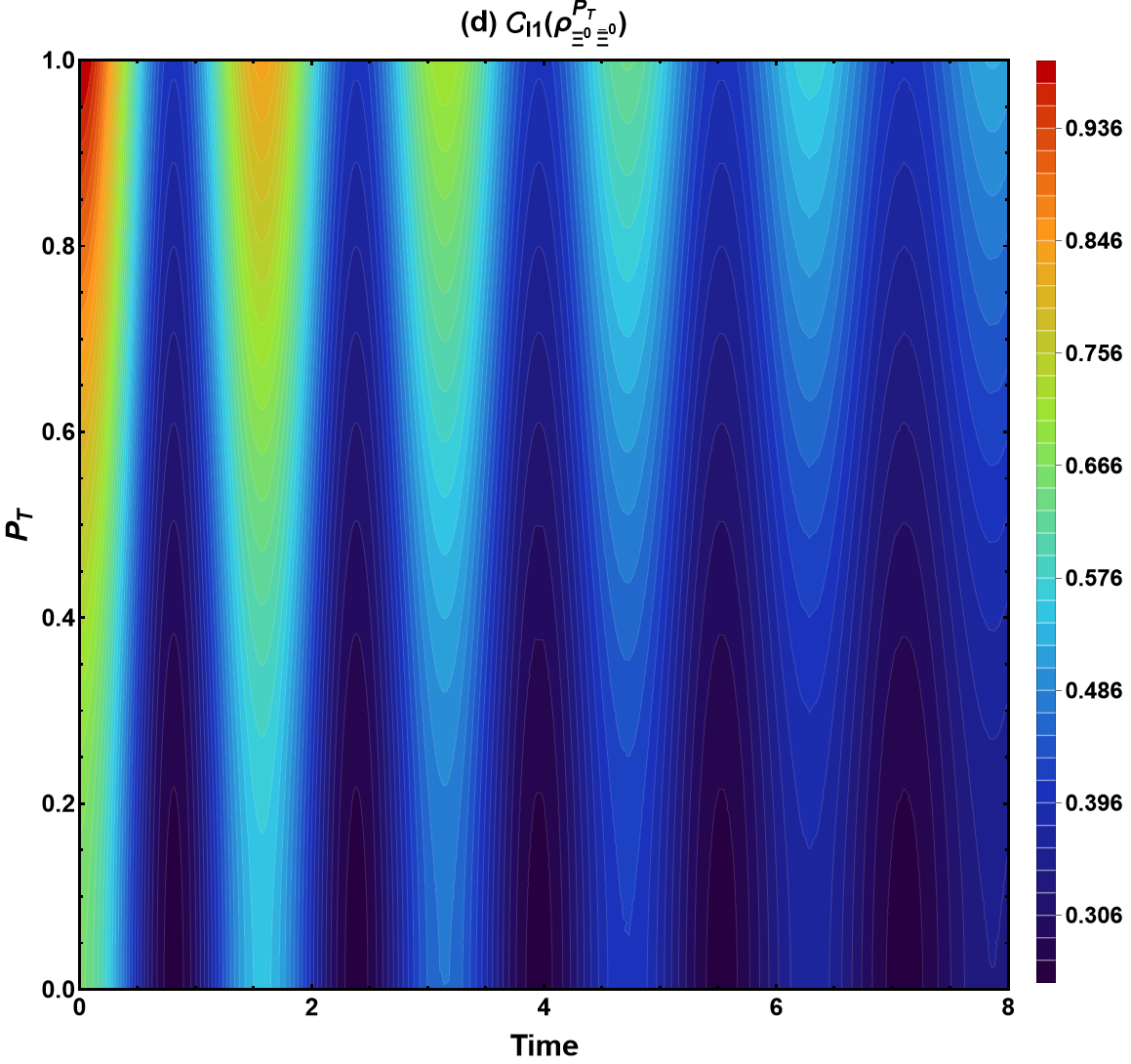}
	\includegraphics[width=0.24\linewidth]{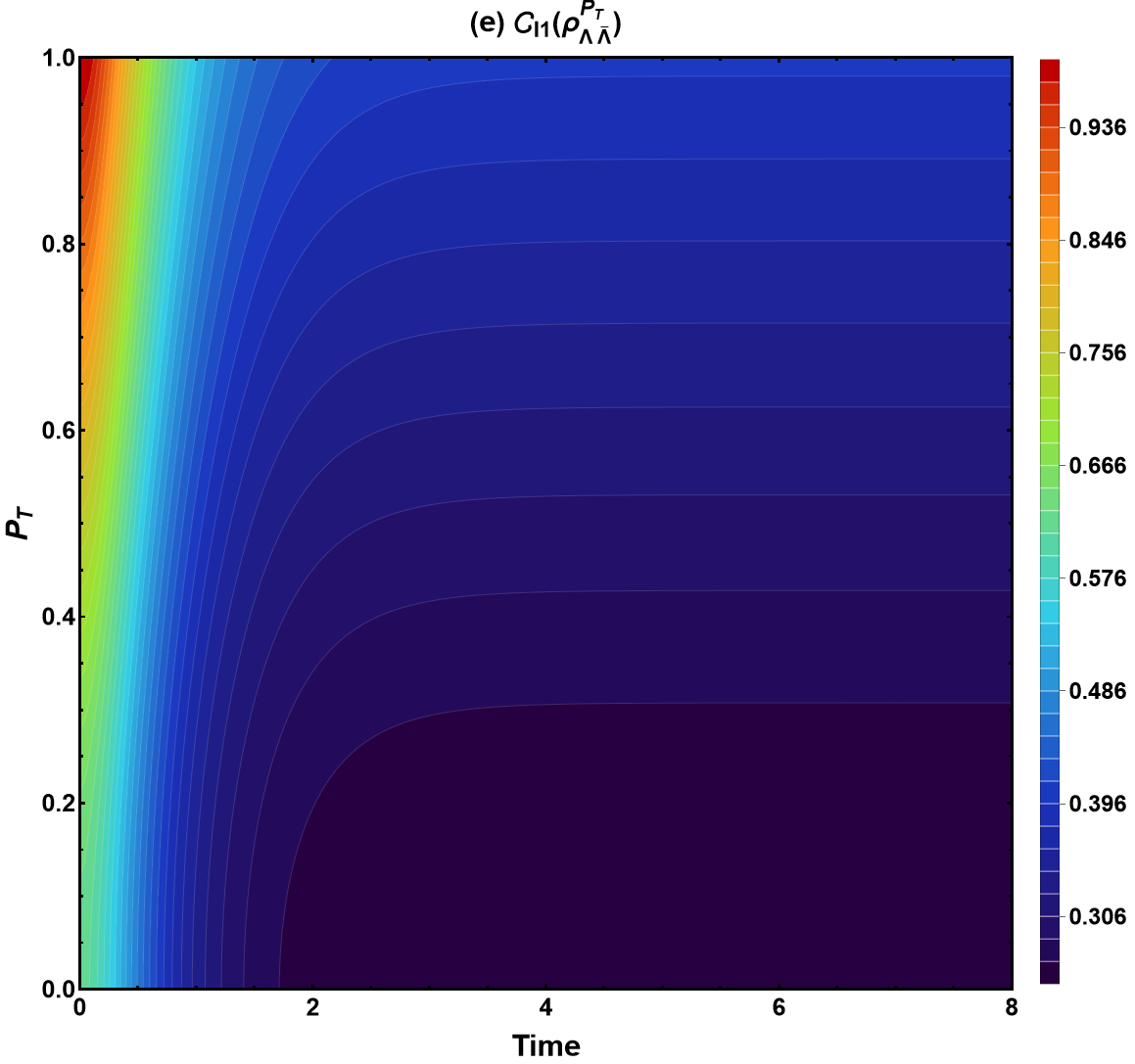}
	\includegraphics[width=0.24\linewidth]{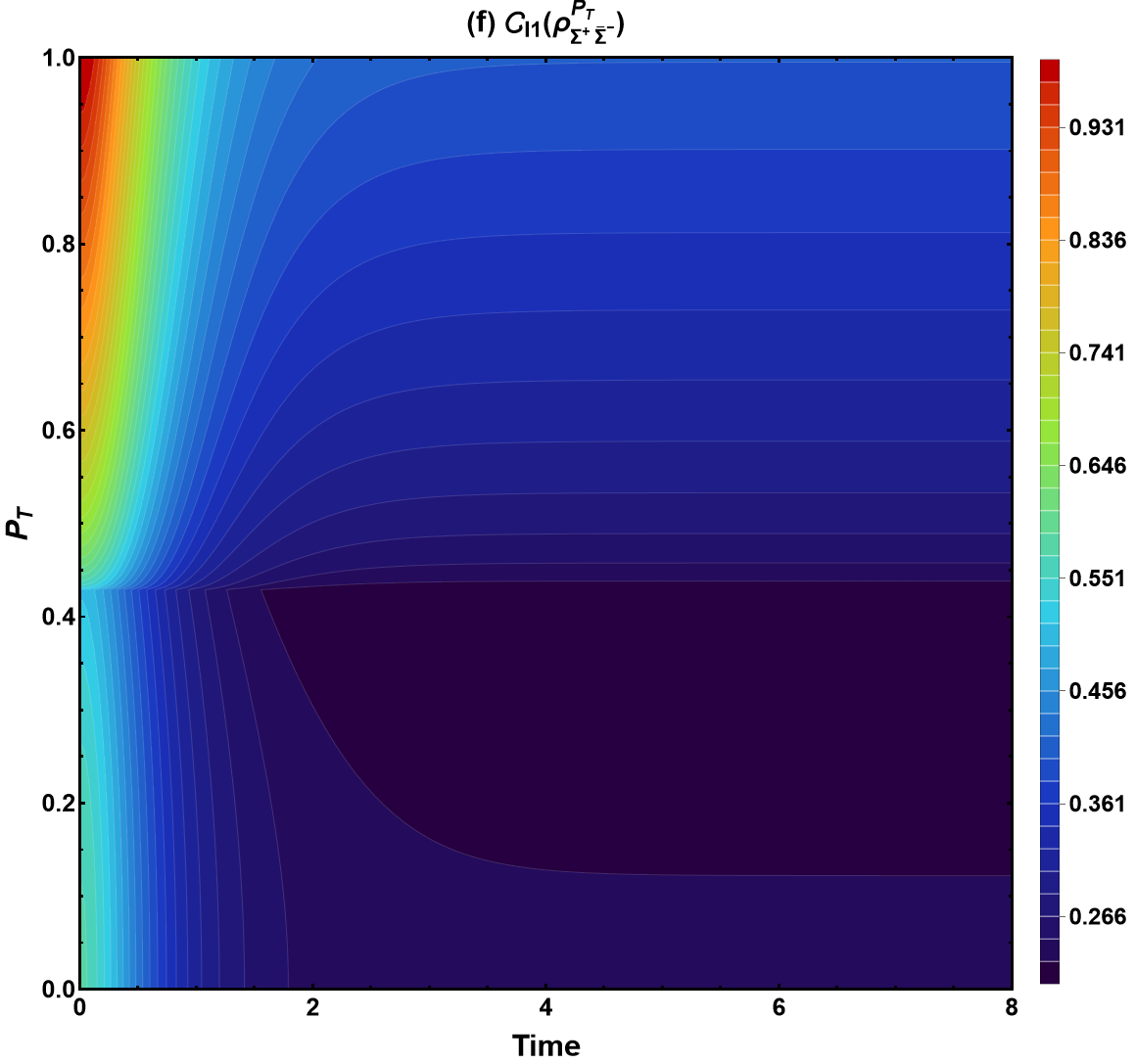}
	\includegraphics[width=0.24\linewidth]{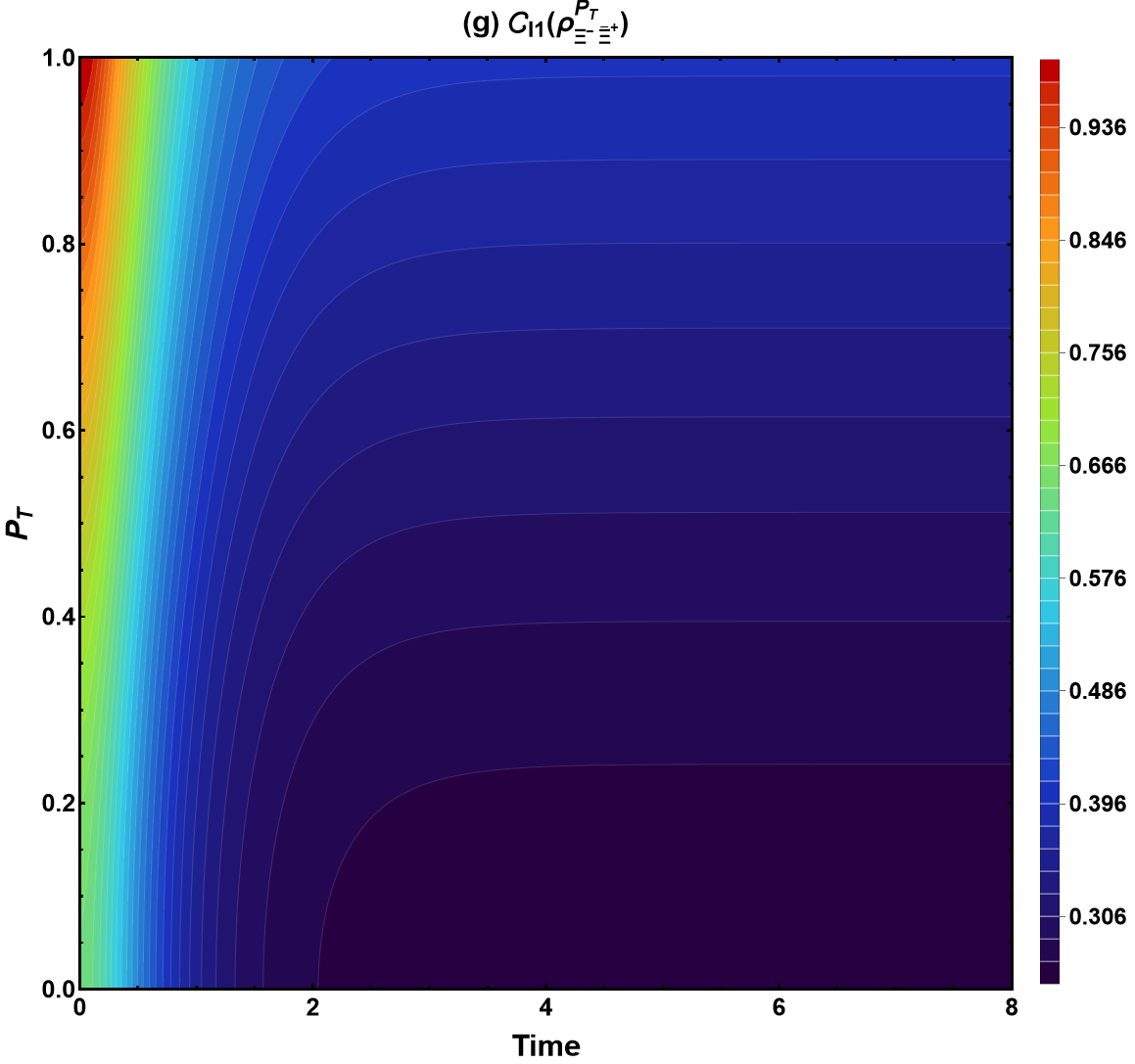}
	\includegraphics[width=0.24\linewidth]{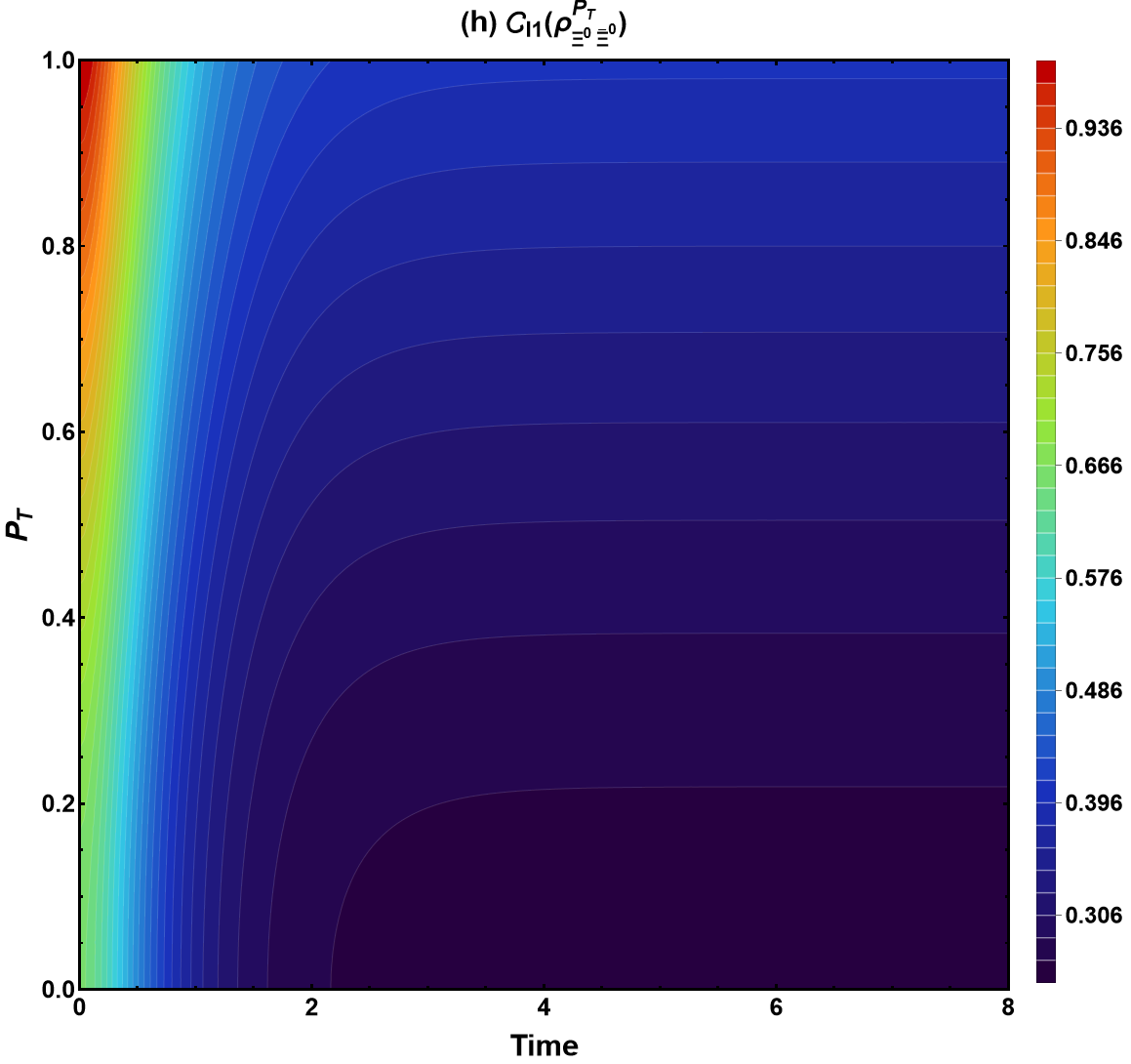}
\caption{
	Dynamical evolution of the $l_1$-norm quantum coherence
	$C_{l_1}(\rho^{P_T}_{Y\bar{Y}})$ as a function of time and the
	transverse polarization degree $P_T$ for
	$J/\psi\rightarrow Y\bar{Y}$ with
	$Y=\Lambda$, $\Sigma^{+}$, $\Xi^{-}$, and $\Xi^{0}$ at
	$\cos\theta=0.5$ and $\phi=0$. Panels (a)--(d) [(e)--(h)] correspond
	to the non-Markovian (Markovian) regime  with $\tau=5$ ($\tau=0.2$) and
	$\mu=0.4$. The experimental parameters are taken from
	Table~\ref{tab:BESIII}.
}
	\label{fig23}
\end{figure}

Figure~\ref{fig23} shows the evolution of the $l_{1}$-norm quantum
coherence $C_{l_1}(\rho^{P_T}_{Y\bar{Y}})$ as a function of the
transverse polarization degree $P_T$ and time for the four hyperon
channels. The upper panels correspond to the non-Markovian regime,
whereas the lower panels represent the Markovian dynamics.
For all channels, the coherence increases with the transverse
polarization, reaching its largest values for highly polarized beams.
In the non-Markovian regime, $C_{l_1}$ exhibits damped oscillations,
which originate from the temporary backflow of information from the
environment to the hyperon--antihyperon system. Although the revival
amplitudes decrease with time, a significant amount of coherence
remains preserved during the evolution.
In the Markovian regime, the oscillatory behavior disappears and the
coherence relaxes smoothly toward stationary values. Nevertheless, the
asymptotic coherence remains relatively large, demonstrating the strong
resilience of quantum superposition against correlated dephasing. The
channels $\Lambda\bar{\Lambda}$, $\Xi^{-}\bar{\Xi}^{+}$, and
$\Xi^{0}\bar{\Xi}^{0}$ exhibit nearly identical coherence patterns,
whereas the $\Sigma^{+}\bar{\Sigma}^{-}$ channel displays a distinct
low-coherence region around intermediate polarization values,
indicating a stronger sensitivity to environmental noise.
Compared with the corresponding results for logarithmic negativity and
geometric quantum discord, the $l_{1}$-norm coherence is the most
robust quantum resource, remaining appreciable even in parameter
regions where entanglement is strongly suppressed.
\begin{figure}[H]
	\centering
	\includegraphics[width=0.24\linewidth]{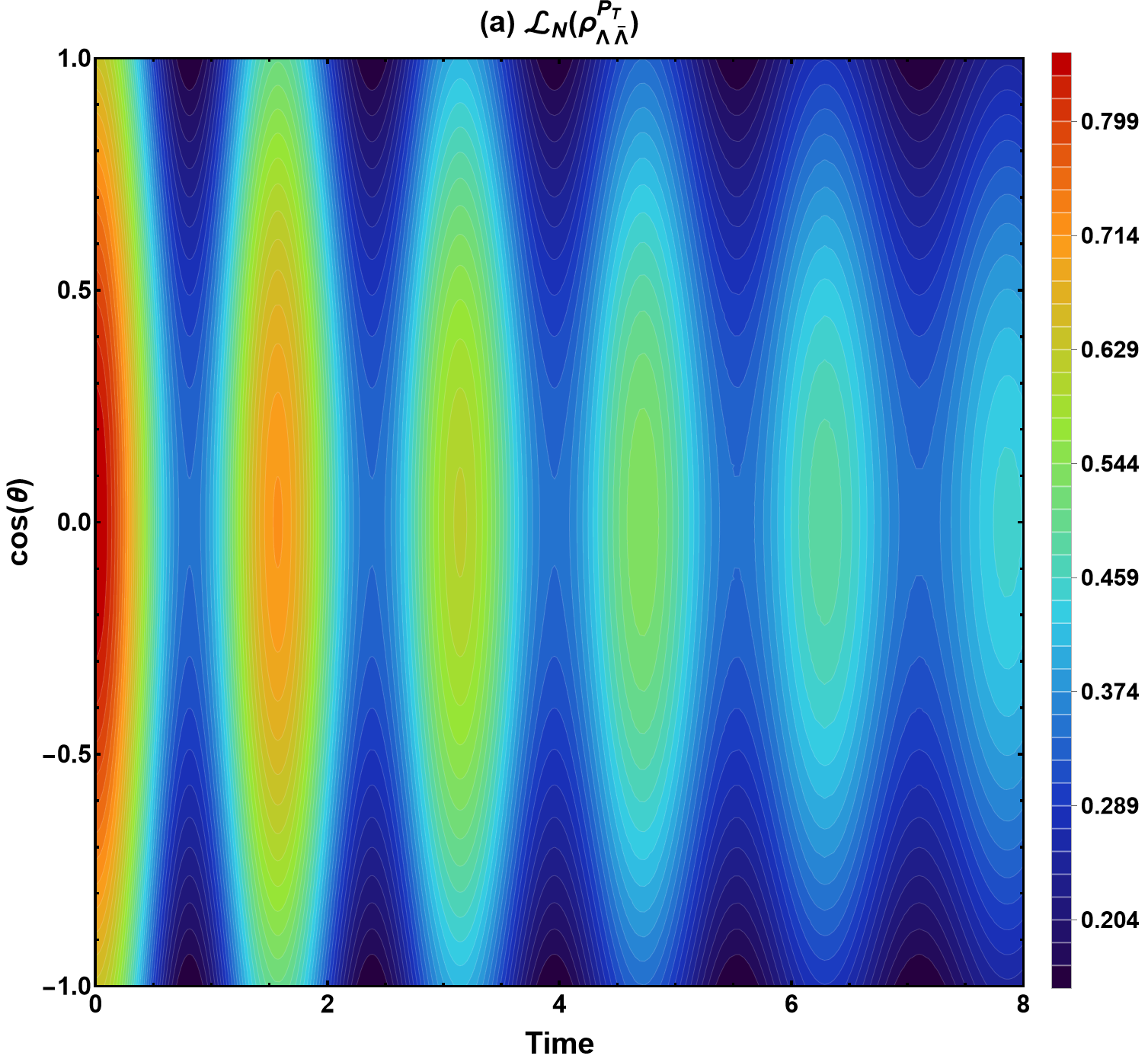}
	\includegraphics[width=0.24\linewidth]{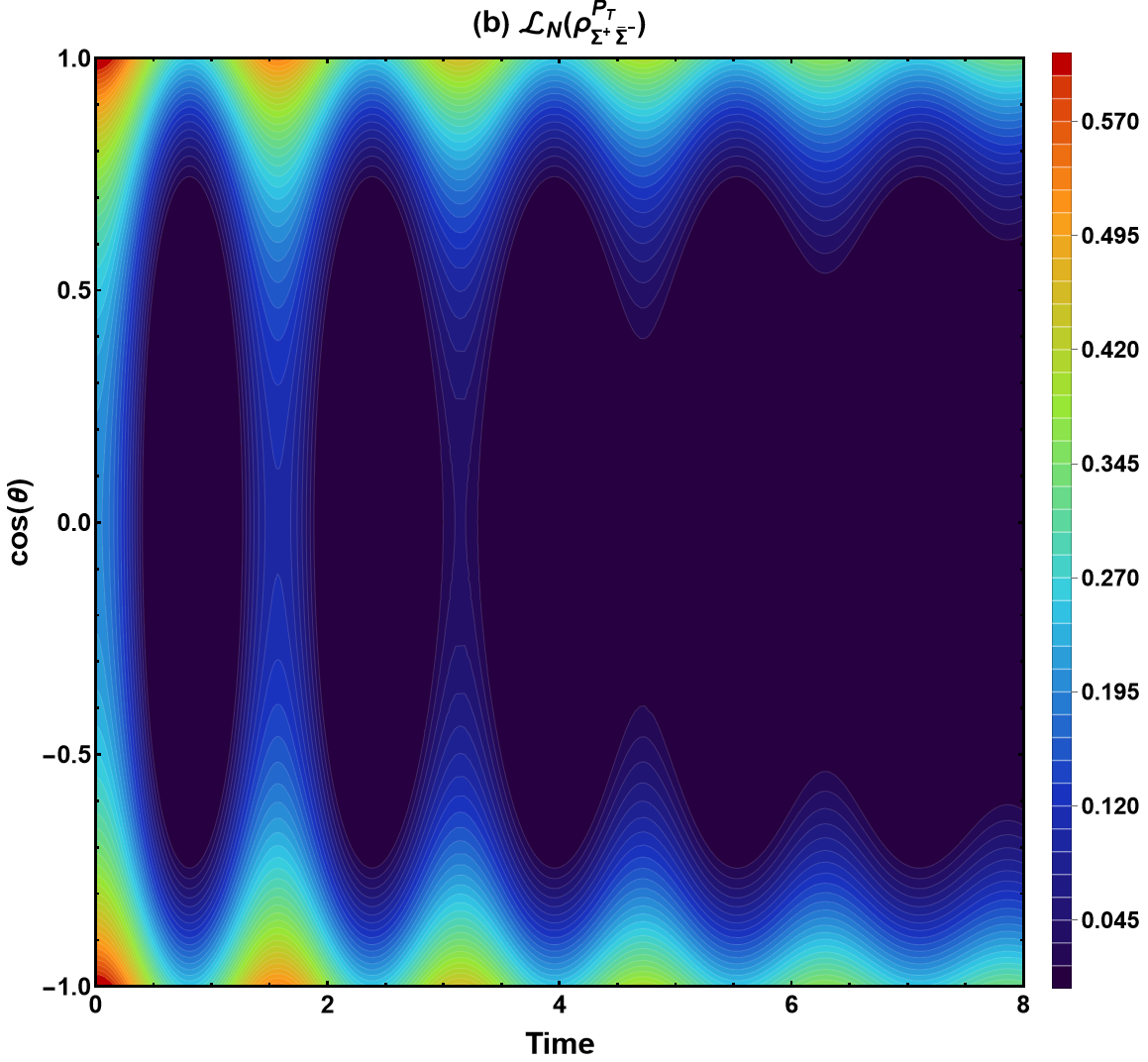}
	\includegraphics[width=0.24\linewidth]{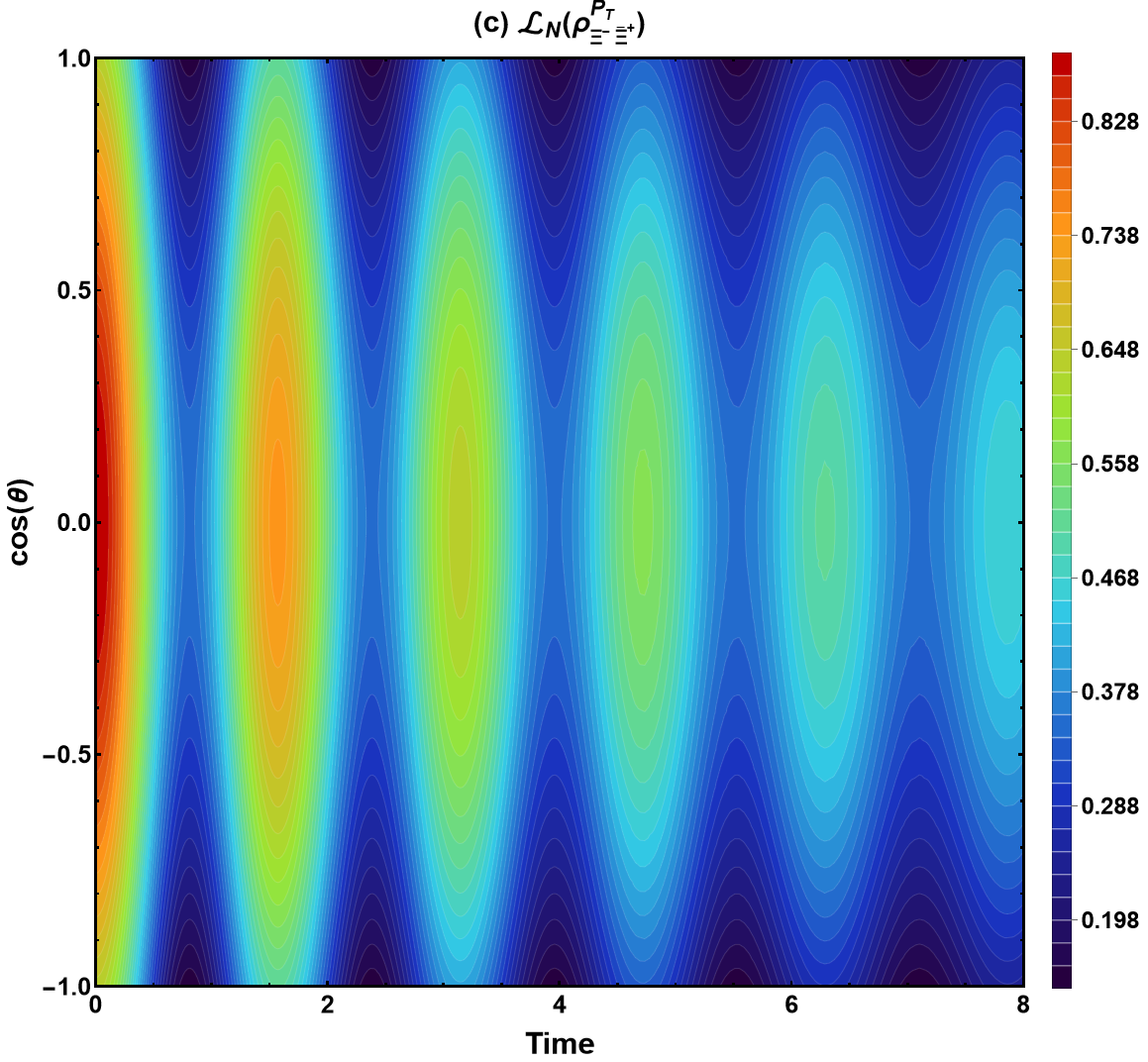}
	\includegraphics[width=0.24\linewidth]{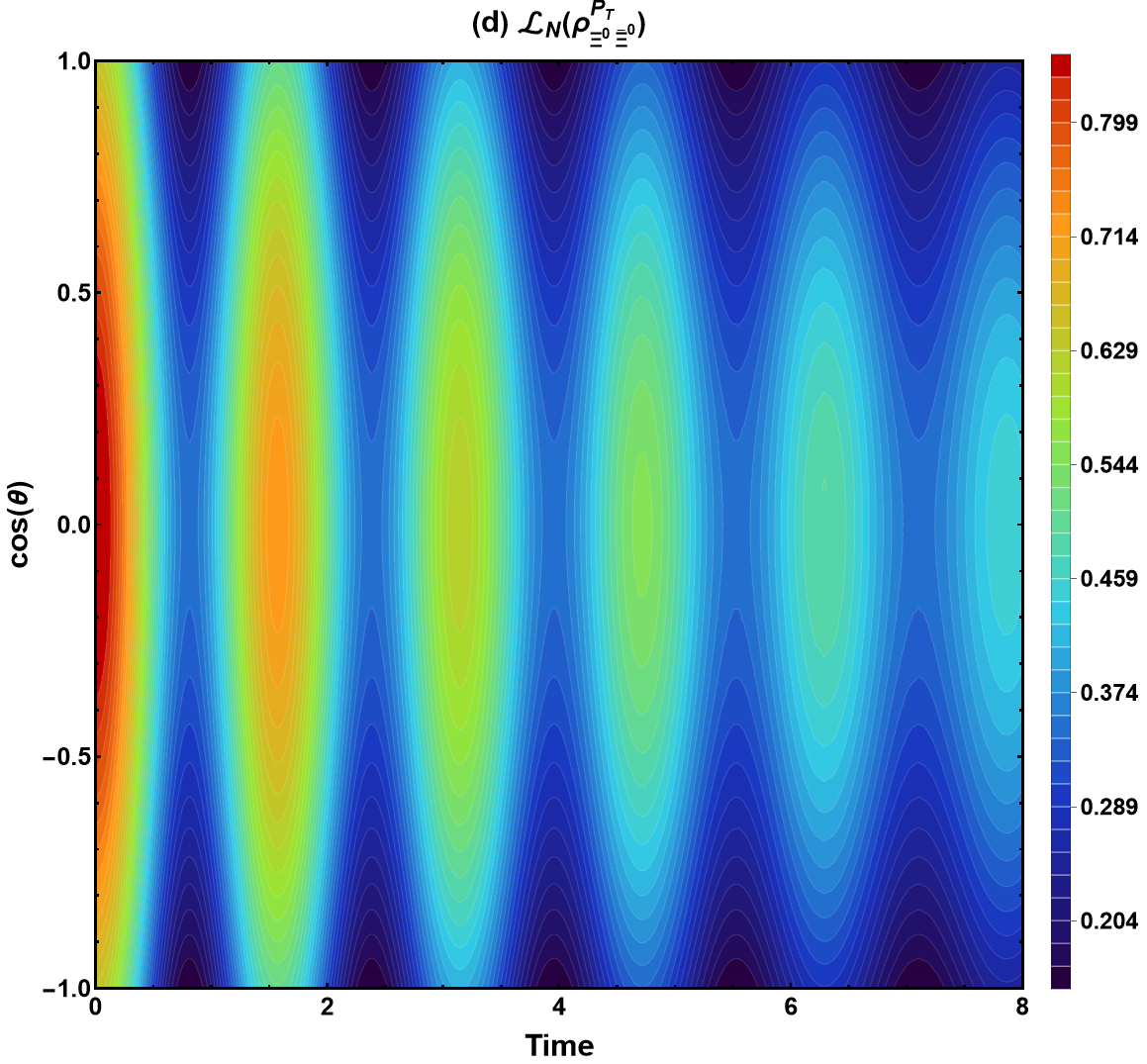}
	\includegraphics[width=0.24\linewidth]{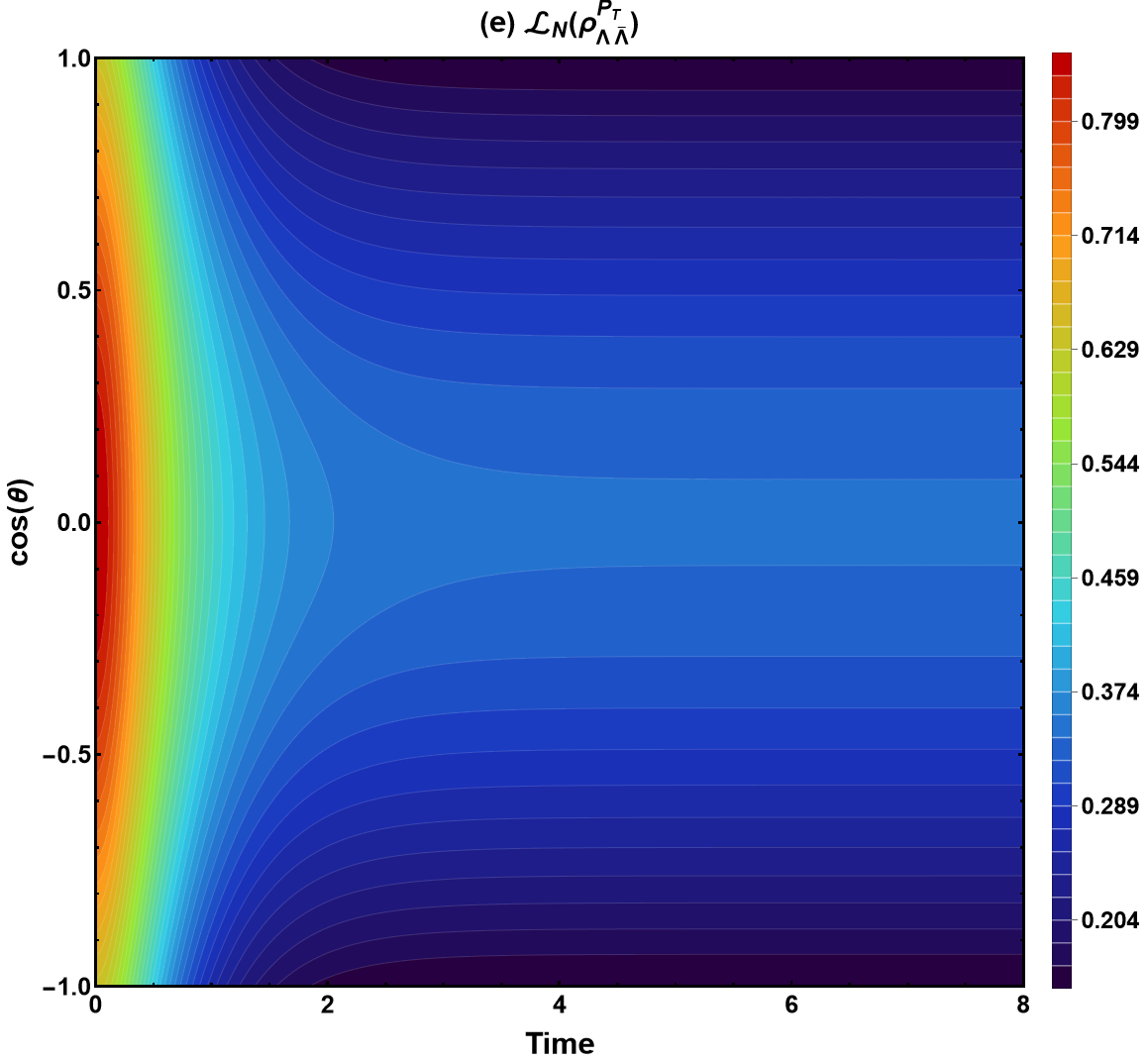}
	\includegraphics[width=0.24\linewidth]{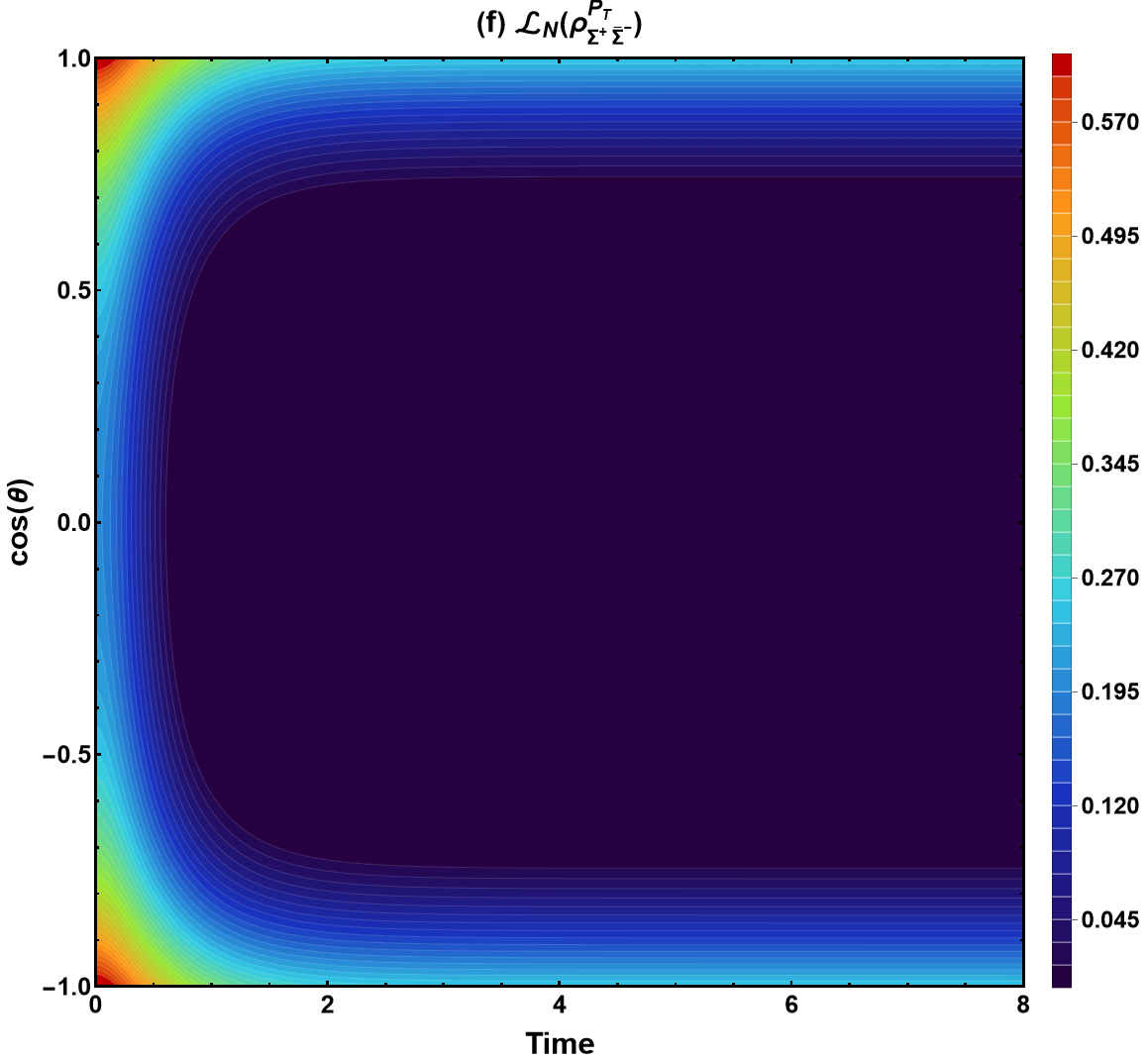}
	\includegraphics[width=0.24\linewidth]{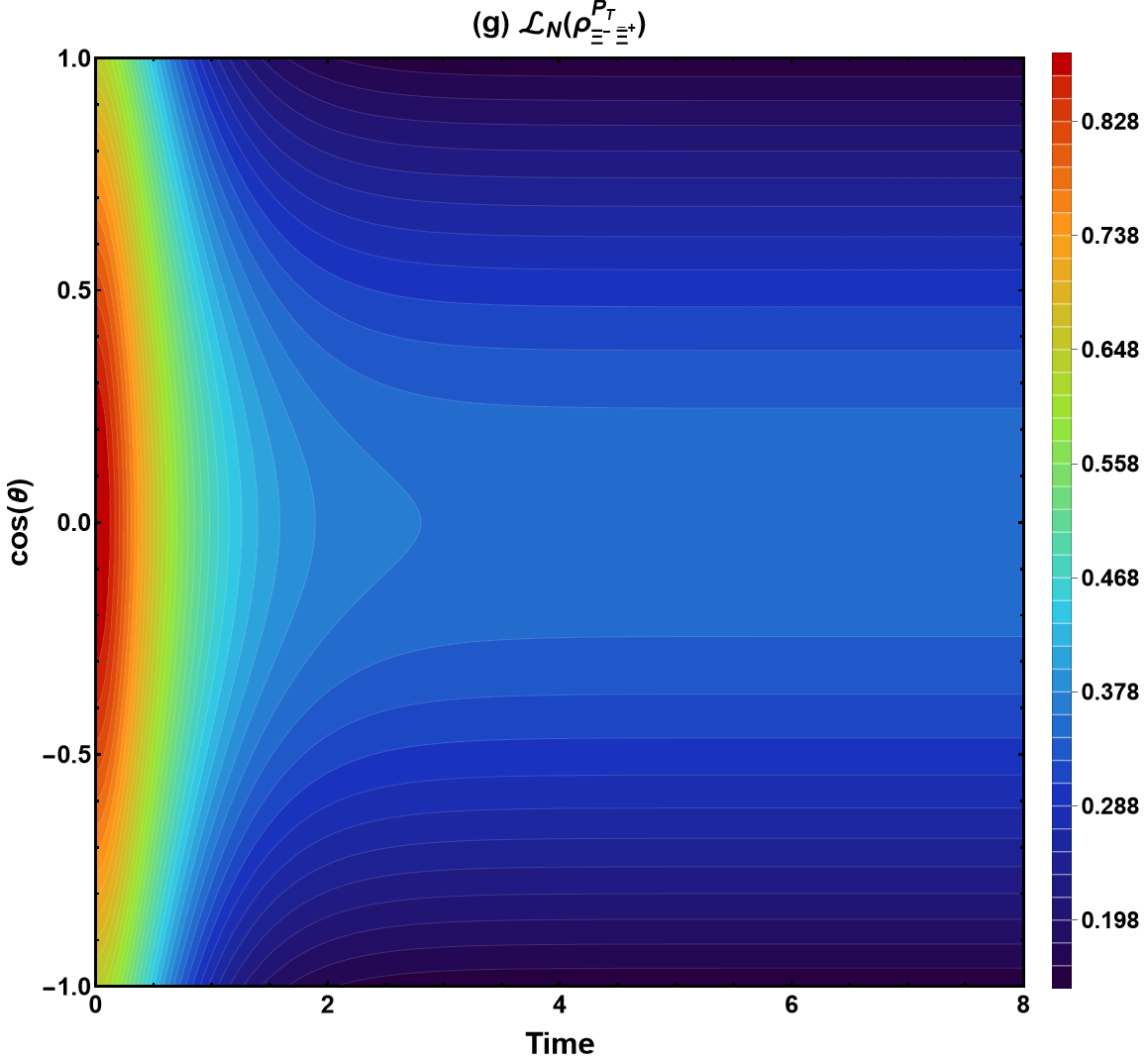}
	\includegraphics[width=0.24\linewidth]{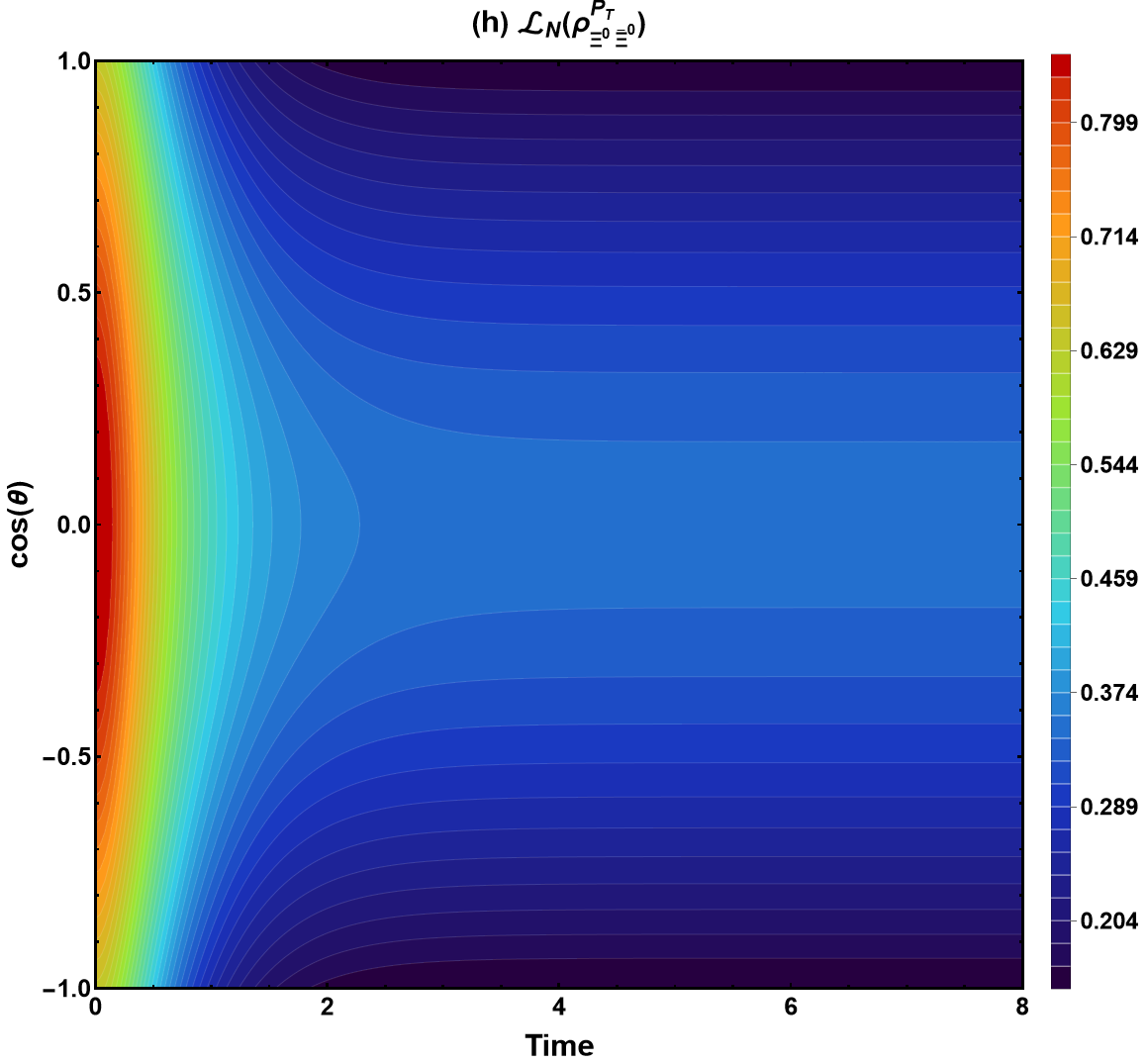}
\caption{
	Dynamical evolution of the logarithmic negativity
	$\mathcal{L}_N(\rho^{P_T}_{Y\bar{Y}})$ as a function of time and the
	production angle $\cos\theta$ for
	$J/\psi\rightarrow Y\bar{Y}$ with
	$Y=\Lambda$, $\Sigma^{+}$, $\Xi^{-}$, and $\Xi^{0}$ at
	$P_T=0.8$ and $\phi=0$. Panels (a)--(d) [(e)--(h)] correspond to the
	non-Markovian (Markovian) regime  with $\tau=5$ ($\tau=0.2$) and
	$\mu=0.4$. The experimental parameters are taken from
	Table~\ref{tab:BESIII}.
}
	\label{fig18}
\end{figure}
The logarithmic negativity $\mathcal{L}_{N}(\rho^{P_T}_{Y\bar{Y}})$ is depicted in Fig.~\ref{fig18} as a function of the production angle $\cos\theta$ and time for the four hyperon--antihyperon channels under transverse beam polarization. The
upper panels correspond to the non-Markovian regime, while the lower
panels represent the Markovian dynamics.
In the non-Markovian regime, all channels exhibit damped oscillatory
behavior characterized by successive entanglement revivals, reflecting
the backflow of information from the environment to the system. The
channels $\Lambda\bar{\Lambda}$, $\Xi^{-}\bar{\Xi}^{+}$, and
$\Xi^{0}\bar{\Xi}^{0}$ display very similar patterns, with the largest
entanglement occurring around the central angular region
$\cos\theta \simeq 0$. By contrast, the
$\Sigma^{+}\bar{\Sigma}^{-}$ channel exhibits a markedly different
distribution, where entanglement is mainly concentrated near
$\cos\theta \simeq \pm1$ and remains strongly suppressed around the
central region.
In the Markovian regime, the revival structure disappears and the
entanglement decays monotonically toward stationary values. Although a
finite amount of entanglement survives at long times for most channels,
its magnitude is substantially reduced compared with the non-Markovian
case. The $\Sigma^{+}\bar{\Sigma}^{-}$ channel remains the most fragile
against decoherence, whereas the $\Lambda\bar{\Lambda}$,
$\Xi^{-}\bar{\Xi}^{+}$, and $\Xi^{0}\bar{\Xi}^{0}$ channels preserve a
larger fraction of their initial quantum correlations.
\begin{figure}[H]
	\centering
	\includegraphics[width=0.24\linewidth]{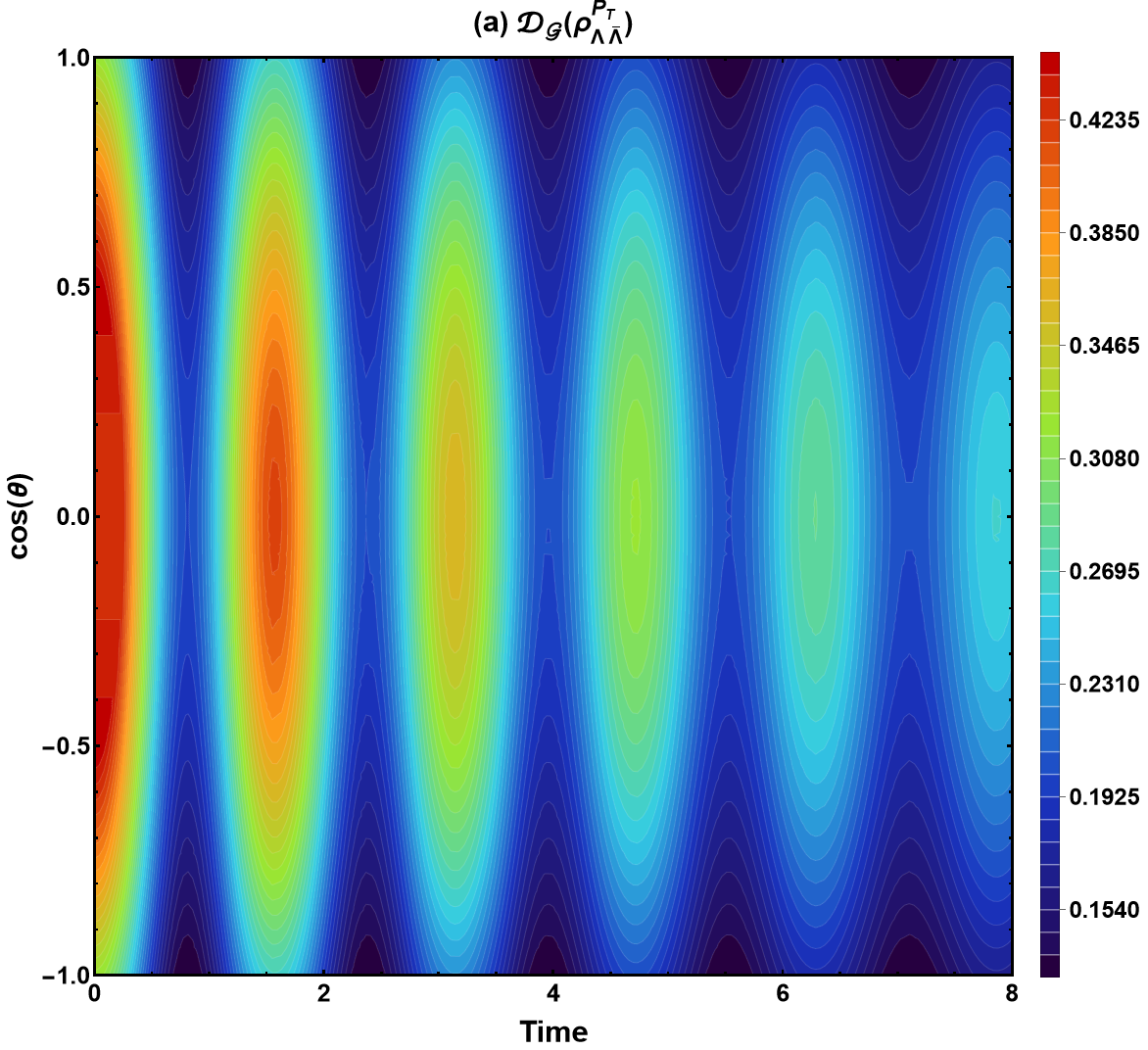}
	\includegraphics[width=0.24\linewidth]{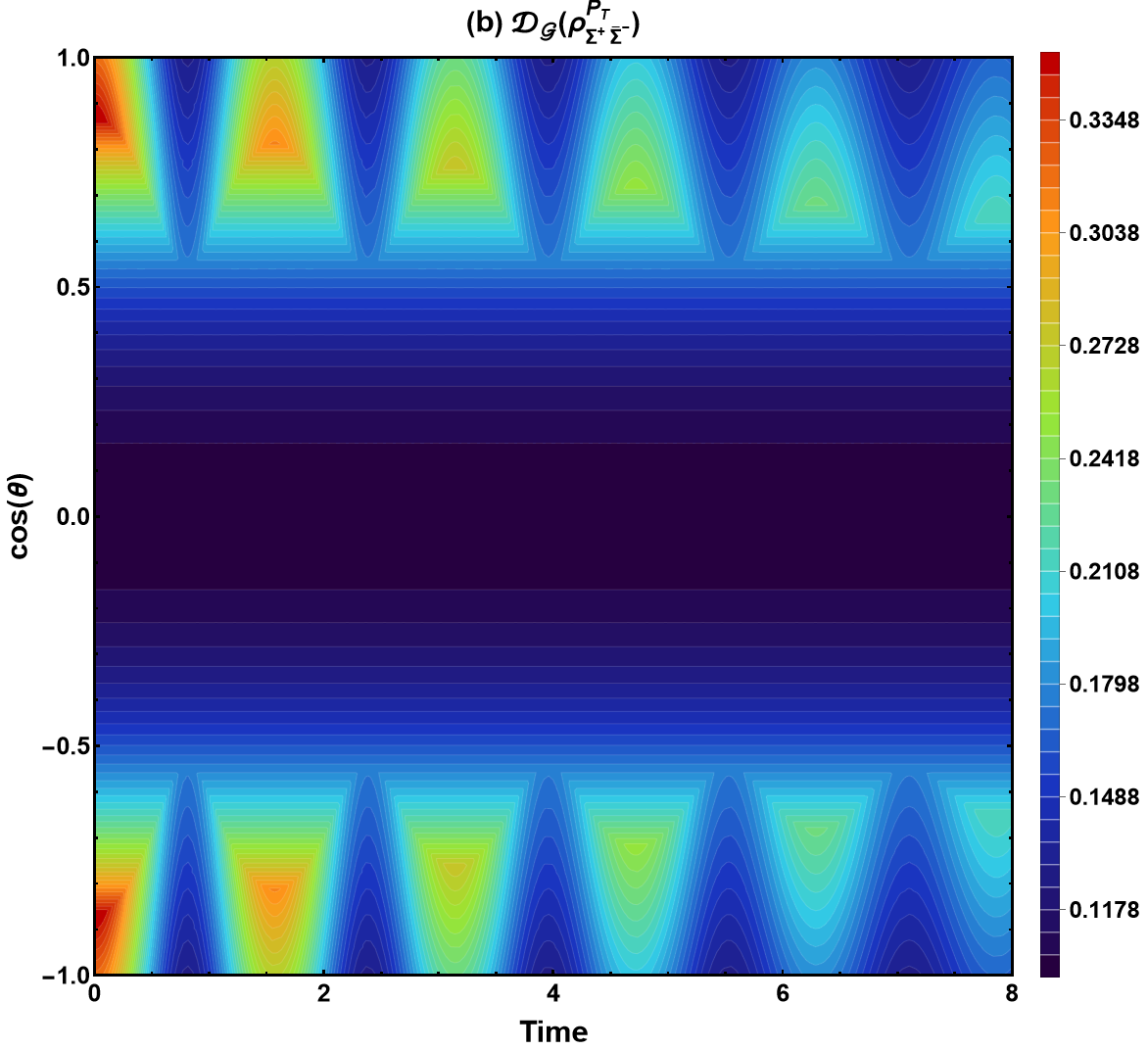}
	\includegraphics[width=0.24\linewidth]{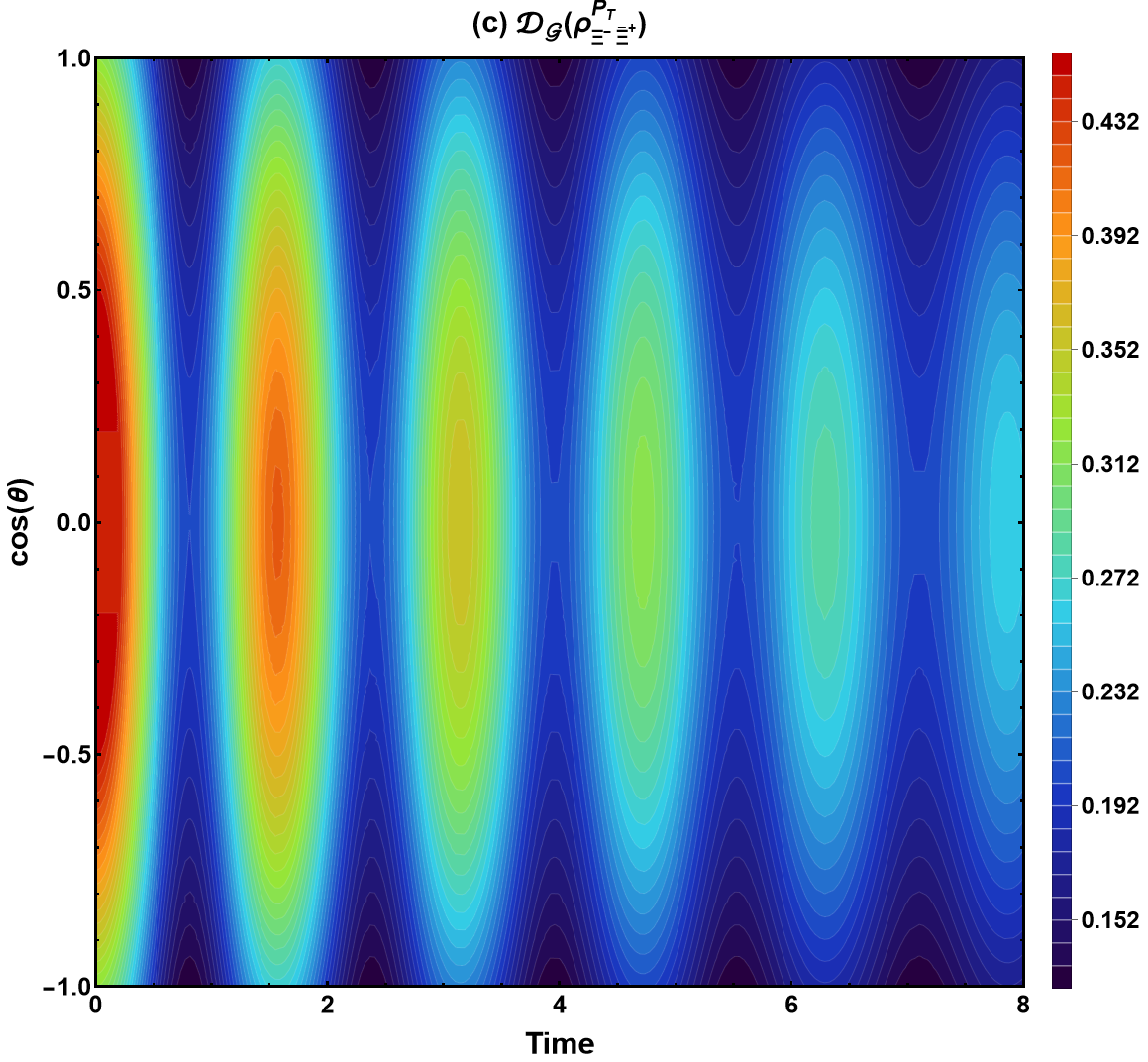}
	\includegraphics[width=0.24\linewidth]{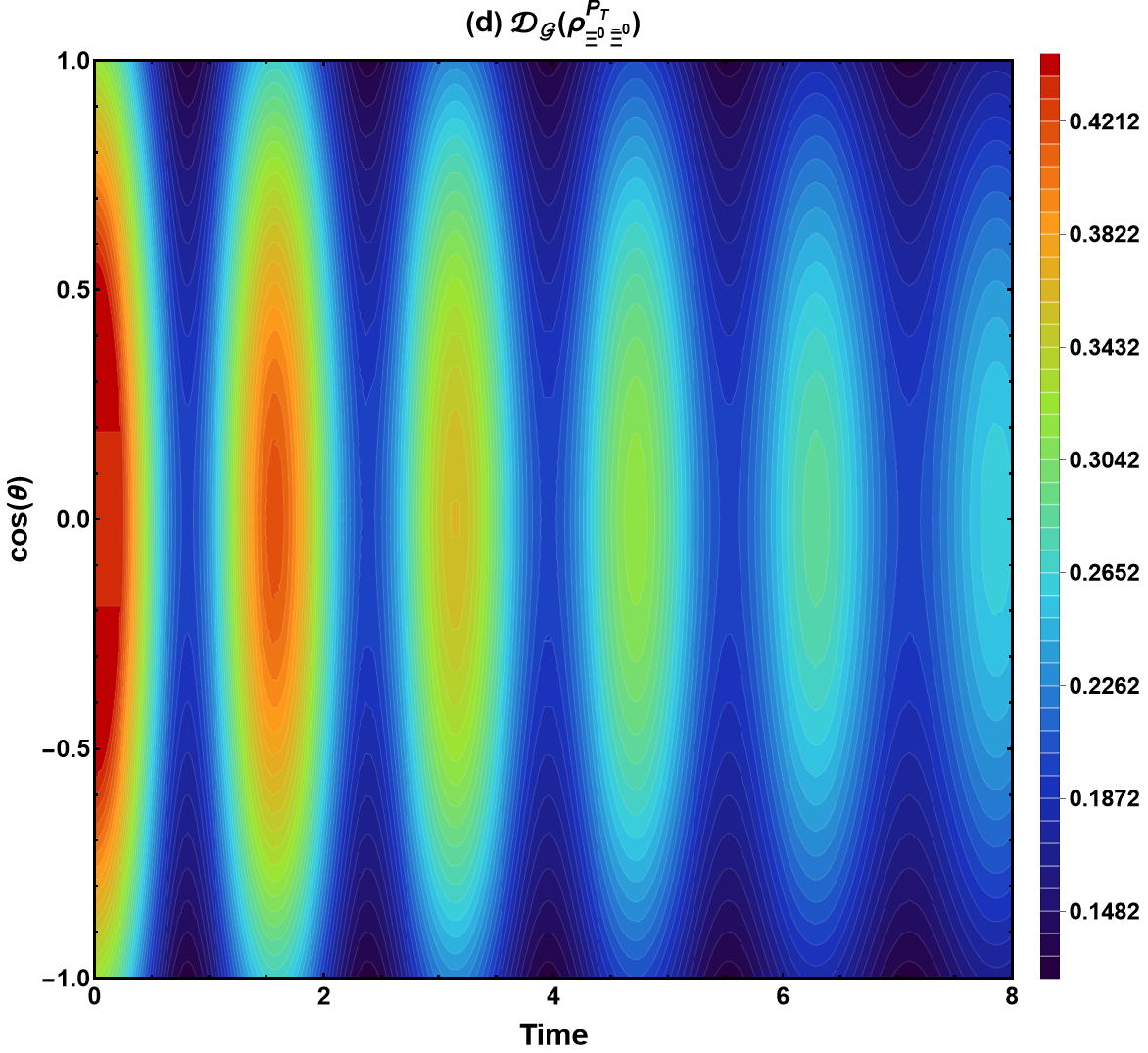}
	\includegraphics[width=0.24\linewidth]{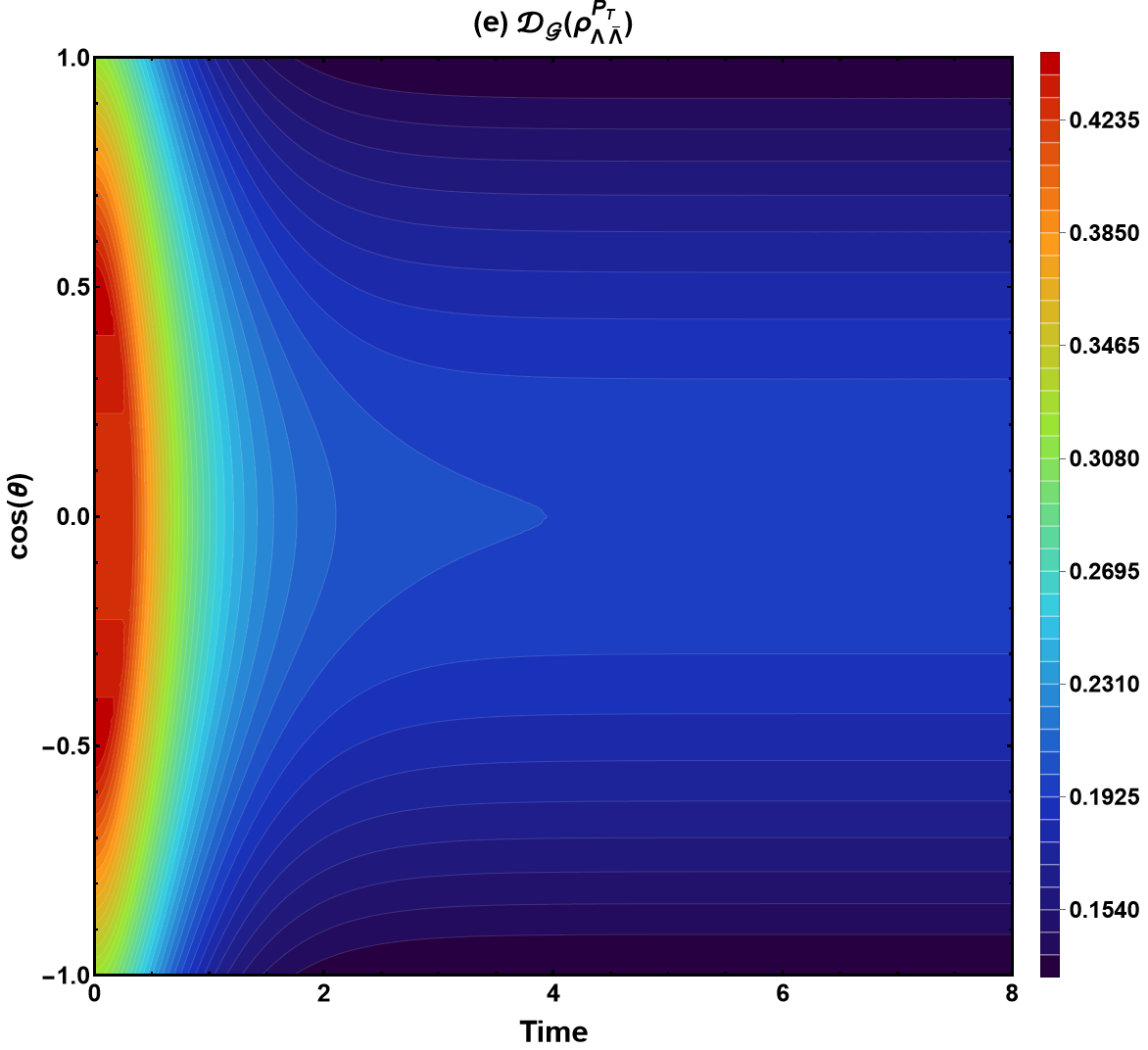}
	\includegraphics[width=0.24\linewidth]{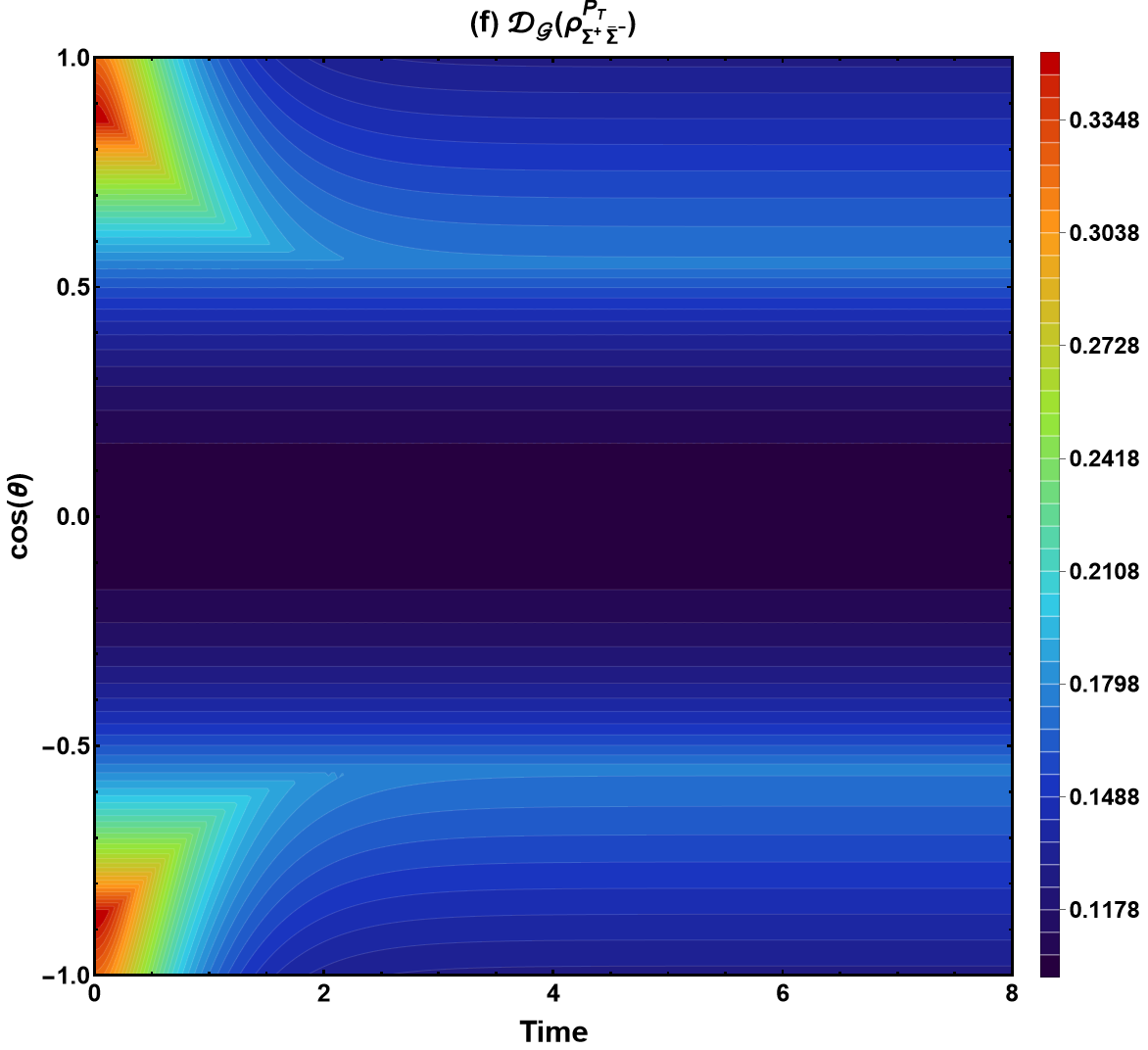}
	\includegraphics[width=0.24\linewidth]{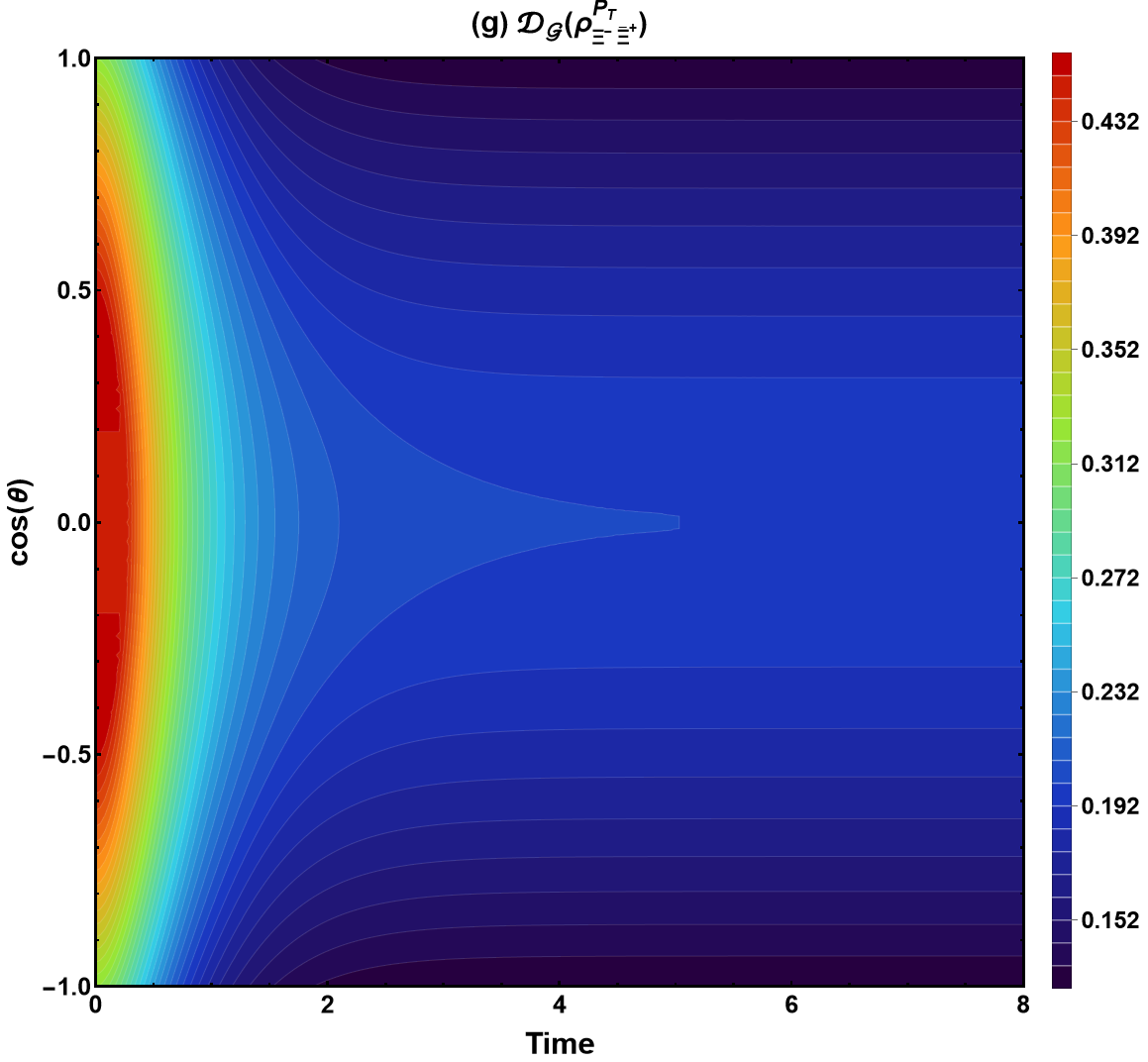}
	\includegraphics[width=0.24\linewidth]{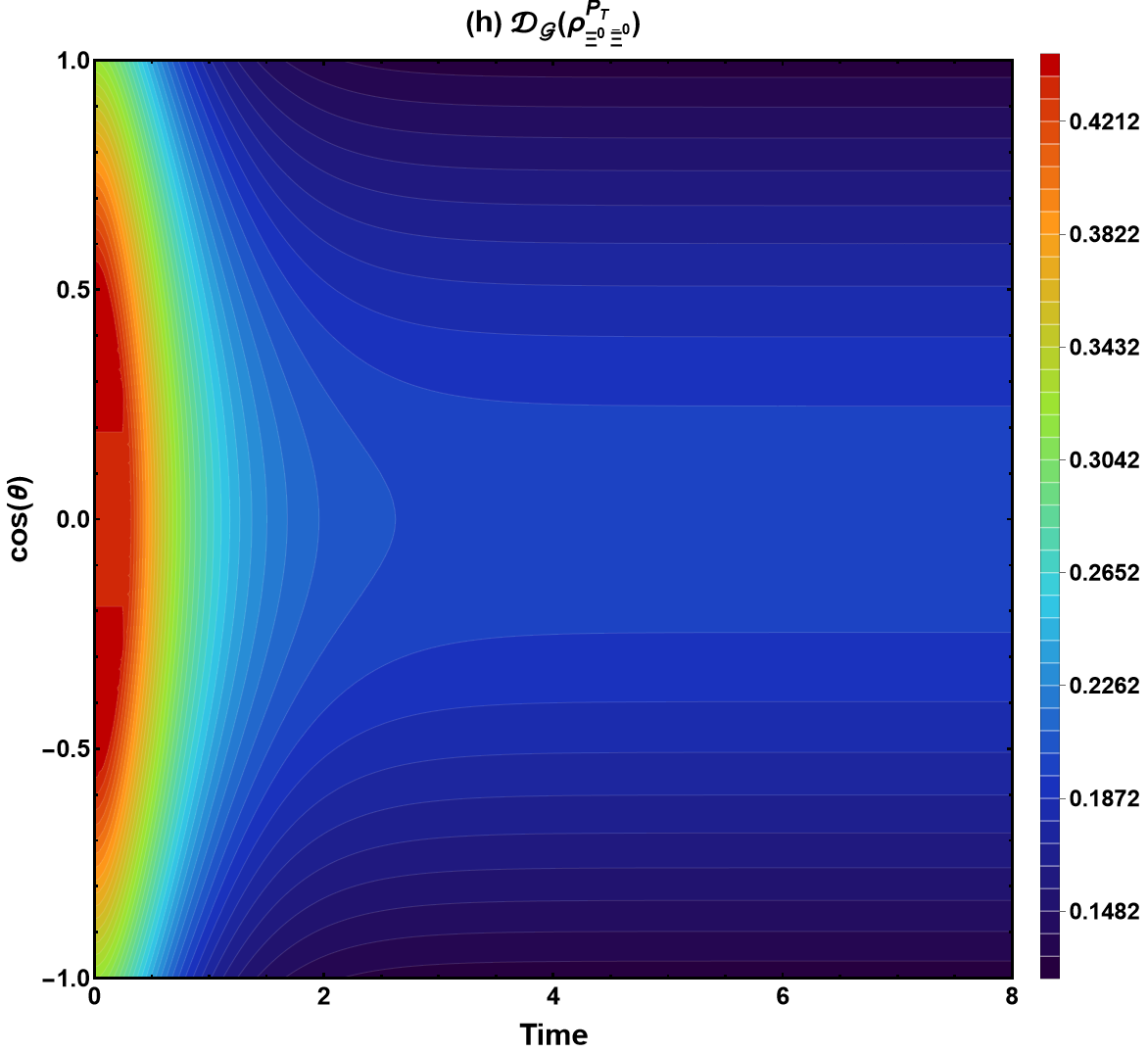}
\caption{
	Dynamical evolution of the geometric quantum discord
	$\mathcal{D}_G(\rho^{P_T}_{Y\bar{Y}})$ as a function of time and the
	production angle $\cos\theta$ for
	$J/\psi\rightarrow Y\bar{Y}$ with
	$Y=\Lambda$, $\Sigma^{+}$, $\Xi^{-}$, and $\Xi^{0}$ at
	$P_T=0.8$ and $\phi=0$. Panels (a)--(d) [(e)--(h)] correspond to the
	non-Markovian (Markovian) regime  with $\tau=5$ ($\tau=0.2$) and
	$\mu=0.4$. The experimental parameters are taken from
	Table~\ref{tab:BESIII}.
}
	\label{fig21}
\end{figure}
The geometric quantum discord $\mathcal{D}_{G}(\rho^{P_T}_{Y\bar{Y}})$ as a function of the production angle $\cos\theta$ and time under transverse beam polarization is presented in Fig.~\ref{fig21}. The upper panels correspond to the non-Markovian regime,
whereas the lower panels represent the Markovian dynamics.
In the non-Markovian regime, the
$\Lambda\bar{\Lambda}$, $\Xi^{-}\bar{\Xi}^{+}$, and
$\Xi^{0}\bar{\Xi}^{0}$ channels exhibit damped oscillatory patterns
associated with information backflow from the environment. The largest
values of geometric discord are concentrated around the central angular
region $\cos\theta \simeq 0$, where pronounced revival structures
persist over long evolution times. In contrast, the
$\Sigma^{+}\bar{\Sigma}^{-}$ channel displays a distinct angular
dependence, with quantum correlations mainly localized near
$\cos\theta \simeq \pm1$ and significantly reduced around the central
region.
For Markovian dynamics, the oscillatory behavior disappears and the
discord relaxes monotonically toward stationary distributions.
Nevertheless, a substantial amount of discord survives at long times,
demonstrating a considerably higher robustness against decoherence than
that observed for logarithmic negativity.  While the
$\Lambda\bar{\Lambda}$, $\Xi^{-}\bar{\Xi}^{+}$, and
\begin{figure}[H]
	\centering
	\includegraphics[width=0.24\linewidth]{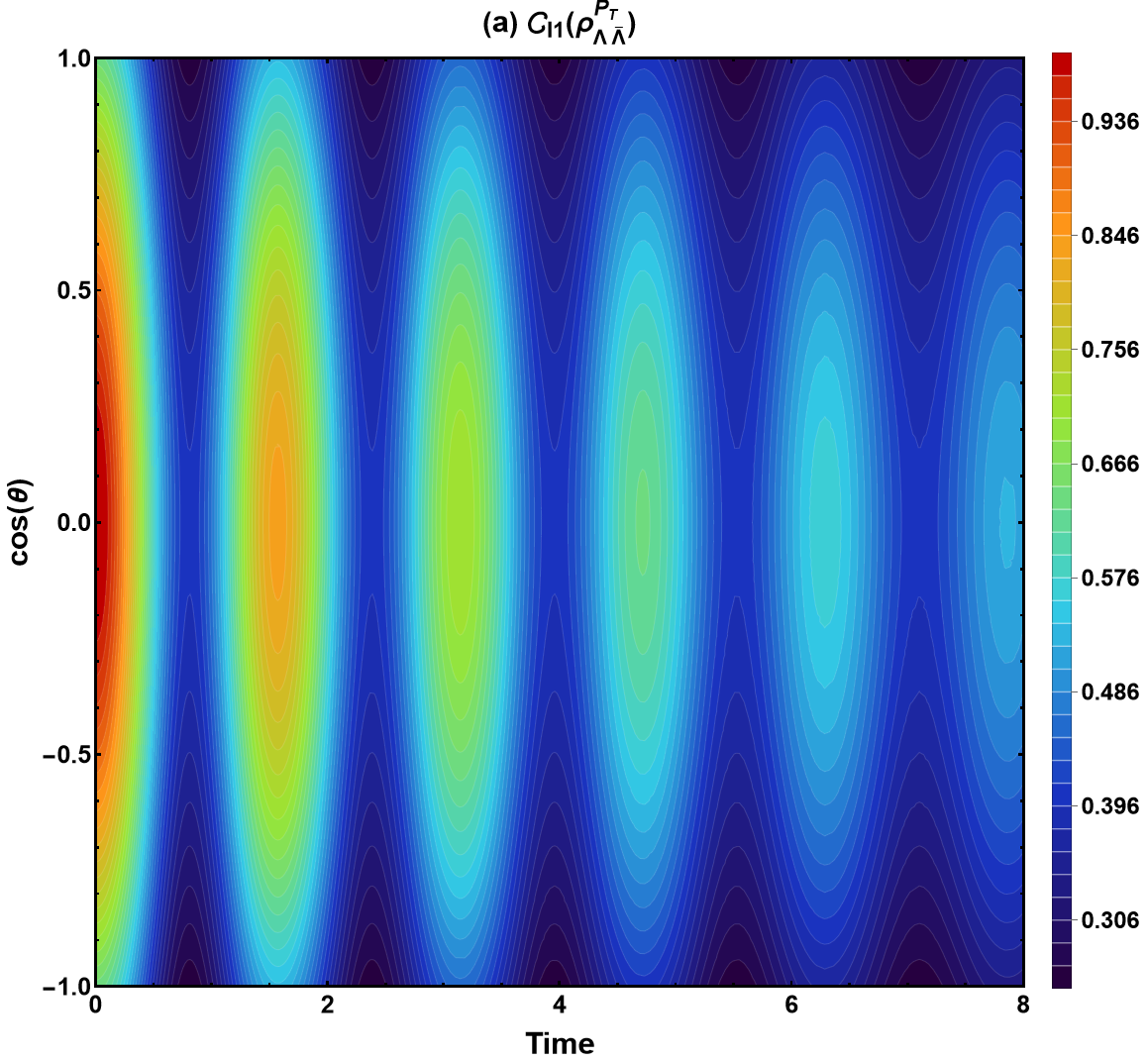}
	\includegraphics[width=0.24\linewidth]{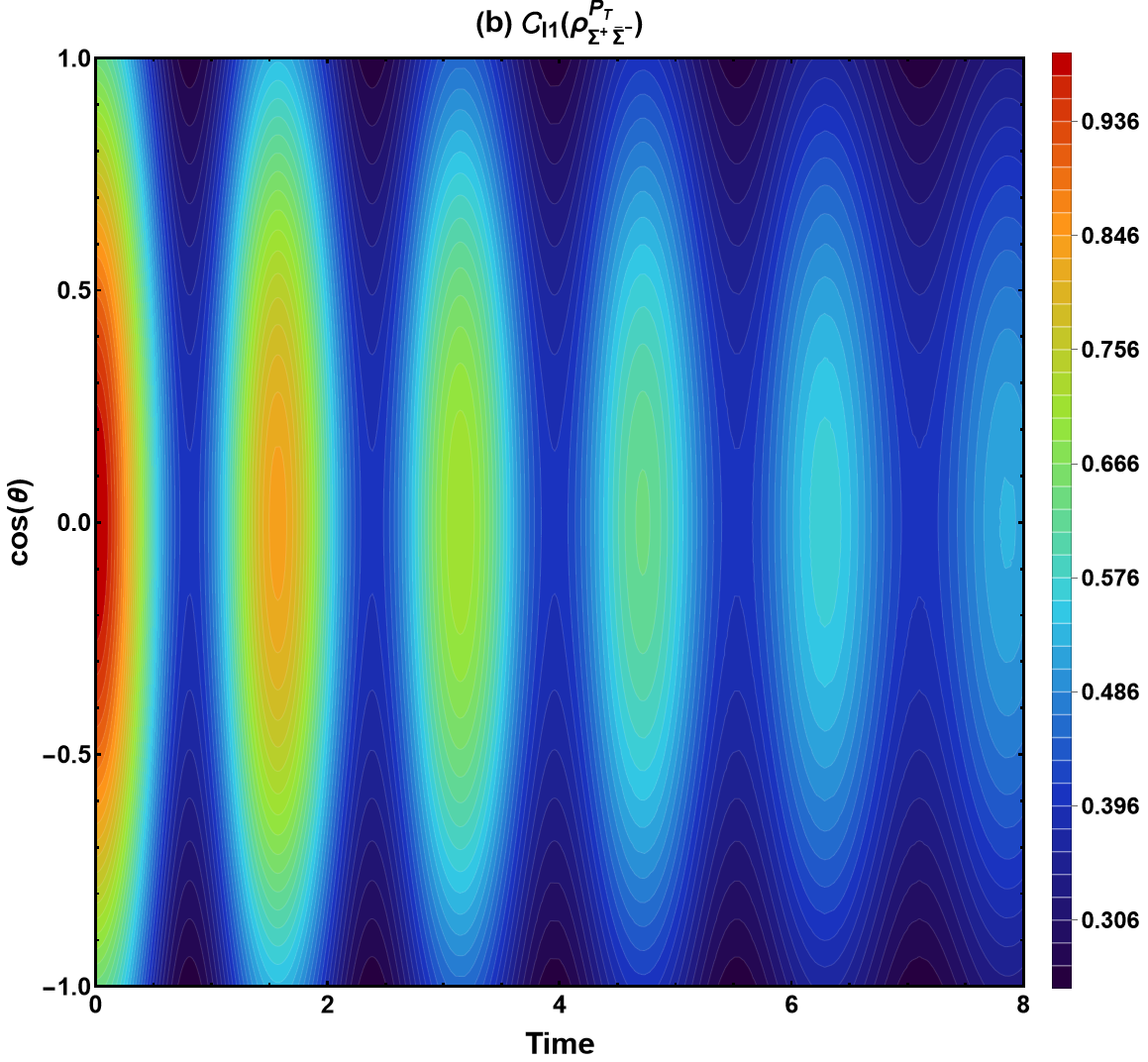}
	\includegraphics[width=0.24\linewidth]{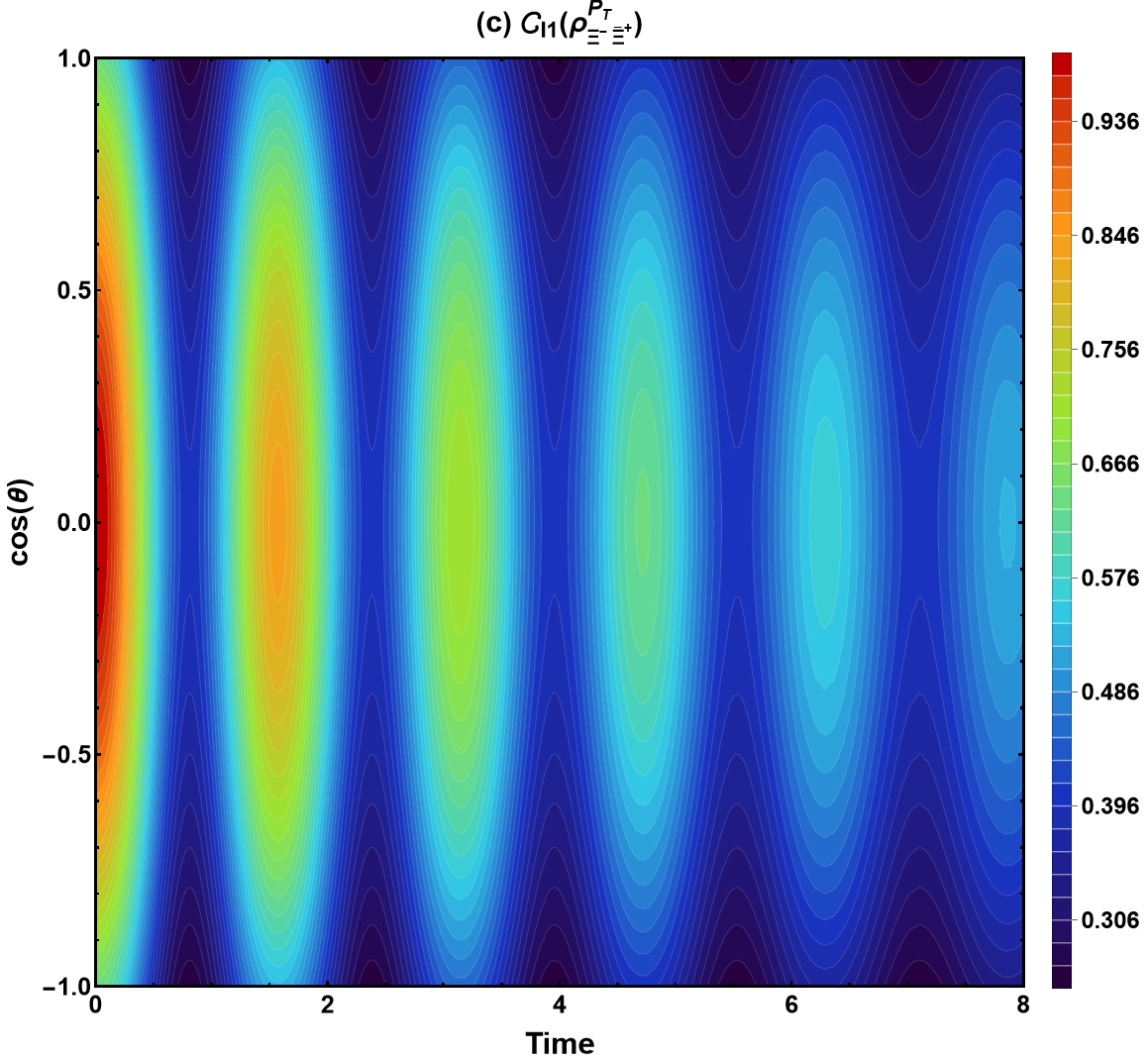}
	\includegraphics[width=0.24\linewidth]{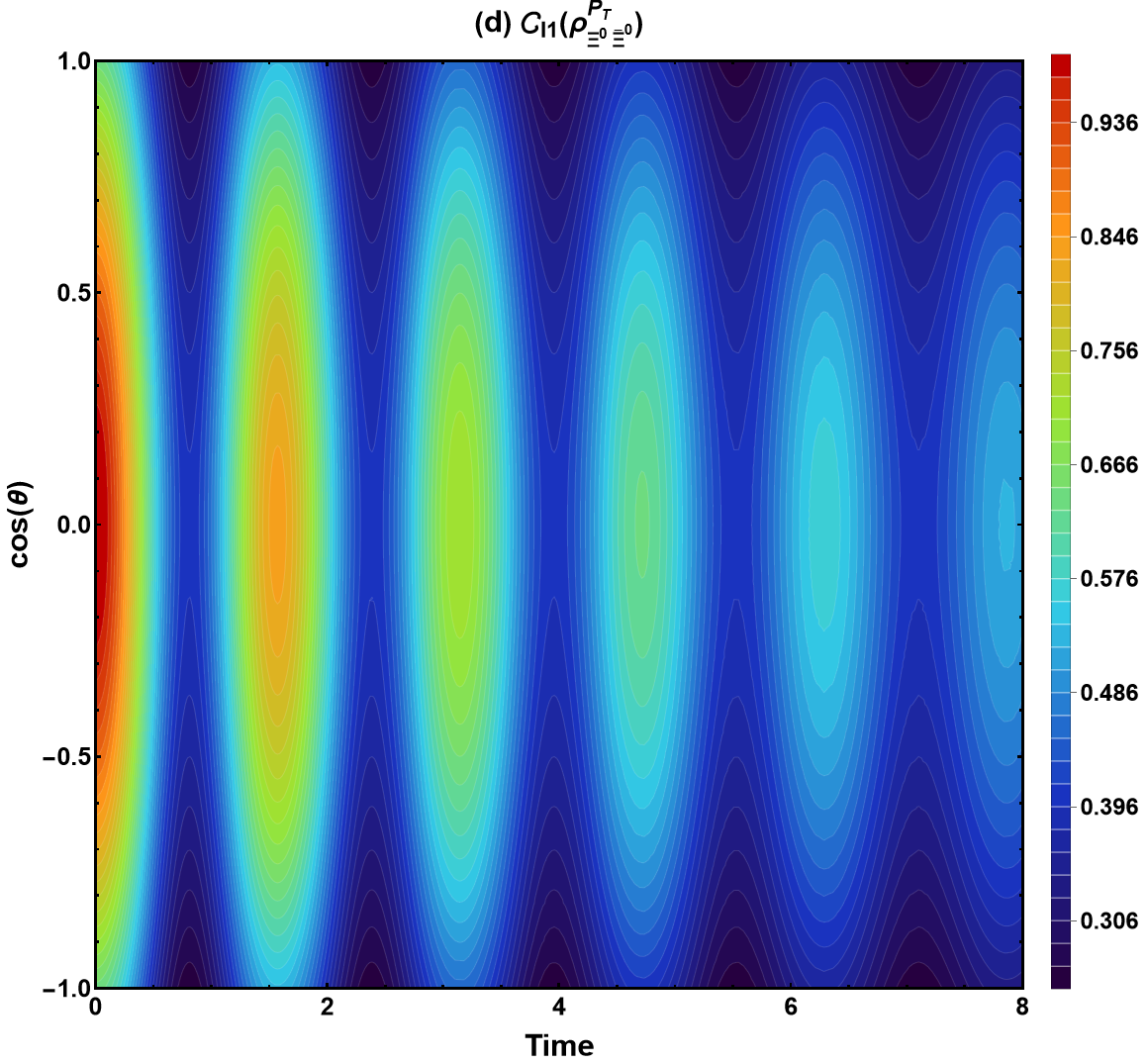}
	\includegraphics[width=0.24\linewidth]{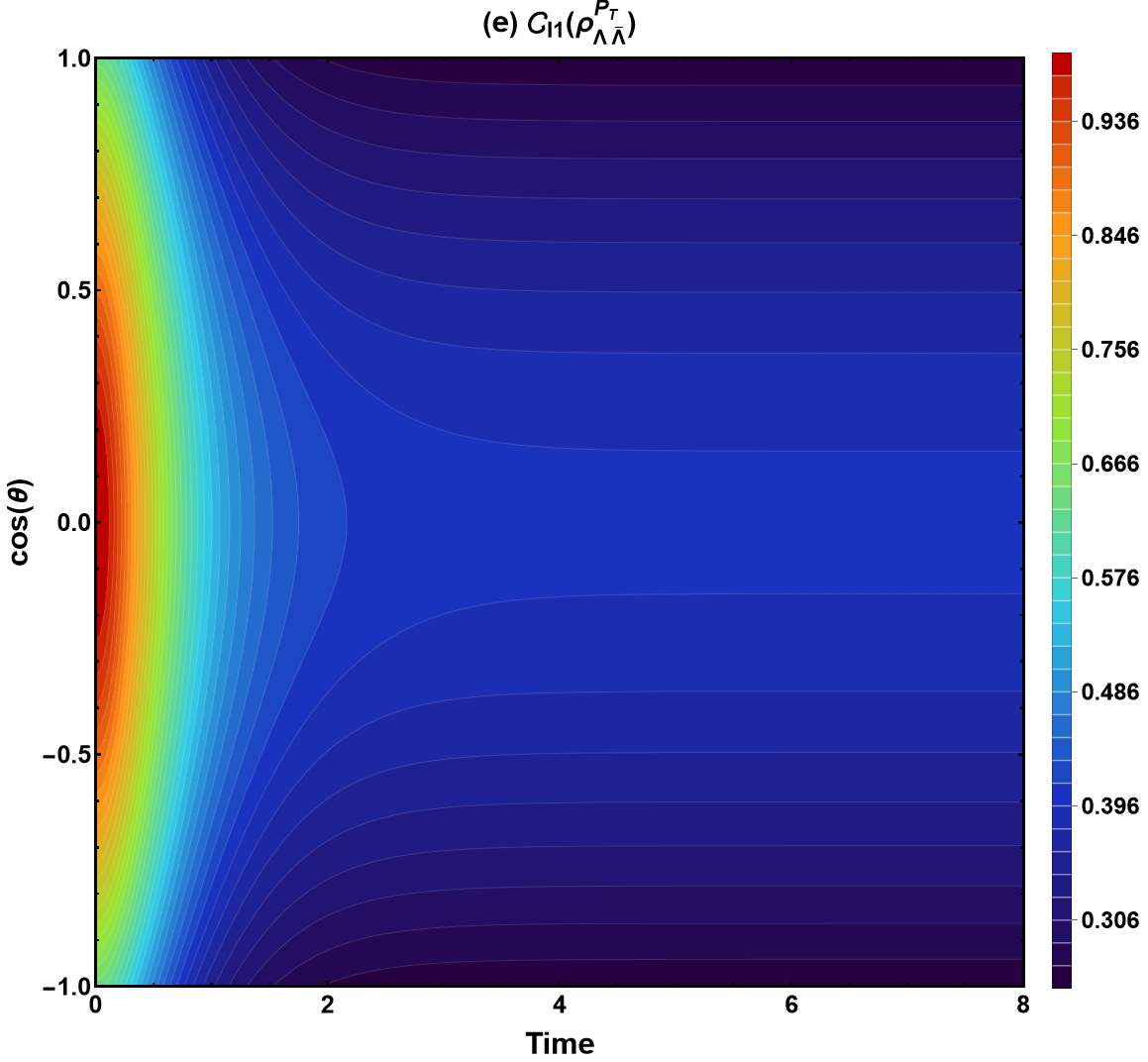}
	\includegraphics[width=0.24\linewidth]{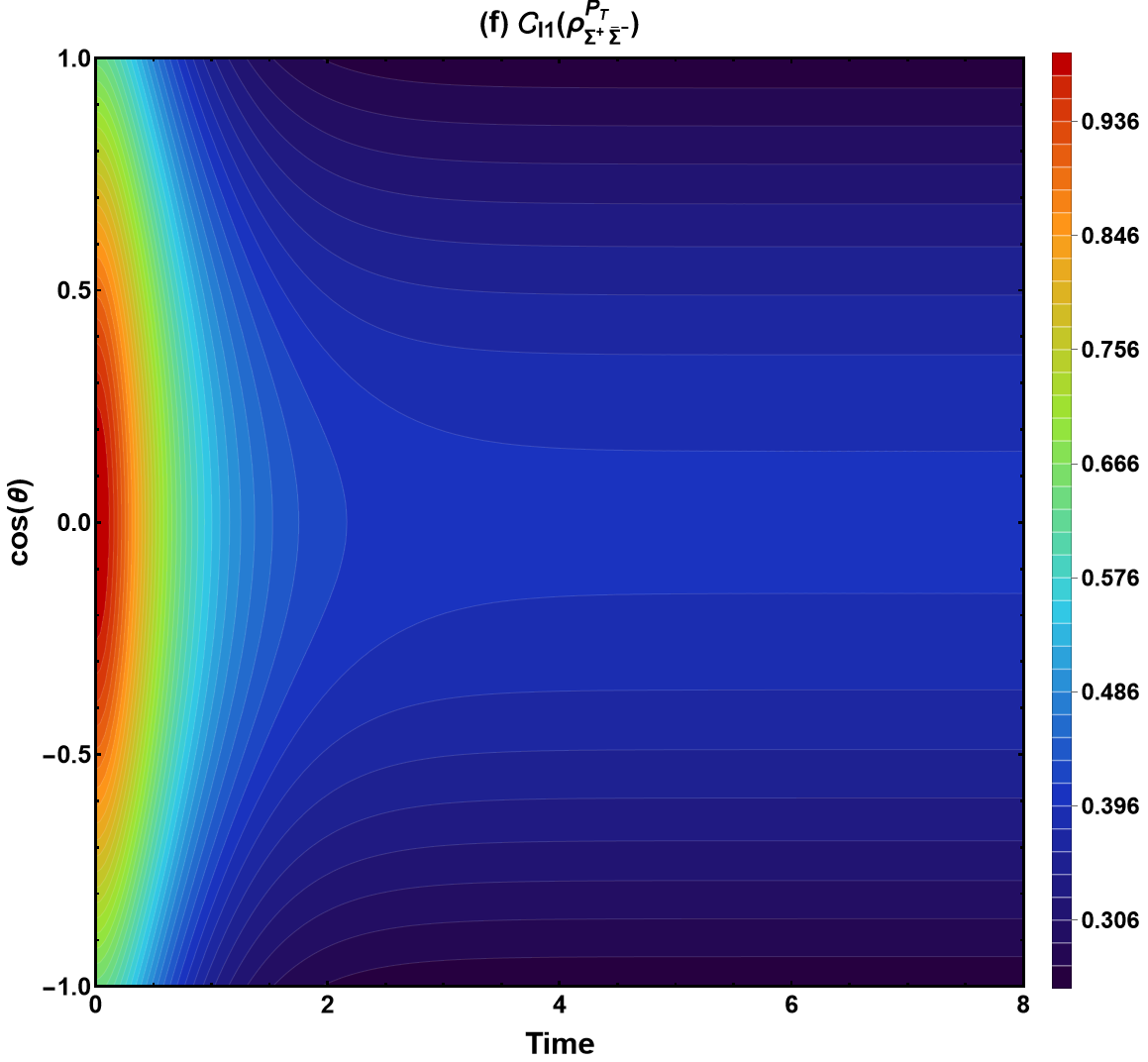}
	\includegraphics[width=0.24\linewidth]{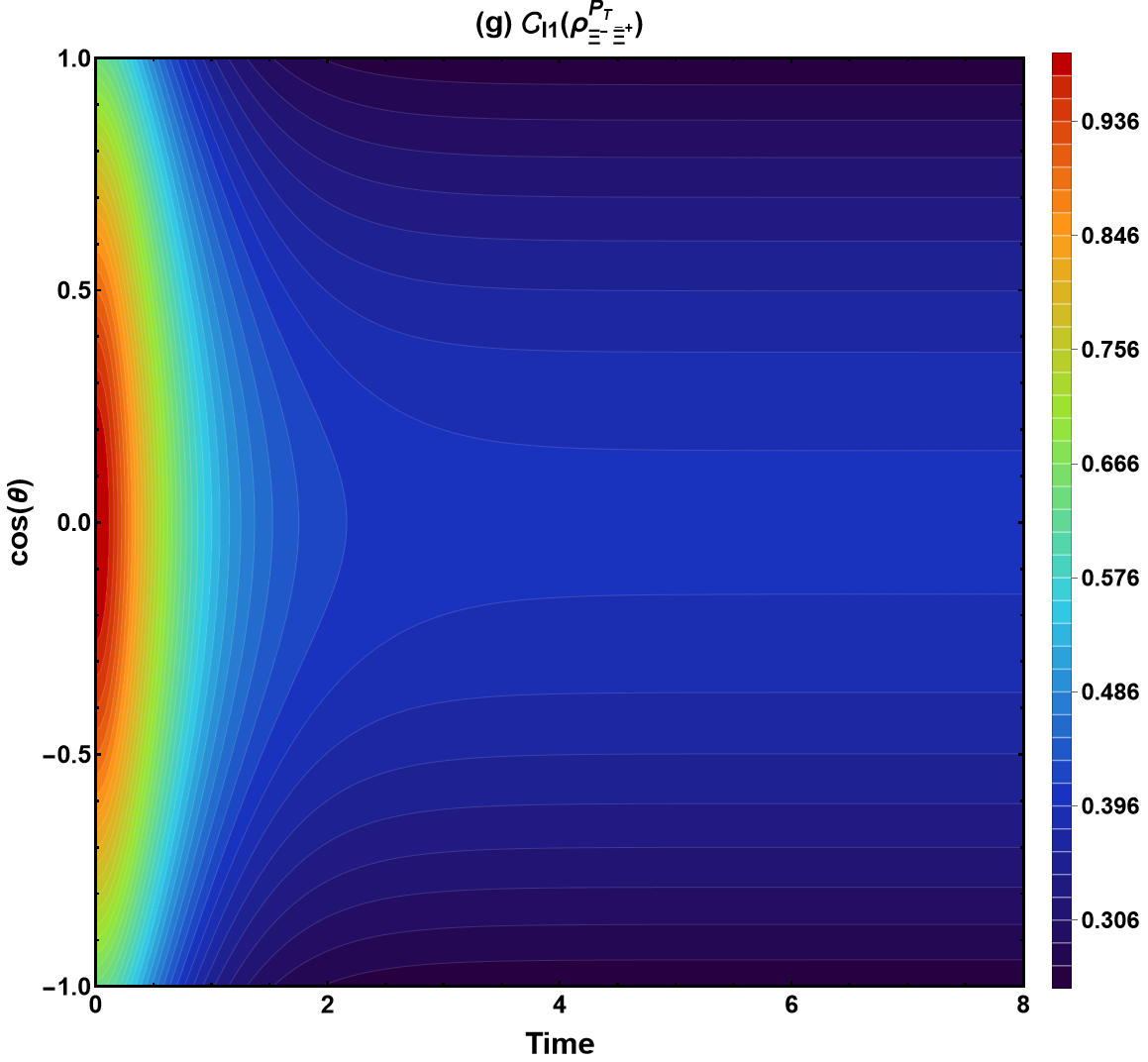}
	\includegraphics[width=0.24\linewidth]{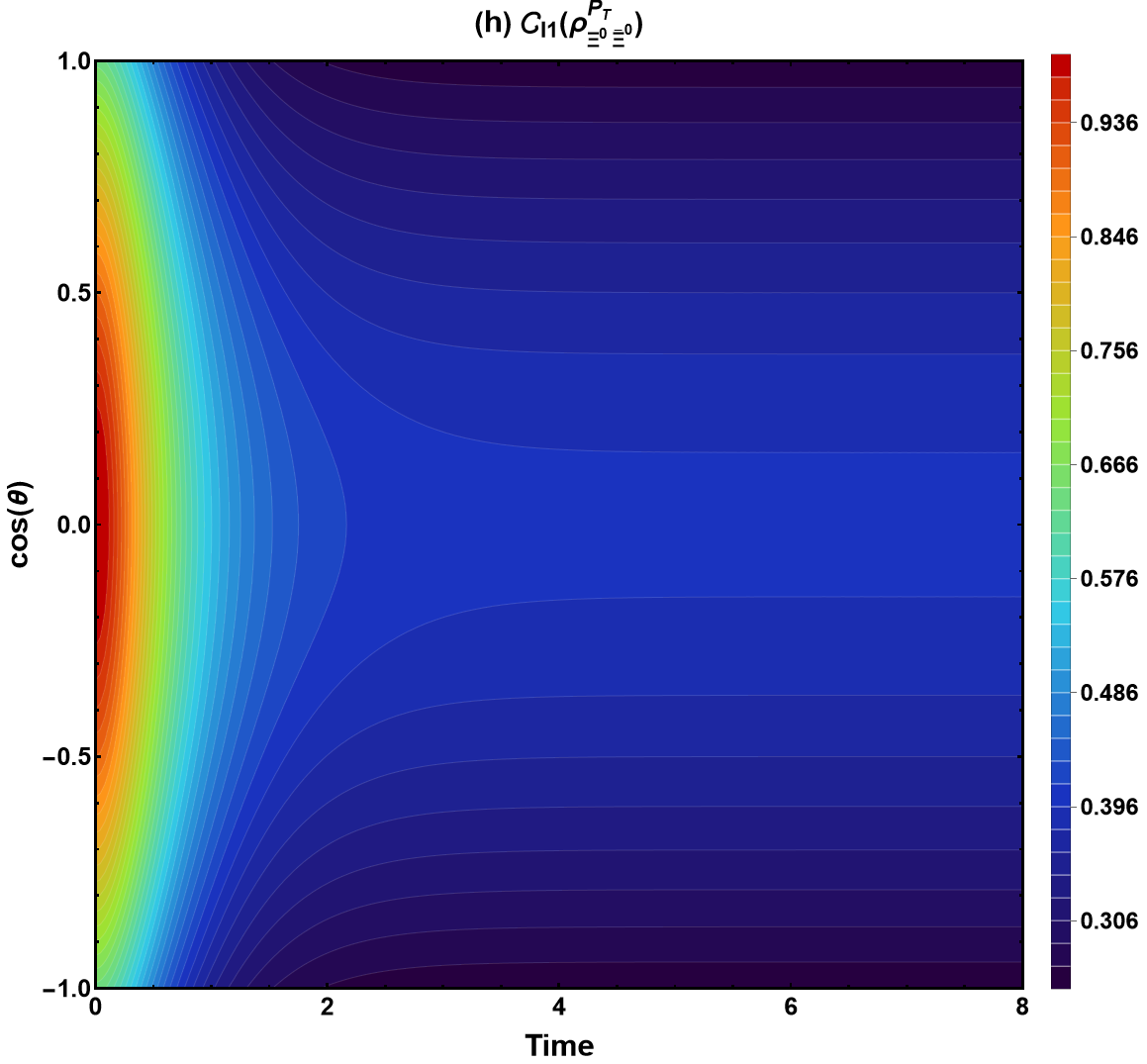}
\caption{
	Dynamical evolution of the $l_1$-norm quantum coherence
	$C_{l_1}(\rho^{P_T}_{Y\bar{Y}})$ as a function of time and the
	production angle $\cos\theta$ for
	$J/\psi\rightarrow Y\bar{Y}$ with
	$Y=\Lambda$, $\Sigma^{+}$, $\Xi^{-}$, and $\Xi^{0}$ at
	$P_T=0.8$ and $\phi=0$. Panels (a)--(d) [(e)--(h)] correspond to the
	non-Markovian (Markovian) regime  with $\tau=5$ ($\tau=0.2$) and
	$\mu=0.4$. The experimental parameters are taken from
	Table~\ref{tab:BESIII}.
}
	\label{fig24}
\end{figure}
$\Xi^{0}\bar{\Xi}^{0}$ channels retain enhanced correlations around
$\cos\theta \simeq 0$, the $\Sigma^{+}\bar{\Sigma}^{-}$ channel
preserves its characteristic edge-dominated structure. Overall, the
results confirm that geometric quantum discord remains a persistent
resource even when entanglement is strongly degraded by environmental
effects.

Shown in Fig.~\ref{fig24} is the dynamical behavior of the $l_{1}$-norm quantum coherence $C_{l_{1}}(\rho^{P_T}_{Y\bar{Y}})$ for varying production angles $\cos\theta$ under transverse beam polarization. In the non-Markovian regime [panels (a)--(d)], the
coherence exhibits damped oscillatory behavior accompanied by revival
phenomena. These revivals originate from the backflow of information
from the environment to the hyperon system, which partially restores
the off-diagonal elements of the density matrix. The largest coherence
values are observed around $\cos\theta \simeq 0$, where spin
interference effects are strongest.
As time increases, the oscillation amplitudes gradually decrease due
to decoherence, but a substantial amount of coherence survives
throughout the evolution. This behavior demonstrates that
non-Markovian memory effects significantly enhance the preservation of
quantum superpositions.
In the Markovian regime [panels (e)--(h)], the revivals disappear and
the coherence decays smoothly toward an asymptotic stationary value,
reflecting the irreversible loss of quantum information into the
environment. Nevertheless, finite coherence persists even at long
times, indicating that coherence is more robust against environmental
noise than entanglement-based correlations.
Another notable feature is the strong similarity among the four
hyperon channels. The nearly identical angular and temporal
distributions suggest that the coherence dynamics is mainly governed
by the common spin structure of the hyperon--antihyperon pairs rather
than by channel-dependent decay parameters. Overall, the results show
that transverse beam polarization generates long-lived coherent spin
superpositions, particularly in the presence of non-Markovian memory
effects.
\begin{figure}[H]
	\centering
	\includegraphics[width=0.24\linewidth]{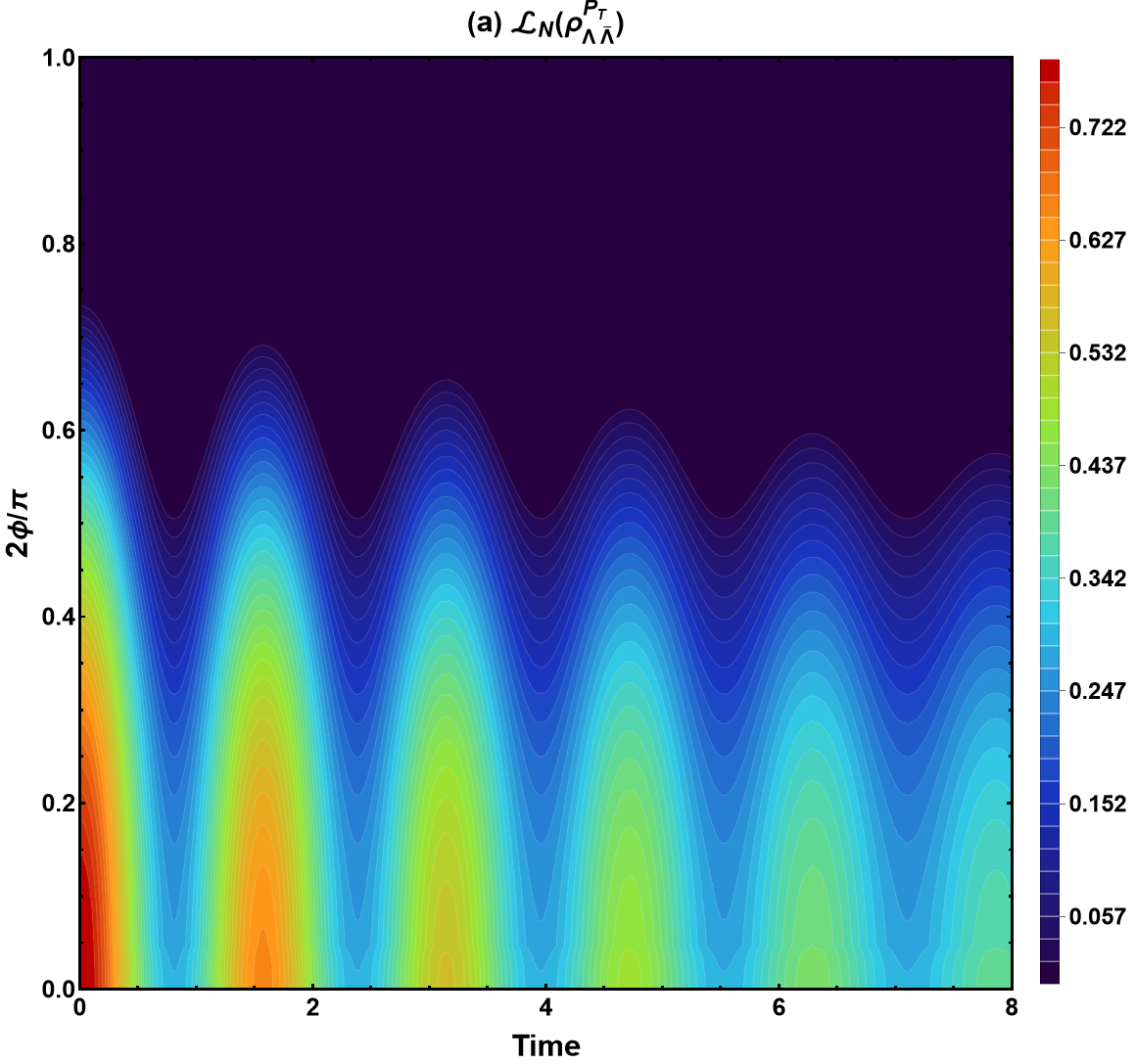}
	\includegraphics[width=0.24\linewidth]{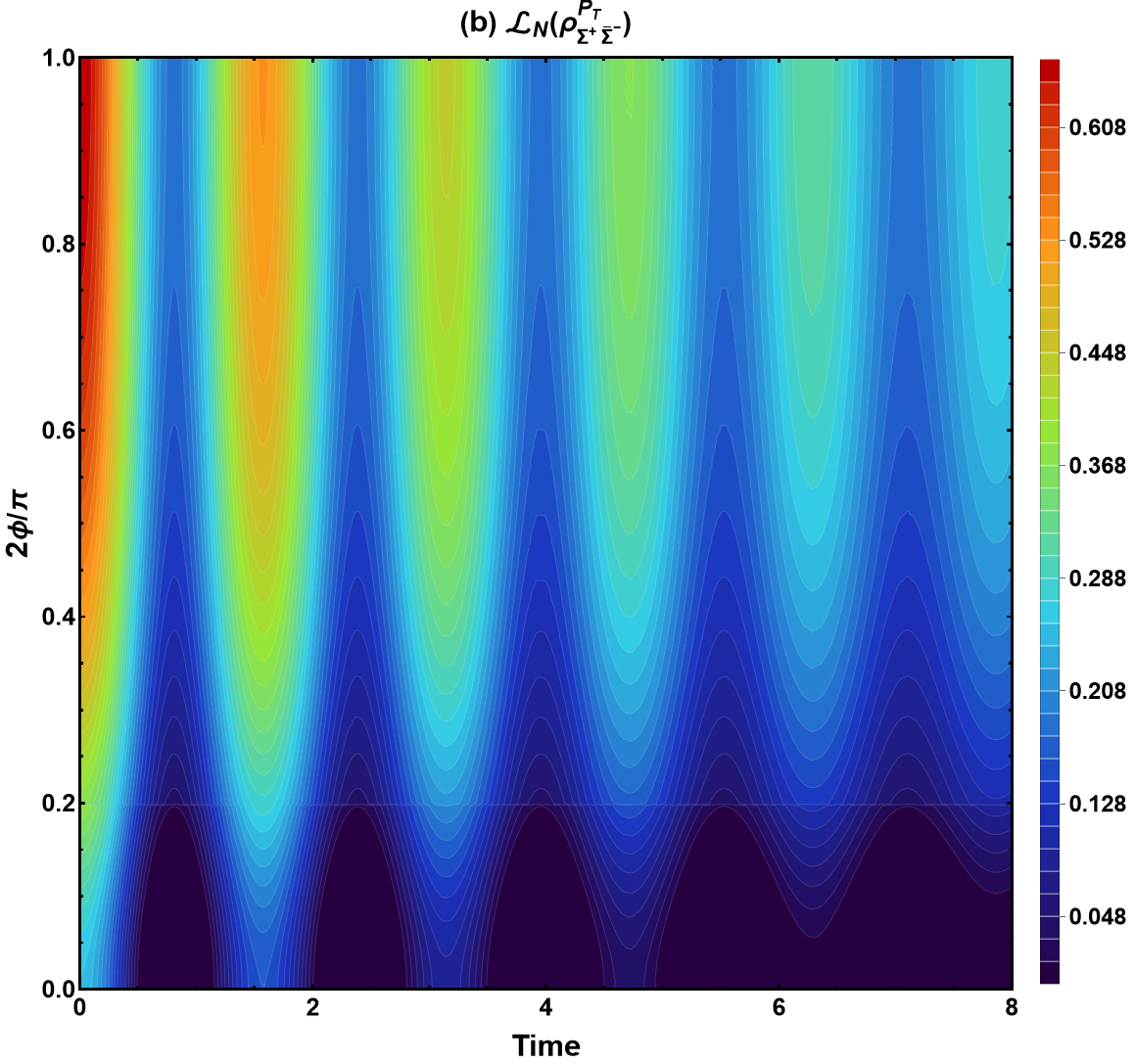}
	\includegraphics[width=0.24\linewidth]{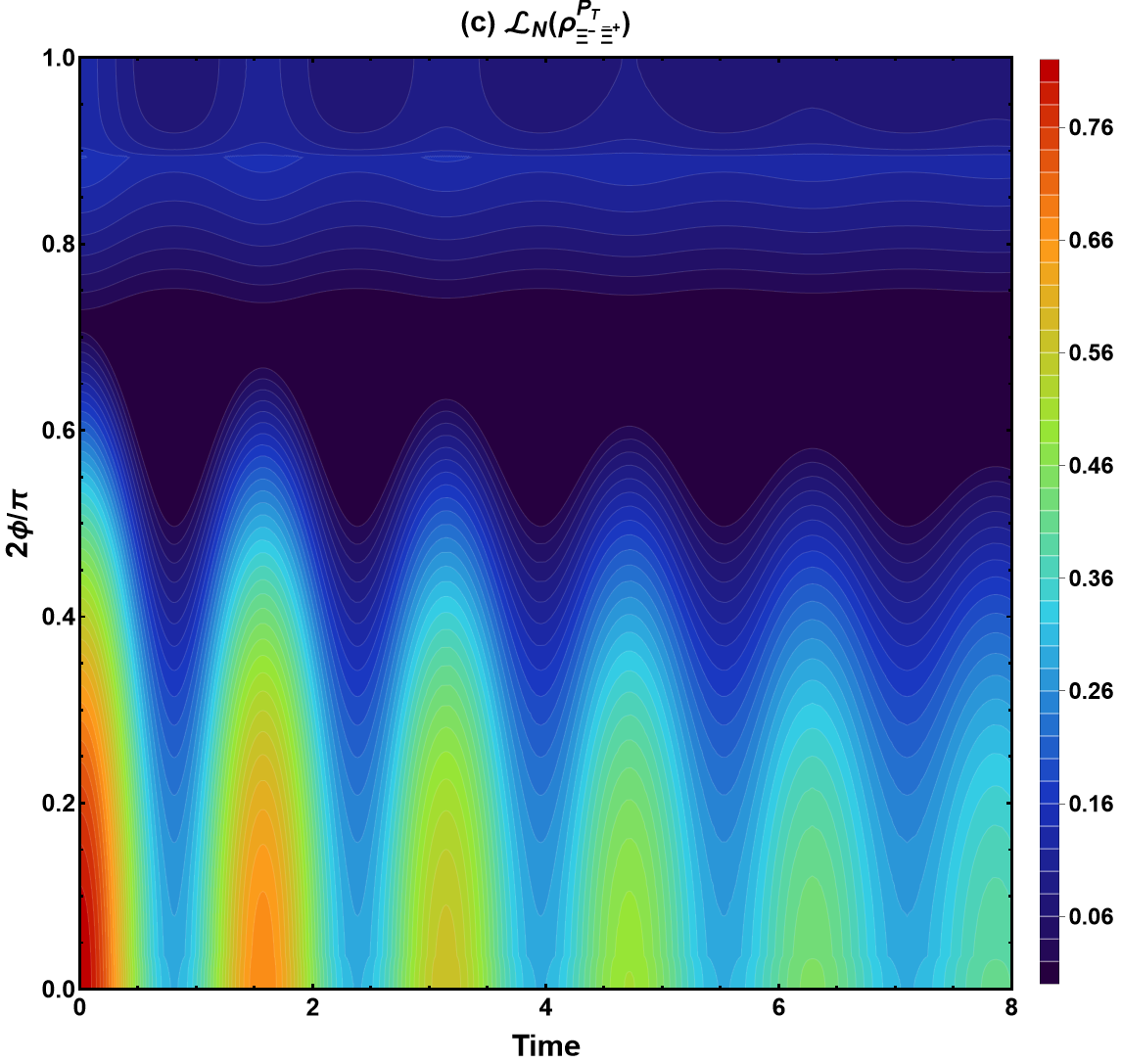}
	\includegraphics[width=0.24\linewidth]{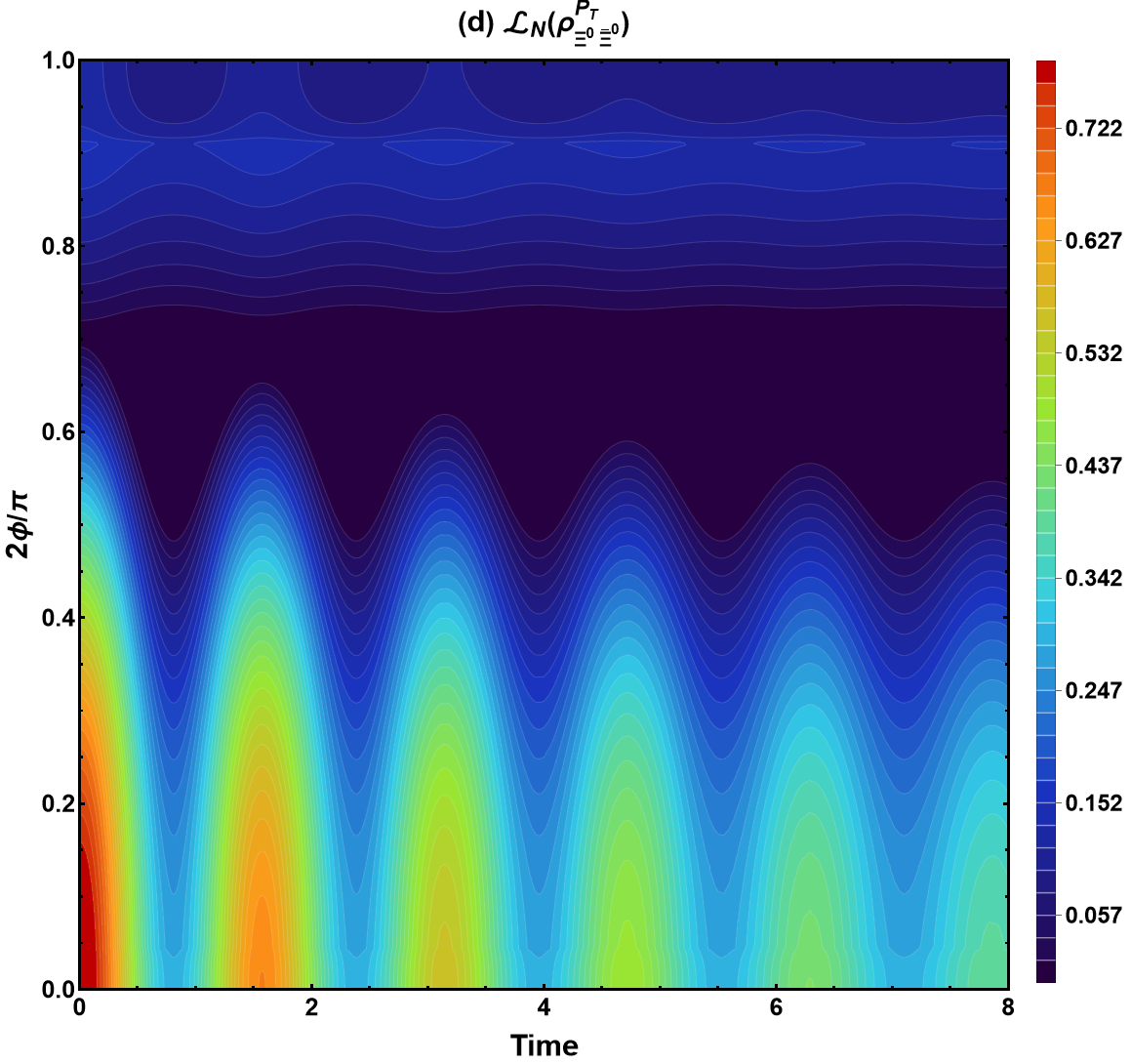}
	\includegraphics[width=0.24\linewidth]{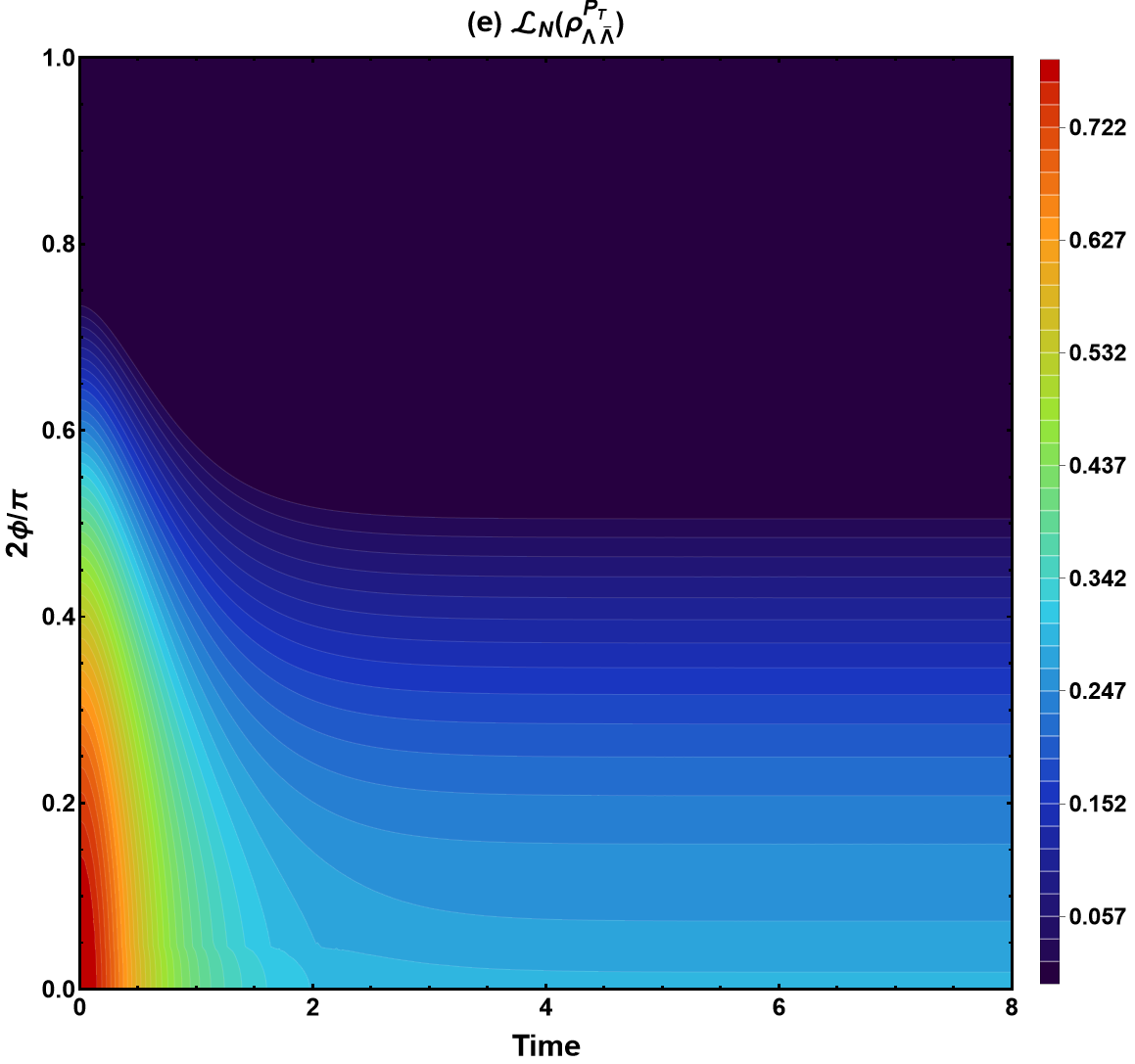}
	\includegraphics[width=0.24\linewidth]{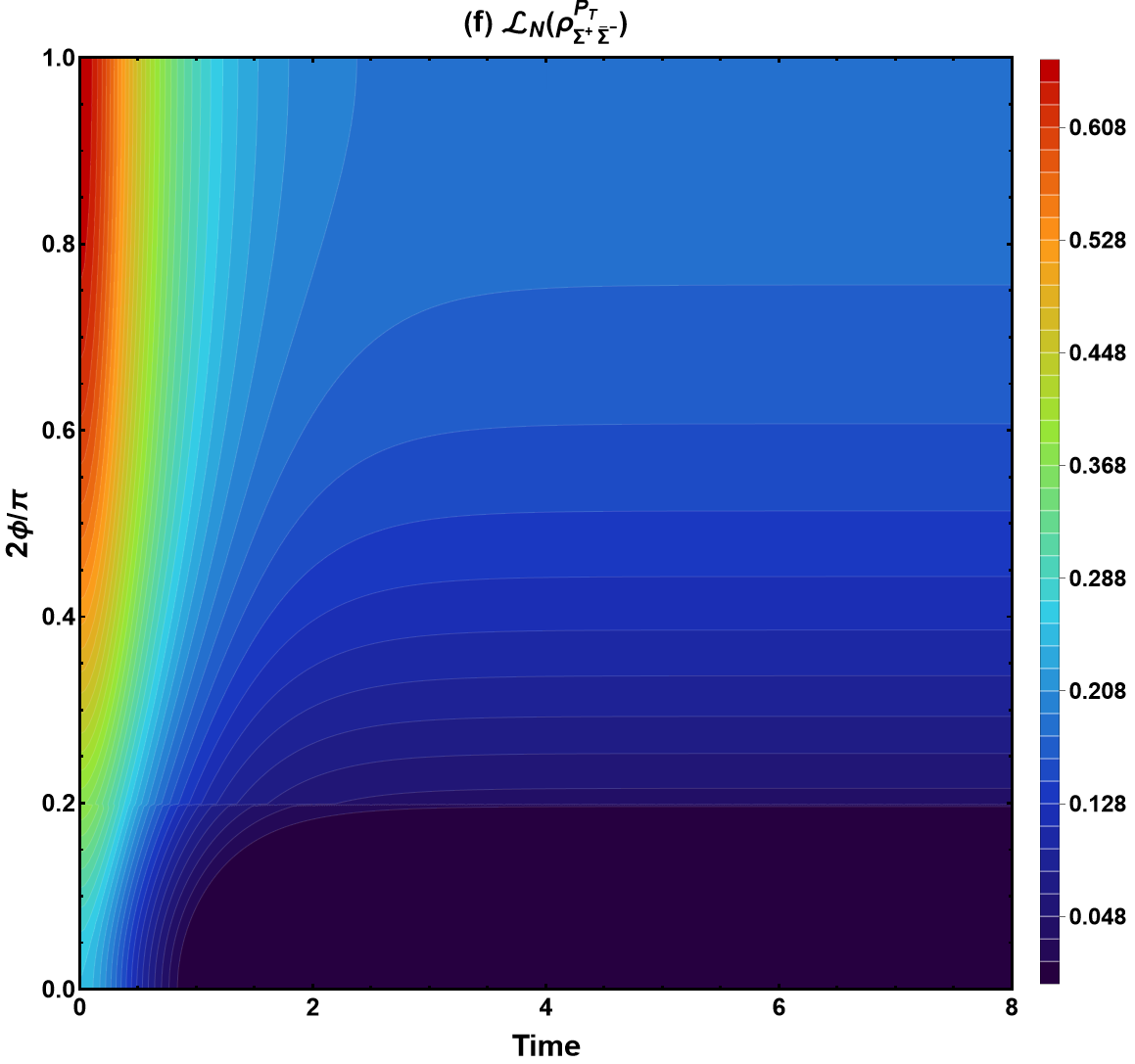}
	\includegraphics[width=0.24\linewidth]{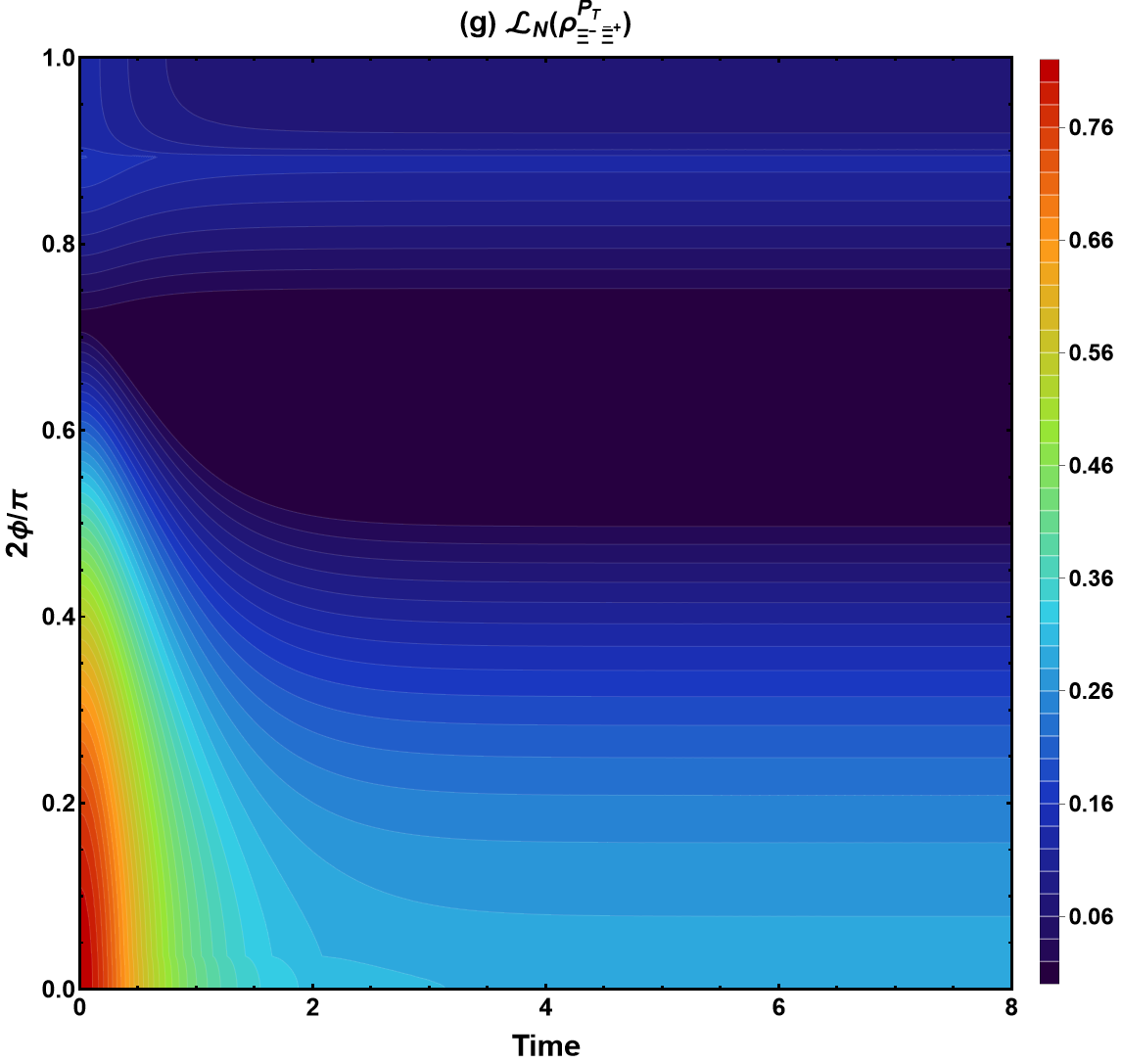}
	\includegraphics[width=0.24\linewidth]{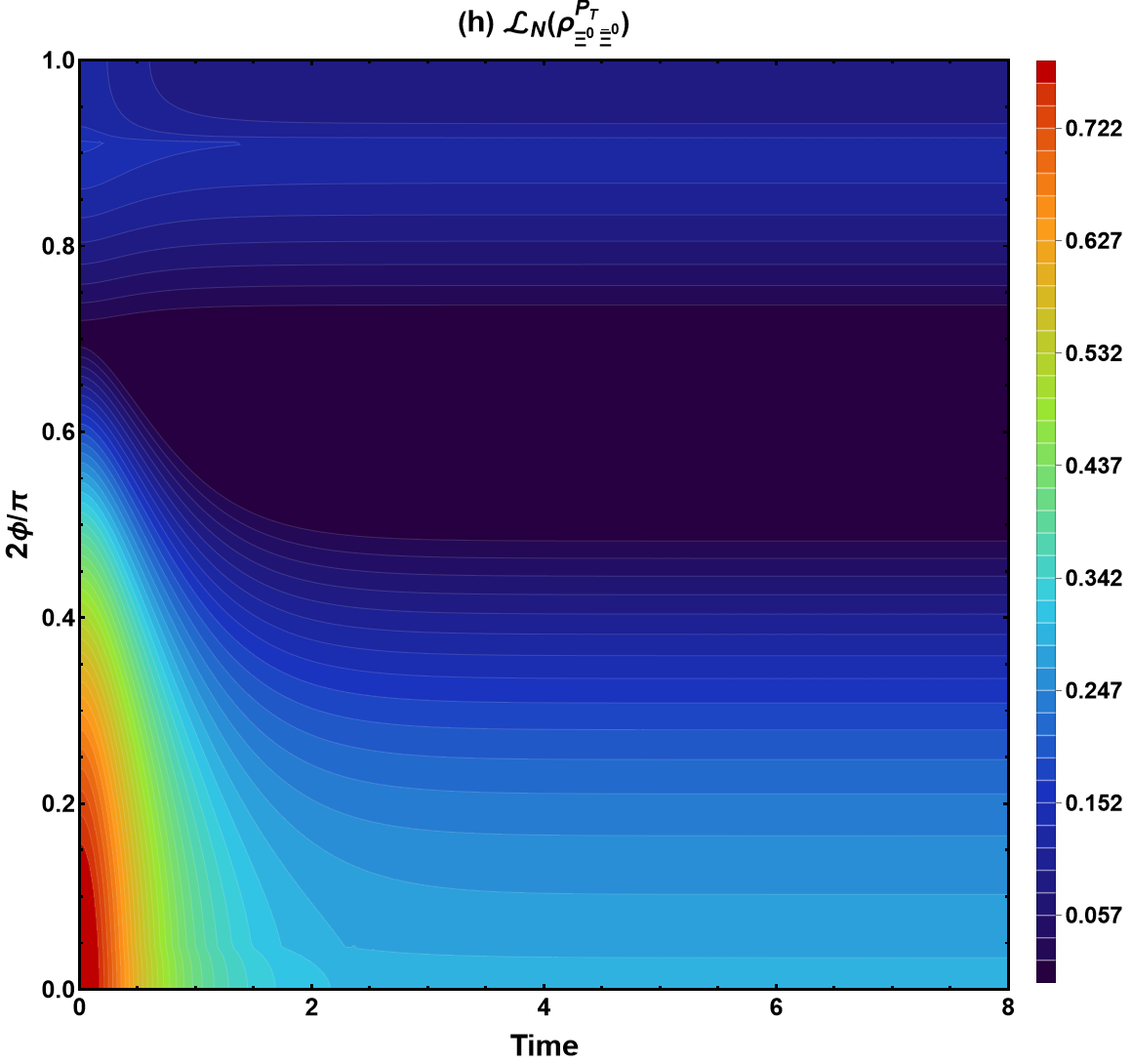}
\caption{
	Dynamical evolution of the logarithmic negativity
	$\mathcal{L}_N(\rho^{P_T}_{Y\bar{Y}})$ as a function of time and the
	azimuthal angle $\phi$ for
	$J/\psi\rightarrow Y\bar{Y}$ with
	$Y=\Lambda$, $\Sigma^{+}$, $\Xi^{-}$, and $\Xi^{0}$ at
	$\cos\theta=0.5$ and $P_T=0.8$. Panels (a)--(d) [(e)--(h)] correspond
	to the non-Markovian (Markovian) regime  with $\tau=5$ ($\tau=0.2$) and
	$\mu=0.4$. The experimental parameters are taken from
	Table~\ref{tab:BESIII}.
}
	\label{fig19}
\end{figure}
Figure~\ref{fig19} presents the dynamics of the logarithmic negativity
$\mathcal{L}_{N}(\rho^{P_T}_{Y\bar{Y}})$ as a function of time and the
azimuthal angle $\phi$ (displayed through the normalized quantity
$2\phi/\pi$) for transversely polarized beams. The upper panels
(a)--(d) correspond to the non-Markovian regime, while the lower
panels (e)--(h) describe the Markovian dynamics.
In the non-Markovian regime, the entanglement exhibits a sequence of
damped oscillatory revivals whose amplitudes gradually decrease with
time. These revivals are a direct consequence of environmental memory
effects, which partially restore the quantum correlations lost during
the evolution. The azimuthal angle strongly influences the
entanglement distribution, indicating that the spin configuration in
the transverse production plane plays an important role in the
generation of quantum correlations. The
$\Lambda\bar{\Lambda}$ and $\Xi^{0}\bar{\Xi}^{0}$ channels display the
most pronounced revival structures, whereas the
$\Sigma^{+}\bar{\Sigma}^{-}$ channel shows a stronger dependence on
the azimuthal angle, with entanglement concentrated in specific
angular regions.
By contrast, the Markovian regime is characterized by a monotonic
suppression of logarithmic negativity and the complete disappearance
of revival patterns. This behavior reflects the irreversible transfer
of information from the hyperon system to the environment and the
absence of memory-induced backflow processes. Although entanglement
decreases significantly, finite residual correlations remain visible
at long times for several channels, demonstrating a certain degree of
robustness against decoherence.
Overall, the results show that the azimuthal angle $\phi$ acts as an
effective control parameter for the distribution of entanglement under
transverse polarization. Furthermore, non-Markovian memory effects
substantially enhance the persistence of quantum correlations by
sustaining repeated entanglement revivals, whereas Markovian dynamics
drives the system toward a stationary mixed state with reduced
entanglement.

\begin{figure}[H]
	\centering
	\includegraphics[width=0.24\linewidth]{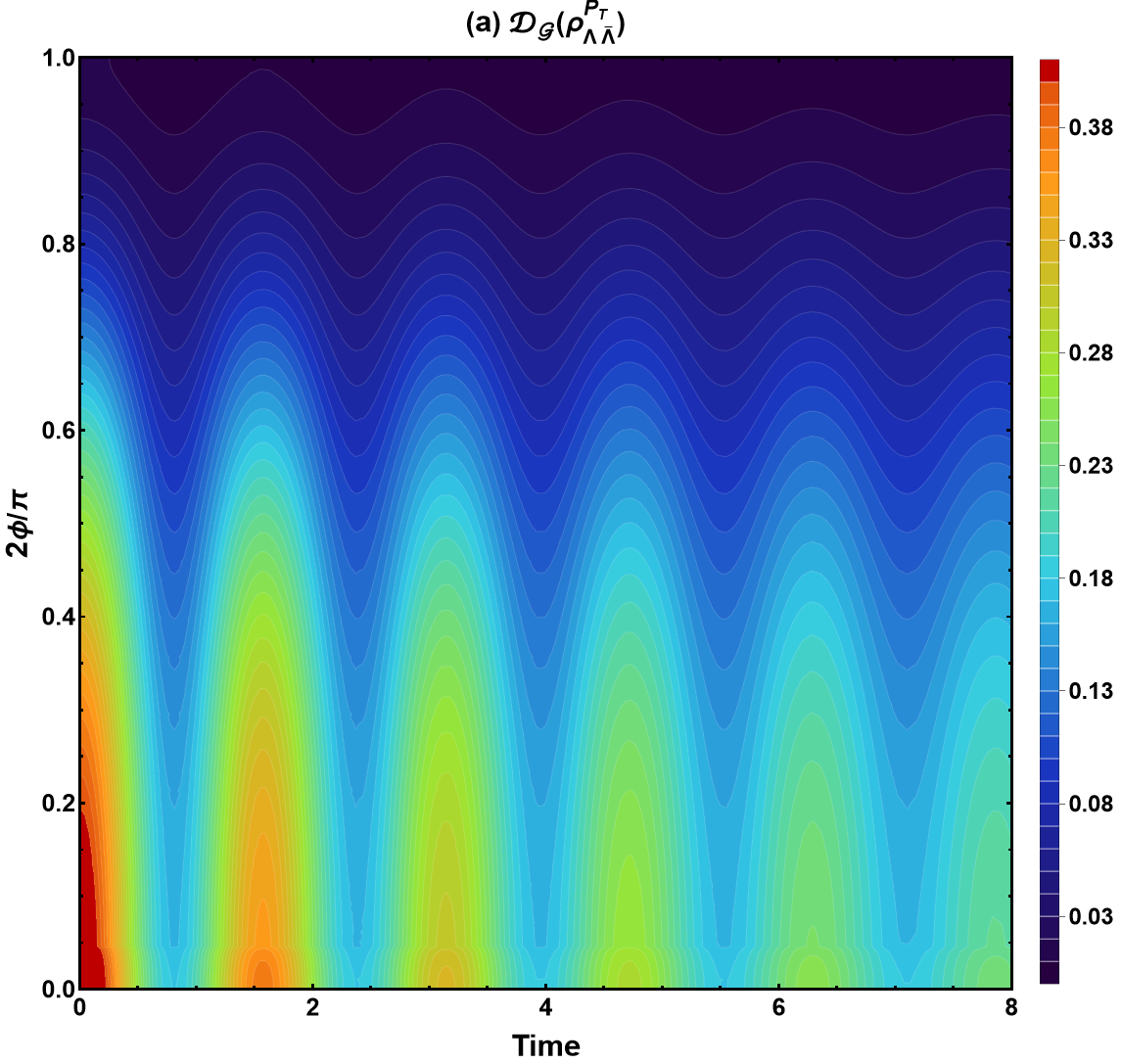}
	\includegraphics[width=0.24\linewidth]{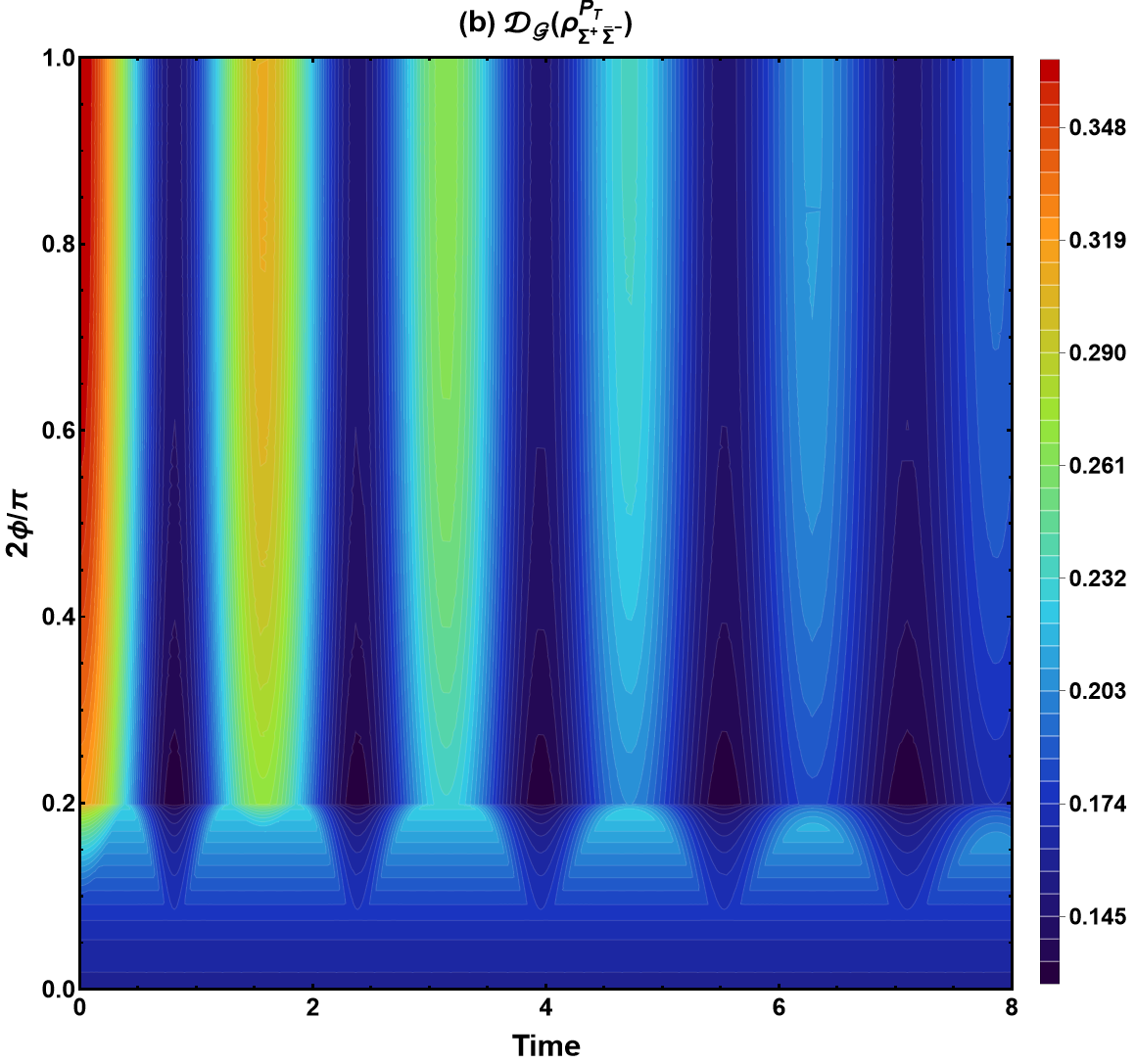}
	\includegraphics[width=0.24\linewidth]{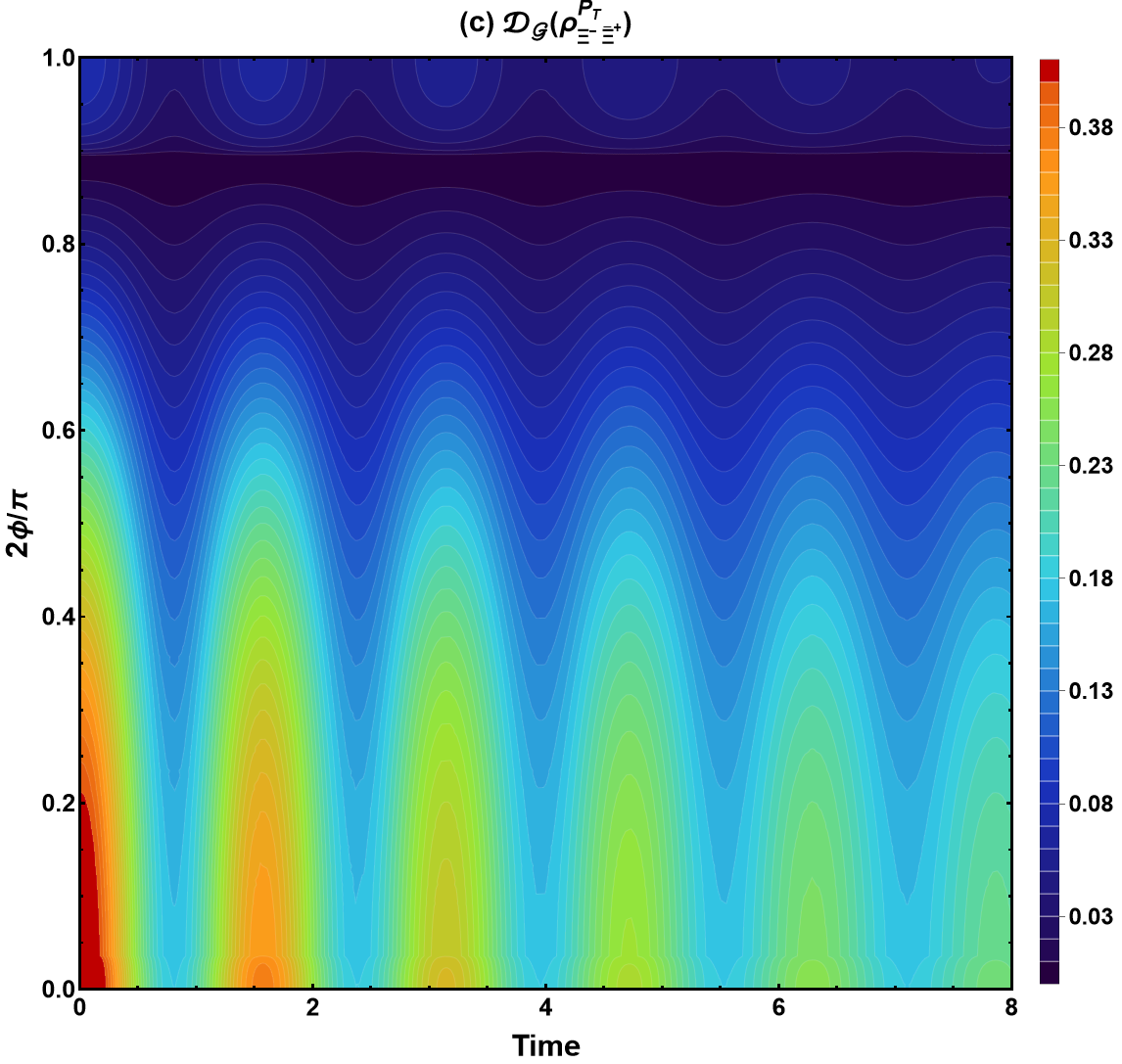}
	\includegraphics[width=0.24\linewidth]{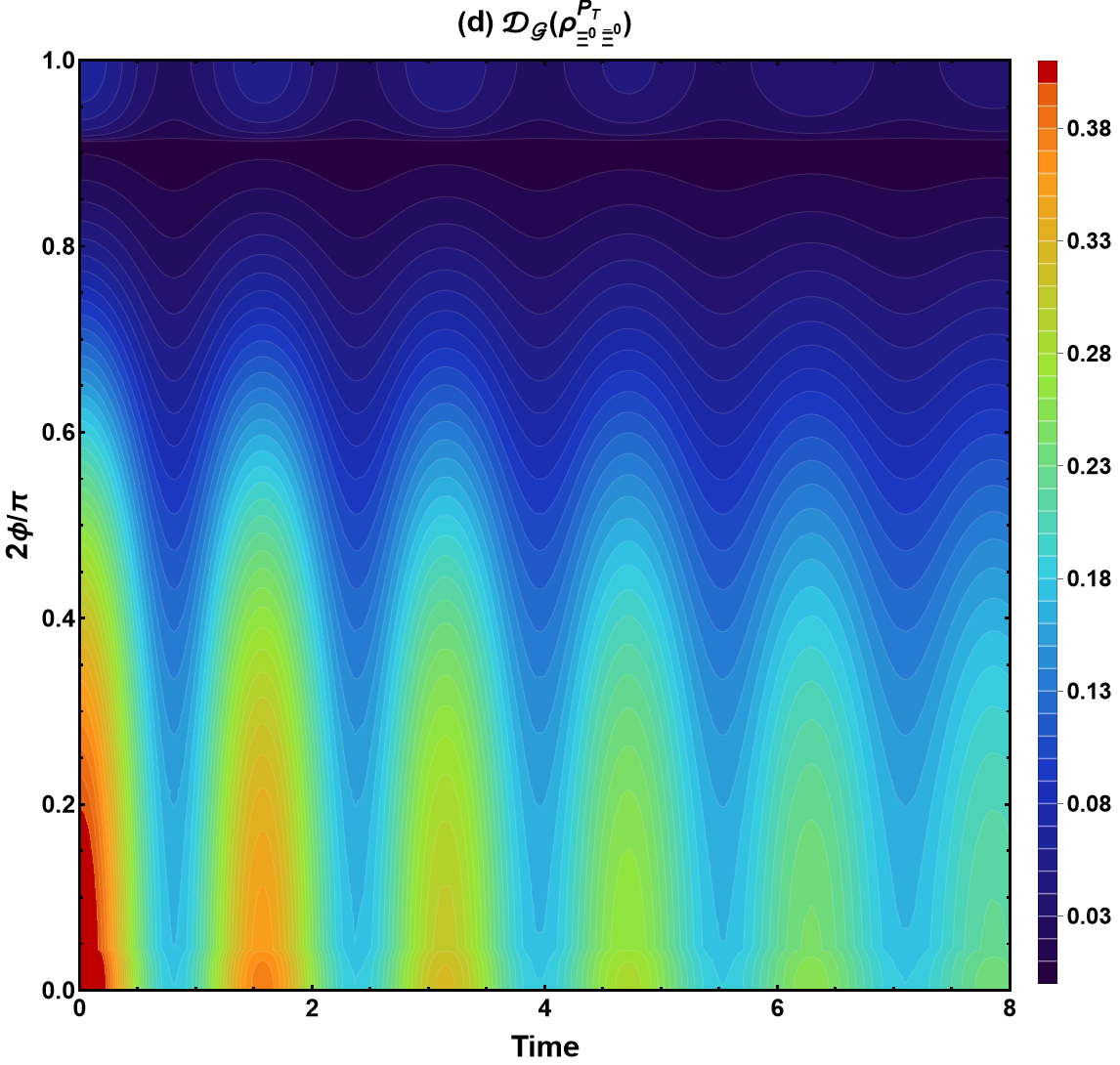}
	\includegraphics[width=0.24\linewidth]{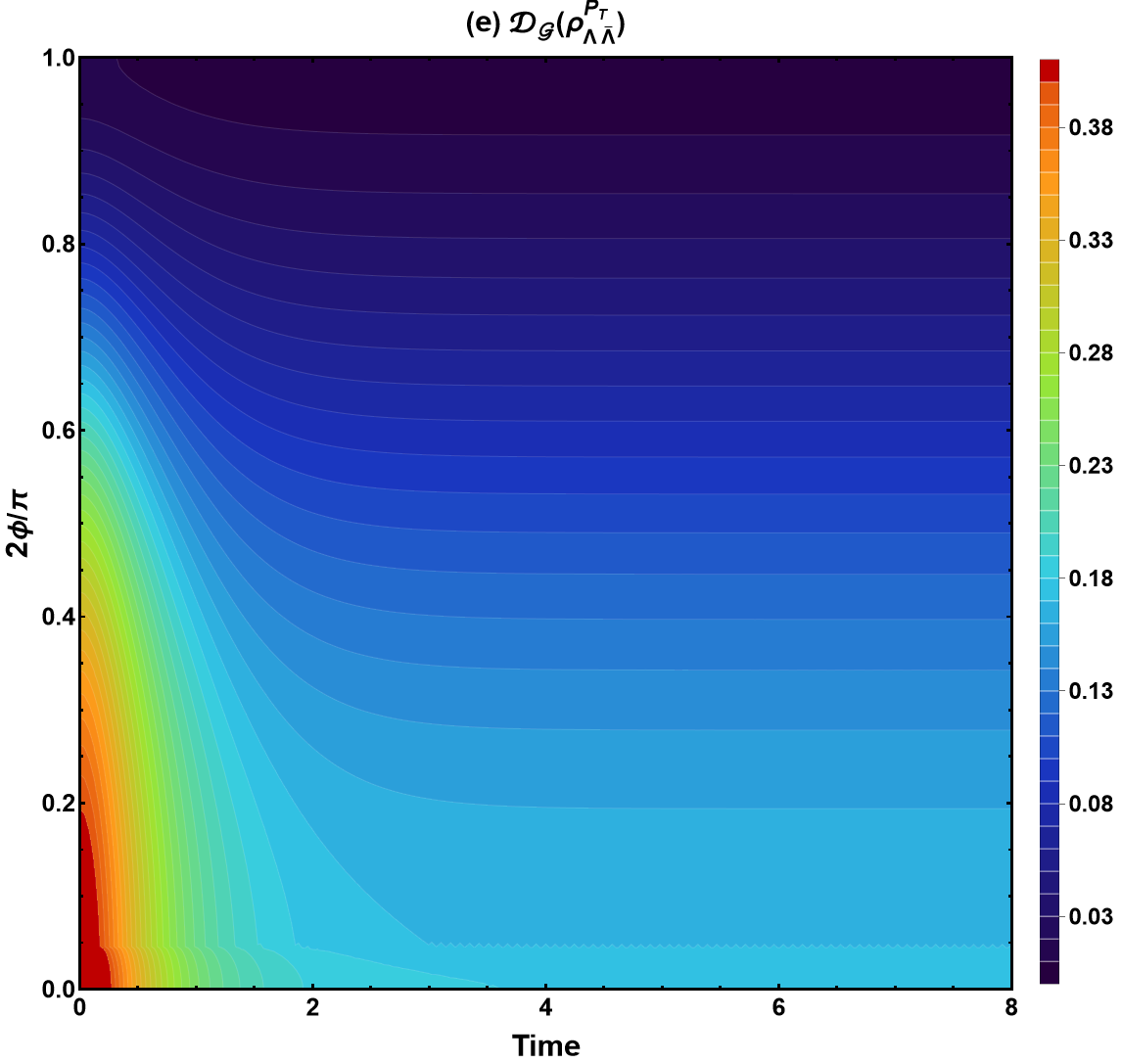}
	\includegraphics[width=0.24\linewidth]{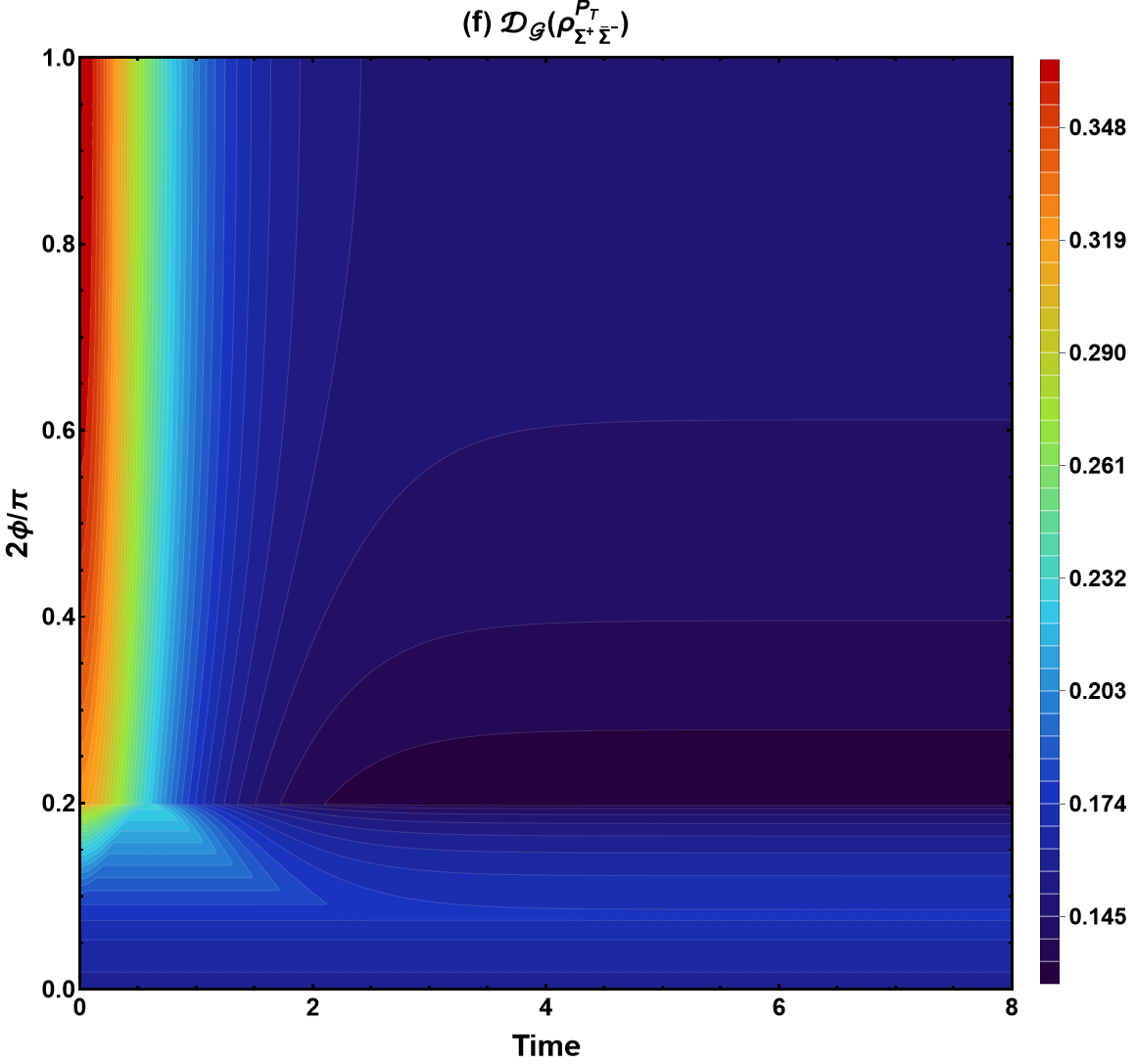}
	\includegraphics[width=0.24\linewidth]{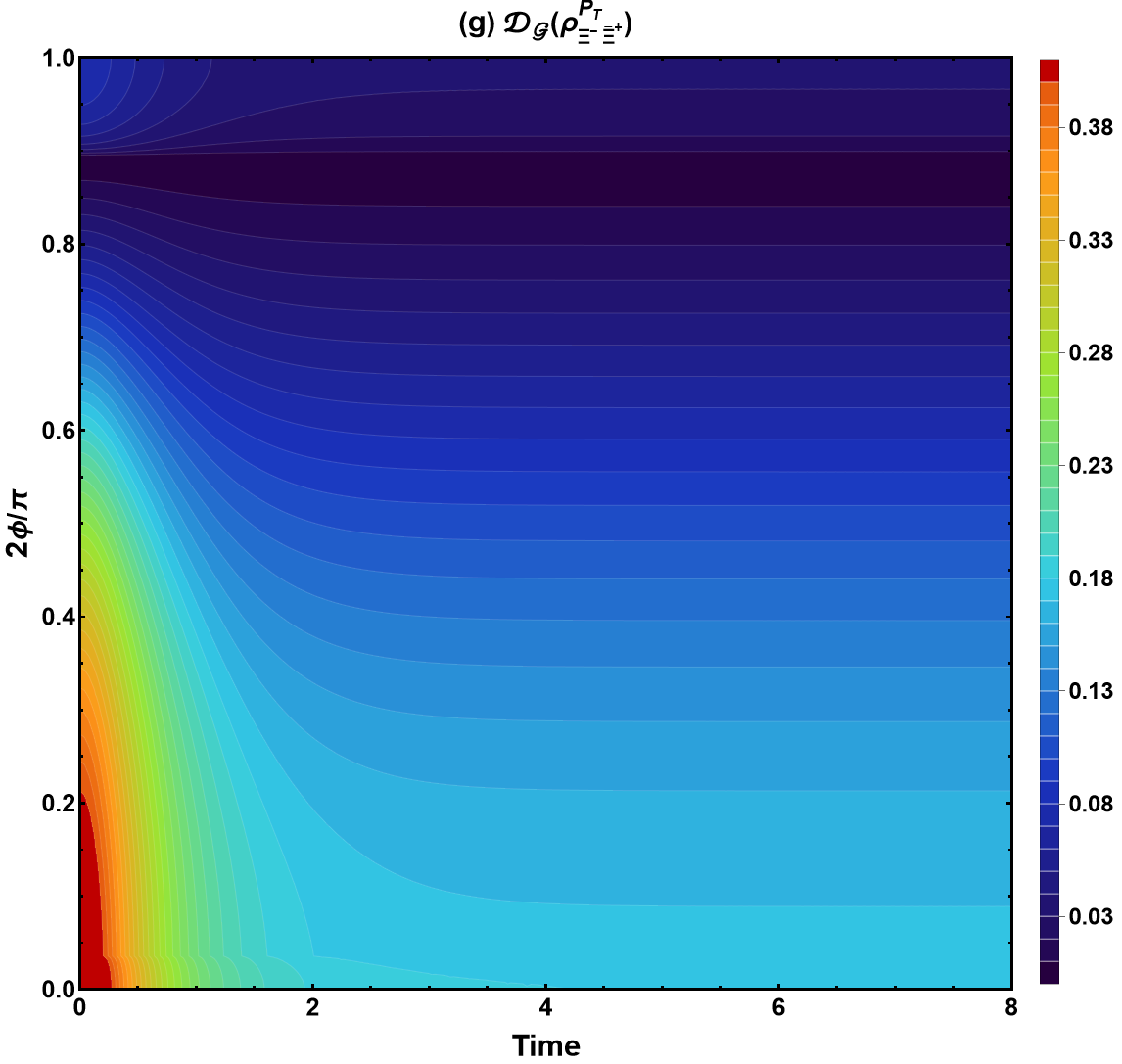}
	\includegraphics[width=0.24\linewidth]{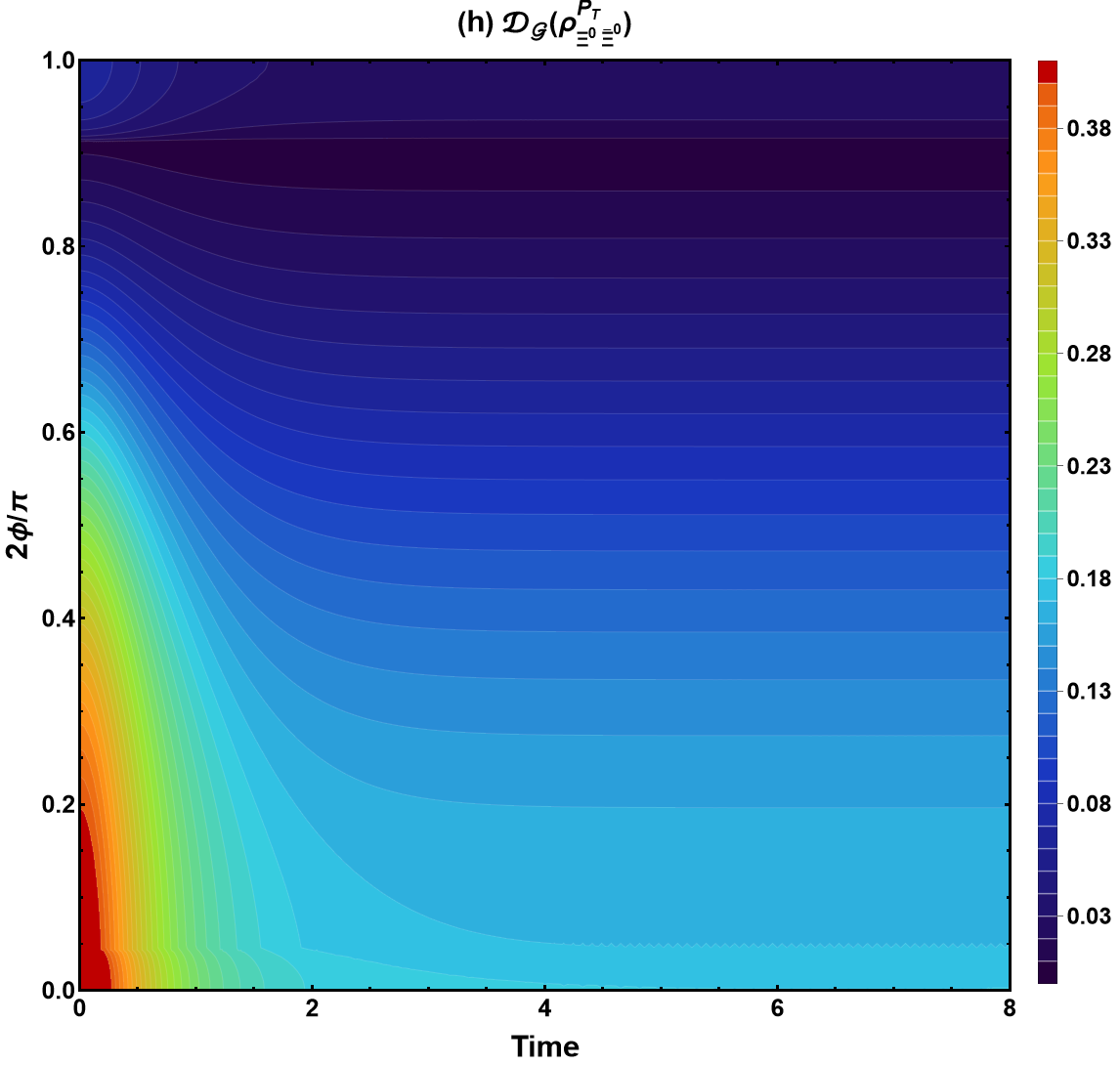}
\caption{
	Dynamical evolution of the geometric quantum discord
	$\mathcal{D}_G(\rho^{P_T}_{Y\bar{Y}})$ as a function of time and the
	azimuthal angle $\phi$ for
	$J/\psi\rightarrow Y\bar{Y}$ with
	$Y=\Lambda$, $\Sigma^{+}$, $\Xi^{-}$, and $\Xi^{0}$ at
	$\cos\theta=0.5$ and $P_T=0.8$. Panels (a)--(d) [(e)--(h)] correspond
	to the non-Markovian (Markovian) regime  with $\tau=5$ ($\tau=0.2$) and
	$\mu=0.4$. The experimental parameters are taken from
	Table~\ref{tab:BESIII}.
}
	\label{fig22}
\end{figure}
The dependence of the geometric quantum discord $\mathcal{D}_{G}(\rho^{P_T}_{Y\bar{Y}})$ on the azimuthal angle $\phi$ and time is illustrated in Fig.~\ref{fig22} for the four hyperon--antihyperon channels under transverse beam polarization. The upper panels (a)--(d) correspond to the
non-Markovian regime, while the lower panels (e)--(h) represent the
Markovian dynamics.
In the non-Markovian regime, all channels exhibit pronounced oscillatory
structures accompanied by gradual damping. These revivals originate from
the backflow of information from the environment to the hyperon system,
which temporarily restores the quantum correlations lost through
decoherence. The $\Lambda\bar{\Lambda}$, $\Xi^{-}\bar{\Xi}^{+}$, and
$\Xi^{0}\bar{\Xi}^{0}$ channels display very similar behaviors, with the
largest discord occurring at small azimuthal angles and early times,
followed by a sequence of damped revivals. In contrast, the
$\Sigma^{+}\bar{\Sigma}^{-}$ channel exhibits a stronger angular
dependence, where the discord is concentrated predominantly in the region
$2\phi/\pi\gtrsim0.2$. This indicates that the azimuthal angle plays a
more significant role in controlling the nonclassical correlations of
this channel.
A qualitatively different behavior is observed in the Markovian regime.
The revival structures disappear completely and the discord decreases
monotonically toward stationary values, reflecting the irreversible loss
of information into the environment. Nevertheless, a
\begin{figure}[H]
	\centering
	\includegraphics[width=0.24\linewidth]{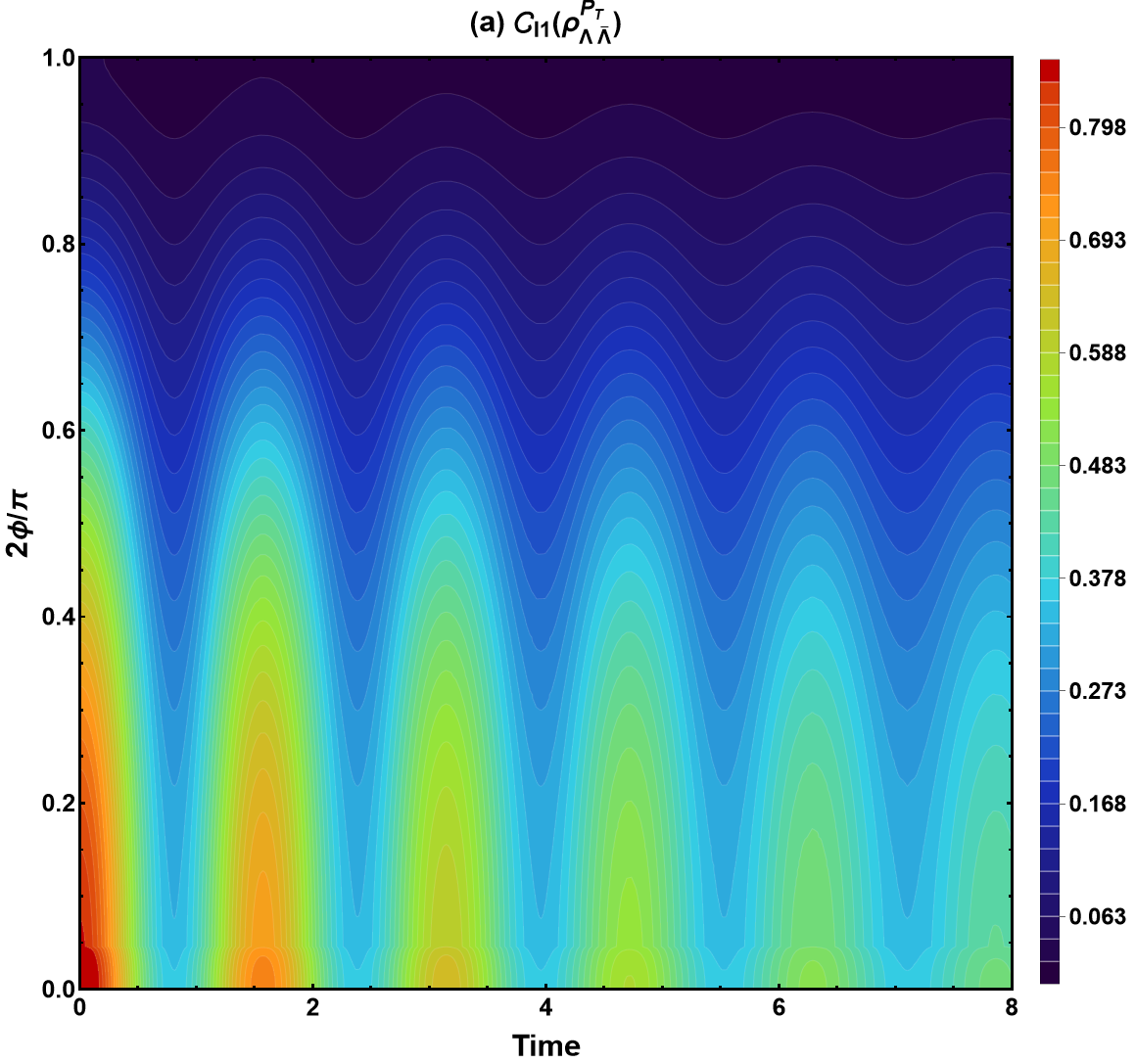}
	\includegraphics[width=0.24\linewidth]{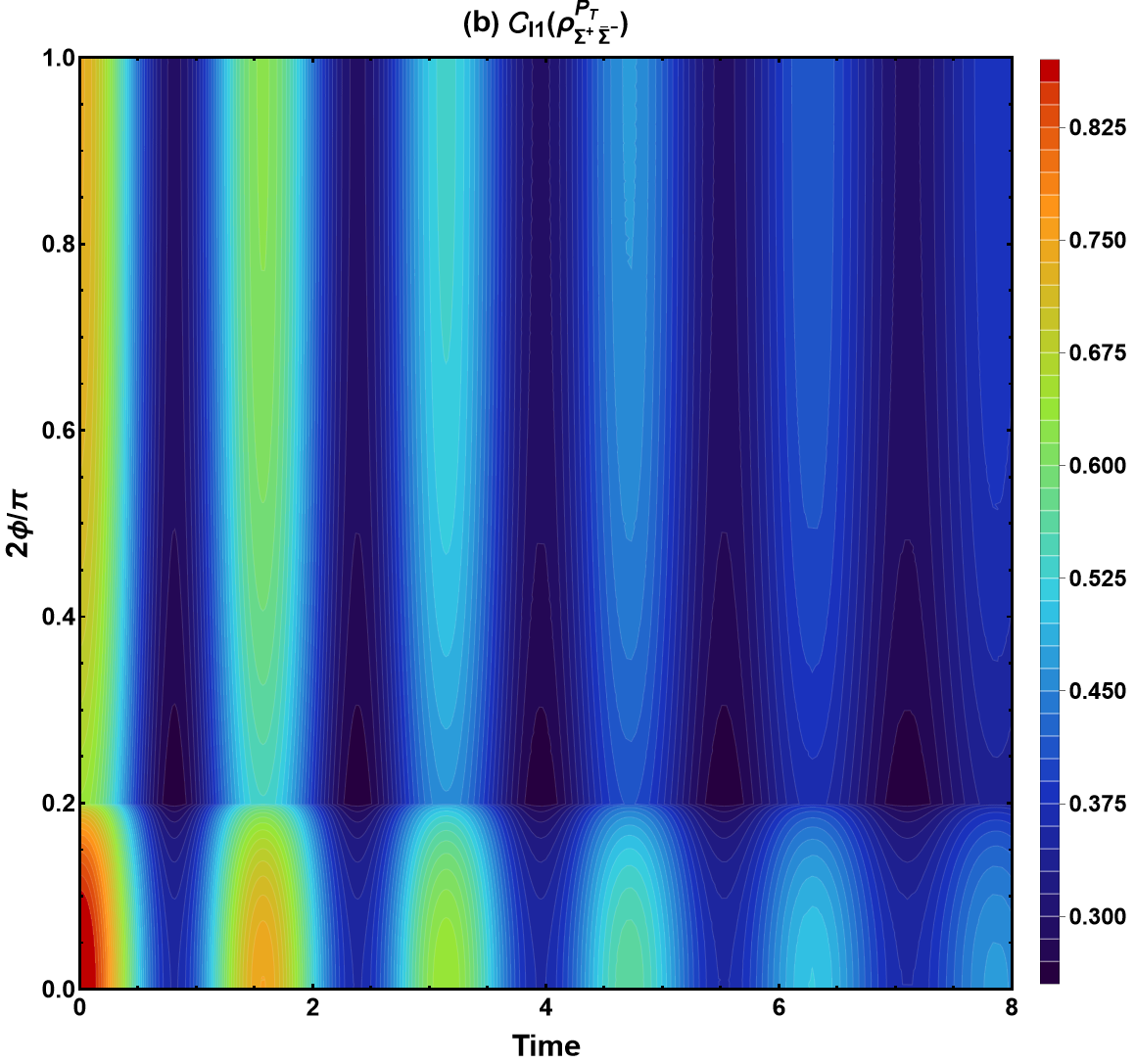}
	\includegraphics[width=0.24\linewidth]{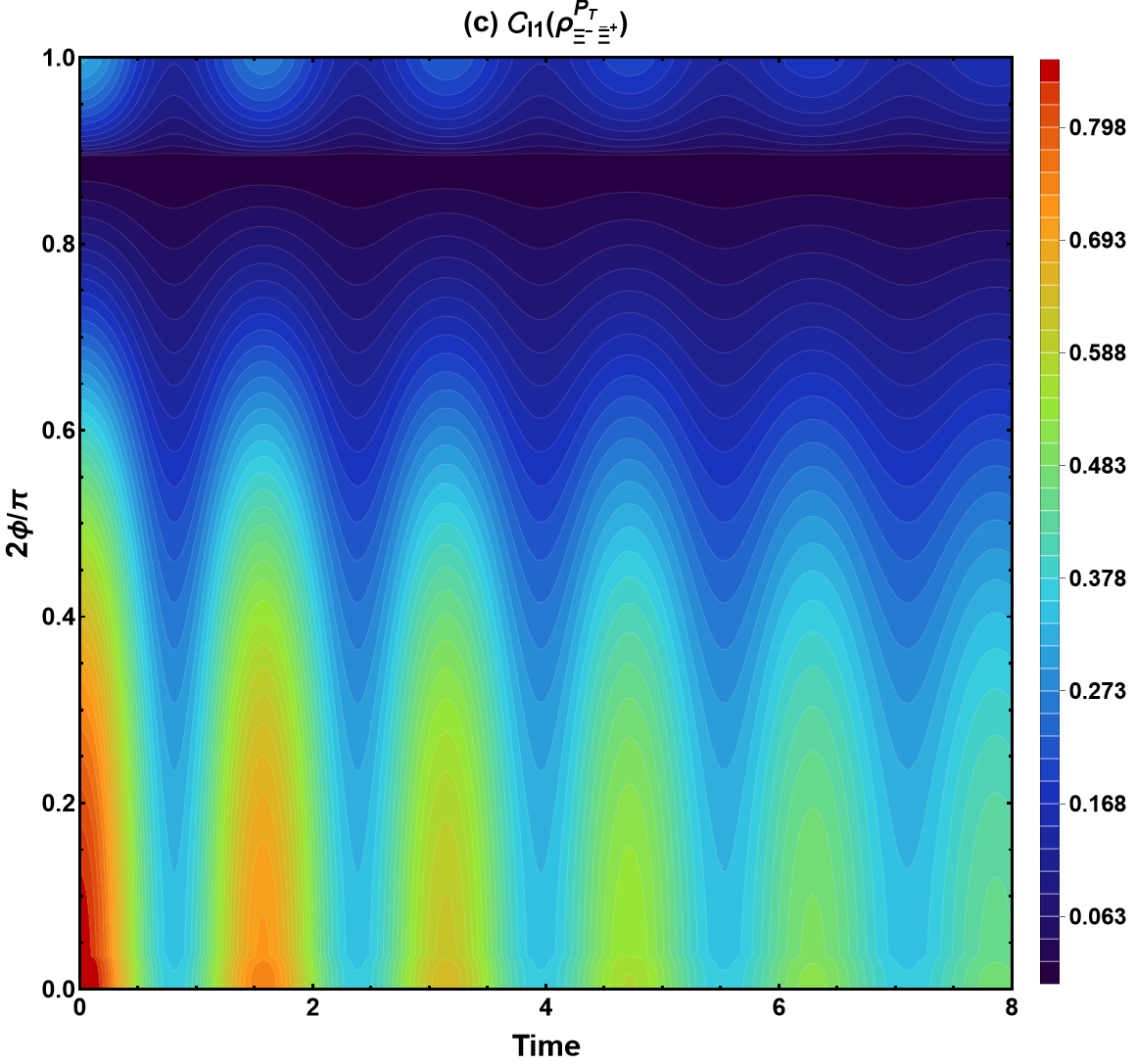}
	\includegraphics[width=0.24\linewidth]{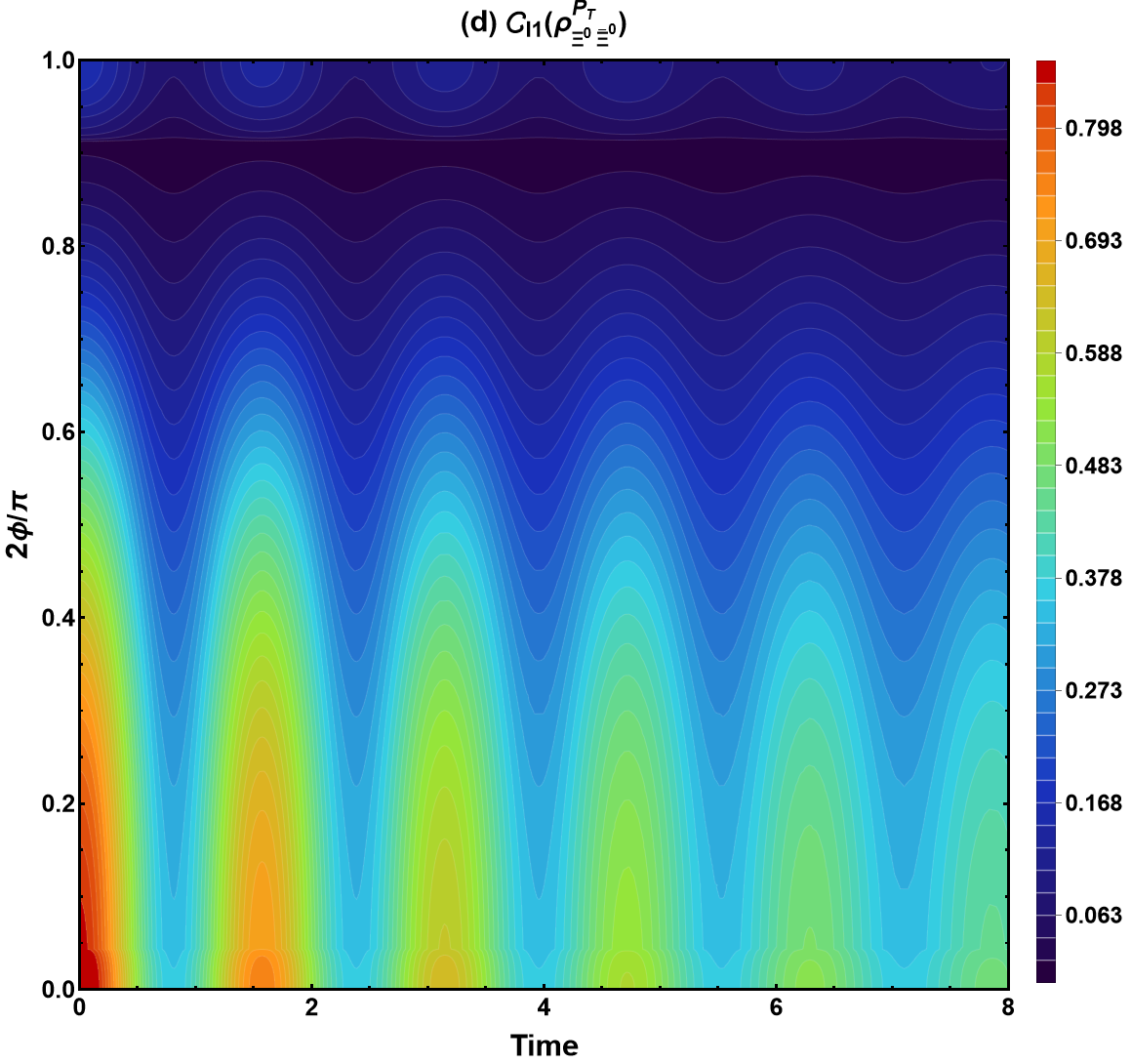}
	\includegraphics[width=0.24\linewidth]{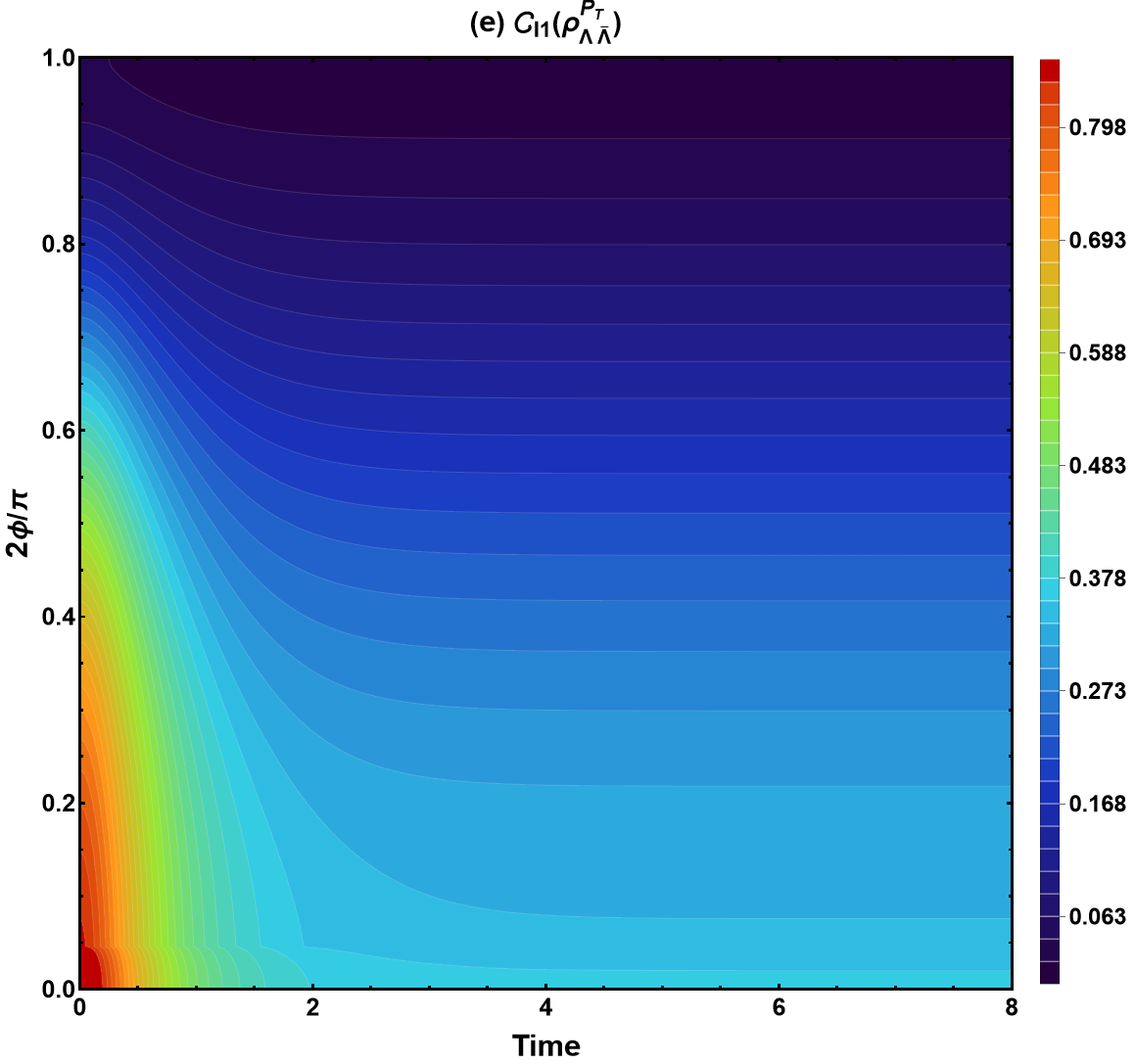}
	\includegraphics[width=0.24\linewidth]{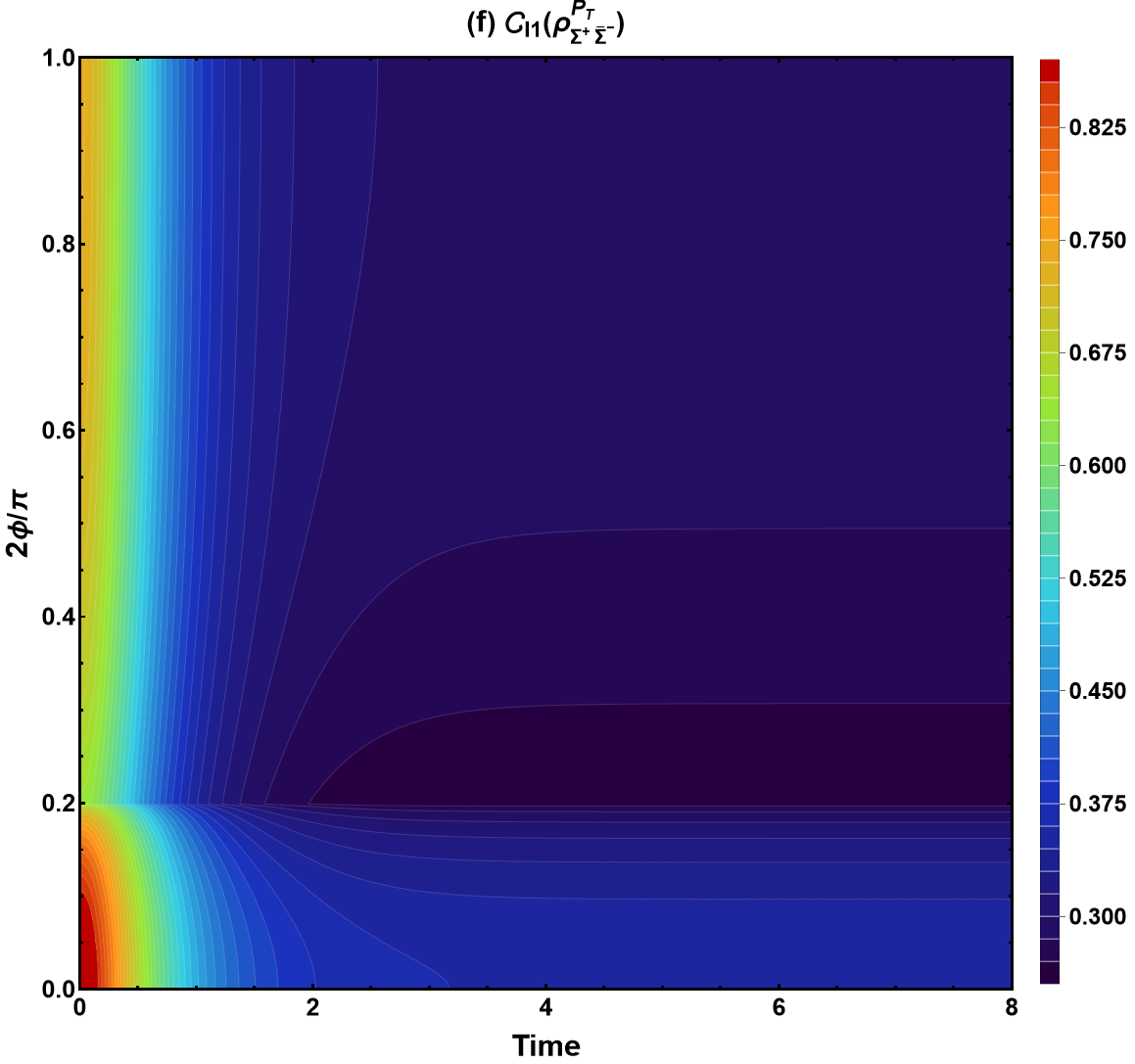}
	\includegraphics[width=0.24\linewidth]{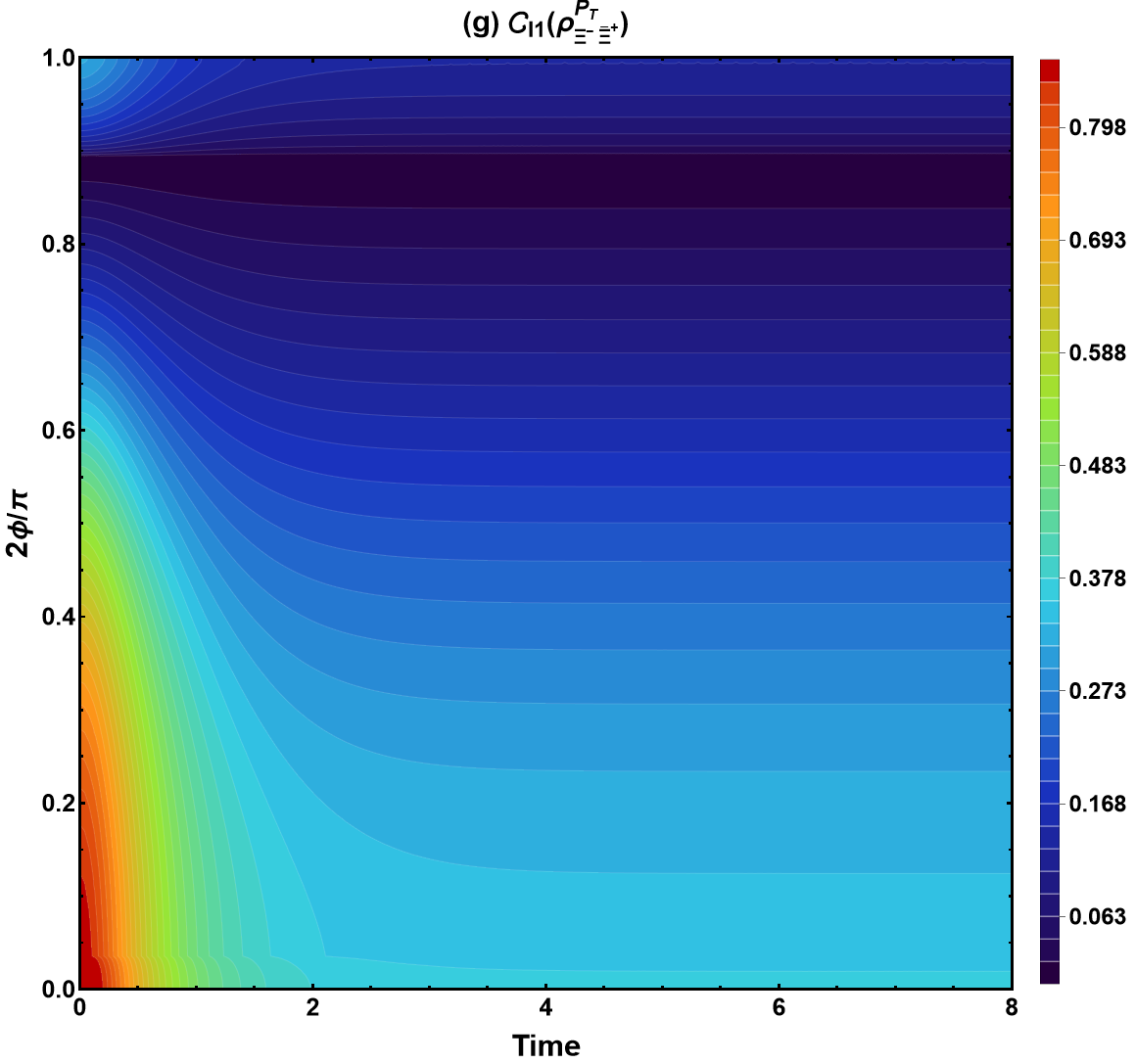}
	\includegraphics[width=0.24\linewidth]{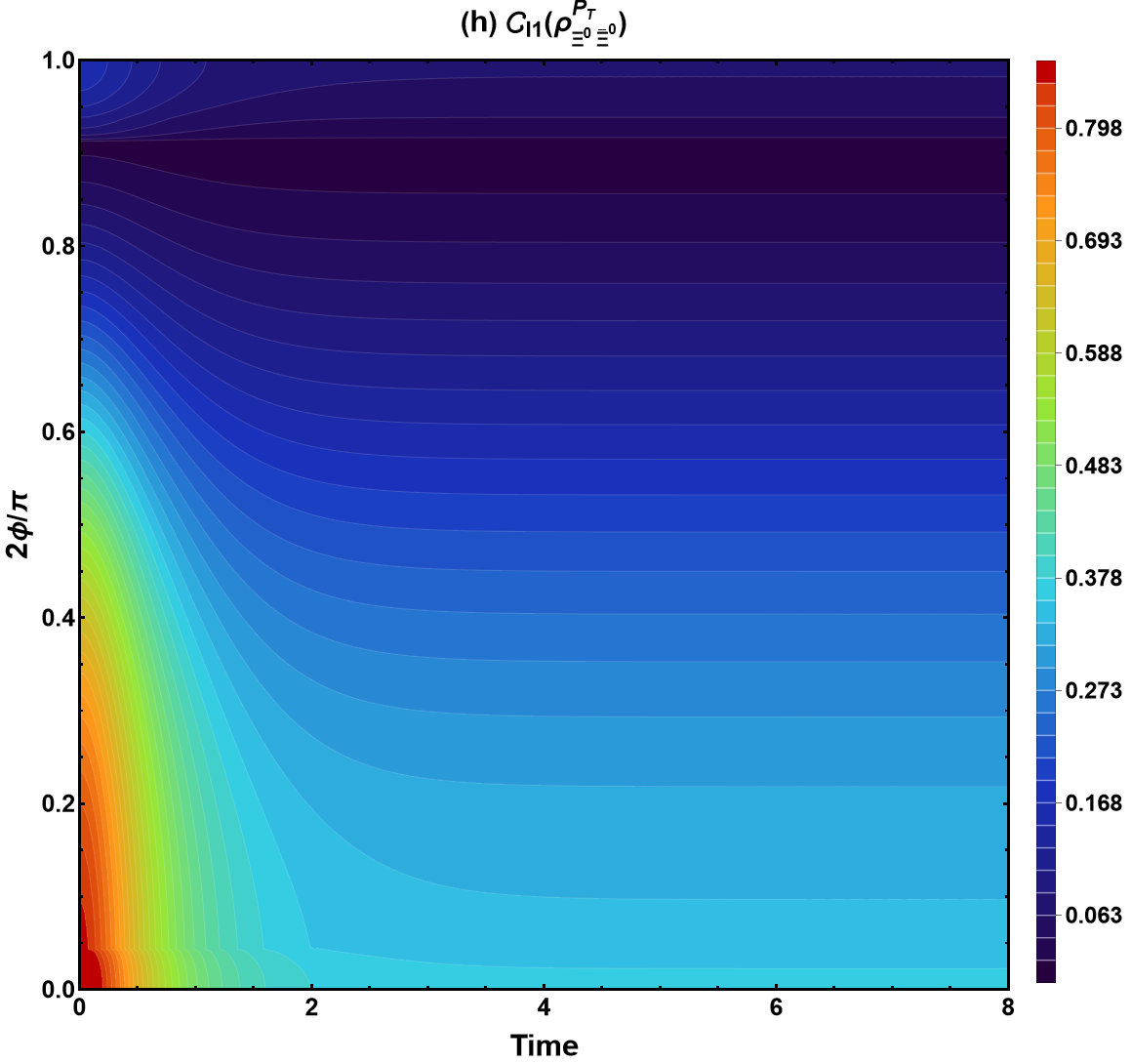}
\caption{
	Dynamical evolution of the $l_1$-norm quantum coherence
	$C_{l_1}(\rho^{P_T}_{Y\bar{Y}})$ as a function of time and the
	azimuthal angle $\phi$ for
	$J/\psi\rightarrow Y\bar{Y}$ with
	$Y=\Lambda$, $\Sigma^{+}$, $\Xi^{-}$, and $\Xi^{0}$ at
	$\cos\theta=0.5$ and $P_T=0.8$. Panels (a)--(d) [(e)--(h)] correspond
	to the non-Markovian (Markovian) regime  with $\tau=5$ ($\tau=0.2$) and
	$\mu=0.4$. The experimental parameters are taken from
	Table~\ref{tab:BESIII}.
}
	\label{fig25}
\end{figure}
 finite amount of
discord survives at long times for all channels, demonstrating the
robustness of quantum correlations beyond entanglement. Compared with the
logarithmic negativity shown in Fig.~\ref{fig19}, the geometric discord
remains nonzero over a broader region of the parameter space, confirming
that quantum discord is more resilient against environmental decoherence.
Overall, the results show that transverse polarization and the azimuthal
angle strongly influence the distribution of quantum correlations, while
non-Markovian memory effects significantly enhance their preservation.

Figure~\ref{fig25} illustrates the dynamics of the quantum coherence
$C_{l_1}(\rho^{P_T}_{Y\bar{Y}})$ as a function of the azimuthal angle
$\phi$ and time for the four hyperon-antihyperon channels under
transverse beam polarization. The upper panels correspond to the
non-Markovian regime, whereas the lower panels describe the Markovian
evolution.
In the non-Markovian regime, the coherence exhibits pronounced damped
oscillations for all channels, reflecting the recurrent exchange of
information between the hyperon system and its environment. The
$\Lambda\bar{\Lambda}$, $\Xi^{-}\bar{\Xi}^{+}$, and
$\Xi^{0}\bar{\Xi}^{0}$ channels display very similar behaviors, with
periodic revivals extending over the entire time interval. These
revivals are accompanied by a gradual reduction of their amplitudes,
indicating that memory effects partially counteract decoherence while
remaining insufficient to prevent the eventual degradation of quantum
coherence. By contrast, the $\Sigma^{+}\bar{\Sigma}^{-}$ channel shows
a stronger dependence on the azimuthal angle, where the coherence is
distributed nonuniformly and exhibits a pronounced transition around
$2\phi/\pi\simeq0.2$.
The Markovian dynamics leads to a qualitatively different behavior.
The oscillatory revival structures disappear and the coherence decays
monotonically toward stationary values, demonstrating the irreversible
loss of information into the environment. Despite this suppression,
a finite amount of coherence survives at long times in all channels,
highlighting the intrinsic robustness of coherence against environmental
noise. Furthermore, the long-time values remain significantly larger
than those observed for entanglement and geometric quantum discord,
confirming the well-established hierarchy of quantum correlations in
which coherence constitutes the most resilient quantum resource. These
results demonstrate that transverse polarization and the azimuthal angle
substantially influence the distribution of coherence, while
non-Markovian memory effects play a crucial role in preserving quantum
features during the dynamical evolution.

In summary, the dynamical analysis reveals a clear interplay between beam polarization, production geometry, and environmental effects in determining the persistence of quantum resources in hyperon--antihyperon systems. Non-Markovian environments substantially enhance the preservation of quantum correlations through recurrent revival phenomena induced by information backflow, whereas Markovian dynamics lead to a monotonic relaxation toward stationary states. Both longitudinal and transverse polarizations generally strengthen the quantum resources and enlarge the parameter regions where nonclassical features remain observable. Among the considered quantifiers, the $l_{1}$-norm coherence is found to be the most robust against decoherence, followed by geometric quantum discord, while logarithmic negativity is the most fragile. Although quantitative differences exist among the various hyperon channels, particularly for the $\Sigma^{+}\bar{\Sigma}^{-}$ system, the overall hierarchy of quantum resources remains unchanged. These results demonstrate that the combined action of beam polarization and environmental memory provides an efficient mechanism for preserving quantum signatures in hyperon production processes, thereby offering promising perspectives for the experimental investigation of quantum correlations in high-energy systems.

\subsection{Effect of memory parameter }
\begin{figure}[H]
	\centering
	\includegraphics[width=0.24\linewidth]{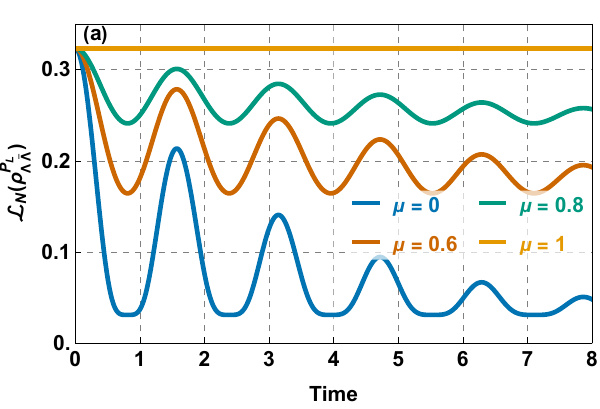}
	\includegraphics[width=0.24\linewidth]{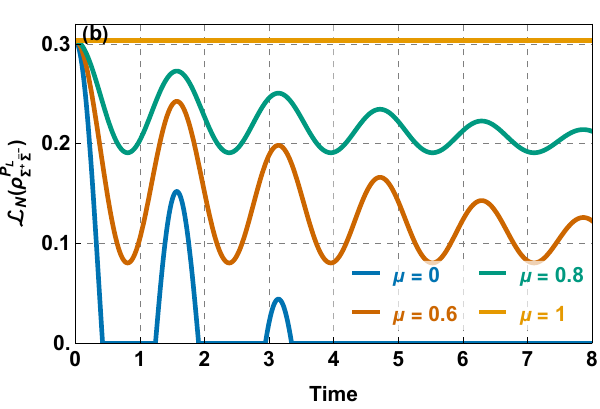}
	\includegraphics[width=0.24\linewidth]{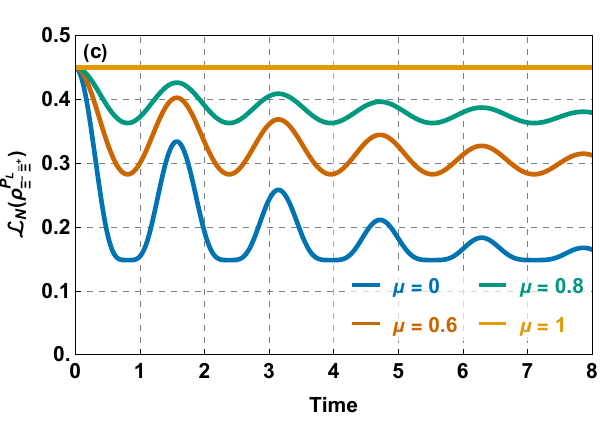}
	\includegraphics[width=0.24\linewidth]{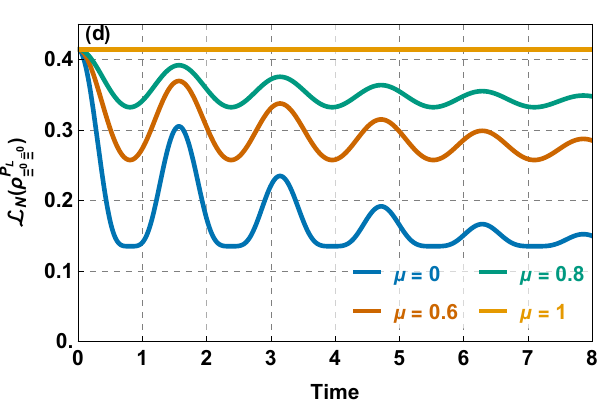}
	\includegraphics[width=0.24\linewidth]{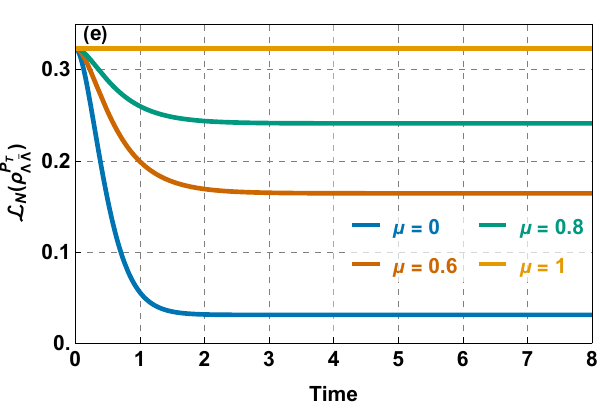}
	\includegraphics[width=0.24\linewidth]{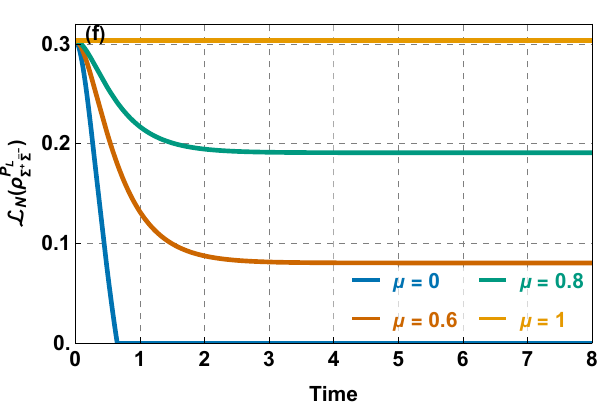}
	\includegraphics[width=0.24\linewidth]{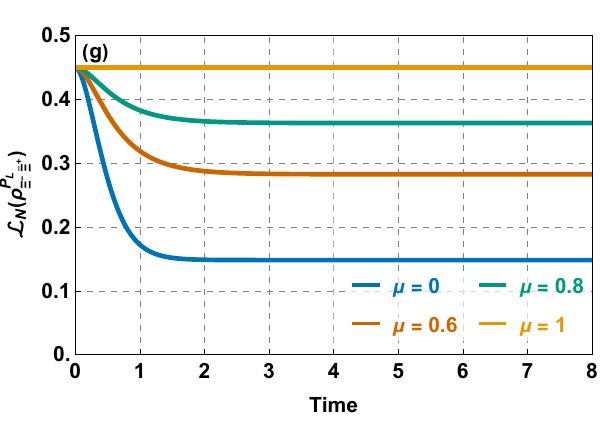}
	\includegraphics[width=0.24\linewidth]{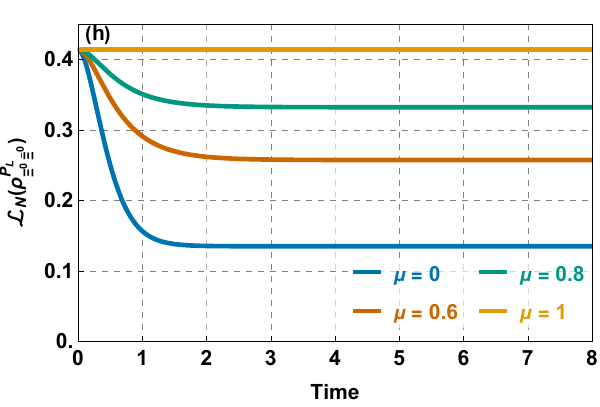}
\caption{
	Time evolution of the logarithmic negativity
	$\mathcal{L}_N(\rho^{P_T}_{Y\bar{Y}})$ for
	$J/\psi\rightarrow Y\bar{Y}$ with
	$Y=\Lambda$, $\Sigma^{+}$, $\Xi^{-}$, and $\Xi^{0}$ for different
	values of the memory parameter $\mu$ at
	$\cos\theta=0.5$, $P_T=0$, and $\phi=0$.
	Panels (a)--(d) [(e)--(h)] correspond to the non-Markovian
	(Markovian) regime  with $\tau=5$ ($\tau=0.2$). The experimental parameters are taken from
	Table~\ref{tab:BESIII}.
}
	\label{fig28}
\end{figure}
The effect of the memory parameter $\mu$ on the dynamical evolution of the logarithmic negativity for the four hyperon--antihyperon channels is illustrated in Fig.~\ref{fig28}. A clear enhancement of entanglement is
observed as $\mu$ increases from $0$ to $1$. For $\mu=0$, corresponding
to an uncorrelated dephasing environment, the entanglement experiences
the strongest degradation. In the non-Markovian regime
[Figs.~\ref{fig28}(a)--(d)], damped oscillations and revivals are
present due to information backflow, but their amplitudes decrease
progressively with time. By contrast, larger values of $\mu$ reduce the
decohering action of the environment, leading to higher minima,
stronger revivals, and larger asymptotic entanglement values.
The limiting case $\mu=1$ is particularly remarkable. In this regime,
the logarithmic negativity becomes essentially time independent,
indicating that the correlated environment preserves the relative
quantum phases responsible for entanglement. The environmental noise
acts collectively on the two particles and therefore cannot efficiently
destroy their shared quantum correlations. As a consequence, the initial
entanglement is almost completely protected during the evolution.
The same tendency is observed in the Markovian regime
[Figs.~\ref{fig28}(e)--(h)], although the oscillatory revivals disappear
because information backflow is absent. Increasing $\mu$ nevertheless
significantly slows down the loss of entanglement and raises the
stationary values reached at long times. These results demonstrate that
environmental memory constitutes an efficient mechanism for protecting
entanglement in hyperon-antihyperon systems, independently of the
particular production channel.
\begin{figure}[H]
	\centering
	\includegraphics[width=0.24\linewidth]{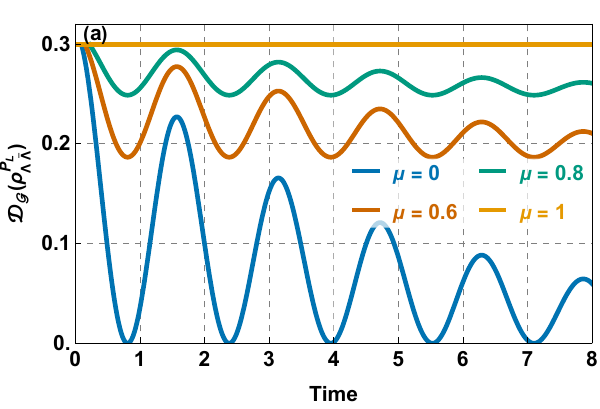}
	\includegraphics[width=0.24\linewidth]{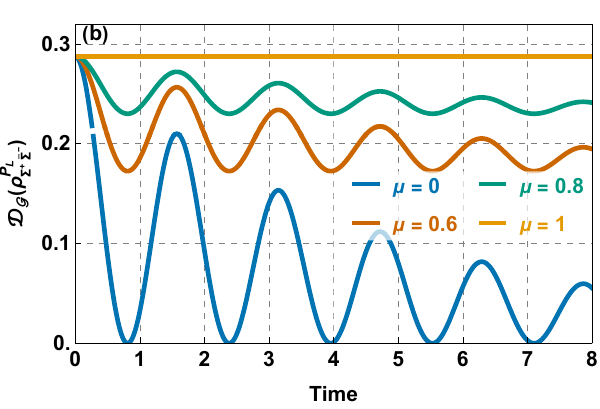}
	\includegraphics[width=0.24\linewidth]{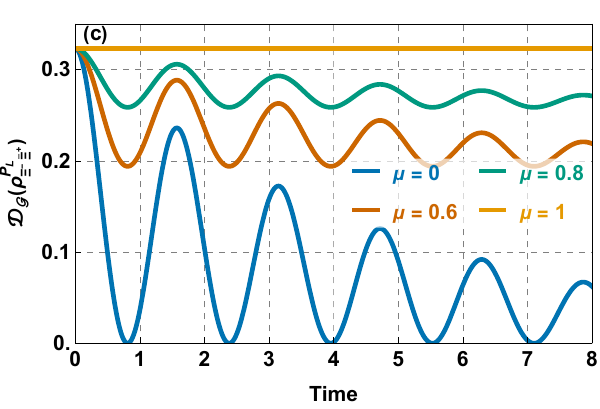}
	\includegraphics[width=0.24\linewidth]{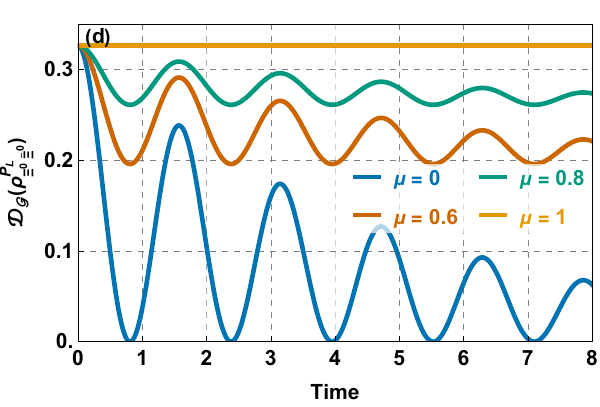}
	\includegraphics[width=0.24\linewidth]{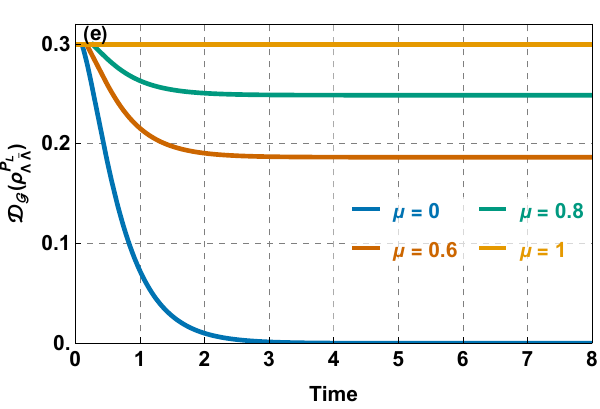}
	\includegraphics[width=0.24\linewidth]{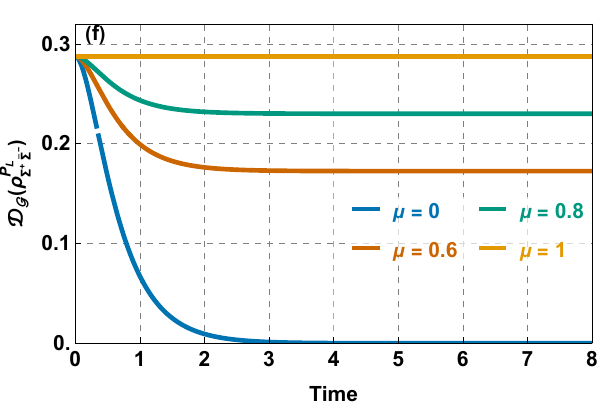}
	\includegraphics[width=0.24\linewidth]{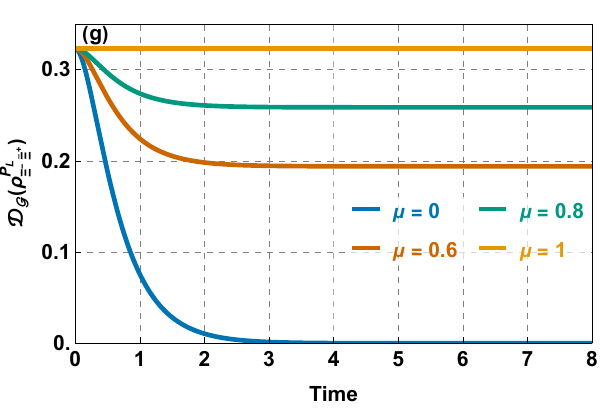}
	\includegraphics[width=0.24\linewidth]{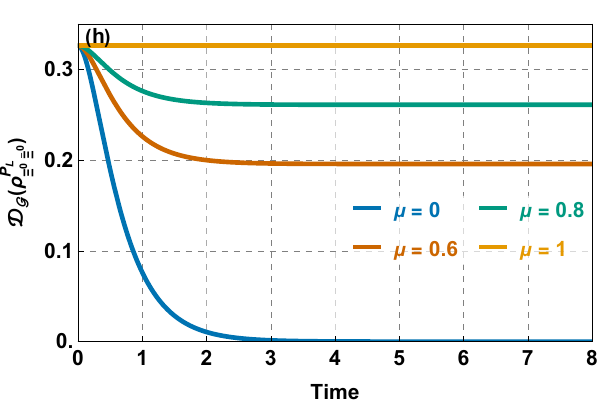}
\caption{
	Time evolution of the geometric quantum discord
	$\mathcal{D}_G(\rho^{P_T}_{Y\bar{Y}})$ for
	$J/\psi\rightarrow Y\bar{Y}$ with
	$Y=\Lambda$, $\Sigma^{+}$, $\Xi^{-}$, and $\Xi^{0}$ for different
	values of the memory parameter $\mu$ at
	$\cos\theta=0.5$, $P_T=0$, and $\phi=0$.
	Panels (a)--(d) [(e)--(h)] correspond to the non-Markovian
	(Markovian) regime  with $\tau=5$ ($\tau=0.2$). The experimental parameters are taken from
	Table~\ref{tab:BESIII}.
}
	\label{fig29}
\end{figure}
Figure~\ref{fig29} illustrates the influence of the memory parameter
$\mu$ on the dynamical evolution of the geometric quantum discord
$\mathcal{D}_{G}$ for the four hyperon--antihyperon channels.
As in the case of logarithmic negativity, increasing $\mu$ strongly
enhances the preservation of quantum correlations. For an uncorrelated
environment ($\mu=0$), the discord undergoes the strongest degradation.
In the non-Markovian regime [Figs.~\ref{fig29}(a)--(d)], damped
oscillations and recurrent revivals are observed as a consequence of
information backflow from the environment. However, the revival
amplitudes decrease progressively with time, reflecting the persistent
action of decoherence. As the memory parameter increases, the damping
becomes weaker and the minima of the oscillations shift toward larger
values, indicating a more efficient protection of quantum correlations.
In the limiting case $\mu=1$, the discord remains constant throughout
the evolution, revealing a complete suppression of decoherence-induced
losses. A similar tendency is observed in the Markovian regime
[Figs.~\ref{fig29}(e)--(h)], where the oscillatory behavior disappears
but increasing $\mu$ significantly raises the stationary discord.
These results demonstrate that environmental correlations efficiently
protect nonclassical correlations and that geometric quantum discord,
being more robust than entanglement, survives over a
\begin{figure}[H]
	\centering
	\includegraphics[width=0.24\linewidth]{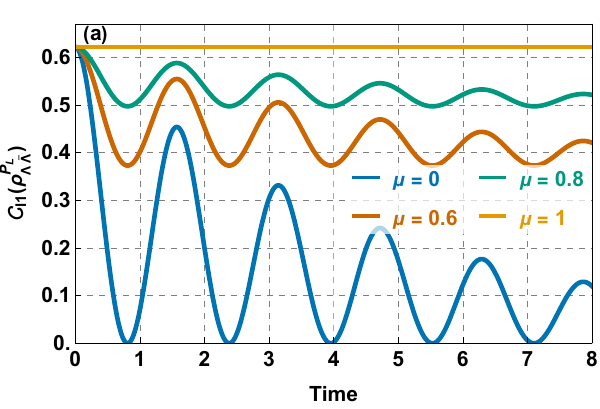}
	\includegraphics[width=0.24\linewidth]{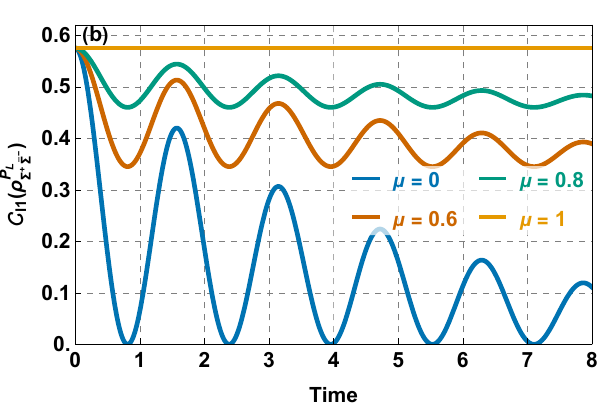}
	\includegraphics[width=0.24\linewidth]{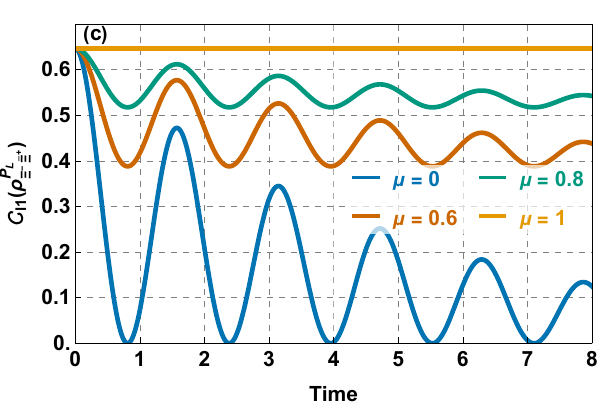}
	\includegraphics[width=0.24\linewidth]{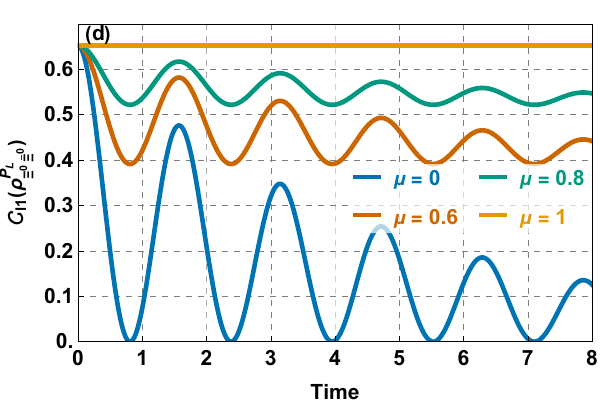}
	\includegraphics[width=0.24\linewidth]{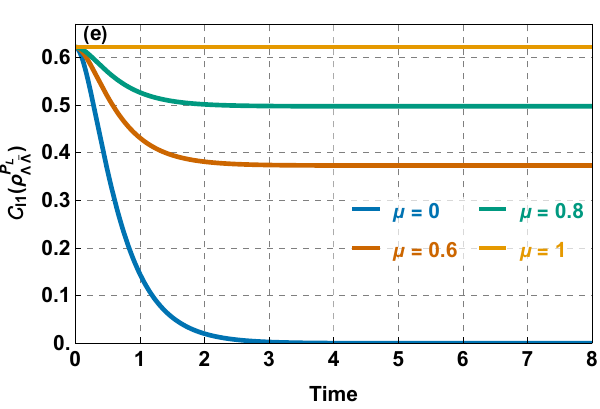}
	\includegraphics[width=0.24\linewidth]{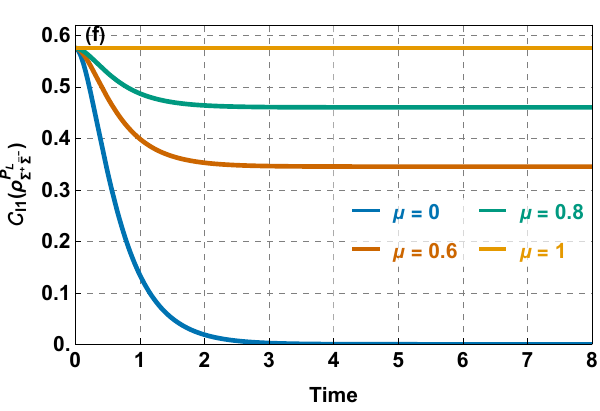}
	\includegraphics[width=0.24\linewidth]{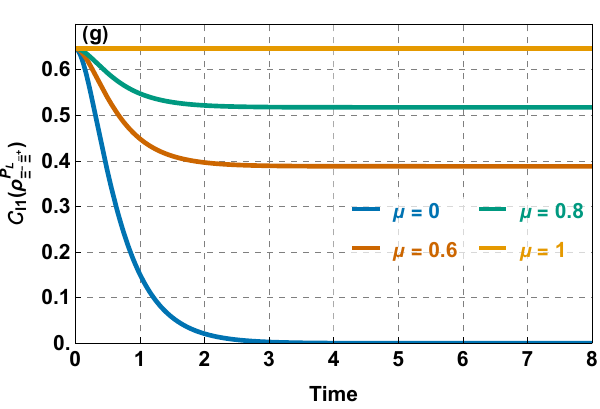}
	\includegraphics[width=0.24\linewidth]{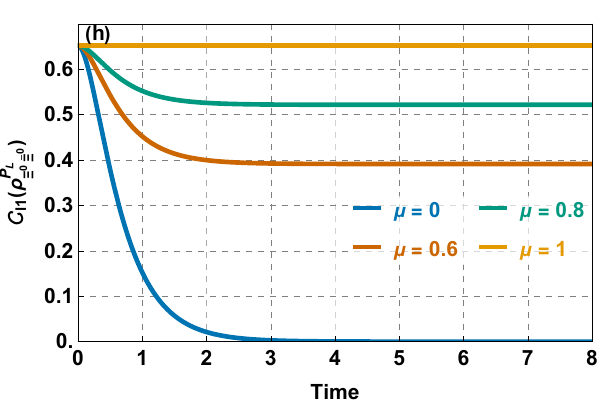}
\caption{
	Time evolution of the $l_1$-norm quantum coherence
	$C_{l_1}(\rho^{P_T}_{Y\bar{Y}})$ for
	$J/\psi\rightarrow Y\bar{Y}$ with
	$Y=\Lambda$, $\Sigma^{+}$, $\Xi^{-}$, and $\Xi^{0}$ for different
	values of the memory parameter $\mu$ at
	$\cos\theta=0.5$, $P_T=0$, and $\phi=0$.
	Panels (a)--(d) [(e)--(h)] correspond to the non-Markovian
	(Markovian) regime  with $\tau=5$ ($\tau=0.2$). The experimental parameters are taken from
	Table~\ref{tab:BESIII}.
}
	\label{fig30}
\end{figure}
 broader range of
memory strengths and evolution times.

Figure~\ref{fig30} presents the influence of the memory parameter
$\mu$ on the dynamical evolution of the $l_{1}$-norm quantum coherence
$C_{l_{1}}(\rho^{P_L}_{Y\bar{Y}})$ for the four hyperon--antihyperon
channels. A common feature of all panels is the strong enhancement of
coherence preservation as the memory parameter increases. For
$\mu=0$, corresponding to an uncorrelated dephasing environment, the
coherence experiences the strongest degradation. In the non-Markovian
regime [Figs.~\ref{fig30}(a)--(d)], this degradation is accompanied by
damped oscillations and recurrent revivals originating from the
backflow of information from the environment to the hyperon pair.
Although the oscillation amplitudes decrease progressively with time,
the revivals remain visible over the entire evolution interval.
As the memory parameter increases from $\mu=0$ to $\mu=0.6$ and
$\mu=0.8$, both the damping rate and the oscillation amplitude are
significantly reduced. This behavior indicates that correlated noise
acts collectively on the two-particle system and therefore becomes less
efficient in destroying the off-diagonal density-matrix elements
responsible for quantum coherence. In the limiting case $\mu=1$, the
coherence remains completely frozen during the evolution, demonstrating
that a fully correlated environment suppresses decoherence-induced
coherence losses.
The same tendency is observed in the Markovian regime
[Figs.~\ref{fig30}(e)--(h)], where increasing $\mu$ substantially raises
the asymptotic stationary coherence. Remarkably, even for intermediate
memory strengths, the long-time coherence remains considerably larger
than the corresponding values of logarithmic negativity and geometric
quantum discord. This confirms that quantum coherence is the most
robust quantum resource in hyperon--antihyperon systems. The results
therefore demonstrate that environmental memory not only protects
quantum correlations but is particularly effective in preserving
coherent spin superpositions, which survive over extended evolution
times even in the presence of dephasing noise.

Since the memory parameter enters through the correlated dephasing
channel rather than through the production mechanism itself, the
qualitative effects observed for longitudinal polarization remain valid
for transverse polarization. In both configurations, increasing $\mu$
reduces decoherence, suppresses damping, and enhances the preservation
of quantum resources. The only noticeable difference lies in the
overall magnitude of the generated entanglement, discord, and
coherence, which is determined by the polarization-dependent spin
correlation structure of the corresponding hyperon-antihyperon channel.
\subsection{Hierarchy of quantum resources}
\begin{figure}[H]
	\centering
	\includegraphics[width=0.24\linewidth]{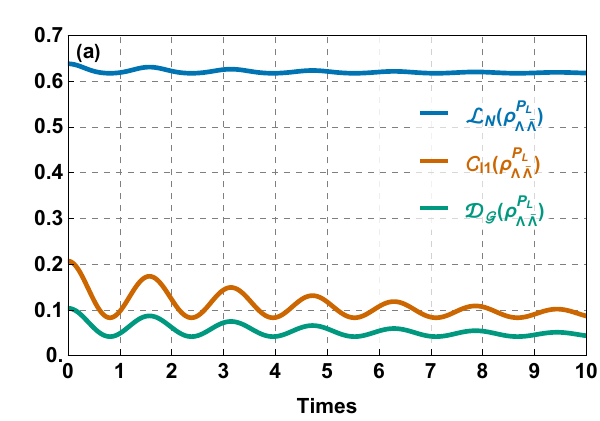}
	\includegraphics[width=0.24\linewidth]{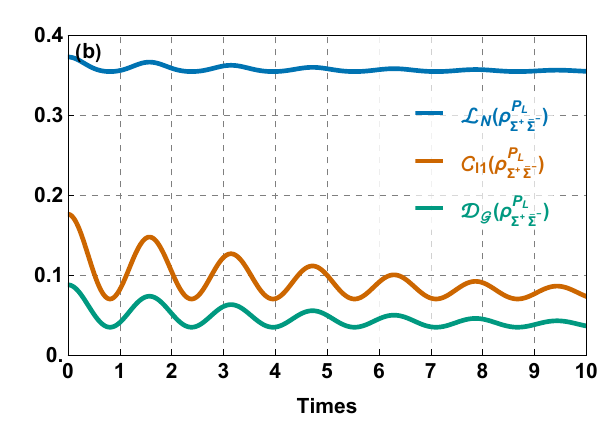}
	\includegraphics[width=0.24\linewidth]{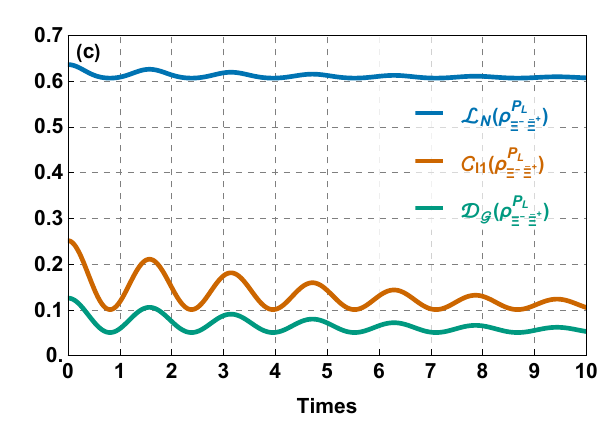}
	\includegraphics[width=0.24\linewidth]{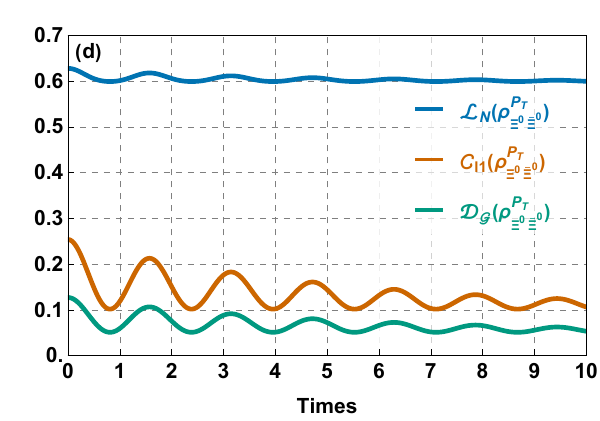}
	\includegraphics[width=0.24\linewidth]{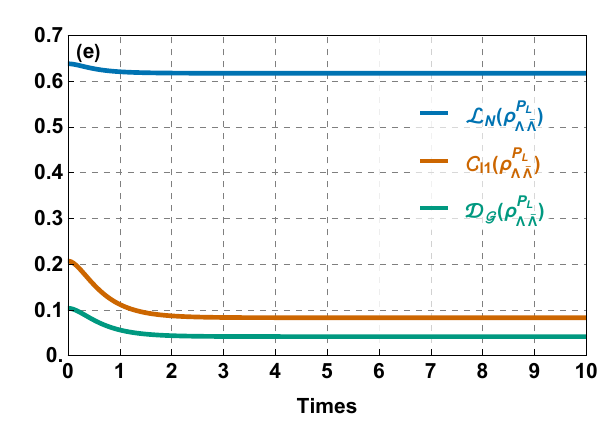}
	\includegraphics[width=0.24\linewidth]{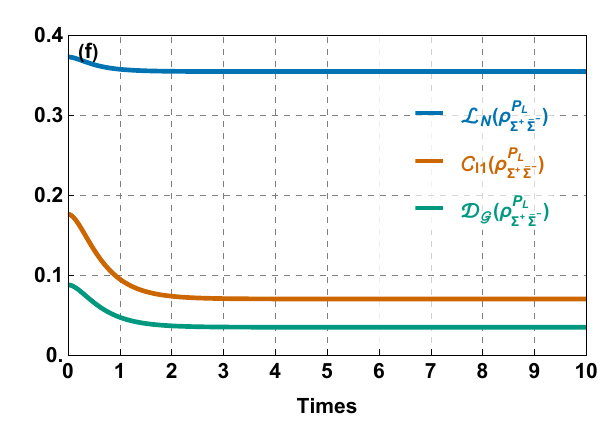}
	\includegraphics[width=0.24\linewidth]{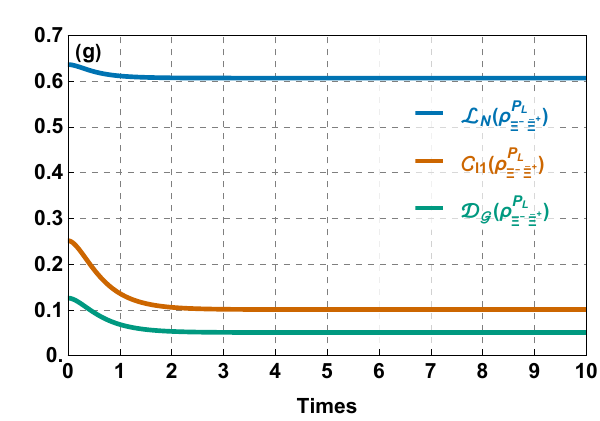}
	\includegraphics[width=0.24\linewidth]{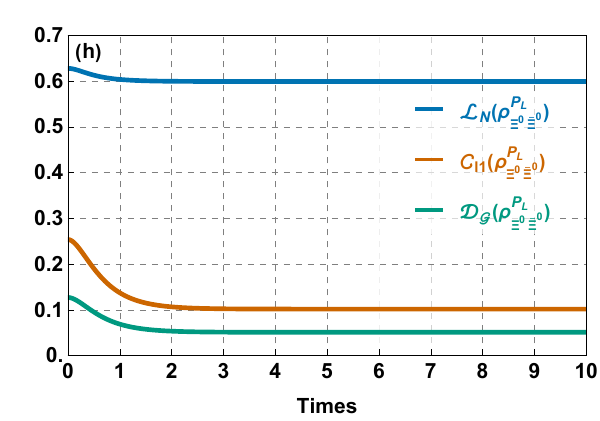}
\caption{
	Hierarchy of quantum resources characterized by the $l_1$-norm quantum coherence
	$C_{l_1}(\rho^{P_L}_{Y\bar{Y}})$, geometric quantum discord
	$\mathcal{D}_G(\rho^{P_L}_{Y\bar{Y}})$, and logarithmic negativity
	$\mathcal{L}_N(\rho^{P_L}_{Y\bar{Y}})$ for
	$J/\psi\rightarrow Y\bar{Y}$ with
	$Y=\Lambda$, $\Sigma^{+}$, $\Xi^{-}$, and $\Xi^{0}$ at
	$\cos\theta=0.5$ and $P_L=0.8$.
	Panels (a)--(d) [(e)--(h)] correspond to the non-Markovian
	(Markovian) regime  with $\tau=5$ ($\tau=0.2$) and
	$\mu=0.4$. The experimental parameters are taken from
	Table~\ref{tab:BESIII}.
}
	\label{fig26}
\end{figure}
A comparison between the logarithmic negativity $\mathcal{L}_{N}$, the $l{1}$-norm coherence $C_{l_{1}}$, and the geometric quantum discord $\mathcal{D}_{G}$ reveals the hierarchy of quantum resources in the four hyperon--antihyperon channels, as shown in Fig.~\ref{fig26}. A
remarkable feature is that the ordering of these resources remains
unchanged throughout the evolution and is independent of both the
hyperon species and the environmental regime. In all cases, the
logarithmic negativity attains the largest values, followed by the
quantum coherence, while the geometric quantum discord constitutes the
smallest contribution, yielding the hierarchy
$\mathcal{L}_{N}>C_{l_{1}}>\mathcal{D}_{G}$. In the non-Markovian
regime, all quantities exhibit damped oscillations and revival
phenomena induced by information backflow, whereas in the Markovian
regime they relax monotonically toward stationary values. Despite these
different dynamical behaviors, the hierarchy is preserved at all times,
indicating that it is an intrinsic property of the spin-correlation
structure generated in the $J/\psi\rightarrow Y\bar{Y}$ process rather
than a consequence of environmental effects. The persistence of this
ordering demonstrates that entanglement constitutes the dominant quantum
resource in the considered hyperon systems, while coherence and discord
provide complementary signatures of nonclassical correlations.
\begin{figure}[H]
	\centering
	\includegraphics[width=0.24\linewidth]{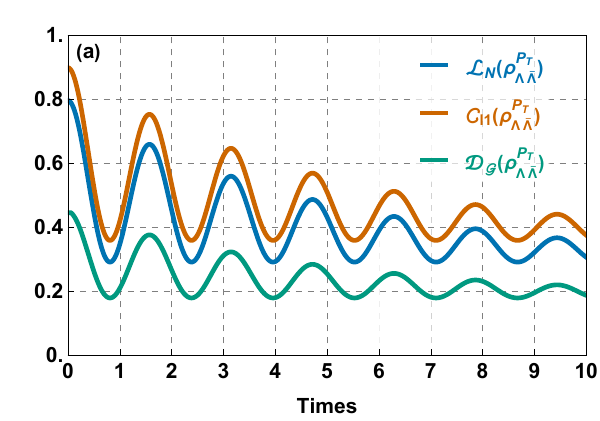}
	\includegraphics[width=0.24\linewidth]{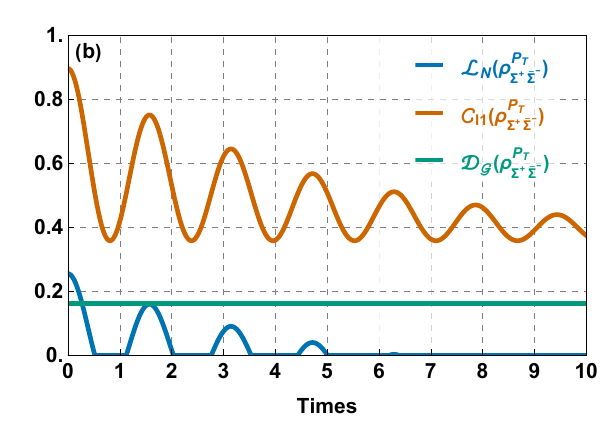}
	\includegraphics[width=0.24\linewidth]{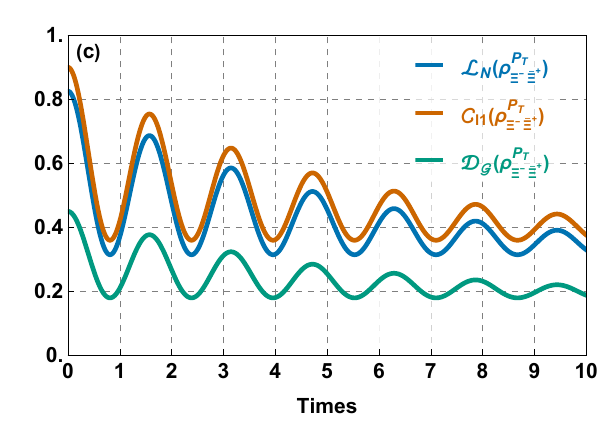}
	\includegraphics[width=0.24\linewidth]{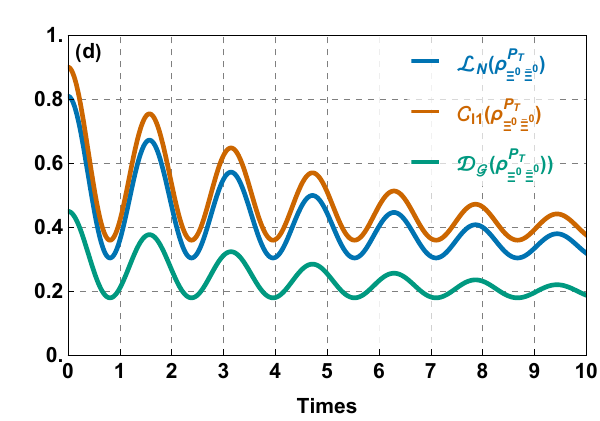}
	\includegraphics[width=0.24\linewidth]{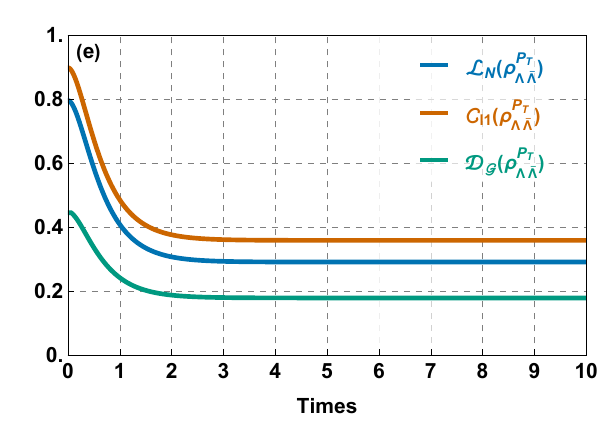}
	\includegraphics[width=0.24\linewidth]{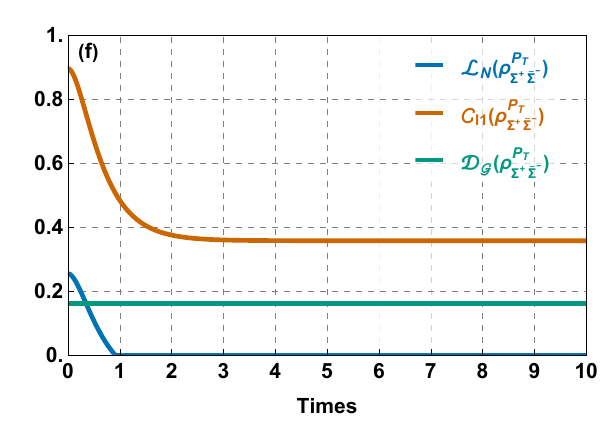}
	\includegraphics[width=0.24\linewidth]{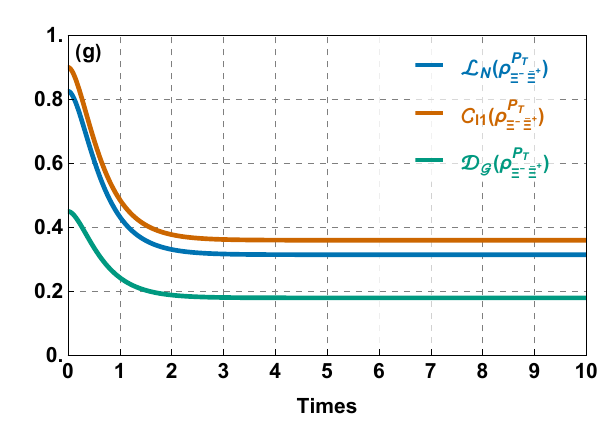}
	\includegraphics[width=0.24\linewidth]{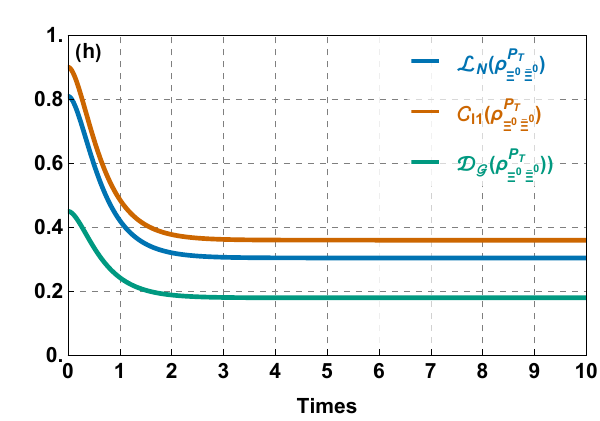}
\caption{
	Hierarchy of quantum resources characterized by the $l_1$-norm quantum coherence
	$C_{l_1}(\rho^{P_T}_{Y\bar{Y}})$, geometric quantum discord
	$\mathcal{D}_G(\rho^{P_T}_{Y\bar{Y}})$, and logarithmic negativity
	$\mathcal{L}_N(\rho^{P_T}_{Y\bar{Y}})$ for
	$J/\psi\rightarrow Y\bar{Y}$ with
	$Y=\Lambda$, $\Sigma^{+}$, $\Xi^{-}$, and $\Xi^{0}$ at
	$P_T=0.8$, $\cos\theta=0.5$, and $\phi=0$.
	Panels (a)--(d) [(e)--(h)] correspond to the non-Markovian
	(Markovian) regime with $\tau=5$ ($\tau=0.2$) and
	$\mu=0.4$. The experimental parameters are taken from
	Table~\ref{tab:BESIII}.
}

	\label{fig27}
\end{figure}
Figure~\ref{fig27} summarizes the hierarchy of quantum resources in the
hyperon--antihyperon systems by comparing the logarithmic negativity
$\mathcal{L}_{N}$, the geometric quantum discord $\mathcal{D}_{G}$, and
the $l_{1}$-norm coherence $C_{l_{1}}$ for the four production channels.
A remarkable feature is the existence of a stable ordering among these
quantities throughout the entire dynamical evolution. Independently of
the considered hyperon channel and of the environmental regime, the
quantum coherence remains the dominant resource, followed by the
logarithmic negativity, while the geometric quantum discord assumes the
smallest values. The resulting hierarchy can therefore be written as
$
C_{l_{1}}>\mathcal{L}_{N}>\mathcal{D}_{G}.
$
In the non-Markovian regime, all resources exhibit damped oscillatory
revivals originating from the backflow of information between the
environment and the hyperon pair. Although the amplitudes decrease with
time, the above hierarchy is preserved during the entire evolution. In
the Markovian regime, the oscillations disappear and the three
quantities monotonically approach stationary values; nevertheless, the
same ordering remains valid in the asymptotic limit. The
$\Sigma^{+}\bar{\Sigma}^{-}$ channel constitutes a special case, where
the entanglement is strongly suppressed and may even vanish for extended
time intervals, while both coherence and geometric discord remain
finite. This behavior highlights the greater robustness of coherence and
discord against environmental decoherence compared with entanglement.
The persistence of the hierarchy
$C_{l_{1}}>\mathcal{L}_{N}>\mathcal{D}_{G}$ indicates that quantum
superposition effects constitute the most resilient manifestation of
quantumness in hyperon systems. Entanglement represents an intermediate
resource that benefits significantly from non-Markovian memory effects,
whereas geometric discord provides a weaker but more universally
surviving form of nonclassical correlation. These results demonstrate
that the different quantum resources respond differently to correlated
dephasing dynamics while preserving a well-defined hierarchical
structure.

In summary, the analysis performed in this section reveals that the
quantum resources generated in polarized hyperon--antihyperon systems
are strongly influenced by both beam polarization and environmental
memory effects. While longitudinal and transverse polarizations mainly
control the magnitude of the quantum correlations, non-Markovian
dynamics significantly enhances their preservation through information
backflow and revival phenomena. The comparative investigation of
logarithmic negativity, geometric quantum discord, and $l_{1}$-norm
coherence further demonstrates the existence of a well-defined
hierarchy of quantum resources, with coherence emerging as the most
robust quantity against decoherence, followed by entanglement and
geometric discord. Overall, these results highlight the potential of
polarized hyperon pairs as a promising platform for exploring quantum
correlations in high-energy physics and for probing the interplay
between spin dynamics, environmental effects, and quantum information
resources.

	\section{Conclusion}\label{sec7}

In this work, we have investigated the behavior of quantum resources in polarized hyperon--antihyperon systems produced through the process
$e^{+}e^{-}\rightarrow J/\psi\rightarrow Y\bar{Y}$
($Y=\Lambda,\Sigma^{+},\Xi^{-},\Xi^{0}$)
in the presence of correlated dephasing environments. By combining realistic BESIII experimental parameters with an open-quantum-system framework, we analyzed both the static and dynamical properties of logarithmic negativity, geometric quantum discord, and quantum coherence under longitudinal and transverse beam polarizations.
Our results reveal that beam polarization plays a crucial role in shaping the distribution of quantum correlations. In the stationary regime, longitudinal and transverse polarizations generate distinct correlation landscapes and significantly modify the amount of accessible quantum resources. While entanglement is strongly dependent on the production kinematics and polarization configuration, geometric discord and quantum coherence remain appreciable over much broader parameter regions, demonstrating a higher resilience against correlation degradation.
The dynamical analysis highlights the fundamental impact of environmental memory effects. Non-Markovian dynamics induce recurrent revivals of quantum resources through information backflow from the environment to the hyperon system, considerably extending the lifetime of quantum correlations. In contrast, Markovian evolution leads to a monotonic and largely irreversible suppression of quantum resources. The enhancement produced by memory effects is observed for all considered hyperon channels and becomes particularly pronounced for highly polarized beams.
A systematic comparison among the different quantifiers establishes a clear hierarchy of quantum resources. Quantum coherence emerges as the most robust manifestation of quantumness, followed by geometric quantum discord and logarithmic negativity. This hierarchy remains stable in both stationary and dynamical regimes, indicating that nonclassical correlations may survive environmental decoherence even when entanglement becomes strongly reduced. Such behavior demonstrates that quantum resources beyond entanglement provide valuable information on the underlying spin-correlation structure of hyperon pairs.
From an experimental perspective, the predicted effects occur within parameter regions accessible to current BESIII measurements and future high-luminosity facilities such as STCF and CEPC. The combination of beam polarization, precise spin-correlation measurements, and large event samples offers a realistic route toward probing decoherence effects, environmental memory, and the hierarchy of quantum resources in high-energy particle systems.
Our findings therefore establish polarized hyperon--antihyperon pairs as a unique laboratory at the interface between quantum information science, open quantum systems, and high-energy physics. They further demonstrate that environmental effects, often neglected in collider-based quantum studies, can profoundly influence the structure and persistence of quantum correlations. Future investigations may extend this framework to more general noise models, relativistic open-system scenarios, and experimental tests of quantum resources in next-generation collider experiments. 

Beyond the specific results obtained for individual quantum resources, the present work establishes a unified framework for investigating open-system effects in polarized hyperon--antihyperon pairs. By combining realistic hyperon density matrices, correlated dephasing channels with memory, and experimentally accessible polarization configurations, our study extends the application of quantum-information concepts to a realistic high-energy environment. The identification of a robust resource hierarchy and the demonstration of memory-assisted protection mechanisms provide new insights into the preservation of quantum correlations in particle-physics experiments.

\end{document}